\documentclass[aps,rmp,reprint,amsmath,amssymb,graphicx,longbibliography]{revtex4-1}

\usepackage{bm}

\usepackage[dvipsnames]{xcolor}
\usepackage{graphicx}
\usepackage{dcolumn}
\usepackage{bm}
\usepackage[colorlinks=true,linkcolor =black,citecolor=blue,urlcolor=blue,anchorcolor =blue]{hyperref}
\usepackage{wasysym}
\usepackage{stmaryrd}
\usepackage{times}
\usepackage{verbatim}
\usepackage{subfigure}
\usepackage{amsmath}
\newcommand{\RNum}[1]{\uppercase\expandafter{\romannumeral #1\relax}}

\newcommand{\bea}{\begin{eqnarray}}
\newcommand{\eea}{\end{eqnarray}}
\newcommand{\be}{\begin{eqnarray}}
\newcommand{\ee}{\end{eqnarray}}
\newcommand{\bw}{\begin{widetext}}
	\newcommand{\ew}{\end{widetext}}

\newcommand{\bs}{\boldsymbol}

\newcommand{\mc}{\mathcal}
\newcommand{\mb}{\mathbb}

\begin{document}

\title{Thermal Hall effects in quantum magnets}

\author{Xiao-Tian Zhang$^2$}
\thanks{These authors contributed equally.}
\author{Yong Hao Gao$^3$}
\thanks{These authors contributed equally.}
\author{Gang Chen$^{1,3,4}$}
\email{chenxray@pku.edu.cn}

\affiliation{$^{1}$International Center for Quantum Materials, 
School of Physics, Peking University, Beijing 100871, China}

\affiliation{$^{2}$Kavli Institute for Theoretical Sciences, University of Chinese Academy of Sciences, Beijing 100190, China}

\affiliation{$^{3}$
Department of Physics and HKU-UCAS Joint Institute for Theoretical and Computational Physics at Hong Kong, The University of Hong Kong, Hong Kong,
China}

\affiliation{$^{4}$
Collaborative Innovation Center of Quantum Matter, Beijing 100871, China}

\date{\today}

\begin{abstract}
In the recent years, the thermal Hall transport has risen as an important diagnosis 
of the physical properties of the elementary excitations in various quantum materials,
especially among the Mott insulating systems where the electronic transports
are often featureless. Here we review the recent development of thermal Hall
effects in quantum magnets where all the relevant excitations are charge-neutral.
In addition to summarizing the existing experiments, we pay a special attention 
to the underlying mechanisms of the thermal Hall effects in various magnetic systems, 
and clarify the connection between the microscopic physical variables and  
the emergent degrees of freedom in different quantum phases. The external 
magnetic field is shown to modify the intrinsic Berry curvature properties    
of various emergent and/or exotic quasiparticle excitations in distinct 
fashions for different quantum systems and quantum phases, contributing 
to the thermal Hall transports. These include, for example, the conventional 
ones like the magnons in ordered magnets, the triplons in dimerized magnets, 
the exotic and fractionalized quasparticles such as the spinons and the magnetic 
monopoles in quantum spin liquids. We review their contribution and 
discuss their presence in the thermal Hall conductivity in different physical contexts. 
We expect this review to provide a useful guidance for the physical
mechanism of the thermal Hall transports in quantum magnets. 
\end{abstract}






\maketitle
\tableofcontents







\section{Introduction}
\label{sec1}

Quantum magnets are well-known correlated spin systems where the local spin moments  
arise from the Coulomb interaction and interact via the exchange interactions. 
Quantum magnetism is one of the most active fields of research in the
modern condensed matter physics, and 
provides the opportunities to bridge the frontier experiments and the fundamental theories. 
There is a significant research interest, 
especially in the low-dimensional and frustrated quantum spin systems~\cite{lacroix2011introduction,Starykh_2015}.  
Such systems have a large number of experimental realizations 
and exhibit a variety of phenomena whose physical origin can be 
attributed to quantum effects, low dimensionality and magnetic 
frustration. The ground state of quantum magnets can be generally 
classified into the magnetically ordered phases and quantum disordered 
phases. For the magnetically ordered phases, the low-energy elementary 
magnetic excitation is the conventional and well-understood magnon 
quasiparticle that carries integer spin quantum numbers. 
For the quantum disordered states, there exist two large families of 
states that are most relevant for the existing experimental interest.
The first one are the valence bond solid and the plaquette ordered states.
These states are composed of the spin singlets and preserve the time 
reversal symmetry. The lattice translation symmetry, however, is 
broken by the formation of spin singlets. The elementary excitations  
are the triplon quasiparticles and are in fact quite similar to the magnons. 
The other disordered state is the more fascinating quantum spin liquid 
state. This state is particularly appealing due to its potential relevance 
to the high-temperature superconductivity~\cite{Anderson1987} and 
quantum-computation applications~\cite{Kitaev2006} and has 
attracted a tremendous attention in the recent years. The elementary 
excitations in the quantum spin liquids are generally exotic but vary 
significantly in different spin liquids~\cite{Savary2016}. 
They can be bosonic, fermionic, 
or even anyonic, and carry the fractional spin and/or lattice symmetry 
quantum numbers. 

Identifying the nature of the ground states and the low-energy excitations  
is probably one of the central questions in the field of quantum magnetism. 
Various experimental probes, such as the thermodynamic ones like the 
magnetic susceptibility and the specific heat, and the spectroscopic ones 
such as the neutron scattering measurements, the muon spin relaxation 
and nuclear magnetic resonance measurements, are commonly employed 
to provide the specific physical information about the system. 
The charge transport, that often provides some crucial clues 
for the electronic structures such as the quantum Hall liquids,
superconductivity and non-Fermi liquids~\cite{RevModPhys.82.1743}, 
turns out to be featureless 
in the Mott insulating quantum magnets, excepting exhibiting an insulating behavior. 
The magnetic excitations here carry no electrical charge 
and thus have no conventional transport response driven by an electrical field. 
Fortunately, the thermal transports open up another window for probing and 
unveiling the nature of these charge-neutral excitations. 
The heat currents in the insulating quantum magnets are then 
carried by the emergent and neutral modes.
For example, the heat-carrying magnetic excitations are simple 
magnons in the ordered phases, are replaced by the 
deconfined quasiparticles in spin liquids.  
These magnetic excitations transport the heat and contribute 
to the longitudinal thermal conductivity $\kappa_{xx}$ 
in the same way as the physical electrons carry and  
 transport electric charges in an electrical conductor.  One major difficulty in the understanding of   
the longitudinal thermal conductivity, however, is that,
many other excitations, most notably phonons, may get involved 
and interact with the magnetic excitations in a complicated fashion~\cite{PhysRevB.106.245139}. 
Hence, it requires a bit scrutiny to extract the precise 
magnetic contribution from the longitudinal thermal   
conductivity measurements. 
To avoid the complication in the interpretation of the longitudinal thermal conductivity, 
one feasible way is to turn to the understanding of the thermal Hall conductivity $\kappa_{xy}$. 
Comparing with the thermodynamic and the spectroscopic measurements 
that detect the physical properties about the density of states 
and the dispersion relations of the magnetic excitations, 
the thermal Hall conductivity is related to the Berry curvature 
and/or the topological properties of the excitations and/or the ground states,
and thus is connected to the wavefunction properties of the underlying many-body 
states. On the experimental side, since the spin degree of freedom 
is directly coupled to the external magnetic field, the thermal Hall    
conductivity can be a bit more selective to unveil the magnetic properties and the Berry curvature
properties of the excitations,
though the phonons could still get involved in the thermal Hall transports via the coupling
to other degrees of freedom and the generation or transferring of Berry curvatures.
In fact, the thermal Hall measurement has proved to be quite successful 
in revealing the intrinsic topological properties of both
integer and fractional quantum Hall liquids, and may even be decisive 
at the current experimental stage to confirm the non-Abelian nature of 
the ${\nu=5/2}$ fractional quantum Hall liquid~\cite{PhysRevB.55.15832,Banerjee_2018}. Quantized or fractional 
thermal Hall effect is an indication of the gravitational anomaly and 
thus provides an important addition to the Hall transports of the electric 
charges~\cite{PhysRevB.85.184503,PhysRevB.85.045104,PhysRevLett.114.016805}. 
In contrast to the thermal Hall measurements on the quantum Hall liquids 
that require the strong magnetic fields, extremely low temperatures and 
high sample quality, more experiments with less constraints have 
been conducted and more data were accumulated in various 
quantum magnets, which calls for the theoretical understanding. 



The thermal Hall effect is the thermal analogue of the Hall effect. 
A temperature gradient is produced across the solid instead of 
an electric field. When a magnetic field is applied, an orthogonal  
temperature gradient develops. In this review,  both theoretical 
and experimental progress on the thermal Hall effects of the 
insulating quantum magnets are discussed in the contexts 
of magnetically ordered states as well as various disordered 
states, where the heat current is totally carried by the 
charge-neutral objects such as the magnons and the deconfined spinons. 
For the ordered magnets, time reversal symmetry is broken already,
and the intrinsic thermal Hall effect is often attributed to the magnonic 
excitations. In this case, as there are no charged currents in the solid, 
the magnetic field cannot directly exert a Lorentz-like force on the magnons. 
The origin of the magnon thermal Hall effect can be traced 
back to the topologically nontrivial magnon bands and/or 
the magnon Berry curvatures. A similar picture applies to 
the triplon thermal Hall effects for the valence bond singlet states 
of the dimerized magnets. In contrast, for the quantum spin liquids 
where the magnetic orders are absent even down to the zero 
temperature limit, the deconfined and fractionalized excitations 
carry the emergent gauge charges, thus could contribute to 
the thermal Hall response via the emergent Lorentz force. 
How the external magnetic field induces the internal Lorentz force
for the exotic quasiparticles is a challenging but key question to be 
addressed, and the mechanism varies widely for different spin liquids 
and different emergent quasiparticles.

\subsection{Formalism of thermal Hall conductivity}

Before proceeding further, we briefly review the theoretical formalism 
for the thermal Hall conductivity of the magnetic excitations, 
which would be beneficial for the main part of this review.  
In the standard linear response theory to an external probe,  
the external probe simply enters as a perturbation. 
To study the thermal Hall transport, one needs to consider     
the temperature gradient that works as the external driving 
force for the thermal transport. Unlike the conventional linear
response theory, here the Hamiltonian stays invariant 
while the distribution function $\exp{(-\beta H )}$ is 
modified~\cite{Luttinger1964,Matsumoto2011,Matsumoto2011B,Shindou2014,Qin2011}.
Thus the theoretical treatment requires some extra care. 
This difficulty is overcome by the introduction of a fictitious 
pseudo-gravitational potential as shown by Joaquin 
Luttinger~\cite{Luttinger1964}. The temperature gradient is 
defined by ${T(\boldsymbol{r})=T_0[1-\eta(\boldsymbol{r})]}$ 
with a constant $T_0$ and a space-dependent small  
parameter $\eta(\boldsymbol{r})$, that can be regarded 
as a space-dependent prefactor to the Hamiltonian,   
${\exp{[-H/(k_BT(\boldsymbol{r})) ]} \simeq 
	\exp{[-(1+\eta(\boldsymbol{r}))H/(k_BT_0)}]}$. 
Then $\eta(\boldsymbol{r})H$ is regarded as a perturbation  
to the Hamiltonian from the temperature gradient. Generally, 
one can incorporate the temperature gradient into the Hamiltonian 
as a perturbation by using the pseudo-gravitational potential. 
In the linear response, one can further assume $\eta(\boldsymbol{r})$ 
to be linear in the position and expand the response in terms of 
$\nabla\eta(\boldsymbol{r})$. The energy current density is derived as follows, 
${ j_{\mu}^{E}(\boldsymbol{r})= j_{0\mu}^{E}(\boldsymbol{r})+j_{1\mu}^{E}(\boldsymbol{r}) }$, 
where $j_{0\mu}^{E}(\boldsymbol{r})$ is independent of 
$\nabla\eta(\boldsymbol{r})$ and $j_{1\mu}^{E}(\boldsymbol{r})$ 
is linear in $\nabla\eta(\boldsymbol{r})$. They both contribute 
to the thermal transport coefficients. 

Following this scheme, the energy current density of the magnetic excitations
in quantum magnets can be derived in the linearly perturbative regime,  
and the thermal Hall conductivity for a non-interacting spinless boson    
Hamiltonian~\cite{Matsumoto2011,Matsumoto2011B,Shindou2014} is 
written as 
\begin{eqnarray}
\kappa_{xy} =  -\frac{k_B^2T}{\hbar V} \sum_{n} \sum_{\bs{k}} c_2\big[n_B(\epsilon_{n{\bs{k}}},T)\big]\Omega_{n{\bs{k}}},
\end{eqnarray}
where $n_B(\epsilon_{n{\bs{k}}},T)$ is the Bose-Einstein distribution given 
by the elementary excitations with the dispersion $\epsilon_{n{\bs{k}}}$, 
and $c_2(x)$ is a weighting function defined by
\begin{eqnarray}
c_2(x)=(1+x) \left(\ln{\frac{1+x}{x}}\right)^2-(\ln x)^2-2{\rm Li}_2(-x),
\label{k_xy_boson}
\end{eqnarray}
with ${\rm Li}_2(x)$ being the dilogarithm. It can be readily verified that $c_2(x)$ 
increases monotonically with $x$ and has a minimum value of 0 in the limit $x\rightarrow0^+$. 
In contrast, it tends to ${\pi}^2/3$ in the opposite limit $x\rightarrow +\infty$. 
Moreover, $\Omega_{n{\bs{k}}}$ is the Berry curvature of the bosonic excitations 
of the $n$-th band, which indicates that the thermal Hall conductivity $\kappa_{xy}$ 
in Eq.\eqref{k_xy_boson} is rooted in a topological origin, since it is directly related 
to the Berry curvature in the momentum space. The integral of the Berry curvature 
$\Omega_{n{\bs{k}}}$ over the Brillouin zone defines the Chern number 
(first Chern index) for the $n$-th boson band
\begin{equation}
\mc{C}_n = \frac{1}{2\pi}\int_{\rm BZ}d{\bs{k}}\Omega_{n\bs{k}}\in \mb{Z} .
\label{chernDef}
\end{equation}

In the zero temperature limit, or for the temperature much smaller than 
the bosonic excitation gap, $n_B(\epsilon_{n{\bs{k}}},T)\approx0$. 
Thus one can simply obtain $\kappa_{xy}/T\rightarrow0$ from the mathematical 
properties of $c_2(x)$. On the other hand, if the temperature is much higher 
than the maximum energy of the boson bands such that 
$n_B(\epsilon_{n{\bs{k}}},T)\gg1$, 
then
\begin{equation}
\frac{\kappa_{xy}}{T} \approx -\frac{\pi k_B^2}{6\hbar}\sum_{n}\mc{C}_n=0,
\end{equation}
where we have used the fact that the sum of the Chern numbers over all particle bands is zero. 
Therefore, for any bosonic excitations the thermal Hall conductivity $\kappa_{xy}/T$ can tend 
to zero both in the zero temperature limit and in the high temperature limit, and the distinct
features for different systems could only emerge from the finite temperature region 
(roughly between the excitation gap scale and the bandwidth of the boson spectra). 
The authors of Refs.~\cite{Matsumoto2011,Matsumoto2011B,Shindou2014} 
originally studied the thermal Hall effect of the non-interacting 
spin waves in the magnetically ordered systems. In fact, this 
scheme can be readily generalized to other bosonic excitations, 
even involving the spin indices.

For the case of fermionic excitations, the result is derived by \citet{Qin2011} 
within the same framework using the pseudo-gravitational potential, 
where a thermal Hall conductivity formula for the fermionic system 
with a nonzero chemical potential $\mu$ was obtained as
\begin{equation}
\kappa_{xy}=-\frac{1}{T}\int d\epsilon \, (\epsilon-\mu)^2
\frac{\partial n_F(\epsilon,\mu,T)}{\partial \epsilon}\sigma_{xy}(\epsilon),
\label{k_xy_fermi}
\end{equation}	
where ${n_F(\epsilon,\mu,T)}$ is the Fermi-Dirac distribution 
and 
\begin{equation}
\sigma_{xy}(\epsilon)=-\frac{1}{\hbar V}\sum_n
\sum_{\bs{k}}\Theta(\epsilon-\epsilon_{n\bs{k}})\Omega_{n{\bs{k}}}
\end{equation}	
is the zero temperature anomalous Hall coefficient
for a system with the chemical potential $\epsilon$ and unit charge.
Here $\Omega_{n{\bs{k}}}$ is the Berry curvature for the fermionic quasiparticles 
of the band indexed by $n$, and $\epsilon_{n\bs{k}}$ is the energy dispersion of this band. 
For the temperature much higher than the maximum energy of the fermion bands, 
all the bands are almost equally populated as determined by the Fermi-Dirac distribution function. 
Then the summation of Berry curvatures of all bands vanishes and $\kappa_{xy}/T\rightarrow0$ 
as for the bosonic system. 
Conversely, in the zero temperature limit, the derivative of the Fermi-Dirac distribution function 
represents a sharp peak and can be expanded as
\begin{equation}
\frac{\partial n_F(\epsilon,\mu,T)}{\partial \epsilon}=-\delta(\epsilon-\mu)-\frac{(\pi k_BT)^2}{6}\frac{d^2}{d \epsilon^2}\delta(\epsilon-\mu)+...
\end{equation}
Thus the thermal Hall conductivity of Eq.~\eqref{k_xy_fermi} can be recast into
\begin{equation}
\kappa_{xy}=\frac{\pi^2 k_B^2 T}{6}\int d\epsilon(\epsilon-\mu)^2\frac{d^2}{d \epsilon^2}\delta(\epsilon-\mu)\sigma_{xy}(\epsilon).
\end{equation}	
Using the relation ${\delta''(x)=2\delta(x)/x^2}$, one can easily obtain
\begin{eqnarray}
\frac{\kappa_{xy}}{T}&=&-\frac{\pi^2k_B^2}{3\hbar V} 
\sum_n
\sum_{\bs{k}}\Theta(\mu-\epsilon_{n\bs{k}})\Omega_{n{\bs{k}}}\\
&=&-\frac{\pi k_B^2}{6\hbar} \sum_n \int_{\rm BZ}\frac{d\bs{k}}{2\pi}\Theta(\mu-\epsilon_{n\bs{k}})\Omega_{n{\bs{k}}}.
\label{fermionkappa}
\end{eqnarray}
It suggests that, for fermionic system, in the zero temperature limit ${\kappa_{xy}/T\neq0}$ 
if $\sum_n\int_{\rm BZ}{d\bs{k}}\Theta(\mu-\epsilon_{n\bs{k}})\Omega_{n{\bs{k}}}$ is non-vanishing, 
which is different from the bosonic system where $\kappa_{xy}/T$ always tends to zero 
in the zero temperature limit. Especially, for the gapped system where the chemical potential 
$\mu$ lies in the gap and the bands are separated from each others, Eq.~\eqref{fermionkappa} 
can be further simplified as 
\begin{equation}
\frac{\kappa_{xy}}{T}=-\frac{\pi k_B^2}{6\hbar}\sum_{n\in{\rm filled~bands}}\mc{C}_n.
\label{fermionkappaC}
\end{equation}
Here $\mc{C}_n$, with the same definition as in Eq.~\eqref{chernDef}, is the Chern number 
of the $n$-th fermion band, and it means that ${\kappa_{xy}}/{T}$ is integer quantized 
in units of ${\pi k_B^2}/{6\hbar}$ at zero temperature. Moreover, for the system with the 
fermion pairing terms (the Hamiltonian can be manipulated into the Bogoliubov–de Gennes form), 
or for the system with Majorana fermionic excitations, an additional $1/2$ factor is needed 
to be multiplied to the right hand side of Eq.~\eqref{fermionkappaC}, which implies that 
${\kappa_{xy}}/{T}$ now is half-integer quantized in the zero temperature limit.

Armed with the above explicit formalism, the theoretical understanding 
of the thermal Hall effects in quantum magnets can be progressed accordingly 
to predict the experiments and/or understand the existing experiments. 
In more complex realistic situations, however, there can be several components 
and degrees of freedom involved in the thermal Hall transports, such as the 
phonons and the emergent gauge fields in quantum spin liquids, 
and more scrutiny in the analysis should be needed.  
It is also worth mentioning that, based on the above formalism, certain 
universal properties about the thermal Hall conductivity may be 
established at the intermediate temperature regime. It was shown 
by Y.~F. Yang et al.~\cite{Fuchun2020}, that at the intermediate 
temperatures comparable to the bandwidth of the excitation spectra 
with the non-trivial topology, the thermal Hall conductivity could 
exhibit an unexpected universal scaling with a simple exponential form. 
This universal scaling behavior depends on whether the magnetic excitations 
are bosonic or fermionic, and may be used to distinguish the bosonic and 
fermionic excitations. At the low temperatures close to the ground states, 
however, no such universal behavior in the thermal Hall conductivity is expected.

\subsection{Physical contents about thermal Hall effect in quantum magnets}

The previous subsection is an overview of the theoretical framework about 
the thermal Hall transports. More specific discussion and explanation
of the microscopic theory and mechanism requires the detailed
understanding of the quantum phases and the correspondence 
between the microscopic variables and the emergent degrees of freedom. 
These will be covered in the main text of this review.  
As for the actual thermal Hall experiments in quantum magnets, 
there are several measurements that are often performed and   
analyzed. It can be a bit illuminating to briefly mention them here. 

1) First of all, the thermal Hall transport is measured as a function of   
temperature. In some of the interesting cases, the thermal Hall signal 
could exhibit the sign reversal as a function of temperature.   
Many times, the thermal Hall signals are originated from the 
thermal activation of the emergent quasiparticles, except 
for the cases with the edge modes that can be related to the ground
state properties such as chiral spin liquids. The thermal activation 
allows one to access the band structure properties such as 
the density of states and the Berry curvature distribution
of the higher energy bands, and these properties could be 
responsible for the temperature dependence.   

2) Secondly, as the external magnetic field is applied to the system, 
the field dependence of the thermal Hall signal is often measured. 
As the field has the ability to modify both the many-body ground state 
and the excited states, the thermal Hall signal with the magnetic field   
allows one to obtain the information about the change of these
many-body states. In particular, the transition from gapless to gapped 
phases, and the topological transition between topologically trivial and 
non-trivial phases can be reflected in the thermal Hall transport. 
Moreover, the field and temperature can be combined. For example,
one can fix the field and then vary the temperature, and 
check the temperature dependence of the field-induced states. 
The reverse can also be performed. 

3) Thirdly, the physical mechanisms for the thermal Hall effect
in many quantum magnets, as we will explain in the main text, 
often require the anisotropic interactions. The variation of 
the thermal Hall signal with respect of the change of the 
field orientation is then related to the anisotropic interactions
and the crystal symmetries of the system. Therefore,  
the geometrical setting of the field-measurement scheme 
such as the planar thermal Hall effect allows one to 
determine the presence of the thermal Hall signal simply 
from the symmetry analysis. One such setting is the planar 
thermal Hall effect where the external field is placed in the plane of the layered materials. 

4) Fourthly, the thermal Hall measurement is often measured and 
compared together with the longitudinal thermal conductivity to
check if there exists some close correlation between the behaviors 
of two conductivities.


As the different quantum magnets belong to distinct sets of quantum phases
and exhibit rather different physical properties, it is difficult for us to place  
them into a single section to explain the experimental backgrounds. 
Thus, in the presentation of this review, we instead bury the introduction 
of the experimental motivations and results on the thermal Hall effects 
and some of the relevant theoretical attempts in each section.
Moreover, we often use specific and concrete examples to deliver
the theoretical understanding and development and hope the readers 
not to localize their thought and vision on these specific examples. 

The remaining parts of this review are organized as follows. 
In Sec.~\ref{sec2}, we start with the thermal Hall effects of 
the simple magnons in the conventional ordered magnets. 
In Sec.~\ref{sec3}, we classify the physical origins of the thermal Hall effects 
for different disordered states. This includes the triplon thermal Hall effect 
for the valence bond solid of the dimerized magnets, and more extensively, 
the thermal Hall effects of various kinds in different spin liquids. 
In Sec.~\ref{sec4}, we explain the physical origin of the thermal Hall effects 
of the U(1) spin liquids with the gapless spinon matter in both weak and strong 
Mott regimes. In Sec.~\ref{sec5}, we explore the physical origin of the 
thermal Hall effects in the three-dimensional U(1) spin liquids with the gapped spinon matter, 
and focus the discussion on the pyrochlore quantum spin ice U(1) spin liquids.
In Sec.~\ref{sec6}, we address the thermal Hall transports in the    
Kitaev materials, and 
in Sec.~\ref{sec11} we 
suggest 
the quantized thermal Hall effect as a diagnosis of the topological and magnetic 
phase transitions. Finally in Sec.~\ref{sec7}, we conclude 
with a summary and some perspectives about the thermal Hall effects.

\section{Magnon thermal Hall effects}
\label{sec2}

In many quantum magnets, one common fate of the system 
is to develop the conventional magnetic orders and then host 
the coherent and collective magnetic excitations at the low 
temperatures below the ordering transitions. In this section, 
we focus on the magnetically ordered states, where the spin 
wave or the magnon is a collective propagation of the 
precessional motions of the local magnetic moments. 
Magnons can be viewed as the quantized spin waves, 
which are the quantized spin fluctuations of the ordered magnets 
and are clearly charge neutral quantum elementary excitations. 
Although the theoretical formalism about the magnonic excitations  
has been well-developed since the mid of the twentieth century~\cite{PhysRev.58.1098},  
the enormous progress on the topological materials in the last decade~\cite{Ando_2013,doi:10.1146/annurev-conmatphys-031016-025458} 
or so provide new insights into the geometry and topology of these 
magnetic excitations. The Berry phase and Berry curvature of the    
magnons play a central role in the study of the magnon band topology, 
and the first Chern number is the corresponding topological invariant. 
The non-interacting topological magnons can propagate without dissipation, 
therefore, considered as a promising candidate for the magnon spintronics. 
For a detailed description on topological magnons and the topological 
properties of the magnons, one can refer to a recent review by McClarty~\cite{McClarty2022}.

On the experimental side, the angle-resolved photoemission spectroscopy is not 
helpful at all in this case and the spin-polarized scanning tunneling spectroscopy 
is severely restricted. The inelastic neutron scattering (INS) measurement  
could directly detect the magnetic excitations in the spin systems by providing 
the detailed information for the magnon dispersion relations in the bulk.   
The presence of the nodal points or nodal lines could be observed directly 
from the INS measurement~\cite{McClarty2022}; yet, the energy resolution 
remains to be improved for a better observation. 
The INS serves as a valuable experimental tool 
in the investigation on the bulk systems with the topological magnons, 
whereas in two dimensional materials the INS suffers from the low signal intensities 
thus can be unreliable. The topological nature of the magnons is essentially the same as 
the fermionic band topology, in the sense that the topological index of the bulk 
corresponds to the number of boundary modes. The boundary modes are 
the hallmarks for the non-trivial topological systems, which remains to be unattainable for the INS.
An alternative route to decipher the topology is to evaluate the Berry curvature of the magnon bands.
Unfortunately, as remarked generically in Sec.~\ref{sec1}, 
the information of the magnonic wavefunction cannot 
be directly deduced from the INS measurements.
These apparent discrepancies suggest that the magnon band topology 
can not be solely determined by the INS experiments at the present stage.
Often, one relies on the combination of theoretical results 
and the neutron scattering measurements to make claims to the topological magnons. 

In addition to the spectroscopic measurements, we consider the heat transport experiment of the magnons
as the magnons are charge neutral and carry energy. Since the thermal transport of the bosons does 
not show any quantized signal, a direct detection of topological boundary modes for the magnons 
is more difficult than that for the electronic counterparts. 
Lacking a clear signature of nontrivial topology, we turn to the magnon thermal Hall transport 
that dictates the Berry curvature 
and thus the wavefunction properties of the magnons, as can be seen from Eq.~\eqref{k_xy_boson}.

\subsection{Thermal Hall effect in ferromagnets}
\label{sec21}

We begin with magnon thermal Hall effect in simple ferromagnets. 
When the magnon is considered in ferromagnets, the time reversal symmetry 
of the system is broken either spontaneously by the magnetic orders and/or 
explicitly by the external magnetic field. 
Breaking the time reversal symmetry alone, however, is insufficient to generate the magnon thermal Hall effect. 
The magnon wavefunction has to be ``twisted'' in order to generate the non-vanishing Berry curvatures. 
A typical ``twisting'' mechanism is facilitated by the Dzyaloshinskii-Moriya (DM) 
interaction~\cite{Onose2010,Lifa2013,Hirschberger2015,Ong2015,Chisnell2015,
PhysRevB.85.134411,PhysRevB.89.134409,PhysRevB.97.081106} 
between the nearby local spin moments and by long-range dipolar 
interaction~\cite{Matsumoto2011,Matsumoto2011B,Shindou2013A,Shindou2013B}.

\subsubsection{Collinear ferromagnets with Dzyaloshinskii-Moriya  interaction}
\label{sec211}

The DM interaction plays a vital role in a majority of studies on the magnon thermal Hall effect in ferromagnets.  
The antisymmetric DM interaction arises from the spin-orbit coupling via a high order perturbation theory 
in the conventional magnets and is thus usually weak compared to the symmetric Heisenberg term. 
In the modern context of the spin-orbit-coupled Mott insulators, the strength of the spin-orbit coupling is 
relatively large; the corresponding DM interaction cannot be treated as a perturbation to the Heisenberg term~\cite{Gang2008},
as well as the pseudodipole interaction that is even more subleading than the DM interaction 
in the conventional case~\cite{Moriya1960}. 
Thus, the role of the DM interaction (and the pseudodipole interaction) 
should be considered more seriously in many such materials. The competition 
between the anti-symmetric DM interaction and the symmetric Heisenberg 
interaction can engender the non-collinear/non-coplanar spin textures with the long-range order.
The DM interaction often generates the Berry curvature for the magnons, 
which leads to the thermal Hall effect for the magnons in the ordered regimes. 
It is theoretically proposed and numerically verified that the ferromagnetic 
insulator with the DM interaction could host the topological magnons, namely, 
the band structure topology of the charge-neutral magnons gives rise to the 
topologically protected magnon modes propagating along the edge or surface.
Even though the discussion about the magnon thermal Hall effects
occasionally involves the topological magnons, the existence of the 
magnon thermal Hall effects does not necessarily require 
the relevant magnon bands to be topological as 
it is related to the summation of the magnon Berry's curvature 
weighted by the thermal distributions [see Eq.~\eqref{k_xy_boson}].

For the materials realization, the pyrochlore ferromagnet Lu$_2$V$_2$O$_7$ 
is suggested to be the representative compound for the magnon thermal Hall effect, 
which has been demonstrated in experiments~\cite{Onose2010,PhysRevB.85.134411}.
While the magnon Weyl point is later suggested~\cite{PhysRevLett.117.157204} 
to occur in the high-energy magnon bands and may become important at high 
temperatures, the low-temperature thermal Hall effect 
is well interpreted by the Berry curvature of the magnon bands. 
In the following we exemplify the DM interaction induced magnon thermal Hall effect 
with two well-known ferromagnets: the 3D pyrochlore ferromagnet Lu$_2$V$_2$O$_7$
and the 2D kagom\'{e} ferromagnet Cu(1-3, bdc).

\subsubsection{3D pyrochlore ferromagnet Lu$_2$V$_2$O$_7$}
\label{sec2111}

\begin{figure}[t] 
	\centering
	\includegraphics[width=8.6cm]{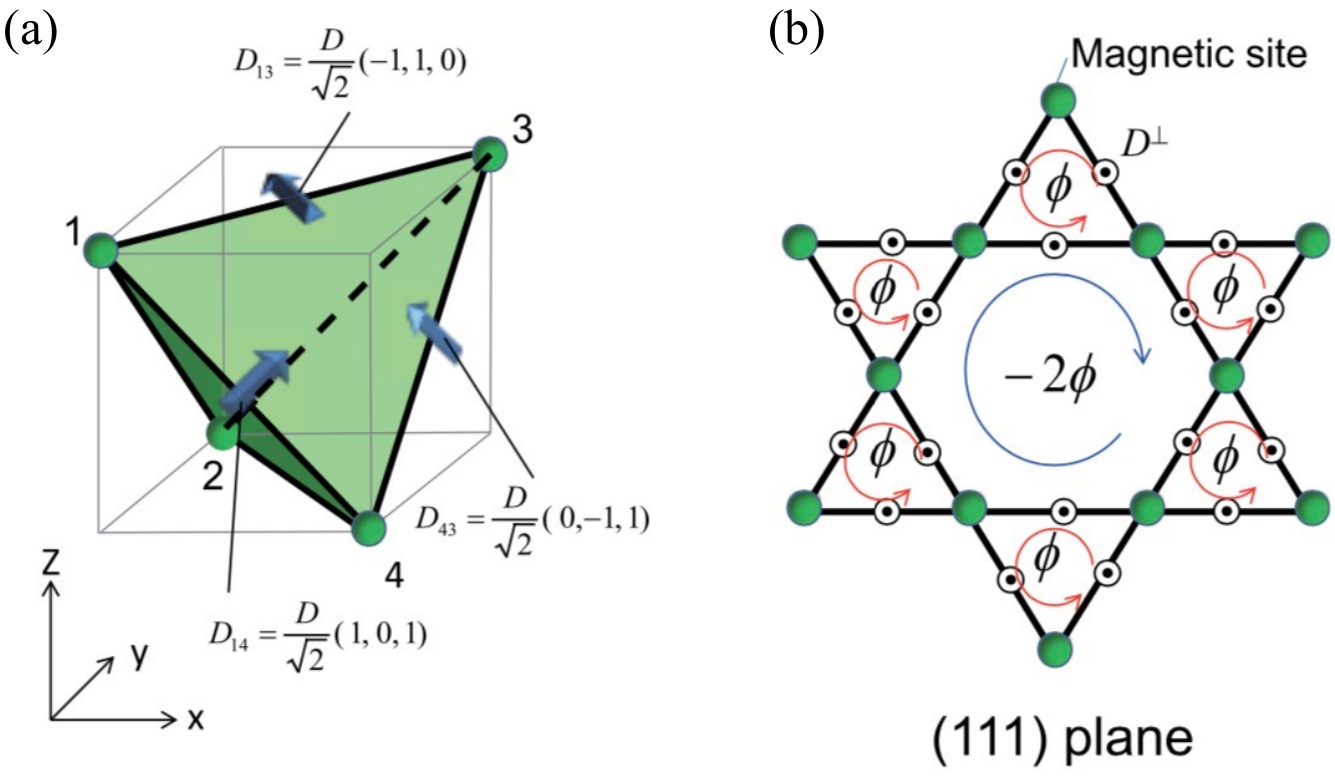}
	\caption{(a) The DM vectors on the pyrochlore lattice and (b) the fictitious U(1) gauge flux for the magnons
		due to the DM interaction in the (111) plane of the pyrochlore lattice (kagom\'e layer). 
		Reprinted from Ref.~\onlinecite{PhysRevB.85.134411}.}
	\label{pyrochloreDM}
\end{figure}

Lu$_2$V$_2$O$_7$ is a ferromagnetic Mott insulator where the 
V$^{4+}$ ions are magnetic with the $3d^1$ electron configuration
and form a pyrochlore lattice~\cite{Onose2010}. 
The pyrochlore structure is composed 
of the corner-sharing tetrahedra in 3D. The lattice can be viewed 
as an alternative stacking of kagom\'e and triangle lattices along  
the [111] crystallographic direction. 
As the $3d$ electron of the V$^{4+}$ ion occupies the lower $t_{2g}$ orbitals, 
the spin-orbit coupling is expected to be relevant despite the presence 
of the splitting among the $t_{2g}$ orbitals. 
In fact, the magnetic moment of 
the V$^{4+}$ ion is about $1.0\,\mu_B$ and is thus strongly 
suppressed compared to the pure ${S=1/2}$ contribution, 
indicating the involvement of the orbital content in the local 
moment~\cite{Su_2019}. The proposed theoretical model for the interacting 
V$^{4+}$ local moments contains a ferromagnetic Heisenberg 
exchange and an antisymmetric DM interaction, \emph{i.e.},
\begin{eqnarray}
	\mc{H} &=&  \sum_{\langle ij\rangle}  {
		- J {\bf S}_i \cdot {\bf S}_j
		+   { {\bf D}_{ij} \cdot ({\bf S}_i \times {\bf S}_j) } } 
	-\sum_i {\bf B}\cdot  {\bf S}_i
	\label{eqham} \\
	& =& 
	\sum_{\langle ij\rangle}   {
		- J {\bf S}_i \cdot {\bf S}_j
		+   { {\bf D}_{ij} \cdot ({\bf S}_i \times {\bf S}_j) } }
 	-   {{\bf B}\cdot \frac{ {\bf S}_i + {\bf S}_j }{6} }  , 
 	\label{Ham_HDMZ}
\end{eqnarray}
where ${\bf S}_i$ refers to the spin-1/2 local moment at the lattice site $i$,
$\langle ij \rangle$ refers to the nearest-neighbor sites, and ${\bf D}_{ij}$ 
is the vector that specifies the DM interaction. In addition, 
a Zeeman coupling to the external magnetic field is included. 
In the second line of the above equation, we have absorbed the 
Zeeman coupling into the bond summation. From the 
Moriya's symmetry rules, the crystal symmetries of the pyrochlore lattice 
can determine the form of the DM interaction, and 
there exists only one component of the DM vector 
[see Fig.~\ref{pyrochloreDM}(a) for the DM vectors 
on a single tetrahedron of the pyrochlore lattice]. 
The simple ferromagnetic Heisenberg model is sufficient to capture 
the collinear ferromagnetic order, but cannot explain the large thermal Hall effect 
in Lu$_2$V$_2$O$_7$. It is thus necessary to introduce the 
DM interaction that would generate the magnon Berry curvature distributions. 
Microscopically, the DM interaction is expected from the spin-orbit coupling~\cite{Moriya1960}.

As the summation of the DM vectors connected to one lattice site vanishes, the DM interaction does 
not modify the ferromagnetic ground state that is demanded by the ferromagnetic Heisenberg part. 
It is then convenient to introduce an orthonormal basis $(\hat{l}, \hat{m}, \hat{n})$ 
with $\hat{n} \equiv {\bf B}/B$, and $\hat{l}$ and $\hat{m}$ 
are two transverse unit vectors normal to $\hat{n}$. Since the ferromagnetic order orients 
along with the external magnetic field, this unit vector $\hat{n}$ also defines the spin ordering 
direction for the ferromagnetism. Both the spin vector and the DM vector are expressed 
in this basis as
\begin{eqnarray}
{\bf S}_i &=& S_i^l \, \hat{l} +  S_i^m \, \hat{m} + S_i^n \,\hat{n}  , 
\\
{\bf D}_{ij} &=& ({\bf D}_{ij} \cdot \hat{l}) \,  \hat{l}  
+ ({\bf D}_{ij} \cdot \hat{m}) \,  \hat{m} 
+ ({\bf D}_{ij} \cdot \hat{n})\,  \hat{n} ,
\end{eqnarray}
where the spin is ferromagnetically ordered in the $S_i^n$ component and 
the $\hat{l}$ and $\hat{m}$ components then become the fluctuations.  
To obtain the magnetic excitations, we first substitute the spin operators  
with the Holstein-Primakoff bosons, 
\begin{eqnarray}
S_i^n &=& S- a_i^\dagger a_i^{} , \label{eq6} \\
S^+_i &=&   S_i^l  + i  S_i^m   = {(2S - a^\dagger_i a_i^{} )^{1/2}} \, a_i^{} , \label{eq7}\\
S^-_i &=&    S_i^l  - i  S_i^m   = a^\dagger_i  {(2S- a^\dagger_i a_i^{} )^{1/2}} \label{eq8},
\end{eqnarray}
where $a^{\dagger}_i$ ($a^{}_i$) refers to the creation (annihilation) 
operator for the Holstein-Primakoff magnon and ${S=1/2}$.  
Due to the cross product structure in the DM interaction, 
the only quadratic term in the spin fluctuation is the one 
contributed from the ${({\bf D}_{ij} \cdot \hat{n})  \hat{n}}$ part. 
The stability of the ferromagnetic state against the DM interaction 
suggests the linear term in the spin fluctuation automatically vanishes. 
Therefore, the spin Hamiltonian becomes
\begin{eqnarray}
\mc{H}& \simeq &
\sum_{\langle ij \rangle }
{-J {\bf S}_i \cdot {\bf S}_j } + D_{ij}^n ( S^l_i S^m_j -S^m_i S^l_j)
- \frac{B}{6} (S_i^n + S_j^n)
\nonumber \\
&=& \sum_{\langle ij \rangle} \Big[
(-\frac{J}{2}  + \frac{i D_{ij}^n}{2}  )S^+_i S^-_j + 
(-\frac{J}{2}  - \frac{i D_{ij}^n}{2}  )S^-_i S^+_j 
\nonumber \\
&& \quad\quad\quad 
- J S_i^n S_j^n  -  \frac{B}{6} (S_i^n + S_j^n) \Big]  \nonumber \\ 
&=& \sum_{\langle ij \rangle}   \Big[
{ -\frac{J_{ij}}{2} } ( e^{-i\phi_{ij}} S^+_i S^-_j  + e^{i\phi_{ij}} S^-_i S^+_j ) 
- J S_i^n S_j^n  \nonumber \\
&& \quad\quad\quad  -  \frac{B}{6} (S_i^n + S_j^n) \Big],
\label{eq16sec2}
\end{eqnarray}
where ${J_{ij} = [J^2+ (D_{ij}^n)^2]^{1/2}}$, ${{\tan \phi_{ij} = D_{ij}^n/J}}$
and $D_{ij}^n$ refers to ${{\bf D}_{ij} \cdot \hat{n}}$. 
It is now straightforward to express the above equation in terms of 
the Holstein-Primakoff bosons from Eqs.~\eqref{eq6}-\eqref{eq8},
\begin{eqnarray}
\mc{H}_{\rm SW} &\simeq & \sum_{\langle ij \rangle} \Big[
{- J_{ij} S} (  e^{-i\phi_{ij}} a^\dagger_i a^{}_j +  e^{i\phi_{ij}} a^\dagger_j a^{}_i )  
\nonumber 
\\
&& \quad\quad
+ (JS + \frac{B}{6}) (a^\dagger_i a^{}_i + a^\dagger_j a^{}_j) 
\Big],
\label{eq10}
\end{eqnarray}
where the standard linear spin-wave approximation is used. 
Here Eq.~\eqref{eq10} essentially describes the hopping of 
the Holstein-Primakoff bosons minimally coupled with the   
fictitious and static U(1) gauge field $\phi_{ij}$ on the pyrochlore lattice. 
Unlike the generic spin-wave Hamiltonian that contains a 
pairing term for the bosons, Eq.~\eqref{eq10} does not have pairing
terms. From Eq.~\eqref{eq10}, one can readily extract the magnon 
dispersions and the eigenstates of the magnon Bloch bands. 
Without the DM interaction, the gauge field ${\phi_{ij} =0}$
and the Hamiltonian is real. The wavefunction of the magnons is real 
and cannot support non-vanishing Berry curvatures.   
The realness of the magnons 
is broken by the DM interaction by introducing the U(1) gauge phase modulation
in the magnon hopping. 
This fictitious U(1) gauge field from the DM interaction
cannot be gauged away because the summation of the DM vectors 
around the triangular or hexagonal plaquettes on the pyrochlore lattice 
is not vanishing and there exist net U(1) gauge fluxes through 
the plaquettes. In Fig.~\ref{pyrochloreDM}(b), the distribution of 
the fictitious U(1) gauge flux on the kagom\'{e} layer is depicted. 
Although the total fictitious U(1) gauge flux is zero in the unit cell,
the Berry curvature for the magnons in the momentum space is 
non-vanishing, which is responsible for the thermal Hall effects 
in Lu$_2$V$_2$O$_7$. 

\begin{figure}[htbp] 
	\centering
	\includegraphics[width=8.6cm]{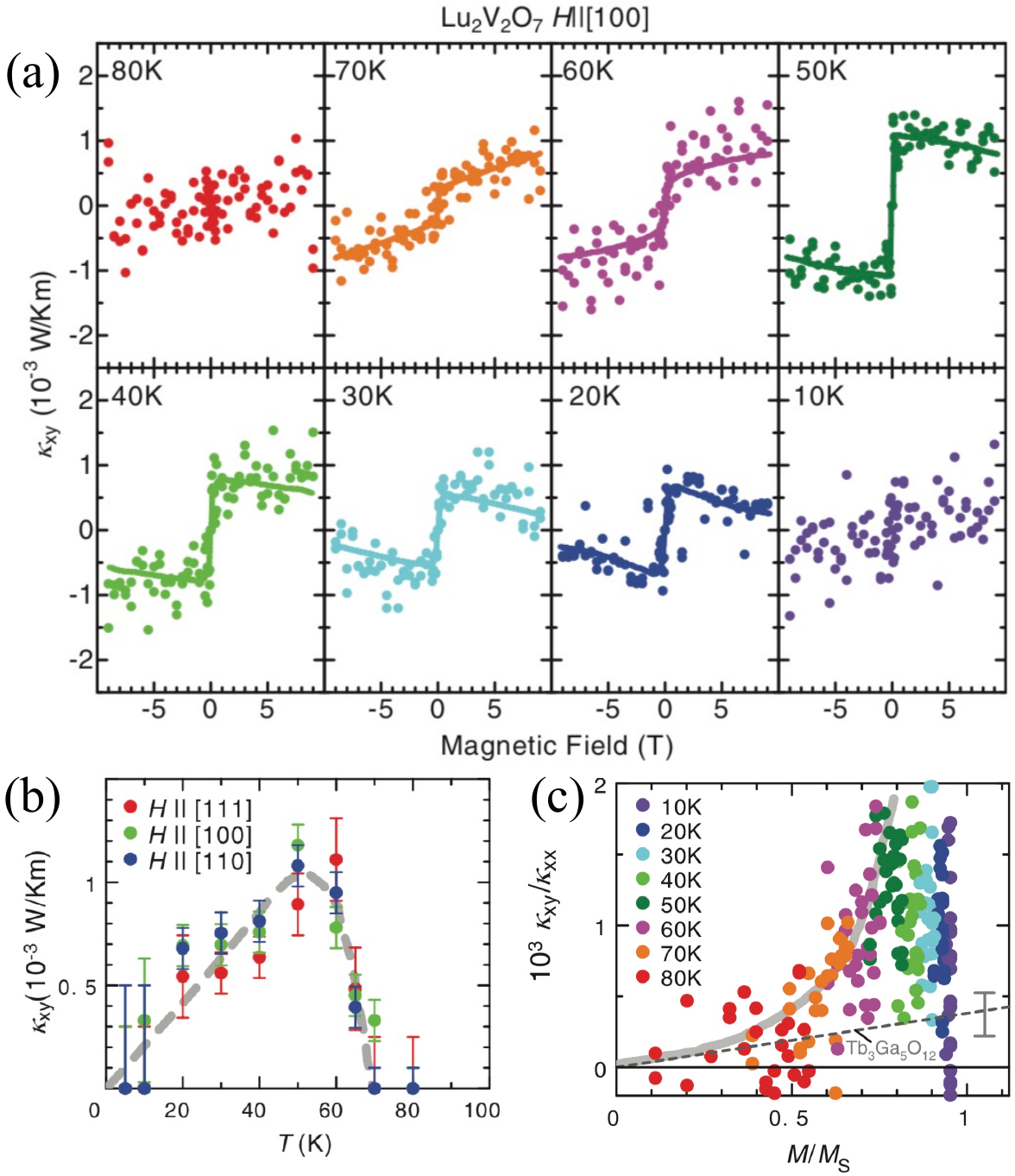}
	\caption{(a) Magnetic field variation of the thermal Hall conductivity 
		in Lu$_2$V$_2$O$_7$ at different temperatures.
		(b) Temperature dependence of the thermal Hall conductivity 
		for the magnetic field aligned along different high symmetry directions.
		(c) The thermal Hall angle $\kappa_{xy}/\kappa_{xx}$ plotted against 
		the magnetization at different temperatures.
		Figures are reprinted from Ref.~\onlinecite{Onose2010}.}
	\label{LuVOthermalHall}
\end{figure}

\begin{figure}[htbp] 
	\centering
	\includegraphics[width=7cm]{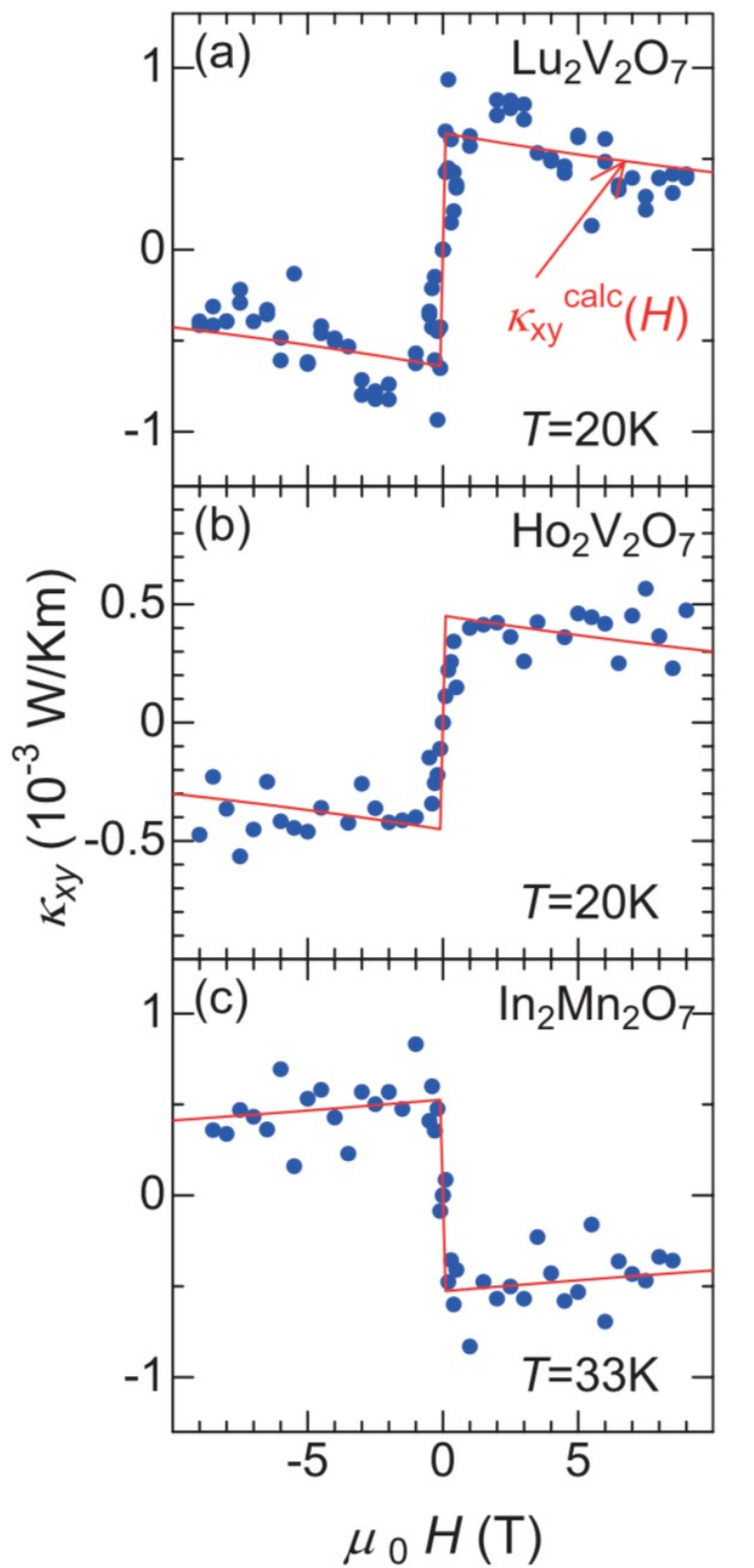}
	\caption{Magnetic field variation of thermal Hall conductivity (a) at 20 K for Lu$_2$V$_2$O$_7$, (b) at 20 K for Ho$_2$V$_2$O$_7$ and 
	(c) at 33 K for In$_2$Mn$_2$O$_7$. The solid (red) line indicates the magnetic field dependence of magnon thermal Hall conductivity 
	given by the theoretical calculation based on the DM interaction. Reprinted from Ref.~\cite{PhysRevB.85.134411}.}
	\label{pyrochloreTHE}
\end{figure}

The measured thermal Hall conductivity for Lu$_2$V$_2$O$_7$ 
with the variation of the magnetic field is plotted in Fig.~\ref{LuVOthermalHall}(a) 
at various temperatures. Below the Curie temperature $T_c$, 
the field dependence of $\kappa_{xy}$ shows a rapid saturation 
at relatively low fields then gradually decrease with higher magnetic fields. 
From the field dependence of the thermal Hall effect, 
the authors found an intimately connection with the spontaneous 
magnetization that indicates an anomalous Hall effect.
The temperature dependence of the anomalous thermal Hall conductivity,
plotted in Fig.~\ref{LuVOthermalHall}(b), are independent of the direction of the field.
This is consistent with the magnon Hamiltonian Eq.~\eqref{eq10} that 
was derived without the pre-assumption of the field direction. 
The phonon contribution to the Hall effect is further excluded  
based on the temperature and field dependence of $\kappa_{xy}$.
The conventional phonon Hall effects were explained from the spin-phonon coupling;
and the phonon thermal Hall effect is expected to be enhanced at higher fields
due to the reduced scattering by the magnetic fluctuation.
The decrease of $\kappa_{xy}$ contradicts with such a scenario.
Moreover, the thermal Hall angle $\kappa_{xy}/\kappa_{xx}$ 
is anticipated to be proportional
to the magnetization for the mechanism based on the phonon scattering by spins;
the experiment in Lu$_2$V$_2$O$_7$ [see Fig.~\ref{LuVOthermalHall}(c)]
shows a steeper decrease with increasing temperature around $T_c$, 
which again contradicts the phonon thermal Hall scenario.

The quantitative comparison between the theoretical calculation from 
the magnon Hamiltonian in Eq.~\eqref{eq10} and the experimental measurements 
are shown in Fig.~\ref{pyrochloreTHE}(a). These results confirmed that the experiment 
conducted in Lu$_2$V$_2$O$_7$ was the first observation of the magnon thermal Hall effect.
Similar measurements and theoretical calculation have been performed on other pyrochlore 
ferromagnets Ho$_2$V$_2$O$_7$ and In$_2$Mn$_2$O$_7$ 
[see Figs.~\ref{pyrochloreTHE}(b) and (c), respectively]. 
The thermal Hall signal under the magnetic field in In$_2$Mn$_2$O$_7$ 
has a different sign from that in Lu$_2$V$_2$O$_7$ and Ho$_2$V$_2$O$_7$, 
and this was suggested to arise from the different sign of the DM interaction 
that leads to the opposite fictitious U(1) gauge flux for the magnons 
on the pyrochlore lattice. 


\subsubsection{2D kagom\'e ferromagnet Cu(1-3, bdc)}
\label{sec2112}

We now turn to the magnon thermal Hall effects in 2D. Unlike the 3D 
counterpart, one could associate a Chern number to the magnon band 
in 2D. When the Chern number is non-zero, there exists a magnon chiral
mode propagating on the edge. Simple theoretical prediction for the 
topological magnons were established for the Heisenberg (anti)ferromagnet 
on the honeycomb lattice with a second-neighbor DM interaction~\cite{Owerre2016B},  
and the system was shown to realize the magnon chiral edge state 
that is analogous to the Haldane model for the quantum anomalous 
Hall effects in the electronic systems but appears at the finite energy. 
To obtain the corresponding magnon thermal Hall effect from the chiral edge modes, 
however, the magnon chiral edge mode should be thermally activated. 

\begin{figure}[t] 
	\centering
	\includegraphics[width=8.6cm]{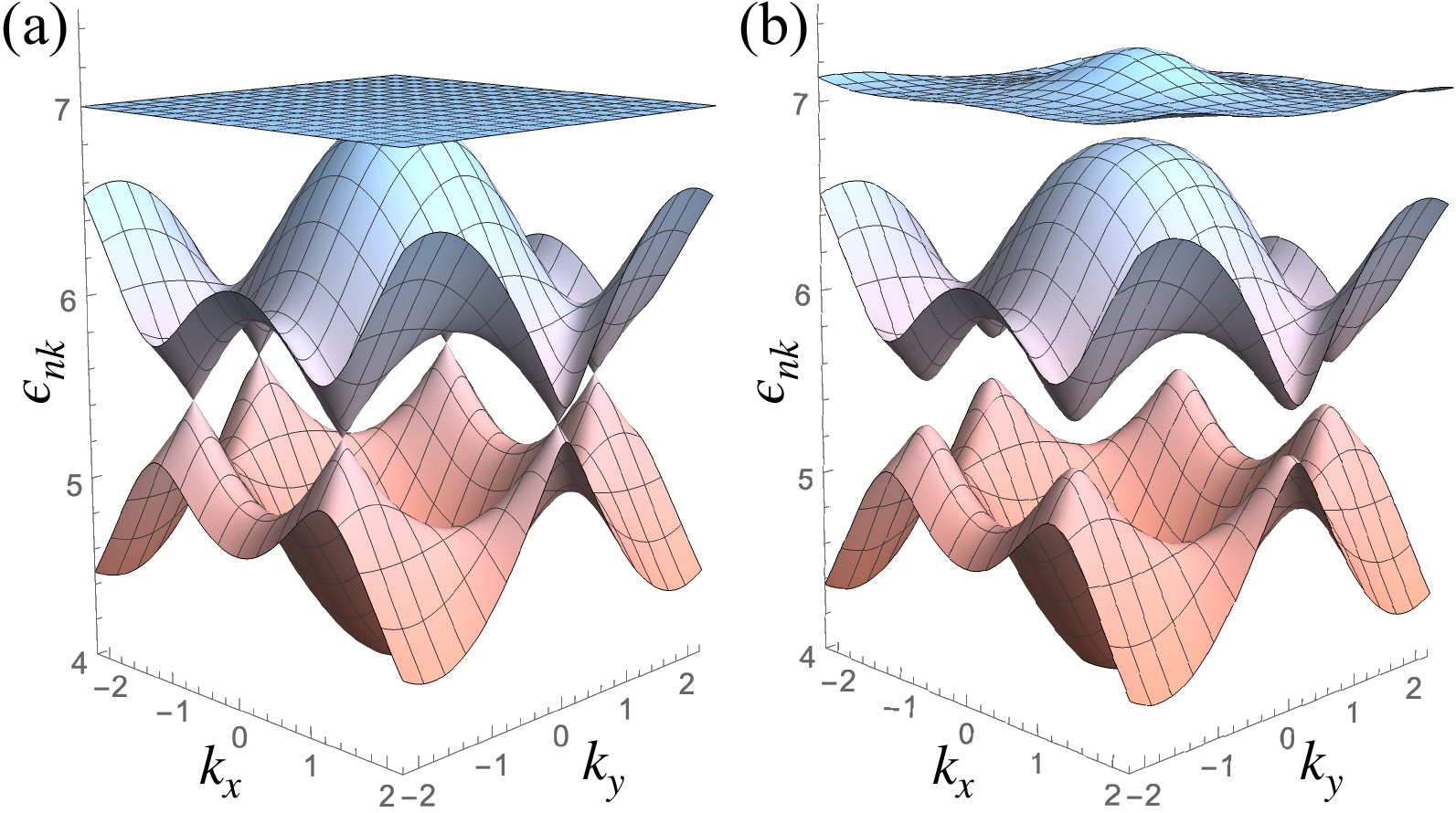}
	\caption{Magnon band structure of kagom\'{e} ferromagnet from linear spin wave theory in the presence of magnetic field $B/J = 1$.  (a) Without DM interaction,  the highest energy band is flat, while the lower two magnon bands form Driac band touching.  (b) With DM interaction (here $D/J=0.4$),  the highest energy band acquires a weak dispersion, and the Dirac band touchings of the lower two magnon bands are gapped out.}
	\label{kagomeMagnon}
\end{figure}

Actually, one representative example of the magnon thermal Hall effect 
in a 2D system that has been experimentally confirmed is 
the collinear kagom\'{e} ferromagnet Cu(1-3, bdc)~\cite{Ong2015,Chisnell2015,Alahmed_2022}
in the presence of the DM interaction. For this 2D ferromagnet, 
the same form of spin model as the one in Eq.~\eqref{eqham} was proposed,
where there exist the ferromagnetic Heisenberg and the DM interactions. 
In this material, the Cu$^{2+}$ ions form a kagom\'e lattice with the ${S=1/2}$ 
local moment at each site. In the ferromagnetically ordered state,  
the magnons for the model in Eq.~\eqref{eqham} were shown to    
have three bands. Due to the geometric reason or the loop structure
for the localized spin-wave mode around the hexagonal plaquette,
the highest energy band is flat in the absence of the DM interaction, 
while the lower two dispersive magnon bands form gapless Dirac band 
touchings, as shown in Fig.~\ref{kagomeMagnon}(a).  
Similar band structures occur widely for the electrons 
in the context of kagom\'{e} metals that is under an active investigation recently~\cite{Ye_2018,Yin_2020}.  
With the introduction of a finite DM interaction, 
the highest magnon energy band develops a weak dispersion, 
and the Dirac band touchings of the lower two magnon bands are 
gapped out [see Fig.~\ref{kagomeMagnon}(b)]. 
Like the case for the pyrochlore ferromagnet, the magnons experience 
a fictitious gauge flux when they hop on the triangular plaquettes of the 
kagom\'{e} lattice. 

Since all the three magnon bands of the kagom\'{e} ferromagnet with the DM interaction
are now well separated and the wavefunctions of the magnons become complex, 
the Chern numbers are found to be non-trivial and turn out to be 
\begin{equation}
\text{Chern number}= +1, \,\, 0, \,\, -1, 
\end{equation}
for the lowest, the middle, and the highest magnon bands in Fig.~\ref{kagomeMagnon}(b), 
respectively. 
Thus, the magnons here can be probably well quoted as \emph{chiral} 
topological magnons. This immediately suggests the presence of the   
magnon thermal Hall effect (and the related orbital magnetization of the chiral magnons~\cite{Alahmed_2022}) 
in this system. Although the finite Chern numbers 
of the magnon bands indicate the presence of the magnon thermal 
Hall effect, the magnon thermal Hall effect is not really a direct measurement 
of the magnon Chern number according to the formula in Eq.~\eqref{k_xy_boson}.
Thus, even if the Chern number of the relevant magnon band is zero,  
as long as the magnon band has a Berry curvature distribution, 
one might still have the thermal Hall effects at finite temperatures. 
This is because the magnon states are thermally activated and the magnon states 
from the zero Chern number band
are not equally populated at a given finite temperature.  
In fact, the middle magnon band in Fig.~\ref{kagomeMagnon}(b) has zero Chern number, 
but the Berry curvature is not actually zero throughout the Brillouin zone,
and this magnon band could still contribute to the thermal Hall signal.

\begin{figure}[t] 
	\centering
	\includegraphics[width=8.6cm]{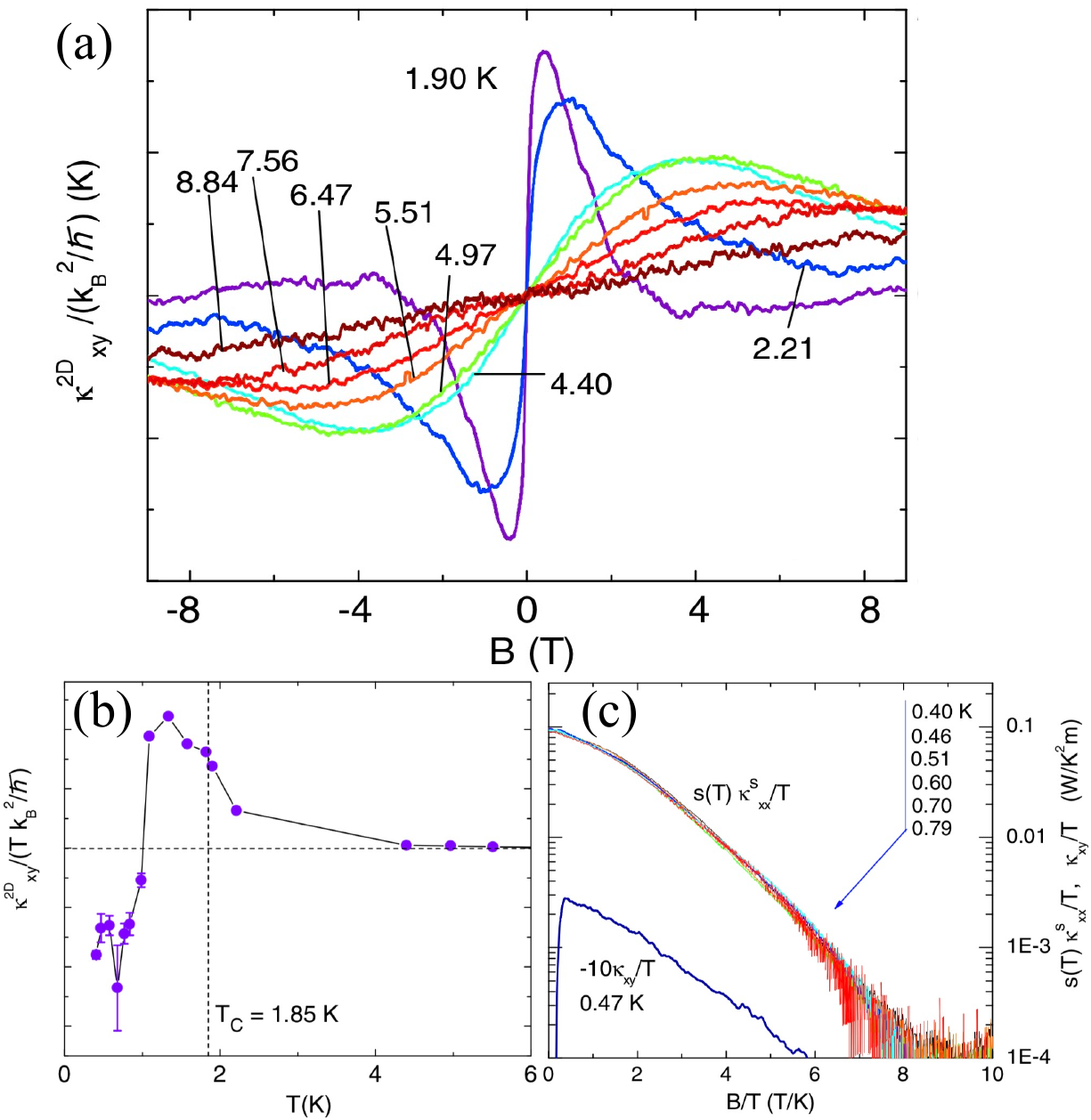}
	\caption{(a) Field dependence of the thermal Hall coefficient in Cu(1,3-bdc).
		(b) Temperature dependence of the quantity $\kappa_{xy}/TB$ in the zero field limit.
		(c) The correlation between the longitudinal and transversal thermal coefficients.
		Figures are reprinted from Ref.~\cite{Ong2015}.}
	\label{CuthermalHall}
\end{figure}

Experimentally, large thermal Hall signals were detected in the 
kagom\'{e} ferromagnet Cu(1,3-bdc) in the ferromagnetic regime 
and persist even up to the paramagnetic regime above the ordering 
temperature at ${T_c=1.8}$K~\cite{Ong2015,Chisnell2015}. 
While the thermal Hall signals were 
suppressed with increasing the magnon gap by the strong magnetic field, 
a non-monotonic dependence of the thermal Hall signal $\kappa_{xy}/T$ on the magnetic 
field was observed in the weak field regime, as shown in Fig.~\ref{CuthermalHall}(a). 
The non-monotonic dependence of $\kappa_{xy}/T$ on the temperature 
was also found as the temperature is varied. Moreover, an interesting
sign change of $\kappa_{xy}/T$ was further revealed under the variation 
of the temperature in the zero field limit [see Fig.~\ref{CuthermalHall}(b)].
The quantity plotted in Fig.~\ref{CuthermalHall}(c) is closely correlated with 
the growth of the longitudinal magnon thermal conductivity $\kappa^S_{xx}$  
below $T_c$, which implies that $\kappa_{xy}$ arises from the spin excitations.
The topological origin, namely the precise role of the magnon Berry curvature,
is examined in a more incisive way by comparing the experimental measurements 
with the theoretical results~\cite{Katsura2010,Lee2015}.
The compelling evidences drawn from the sign change feature of 
the thermal Hall conductivity and its intimate correlation 
with the Berry curvature distribution of the 
magnon bands preclude the phononic origin.
The behaviors of the thermal Hall signal in the ordered regime 
were well captured by the theoretical calculation within
the Holstein-Primakoff spin wave theory for the magnons.

Besides the fundamental mechanism of short-ranged DM interaction combined 
with the collinear ferromagnetic order, the magnon thermal Hall effects have been 
discussed in the magnetic thin films and magnonic crystals, where 
the magnetic dipolar interaction was brought into consideration. 
Like the DM interaction, the magnetic dipolar interaction also belongs 
to the anisotropic spin interactions. 
The magnon thermal Hall effect and the magnon Berry curvatures
in these magnetic systems with the dipolar interactions were shown to be present~\cite{Shindou2014,Shindou2013A,Shindou2013B}.
Shindou et al. used a linearized Landau-Lifshitz equation to 
establish a generic Bogoliubov de Gennes (BdG) Hamiltonian for the magnons.  
It was shown that the magnonic crystal with the dipolar interaction 
acquires a spin-wave bulk band with a nonzero Chern number,
and supports the topological chiral magnonic 
modes on the edges and thus the magnon thermal Hall effects. 
Other types of exchange interactions in the collinear ferromagnetic order 
could also result in the magnon Berry curvature and the corresponding magnon thermal Hall effect. 
For example, the Kitaev and off-diagonal symmetric exchanges 
(also known as pseudo-dipole interaction in the old literature) 
in the field-polarized collinear order could result in a magnon Chern 
insulator~\cite{PhysRevB.98.060405,PhysRevB.98.060404,PhysRevLett.126.147201,PhysRevB.103.174402}, 
which has been employed to partly understand the thermal Hall effects 
in Kitaev materials, and will be discussed in Sec.~\ref{sec6}. 
In addition to the collinear ferromagnets, the magnon thermal Hall effect 
could also emerge from the weak ferromagnets with the noncollinear 
ground state magnetizations, such as the skyrmion spin texture that arises
from the completion of ferromagnetism and DM interaction and supports 
a finite scalar spin chirality~\cite{PhysRevLett.122.057204,PhysRevB.87.024402,PhysRevResearch.4.043085}. 
Such a scalar spin chirality mechanism overlaps with the ones in the ordered antiferromagnets,
and we will give a discussion there.


\subsection{Thermal Hall effect in antiferromagnets}
\label{sec22}

In the above examples we have mainly discussed the magnon thermal Hall effect in the ferromagnets, 
which has been intensively studied both in theories and experiments. Meanwhile, the magnon 
thermal Hall effects in the antiferromagnetic Mott insulators are quite common as well. 
In this subsection we turn our attention to the magnon thermal Hall effects in the antiferromagnets  
with the dominant antiferromagnetic couplings. 
We first discuss the noncollinear and noncoplanar spin configurations  
with a non-vanishing scalar spin chirality, 
next we turn to the coplanar and collinear antiferromagnets
where the net ferromagnetic moment is absent.

\begin{figure}[htbp] 
	\centering
	\includegraphics[width=5cm]{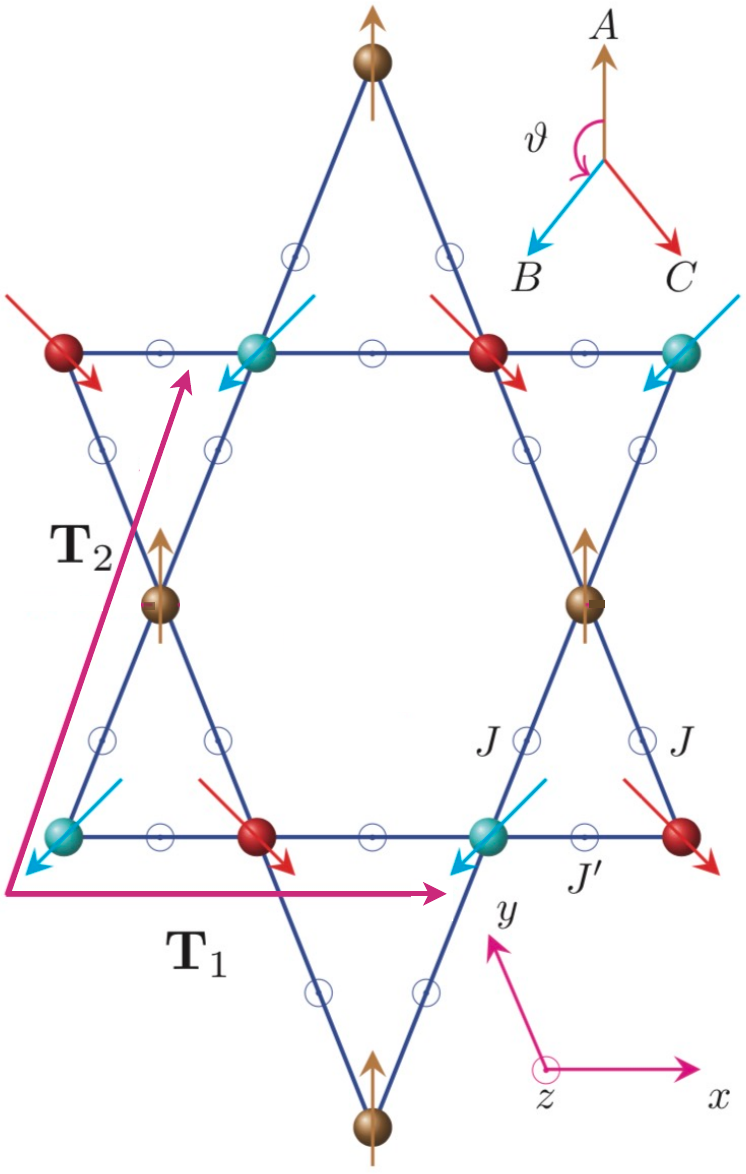}
	\caption{The canted noncollinear/coplanar magnetic order with the positive vector chirality 
	on the distorted kagom\'{e} lattice. 
	The out-of-plane DM interaction lies at the midpoints between two magnetic ions as indicated by small circles. 
	The spin triads are separated by an angle $\vartheta \ne 120^{\circ}$ 
	on each isosceles triangle with $J\ne J^\prime$ as shown in the top figure.
	Figures are reprinted partly from Ref.~\onlinecite{Owerre2017B}.}
	\label{Distorted_Kagome}
\end{figure}

\subsubsection{Ordered antiferromagnets with scalar spin chirality}
\label{sec221}

The scalar spin chirality in the ordered antiferromagnets provides an alternative route 
for the magnon Berry curvatures and the thermal Hall effects. 
Like the fictitious U(1) gauge flux generation for the magnons via the DM interaction 
in the collinear ferromagnet, the scalar spin chirality could manifest itself as an
effective U(1) gauge flux for the magnon hopping in the ordered antiferromagnets, 
which leads to nothing but finite Berry curvatures for the magnon bands
 and the magnon thermal Hall effects. 
It has been shown~\cite{Owerre2016C,Owerre2017,Owerre2017B,PhysRevB.99.054409,PhysRevB.98.094419,PhysRevLett.118.177201} 
that the scalar spin chirality arising from the non-coplanar chiral spin configuration could
yield both the magnon thermal Hall effects and the topological magnons.

As an example, one can consider an antiferromagnetic spin model   
similar to Eq.~\eqref{eqham} on a 2D distorted kagom\'{e} lattice 
with 
\begin{eqnarray}
	{\mathcal H} &=&  \sum_{\langle ij\rangle}  {
		J_{ij} {\bf S}_i \cdot {\bf S}_j
		+   { {\bf D}_{ij} \cdot ({\bf S}_i \times {\bf S}_j) } } 
	-\sum_i {\bf B}\cdot  {\bf S}_i .
\label{Ham_HDMZK}
\end{eqnarray}
The nearest-neighbor spins are coupled by the Heisenberg exchange interaction and the DM interaction.
The exchange couplings are anisotropic due to the lattice distortion.
Here, ${J_{ij}=J>0}$ for the diagonal bonds and ${J_{ij}=J' }$ for the horizontal bonds
with ${J'/J \neq 1}$ as illustrated in Fig.~\ref{Distorted_Kagome}.
The model in the absence of the DM interaction and the Zeeman coupling
is a typical model for frustrated quantum magnetism~\cite{PhysRevB.76.094421}
and can have a subextensive ground state degeneracy along each bond direction 
in the classical regime due to the weathervane modes for ${J'/J>1/2}$. 
Despite the frustration, the out-of-plane 
DM interaction, ${{\bf D}_{ij} = -|D_z| \hat{z}}$, 
could stabilize the coplanar magnetic orders like 
Fig.~\ref{Distorted_Kagome} in certain parameter regime. 
In the presence of the Zeeman coupling with ${{\bf B} = h\hat{z}}$,
due to the magnetic polarization, a finite scalar spin chirality
is induced on the triangular plaquette with
\begin{equation}
 \chi_{ijk} ={{\bf S}_i \cdot ({\bf S}_j \times {\bf S}_k)} \neq 0. 
\end{equation}
This quantity is the triple product of the three neighboring spin vectors on the sites ${i,j,k}$   
from the elementary triangular plaquettes on the lattices. 
The scalar spin chirality constitutes an important type of magnetic correlation 
in addition to the conventional magnetic orders defined by the linear spin operator,
and is further shown to play an important role in the thermal Hall effect 
in several types of spin liquids, which we will revisit and clarify in Sec.~\ref{sec4}.



With a non-vanishing scalar spin chirality, the magnons in the spin-wave Hamiltonian for 
the resulting noncoplanar state experience a fictitious U(1) gauge flux when they hop
on the kagom\'{e} lattice~\cite{Owerre2017B,PhysRevB.99.054409}. According to
Ref.~\cite{Owerre2017B}, the spin-wave Hamiltonian for the linear spin-wave theory
is given as 
\begin{eqnarray}
 {\cal H}_{\rm SW}&= & \sum_{\langle ij\rangle} \frac{1}{2} \big[ t_{ij} ( e^{-i\phi_{ij}} a^\dagger_i a_j^{} + { h.c.}) 
+t_{ij}^0 ( a^\dagger_i a_i^{} + a^\dagger_j a_j^{} ) 
\nonumber \\
&& + \lambda_{ij}  (a^\dagger_i a^\dagger_j + { h.c.})\big] + h\cos\phi \sum_i a^\dagger_i a_i^{},
\label{SWH}
\end{eqnarray}
where 
\begin{eqnarray}
&& t_{ij} = \sqrt{(G_{ij}^R)^2+(G_{ij}^M)^2} ,\\
&& \tan\phi_{ij} = G_{ij}^M/G_{ij}^R, 
\end{eqnarray}
with $G_{ij}^R=J_{ij}[ \cos\theta_{ij}+\sin^2\phi \sin^2(\theta_{ij}/2)] 
 + | D_z| \sin\theta_{ij}  (1 -\sin^2\phi /2 )$ ,
$ G_{ij}^M=\cos\phi (J_{ij} \sin\theta_{ij} - |D_z | \cos\theta_{ij})$.
Here, $t_{ij}^0$ and $\lambda_{ij}$ are unimportant couplings that 
do not have the phase coupling, and
$\theta_i$ and $\phi$ characterize the noncoplanar magnetic order
with ${{\bf S}_i = S (\sin \phi \cos \theta_i, \sin \phi \sin \theta_i, \cos \phi) }$. 
We have used the abbreviation ${\theta_{ij}=\theta_{i}-\theta_{j}}$. 
The emergent gauge flux experienced by the magnons on the triangular plaquette
is related to the solid angle subtended by three noncoplanar spins, i.e. 
the scalar spin chirality~\cite{Owerre2017B}.
Due to the real space Berry phase from the U(1) flux, 
the three magnon bands are now topological~\cite{PhysRevB.99.054409} 
and are characterized by the Chern numbers~\cite{Owerre2017B},
\begin{equation}
{{\cal C}_1=0}, \quad  {{\cal C}_2=-{\rm sgn}(\chi)},\quad  {{\cal C}_3=+{\rm sgn}(\chi)},
\end{equation}
where $\chi$ is the scalar spin chirality on the elementary triangular plaquette
of the kagom\'{e} unit cell.


In the above example, the 
 fictitious gauge flux experienced by the magnons
 arises from the ordered moments or interactions 
 via the scalar spin chirality that 
 drives the thermal Hall transports. 
 In this respect, it is identified that the scalar spin chirality as one important origin for the magnon thermal Hall effect. 
In fact, the relation between the scalar spin chirality and Hall transports have been studied in various itinerant magnets
especially those with the magnetic skyrmion lattice
 where the scalar spin chirality of the local magnetic moments from the skyrmion lattice serves as a fictitious 
gauge flux for the itinerant electrons and generates the so-called
topological Hall effects~\cite{Taguchi2001,PhysRevLett.98.057203,Machida2009}. 
The magnon version for the spin systems would then be naturally dubbed ``topological thermal Hall effect''. 

To generate the scalar spin chirality, the DM interaction seems to play an important role in the ordered antiferromagnets,
though frustrated spin interactions can sometimes play similar roles in stabilizing noncollinear spin configurations. 
For the antiferromagnets, the DM interaction can many times stabilize the coplanar magnetic orders
in the planar geometries or the frustrated systems with the planar plaquettes, 
and the coplanar spin texture can also be equally stabilized by other non-chiral anisotropic interactions, 
such as an easy-plane anisotropy that also has a spin-orbit coupling origin. 
The external field introduces an out-of-plane component for the magnetic ordering,
which then amounts to a noncoplanar chiral or conical-like spin configuration~\cite{Grohol2005}. The noncoplanar 
spin configuration is further related to the emergence of the scalar spin chirality.
The combined effect of coplanar spin orders and external magnetic field could yield 
both thermal Hall effect and topological magnons.

As for the physical applications, the magnetic skyrmion 
textures immediately provide a natural arena for the magnon thermal Hall effects in ordered antiferrromagnets.
The recently discovered skyrmion lattice system 
in the polar magnet GaV$_4$Se$_8$ shows the magnon thermal Hall effects and were
well interpreted as the fictitious U(1) gauge flux of the magnons due to the spin textures~\cite{PhysRevResearch.4.043085}.
A bit more sophisticated version is 
the frustrated diamond lattice antiferromagnet MnSc$_2$S$_4$~\cite{Gao_2020} where 
the magnon thermal Hall transport is observed and is attributed to the skyrmion textures~\cite{takeda2023emergent}.
Another prominent example of frustrated antiferromagnets 
with a planar geometry is the 2D kagom\'{e} lattice antiferromagnets with high spin moments,
such as the jarosites KCr$_3$(OH)$_6$(SO4)$_2$ and KFe$_3$(OH)$_6$(SO4)$_2$ 
where the magnetic orders were found
and the DM interaction is crucial to lift the ground state degeneracy and induces the coplanar 
spin state in the kagom\'{e} plane~\cite{BALLOU2003465}. The scalar spin chirality is 
immediately induced when the out-of-plane field is applied~\cite{Grohol_2005}. 
For the more quantum spin-1/2 Heisenberg model on the kagom\'{e} lattice, however,
 the ground state is still under debate~\cite{Ran2007,Yan_2011,Zhu_2019,Jiang_2019}. 
In fact, the DM interactions suppress the candidate spin liquid ground state 
up to a critical value and are responsible for the formation of the ${{\bs q}=0}$ 
coplanar or noncollinear magnetic order~\cite{Lacroix2002,Grohol2005,PhysRevLett.96.247201}.
While the candidate spin-1/2 kagom\'{e} antiferromagnets, the volborthite and the 
kapellasite compounds both show the signatures of spin liquid physics 
and the interesting behaviors of the thermal Hall effects, 
the precise nature of the driving force is still unclear at this stage. 
We defer the discussion to the section of the thermal Hall effects in the spin liquids.

\subsubsection{Symmetry arguments for magnon thermal Hall effects}
\label{sec222}

At this stage, we have discussed two well-known mechanisms for the magnon thermal Hall effects,
i) ferromagnets with the DM interaction, dipolar or other related interactions 
(including the weak ferromagnetism with the noncollinear spin texture); 
ii) frustrated antiferromagnets with the field-induced scalar spin chirality. 
For both cases, the ferromagnetic moments and/or external magnetic fields were present. 
 To have a better theoretical understanding between the magnetic structure and 
the magnon thermal Hall effect, 
Mook et al. theoretically provided a general symmetry argument 
for the magnon thermal Hall effect~\cite{Mook2019}. 
They argued that the necessary conditions for the magnon thermal Hall effects are
the effective time-reversal symmetry breaking and 
the magnetic point group compatible with ferromagnetism, instead of simply
having ferromagnetic moments and/or external magnetic fields. 
These necessary conditions for a finite thermal Hall conductivity 
were discussed in terms of that rigorous symmetry analyses
 that is in line with the development of the electronic anomalous Hall effect 
in the noncollinear antiferromagnets~\cite{PhysRevB.92.155138,PhysRevResearch.3.033156,PhysRevLett.93.096806,PhysRevLett.83.3737}.
Although it is not directly about the actual physical mechanism of the thermal Hall effect,
the symmetry-based necessary condition provides a useful thumb 
of rule guidance and insights for the magnon thermal Hall effect in an ordered magnet.
Here, we follow Ref.~\cite{Mook2019} to explain these necessary conditions for the magnon
thermal Hall effects.

We start with the effective time-reversal symmetry breaking.  
The coplanar antiferromagnet often has an additional ``effective time-reversal symmetry'' 
${\cal T}^\prime$ under the operation of which the spin configuration is mapped into itself.
By ``effective'' we refer to the simultaneous operations involving the time reversal 
and the spatial symmetry operations. For the coplanar antiferromagnet, 
the effective time reversal is a combination of the time reversal ${\cal T}$
and the $\pi$-rotation of the spins about the axis normal to the ordered plane.
If the ${\cal T}^\prime$ symmetry is preserved, the Berry curvature of the magnon bands
is an odd function with respect to the momentum, i.e.
\begin{eqnarray}
\Omega_{n,{\bf k}} = - \Omega_{n,-{\bf k}} ,
\end{eqnarray}
which renders the thermal Hall response vanishing as dictated by Eq.~\eqref{k_xy_boson}. 
Thus, the breaking of the effective time-reversal symmetry is one of the mandatory conditions
for the magnon thermal Hall effect in the coplanar antiferromagnets.
The ${\cal T}^\prime$-symmetry breaking is then demonstrated 
for different magnetic ordered states on a 2D kagom\'{e} lattice~\cite{Mook2019}.
Having this in mind, one can return to the examples of the previous sections and 
examine the ${\cal T}^\prime$ breaking.

To illustrate the effective time-reversal symmetry breaking, one can consider
 a conventional spin Hamiltonian with the Heisenberg and DM interactions on a kagom\'{e} lattice, 
\begin{equation}
\mc{H} = \frac{1}{2}  \sum_{\langle ij\rangle}[ -J {\bf S}_i \cdot {\bf S}_j + {\bf D}_{ij} \cdot  ({\bf S}_i \times {\bf S}_j)].
\label{AFM_cop}
\end{equation}
This model captures both the kagom\'{e} lattice ferromagnet and antiferromagnet in the previous sections,
as well as the coplanar antiferromagnet. 
Here the DM vector on the bond $ij$ has the in-plane (along the bond) and the out-of-plane components 
${\bf D}_{ij}= D_\parallel \hat{n}_{ij} + D_z \hat{z}$
with $D_\parallel$ and $D_z$ being the respective DM interaction strength. 
The out-of-plane component is allowed by symmetry, whereas the in-plane components survive when the 2D plane is not a mirror plane. 

Let us first consider the mirror reflection symmetric case with ${D_\parallel =0}$
and the DM term is simply given by $ D_z (S_i^x S_j^y - S_i^y S_j^x)$.
A collinear ferromagnet is realized with ${J>0}$ and even with a moderate positive $D_z$. 
The magnetization plane can be chosen arbitrarily since the ferromagnetic moments are collinear.
The ${\cal T}^\prime$ symmetry is broken on the Hamiltonian level 
since the $\pi$-rotation along any in-plane axis can not map the DM term into itself. 
This could lead to a finite thermal Hall effect is expected for the magnons, 
which is the scenario adopted for 2D kagom\'{e} ferromagnet Cu(1,3-bdc) in Sec.~\ref{sec211}. 
On the other hand, non-zero ${D_\parallel \ne 0}$ together with
an antiferromagnetic interaction ${J<0}$ leads to a non-coplanar magnetization configuration.
${D_z>0}$ corresponds to a positive scalar spin chirality.
Due to the out-of-plane canting,
the actual time reversed spins can not be mapped to itself by rotating along any axis.
Therefore, the effective time-reversal symmetry is also broken and the magnon thermal Hall effect is possible.
This is the mechanism that we discussed for the scalar spin chirality induced thermal Hall effect
 in Sec.~\ref{sec221}. Finally, let us discuss the ${\cal T}^\prime$ symmetry breaking 
 for the coplanar antiferromagnet. The coplanar antiferromagnet can be realized 
 by the above Hamiltonian with ${J<0, D_z<0}$ and a moderate out-of-plane canting 
 given by $D_\parallel$. It was pointed out that the coplanar ground state survives from 
 a finite $D_\parallel$ below a critical value~\cite{Lacroix2002}.
The mapping of the actual time reversed spins is facilitated by $\pi$-rotation along
the perpendicular $\hat{z}$-axis which is denoted as ${\cal R}^z_\pi$.
The spin vector is transformed under the rotation as: 
${\cal R}^z_\pi (S_i^x,S_i^y,S_i^z) = (-S_i^x,-S_i^y,S_i^z)$
which leaves the out-of-plane component of the DM term intact;
While, the in-plane components $S_i^y S_j^z-S_i^z S_j^y$ and $S_i^z S_j^x-S_i^x S_j^z$
acquire minus signs since they depend linearly on $S_i^z$.
The ${\cal T}^\prime$ symmetry is broken with a finite $D_\parallel$
when the 2D kagom\'{e} plane is not a mirror plane, and thermal Hall effect is expected~\cite{Mook2019}.

In addition to the effective time-reversal symmetry, the magneto-spatial symmetries ($X$) of 
the magnetic point group can still diminish the thermal Hall effect by relating the Berry curvature
in the reciprocal space if 
\begin{eqnarray}
{\Omega_{n,{\bf k}} = - \Omega_{n,X{\bf k}}}.
\end{eqnarray} 
Thus,
the additional condition to have the thermal Hall current is that 
the magnetic point group of the underlying crystal must be compatible with ferromagnetism~\cite{Mook2019}.
This condition 
follows directly from the understanding of the magneto-spatial symmetries in the electronic anomalous Hall effect~\cite{PhysRevB.95.094406}.
It is important to note that the magnetization vector ${{\bf M}=(M_x,M_y,M_z)}$ 
and the (transverse) heat conductivity vector $\boldsymbol{\kappa}=(\kappa_{yz},\kappa_{zx},\kappa_{xy})$
transform in the same way under the magnetic point group symmetry.
This does not mean that a finite and large ferromagnetic moment is secured as in the collinear ferromagnet cases. 
In fact, certain candidate states with an inverse chiral texture in the coplanar kagom\'{e} antiferromagnets
meet these necessary conditions for a magnon thermal Hall effect~\cite{Mook2019}
where the ferromagnetic moment could be quite small.

\section{Disordered states}
\label{sec3}

For the ordered states in the previous section, things are more-or-less simple. Since the system orders magnetically, 
the magnon excitations and their spectra are often quite clear, and the formalism for the thermal Hall effect is then straightforward. 
One can always sit 
 down and perform the routine calculation for the magnons to obtain the thermal Hall conductance, 
 if the more qualitative symmetry type of arguments are not the necessity. 
 For the disordered states especially quantum spin liquids, however, the origins of the thermal Hall effect are not quite clear by comparison. 
In this section, we turn to discuss the thermal Hall effects in the disordered states. 
As we have mentioned in Sec.~\ref{sec1}, our choices 
of the disordered spin states include the valence bond singlet states, 
the plaquette singlet states, and the quantum spin liquids. 
The first two are actually simple product states and do not have 
the long-range quantum entanglement that is present in the quantum 
spin liquids~\cite{Savary2016}. The elementary excitations of the valence bond singlet 
state for the dimerized magnets are simple triplons~\cite{Giamarchi_2008}. They arise from
the breaking of the spin singlets of the valence bond singlet. Locally,
they correspond to the spin triplet excitations of the valence bonds.
They gain the dispersion from the inter-valence-bond couplings.  
Like the magnons, these triplons are bosonic and carry the interger 
spin quantum numbers.

\begin{figure*}[t] 
	\centering
	\includegraphics[width=17cm]{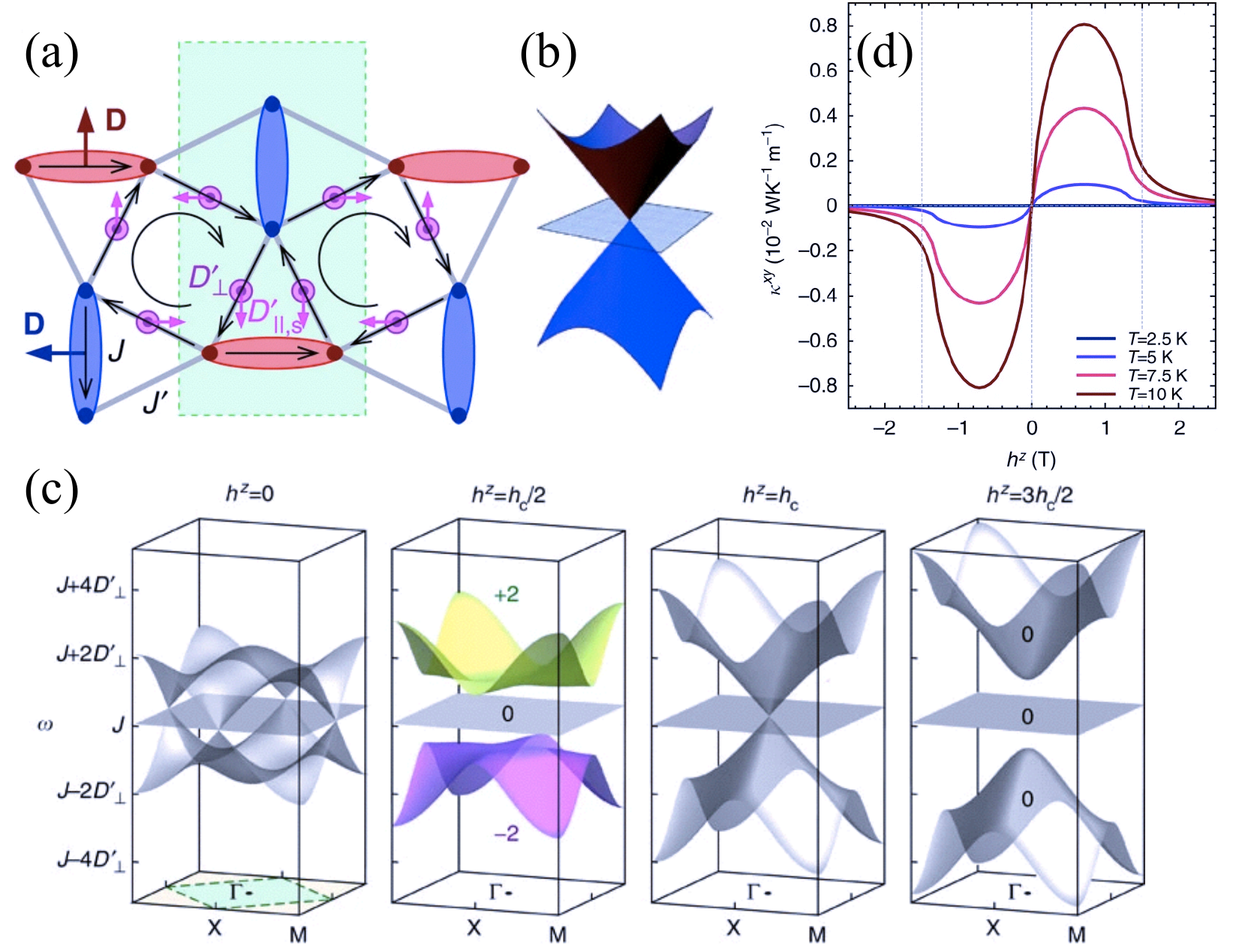}
	\caption{ 	
	     (a) The SrCu$_2$(BO$_3$)$_2$ Sutherland-Shastry lattice with the Heisenberg and DM interactions;
		(b) Spin-$1$ Dirac cone with three bands touching;
		(c) Evolution of the triplon bands and Chern numbers upon tuning the magnetic field;
		(d) Thermal Hall conductivity versus the external magnetic field at different temperatures.
		Figure is reprinted from Ref.~\onlinecite{Romhnyi2015}.}
	\label{triplon_fig}
\end{figure*}

The plaquette singlet states are quite similar to the valence bond singlet states, 
except that the spin singlets are formed on the multiple-site plaquettes rather 
than the two-site bonds~\cite{PhysRevB.54.9007,PhysRevLett.78.2216,Zayed_2017}. 
The properties of the magnetic excitations, however, 
have not been carefully studied before. The underlying theoretical framework 
would be quite analogous to the one for the valence bond solid states. 
One essentially needs to rearrange the local physical Hilbert space
of the plaquette and applies the flavor wave theory by defining
the bosonic flavor operators on the eigenstates of the local 
plaquette. The bosonic flavor that corresponds to the plaquette 
singlet is condensed, and the remaining bosonic flavors then 
turn into the elementary excitations and gain the dispersion 
from the inter-plaquette interactions. Here we do not expand 
the discussion about the thermal Hall effects for the plaquette 
singlet states.

\subsection{Triplon thermal Hall effect in dimerized magnets}

The valence bond solid state and the associate triplon excitations
have been discovered in an archetypal dimerized quantum magnet 
SrCu$_2$(BO$_3$)$_2$~\cite{Miyahara1999}. 
The material comprises layers of strongly interacting ${S =1/2}$ local 
moments from the Cu$^{2+}$ ions that form a lattice known as 
the Shastry-Sutherland lattice (see Fig.~\ref{triplon_fig}(a))~\cite{Kageyama1999,SS1981}.
The Heisenberg exchange interaction on the colored bonds is 
much stronger than the uncolored ones. As a consequence, 
the ground state of the system at the lowest order approximation is the simple
dimer covering of the spin singlets on the colored bonds. 
The low-energy magnetic excitations correspond to breaking 
the spin singlet into the spin triplets and are thus  
the triplons~\cite{Sachdev1990}. 
The triplons are defined on the spin dimers that can be considered 
as the supersites. The triplons gain the dispersion 
from the inter-dimer interaction via the exchange interactions on 
the weak bonds. Mathematically, this can be formulated from the bond operator 
method that replaces the spin singlet and triplets with the bosonic operators via 
the state-operator correspondence.

If the compound SrCu$_2$(BO$_3$)$_2$ can be described by the
Heisenberg model on the Shastry-Sutherland lattice, 
the triplons would have three-fold degenerate bands due to 
the SU(2) symmetry~\cite{Kageyama1999,Gaulin2004}.
In reality, a weak splitting of the bands has been detected 
and is attributed to the weak antisymmetric DM interaction
that breaks the spin rotational symmetry explicitly~\cite{Cepas2001,Cheng2007,Romhanyi2011,Romhnyi2015}.
The proposed spin model is then given as (see Fig.~\ref{triplon_fig}(a))
\begin{eqnarray}
\mc{H} &= & J  \sum_{\langle ij \rangle } {\bf S}_i \cdot {\bf S}_j + J' \sum_{\langle\langle ij \rangle\rangle }  {\bf S}_i \cdot {\bf S}_j - g_z h^z\sum_i {S}_i^z
\nonumber \\
&& + \sum_{\langle ij \rangle } {{\bf D}_{ij} \cdot ({\bf S}_i \times {\bf S}_j) }+ \sum_{\langle\langle ij \rangle\rangle }  {{\bf D}'_{ij} \cdot ({\bf S}_i \times {\bf S}_j)},
\end{eqnarray}
where $J$ is the strength of the exchange coupling on intra-dimer bonds, 
and the intra-dimer DM coupling $D$ is allowed by symmetry below
a structural phase transition at $T\sim 395$K. On the inter-dimer bonds, 
$J'$ and $D'$ are the exchange and DM couplings, respectively. 
In the bond operator treatment and the triplon analysis, one introduces the 
bond operators for the dimer Hilbert space with
\begin{eqnarray}
s^\dagger { |0\rangle}  = { |s\rangle} &=& \frac{1}{\sqrt{2}} ({ | {\uparrow \downarrow }\rangle} - {| {\downarrow \uparrow }\rangle }) ,
\label{eqsinglet}
\\
t_x^\dagger  { |0\rangle}  ={ | t_x \rangle} &=& \frac{i}{\sqrt{2}} ({ |{ \uparrow \uparrow }\rangle} -{ | {\downarrow \downarrow }\rangle}) ,
\label{eqtx}
\\
t_y^\dagger { |0\rangle}  = { | t_y \rangle} &=& \frac{1}{\sqrt{2}} ({ | {\uparrow \uparrow }\rangle }+ {| {\downarrow \downarrow }\rangle} ) ,
\label{eqty}
\\
t_z^\dagger  { |0\rangle}  = { | t_z \rangle} &=&  \frac{-i}{\sqrt{2}} ({ |{ \uparrow \downarrow} \rangle} +{ | {\downarrow \uparrow} \rangle }) ,
\label{eqtz}
\end{eqnarray}
where $s^\dagger, t_x^\dagger, t_y^\dagger, t_z^\dagger$ 
are the boson creation operators for the spin singlet and the spin triplets. In the simple valence bond solids formed
by these singlets on the strong bonds, the singlet boson $s$ is simply condensed, and the triplet bosons (triplons) are
the magnetic excitations. In the presence of the DM interactions that break
the spin-rotational symmetry, the singlets would be weakly hybridized with the triplets with the renormalized
triplons $\tilde{t}_{\mu}$ ($\mu=x,y,z$)~\cite{Romhnyi2015}. 
For the same reason,
the hopping of these renormalized triplons between the spin dimers is ``twisted'' by the antisymmetric DM interactions that
are responsible for the complex hopping amplitudes of 
the triplons, and the triplon Hamiltonian is given as  
\begin{equation}
{\mathcal H}_{\text{triplon}} =\sum_{\boldsymbol k} \sum_{\mu,\nu=x,y,z} \tilde{t}^\dagger_{\mu,{\boldsymbol k}} M_{\mu\nu} ({\boldsymbol k}) \tilde{t}_{\nu,{\boldsymbol k}},
\end{equation}
where $M({\boldsymbol k})$ is given as
\begin{eqnarray}
M({\boldsymbol k}) =\left[
\begin{array}{ccc}
J & ih^z g_z + i D'_{\perp} c_{xy} & \tilde{D}_{\parallel} s_y 
\vspace{2mm}
\\
{- ih^z g_z - i D'_{\perp} c_{xy}} & J & {- \tilde{D}_{\parallel} s_x} 
\vspace{2mm}
\\
\tilde{D}_{\parallel} s_y &  {- \tilde{D}_{\parallel} s_x}  & J 
\end{array}
\right],
\nonumber\\
\end{eqnarray}
where ${s_x=\sin k_x}, {s_y=\sin k_y}, {c_{xy} = \cos k_x + \cos k_y }$, and ${\tilde{D}_{\parallel}={D'_{\parallel,s} - |{\boldsymbol D} |J'/(2J)}}$
is a renormalized inter-dimer DM interaction. 
Remarkably, since $M({\boldsymbol k})$ is a $3\times 3$ hermitian matrix, 
the triplon bands can then be effectively described by a spin-$1$ spinor 
coupled to a pseduo-magnetic field in the momentum space due to the DM interaction,
and this defines a mapping from the momentum space to the effective spinor space. 
The momentum space pseduo-magnetic field (or vector) contains all the information
about the triplon band structure, and its skrymion texture gives rise to
the triplon band structure topology~\cite{Romhnyi2015,PhysRevB.95.195137}. 
The triplon dispersion is Dirac cone-like with three bands touching as demonstrated 
in Fig.~\ref{triplon_fig}(b). The DM interactions renders the triplon with a non-zero 
Berry phase when it hops around the closed paths~\cite{McClarty2017},
and this real space Berry phase is transmitted to the band structure topology in the momentum space
for the triplons.  

As the triplons are magnetic, applying a small magnetic field could further split the Dirac band touching 
of the triplon bands. The splitted triplon band acquires a 
nontrivial topological invariant, namely the Chern number.
The combined effect of the DM interaction and the weak magnetic field endows 
the lowest band with a finite Chern number as illustrated in Fig.~\ref{triplon_fig}(c).
As these triplon bands are gapped bosonic excitations at finite energies, 
the appearance and disappearance of the non-trivial
Chern numbers are not really topological phase transitions. 
Upon further increasing the magnetic field beyond a critical value, 
a trivial gap opens with all three Chern numbers being zero.

The physical property of the triplons is quite analogous to the 
magnons in the ordered magnets, and one can compare 
the bond operator expansion and the triplon analysis with the 
Holstein-Primakoff transformation and the spin-wave analysis. 
As the triplons carry energy but no electrical charge, 
the thermal Hall effect of the triplons is expected upon applying a temperature gradient
once the Berry curvature of the triplon bands exists. 
At the low temperatures, the triplon bands are weakly populated 
rendering the triplon-triplon interactions unimportant, and the free triplon
band theory should be a good approximation.

In Fig.~\ref{triplon_fig}(d), the thermal Hall conductivity as a function of 
the external magnetic field was calculated following the single-particle 
treatment reviewed in Sec.~\ref{sec1}. 
Owing to the bosonic nature of 
the triplon excitations, the triplon quasiparticle population is affected by the 
temperature as well as the applied magnetic field. This renders the 
controllability of the thermal Hall signals in experiments.
Theoretically, a two-dimer model for the frustrated Shastry-Sutherland model with 
the sublattice pseudo-spin is constructed~\cite{Sun2021}.
Varying the angle of the magnetic field, the authors demonstrate the tunable 
experimental signatures for the thermal Hall effect, including a sign changing feature.
On the experimental side, the triplon thermal Hall effect is still out of reach~\cite{Cairns2020}. 
This is possibly due to the triplon-triplon interactions playing a more significant role 
than anticipated in the temperature range under investigation.

\subsection{Generalized ``triplons'' and their thermal Hall transports}

We have already mentioned that the 
bosonic magnetic excitations from the plaquette singlet states 
are very much like the triplons for the dimerized states, except that
 one needs to apply the generalized flavor-wave theory to study them~\cite{PhysRevB.60.6584}.
 In fact, many uprising quantum magnets 
 can be viewed from the perspectives of generalized ``triplon'' excitations
 with respect to the generalized ``singlet'' ground states,
 where the ``singlets'' can appear in different forms. 
  These singlets can be the simple on-site singlet state due to the single-ion spin anisotropy~\cite{PhysRevLett.109.016402,PhysRevResearch.2.033260,PhysRevB.96.020412}, 
 can be the local cluster singlet in the cluster magnets or molecular magnets~\cite{PhysRevB.93.245134,PhysRevLett.113.197202,PhysRevB.97.035124,PhysRevLett.93.126403}, 
 and can be 
so-called the spin-orbital singlet (SOS) that appears in many literature. 
 If the Hamiltonian for the generalized ``triplons'' is ``twisted'' by the complex interactions with lower symmetries,
 one could obtain the Berry curvatures for the generalized ``triplons'' and thus
 the thermal Hall effects of the generalized ``triplons''.
We here give a brief description of the SOS and the magnetic excitations. 
The SOS can appear in different forms, and we describe three of them in the following.

\begin{figure}[t] 
	\centering
	\includegraphics[width=8.5cm]{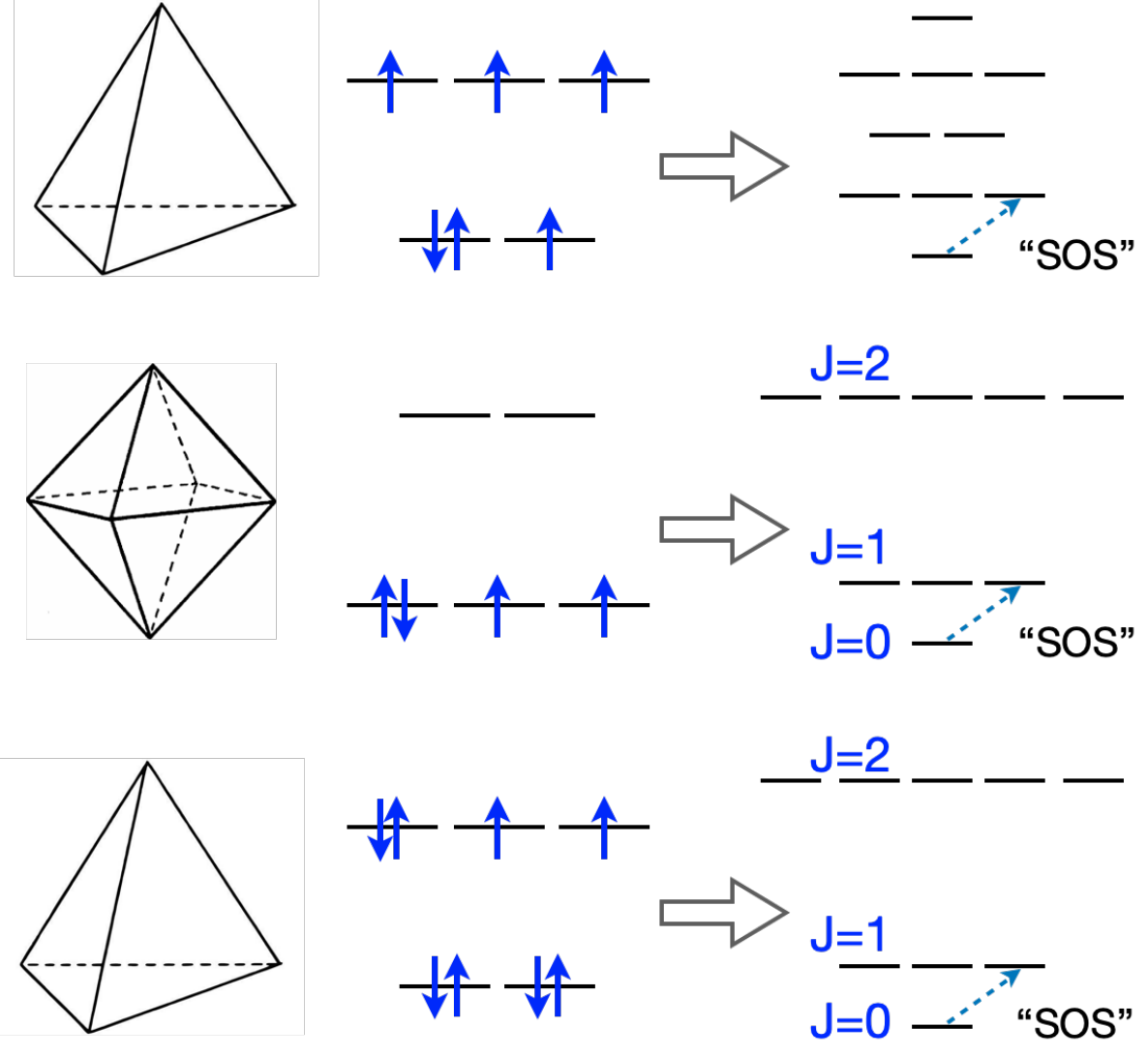}
	\caption{ 	
	    The electron configurations for different spin orbital singlets (``SOS'')
	    and their local triplons. Upper: $3d^6$ in the tetrahedral environment.
	    Middle: $4d^4$/$5d^4$ in the octahedral environment.
	    Lower: $3d^8$ in the tetrahedral environment. 
	   The dashed arrows refer to the local ``triplon''-like excitations.
	   The big arrow indicates the role of the spin-orbit coupling. }
	\label{singlets}
\end{figure}

1) The first one is the SOS with the active $e_g$ orbitals that is relevant for the Fe$^{2+}$ ion 
with the $3d^6$ electron configuration in the tetrahedral crystal field environment. Here, the $e_g$ doublets
are lower in energy than the $t_{2g}$ triplets and are partially filled with the $3d^6$ electrons (see Fig.~\ref{singlets}). 
The spin-orbit coupling is quenched at the linear order and becomes active in the second order.
The single-ion ground state is an SOS between the total spin ${S=2}$ and the active $e_g$ orbitals,
and this SOS state was found to be particularly relevant for the diamond lattice 
antiferromagnet FeSc$_2$S$_4$~\cite{PhysRevLett.102.096406,PhysRevB.80.224409}. 

2) The second one is the SOS with the active $t_{2g}$ orbitals that is relevant for the 
$4d^4$/$5d^4$ electron configurations in the octahedral crystal field environment~\cite{PhysRevLett.111.197201,PhysRevB.84.094420}. 
Here, the $t_{2g}$ triplets
are lower in energy than the $e_{g}$ doublets and are partially filled with the $4d^4$/$5d^4$ electrons (see Fig.~\ref{singlets}). 
Due to the partially filled $t_{2g}$ orbitals, the spin-orbit coupling is active at the linear order. 
The single-ion ground state is an SOS between the total spin ${S=1}$ and the active $t_{2g}$ orbitals. 
This SOS was suggested to be relevant for the Ru$^{4+}$ ion in the square lattice antiferromagnet 
Ca$_2$RuO$_4$~\cite{PhysRevLett.111.197201}, the Ru-based and Mo-based pyrochlore magnets 
A$_2$Ru$_2$O$_7$ and A$_2$Mo$_2$O$_7$~\cite{PhysRevB.98.045109}, 
and other $5d$ spin-orbit-coupled magnets such as the Os$^{4+}$ ion in 
Y$_2$Os$_2$O$_7$~\cite{PhysRevB.93.134426}. A closely related SOS 
is the tetrahedral Ni$^{2+}$ ion with $3d^8$ electron 
where the lower $e_g$ doublets are fully filled and the upper $t_{2g}$ is partially filled with 4 electrons (see Fig.~\ref{singlets}). 
The SOS is also expected here and is likely to be relevant for the diamond lattice antiferromagnet  
NiRh$_2$O$_4$ and the Ni$_2$Mo$_3$O$_8$~\cite{PhysRevB.100.045103,PhysRevMaterials.3.014410}. 

3) The third one is the SOS in the rare-earth magnets where the combination 
of the spin-orbit coupling and the crystal field generates a local SOS~\cite{PhysRevB.99.224407,PhysRevResearch.2.043013,s0217979224500401}. 
This is quite 
common in the rare-earth magnets with an even number of $4f$ electrons and hence
an integer $J$ total moment. If the crystal has a relative low symmetry, all the point group
representations are singlets, and the SOS is unavoidable~\cite{PhysRevResearch.1.033141}. 
In fact, even if the lowest crystal field state 
is not a singlet state, as long as the excited states are not very high in energy, these local excited states
would be involved in the low-temperature magnetic properties~\cite{PhysRevB.99.224407}.

The ``triplon''-like excitations of the above three cases are the magnetic excitations from 
the local SOS to the spin-orbital excited states (see Fig.~\ref{singlets}), and acquire the dispersion from 
the spin-orbital superexchange interaction. A straightforward extension of the bond operator formulation
in Eqs.~\eqref{eqsinglet}, \eqref{eqtx}, \eqref{eqty}, \eqref{eqtz}
can be established for these spin-orbital states.
Due to much complicated local states than the 
dimerized magnets, these ``triplon''-like excitations are not necessarily the singlet-triplet excitations,
and sometimes referred as the spin-orbital excitons in various contexts~\cite{PhysRevLett.111.197201}. 
Due to the orbital involvement and the spin-orbit coupling, the exchange coupling has a relatively low symmetry,
and the effective ``triplon'' hopping can be quite ``twisted'' and often {\sl much more} ``twisted''
than the simple DM interaction~\cite{PhysRevB.82.174440,PhysRevB.84.094420}. When the time-reversal symmetry is broken
either by the external magnetic field or by the ``triplon'' condensation, the resulting magnetic excitations  
could support nontrivial Berry curvature distributions and thermal Hall effects. 
The study of the topological or Berry curvature properties and the thermal Hall transports of these generalized ``triplons'' 
is still in the infancy stage, and there are not many works in this direction yet~\cite{PhysRevLett.122.177201}. 
These systems including these cluster or molecular magnets generally have a much larger local Hilbert space and 
allow more possibilities for unconventional interactions and physical models~\cite{chen2021mott,Li_2022,Balents2014}. 
We expect that the candidate systems could provide interesting and rich phenomena 
for both topological structures and thermal Hall effects.

\subsection{Overview of thermal Hall effects in spin liquids}

We now switch to quantum spin liquid that is a big subject on its
own~\cite{Savary2016,Knolle2019}. 
In the following, we sketch the basic properties about the 
thermal Hall transports for different spin liquids. 
We start from the $\mathbb{Z}_2$ spin liquid. 
Here $\mathbb{Z}_2$ spin liquid merely means that the gauge 
sector is $\mathbb{Z}_2$ and does not really mean that the system 
is fully gapped. The spinon matter sector of the $\mathbb{Z}_2$ 
spin liquids can be gapped or gapless. As the spinon is coupled to 
the external magnetic field and its band structure can be made topological by the
external magnetic field, 
the thermal Hall effect under the field is then controlled by the spinons. The physics then
is not very much different from the anomalous Hall effect of the electron if one only focuses on the Zeeman
effect of the magnetic field. One could readily imagine a Dirac spectrum of spinons 
that become topological under the perturbation of external magnetic fields. 
The gauge flux sector of the $\mathbb{Z}_2$ spin liquid is fully gapped 
so that it cannot be strongly influenced by the weak external magnetic fields. 
Nevertheless, there still exist
some cases that the visons can acquire a nontrivial fractional flux through the small sub-plaquette of the system 
while preserving the 0 or $\pi$ flux through the unit-cell plaquette as demanded by the 
vison projective symmetry group~\cite{PhysRevB.107.045114,PhysRevX.12.041004,song2022translationenriched}. 
Under these special circumstances, the vison can acquire the topological bands with nontrivial
Chern numbers and generate a thermal Hall effect.  
For both gapped or gapless spinons 
in the $\mathbb{Z}_2$ spin liquids, we mostly think about the spinon thermal Hall transports,
and the spinon thermal Hall effect 
is very much analogous to the Hall effect of the band electrons and 
is simply accounted by the Berry curvature properties of the spinons. 
For the chiral spin liquid whose effective theory is a Chern-Simon 
field theory~\cite{Zee1989}, the system is fully gapped in the bulk but  
supports a chiral gapless edge state that automatically generates the 
quantized thermal Hall effects. In fact, this quantized thermal Hall effect 
can be an important and characteristic diagnosis of the chiral spin liquid 
phase. 


For the U(1) spin liquids, the mechanism for the thermal Hall effect is a bit more
complicated than the $\mathbb{Z}_2$ spin liquids and the chiral spin liquids
due to the richer varieties of excitations and the continuous U(1) gauge structure. 
In 2D, the U(1) spin liquids with gapped matters cannot exist due to the proliferation
of the instanton events that leads to the confinement~\cite{Polyakov1977}. 
It is suggested that, 
the gapless spinon matters could suppress the instanton events and support 
a deconfined spin liquid phase~\cite{Hermele2004,SSLee2008}. 
As the gapless bosonic spinons would immediately
condense and generate the magnetic order~\cite{PhysRevB.45.12377}, the gapless spinons have to be 
fermionic. For the 2D U(1) spin liquids, the gapless spinon matter can either 
have a spinon Fermi surface or a Dirac spectrum. If we only consider the direct
coupling between the spinon matter and the external magnetic field, 
then the mechanism would be quite similar to the discussion about the 
$\mathbb{Z}_2$ spin liquids above, and this would be the end of the story. 
Nevertheless, the U(1) spin liquids provide more interesting aspects. 
As the gauge sector here is continuous, the magnetic field could modify 
the gauge sector continuously, which then twists the motion 
of the spinons and generates the thermal Hall effect of the spinons. 
For the 2D U(1) spin liquids, the U(1) gauge flux is related to the scalar spin chirality, 
${\bf S}_i \cdot ({\bf S}_j \times {\bf S}_k)$~\cite{Wen1989B,LeeNagaosa1992,LeeNagaosa2013}. 
Since the physical degrees of freedom in a magnetic Mott insulator are spins
and their coupling to the external magnetic field is the simple Zeeman coupling,
thus the coupling to the scalar spin chirality is usually quite weak. 
In Sec.~\ref{sec4}, we explain two microscopic mechanisms for
the coupling to the scalar spin chirality and the corresponding thermal Hall
effects. One is via the strong charge fluctuations in the weak Mott regime.
The other is from the antisymmetric DM interaction
and works for both weak and strong Mott regimes. Moreover,
both mechanisms are expected to apply to 
the U(1) spin liquids with the gapless fermionic 
spinons in both 2D and 3D.

For the U(1) spin liquids in 3D, as the U(1) lattice gauge theory can be 
stabilized on its own, there is no constraint for the spinon matter sector 
to be gapless~\cite{Senthil2002,Hermele2004B}. 
For the 3D U(1) spin liquids, some of the above discussion 
about the 2D U(1) spin liquids could be carried over to here.  
The way that the external magnetic field generates the internal U(1)
gauge flux is strongly tied to the relationship between the emergent U(1) 
gauge variables and the microscopic spin variables. 
In the context of the 2D U(1) spin liquids that are mentioned above, the 
internal U(1) gauge flux is the scalar spin chirality. For the 3D U(1) spin liquids,
we can have other possibilities. One of the most popular 3D U(1) spin liquids,  
that is also most experimentally relevant, is the pyrochlore spin ice U(1) spin liquid~\cite{Savary2016,Gingras2014,Knolle2019}. 
Over there, the emergent electric field of the U(1) gauge field 
is most often the Ising component of the local spin moment. Quite remarkably, 
the linear Zeeman coupling is able to generate the so-called dual U(1) 
gauge flux and then twists the motion of the ``magnetic monopoles''
that carry the dual U(1) gauge charge. This would immediately generate 
the thermal Hall effects for the ``magnetic monopoles''.
The bosonic spinons may also contribute to the thermal Hall signal
once the high order effect on the direct U(1) gauge flux is considered. 
The thermal Hall effect of the pyrochlore spin ice U(1) spin liquid 
is discussed in details in Sec.~\ref{sec5}.


Here we turn to Kitaev spin liquids and Kitaev materials. 
In last decade or so, this topic has grown into a rather large field~\cite{Trebst2022,Hermanns2018}. 
One of the central questions in this field is whether any 
of the Kitaev materials realizes the Kitaev spin liquids with and without 
the external magnetic field. 
In the original exact solution using the Majorana representation 
by A. Kitaev, the Kitaev model 
without the magnetic field supports both gapped and gapless 
$\mathbb{Z}_2$ spin liquids~\cite{Kitaev2006}. In the gapless ${\mathbb{Z}_2}$
spin liquid, there exist gapless Majorana fermions with 
the Dirac dispersion. When the magnetic field along the $[111]$
direction is applied as a perturbation, the Zeeman coupling generates 
a Dirac mass for the Majorana fermions, and the Majorana part behaves 
more like a quantum anomalous Hall state and can be regarded as
a Chern insulator for the Majorana fermions. 
The vison remains gapped and does not influence the low-energy physics.  
Due to the finite Chern number, the 
system then supports the chiral propagating majorana edge 
mode. Besides the non-Abelian topological properties, this state, 
as Kitaev predicted, supports a half-quantized thermal Hall conductance~\cite{Kitaev2006}. 
More precisely, the magnetic field here drives a 
topological phase transition for the majorana spinons. 
Thus, the thermal Hall property of the Kitaev spin liquid
can be placed onto the generic behaviors of $\mathbb{Z}_2$ spin liquids 
under the magnetic field that is discussed above. 
Moreover, the quantized thermal Hall signal can be used to diagnose
the topological quantum transition of the exotic quasiparticles, which
is the topic for Sec.~\ref{sec11}.
The direct observation of this half-quantized thermal Hall 
conductance in the Kitaev material is a direct and sharp 
confirmation of the magnetic Kitaev spin liquid. 
This is part of the underlying reason for the interest of the
thermal Hall effects in the Kitaev materials. 
Due to the intensive interest and activities in this topic, we single out
Sec.~\ref{sec6} to present some theoretical and experimental
developments on the thermal Hall effects in the Kitaev materials.


\section{Thermal Hall effect in U(1) spin liquids with gapless matter}
\label{sec4}

As we have remarked in Sec.~\ref{sec3}, the driving mechanism of 
the thermal Hall effects in $\mathbb{Z}_2$ spin liquids and chiral spin 
liquids are relatively clear compared to the U(1) spin liquids. We devote 
this section to the discussion about the thermal Hall effects in the 
U(1) spin liquids with the gapless (fermionic) matter. These include, 
for example, the spinon fermi surface U(1) spin liquid
~\cite{Lee2005,Motrunich2005,Motrunich2006,Gang2017,Gang2017B,Lee2018}
and the Dirac spin liquid 
(sometimes dubbed algebraic spin liquid)~\cite{Hermele2005,Ran2007,Hermele2008}.
In 2D, the gapless fermionic matter is believed to be a necessity to 
stabilize the U(1) spin liquids by suppressing the instanton events. 
As the presence of such a state is a prerequisite  
for the discussion of the thermal Hall transports, 
we further elaborate this point briefly below.

The emergent degrees of freedom at the low energies of the 2D U(1) spin liquids 
are the fermionic spinons and the U(1) gauge fields. 
Unlike the $\mathbb{Z}_2$ spin liquids whose existence have been supported by the exactly solvable models 
and Wegner's $\mathbb{Z}_2$ lattice gauge theory~\cite{Fradkin1979,Savary2016,Kitaev2003}, 
the 2D U(1) spin liquids and the associate compact U(1) lattice gauge theory, 
however, face the instability against confinement. 
The compact U(1) lattice gauge theory allows the instanton events
that correspond to the change of the gauge flux by $2\pi$,
and their proliferation would confine 
the gauge charge carried by the spinons~\cite{Polyakov1977}. 
Therefore, the 2D U(1) spin liquids with the gapped spinons cannot avoid the confinement.
Then, the stable 2D U(1) spin liquids are hoped to survive
with the assistance of the gapless (fermionic) spinons.
The stability of these spin liquids undergoes a long debate. 
It was argued that the instantons interact via a 
logarithmically decaying potential in spacetime, thus are 
suppressed~\cite{Ioffe1989,Wen2002,Wen2002B}.
These studies, however, neglect the screening of the instanton-instanton coupling,
as pointed out by further numerics~\cite{Marston1990,Murthy1991}.
The disruption of the instanton binding again leads to the permanent confinement.
The absence of 2D U(1) spin liquid was claimed by Herbut et al. 
for both relativistic~\cite{Herbut2003} and non-relativistic~\cite{Herbut2003B} 
matter fields. Not until a few years later could people realize that
the usual random phase approximation is not 
appropriate due to the spacetime monopoles~\cite{Hermele2004}.
The instanton events are drastic fluctuations to the classical background 
configuration that deserve a specific functional integral for each configuration. 
By treating the monopoles and gapless fermionic matter on 
the equal footing, the stability of these 2D U(1) spin liquids 
was arrived, and a non-compact U(1) gauge theory remains to be a 
good low-energy description~\cite{Kim2005,Hermele2004,SSLee2008}.

Given the stability of the 2D U(1) spin liquids with the gapless fermionic 
matter, we proceed to discuss the thermal Hall effects for this kind of critical spin 
liquids. The microscopic mechanism for the thermal Hall effects involves 
an interplay between the external magnetic field and the emergent 
charge-neutral degrees of freedom, and thus depends on the actual 
microscopic realization of the U(1) spin liquids
and the emergent internal degrees of freedom. 
From a more semiclassical point of view, 
switching on some external fields such as the magnetic field \emph{indirectly} 
exerts an emergent Lorentz force on the emergent exotic quasiparticles. 
The twisted motion of the exotic quasiparticles under a temperature 
gradient gives rise to a transverse heat flow that is nothing 
but the thermal Hall current. On the quantum mechanical ground,
the external magnetic field could induce an internal U(1) gauge flux
that is experienced by the emergent exotic quasiparticles.
The hopping Hamiltonian of the emergent exotic quasiparticles
thus encodes a finite Berry curvature, which is responsible for 
a non-vanishing thermal Hall conductivity. 
In the following, we explain the physics separately
for the 2D U(1) spin liquid in the weak and strong Mott insulating
regimes.

\subsection{Weak Mott insulating U(1) spin liquids}
\label{sec31}

The celebrated examples of the weak Mott insulating spin    
liquids are found in the spin-1/2 triangular-lattice magnets, 
such as the organic compounds $\kappa$-(ET)$_2$Cu$_2$(CN)$_3$~\cite{Saito2003,Saito2005,Yamashita2008,Yamashita2008B}, 
EtMe$_3$Sb[Pd(dmit)$_2$]$_2$~\cite{Yamashita2010,Itou2009},
and more recently, the cluster Mott insulator 1T-TaS$_2$ and 
LiZn$_2$Mo$_3$O$_8$~\cite{Lee2018,PhysRevB.93.245134,PhysRevLett.129.017202}. 
The investigation on the organic salt EtMe$_3$Sb[Pd(dmit)$_2$]$_2$
underwent a controversial debate. Initially, a residual longitudinal 
thermal transport with a finite $\kappa_{xx}/T$  
was observed down to the zero temperature limit  
indicating the presence of the itinerant gapless 
excitations~\cite{Yamashita2010}. The presence of 
the itinerant gapless excitations was then questioned 
by later experiments~\cite{PhysRevX.9.041051,Shiyan2019},
eventually, is further clarified and supported by comparing 
different cooling process carefully~\cite{Yamashita2020}. 
Despite these experimental controversies, the physical 
mechanism for the presence of these weak Mott insulating 
spin liquid candidates is generally believed to be
the strong charge fluctuations in the weak Mott regime~\cite{Motrunich2005,Motrunich2006}.

In the weak Mott insulating regime, 
it is more natural to start from the parent microscopic model, 
i.e. the Hubbard model. 
A stereotype model for the weak Mott insulating spin liquid is 
 the single-band Hubbard model 
for the electrons on the triangular lattice with
\begin{eqnarray}
\mc{H} = -t \sum_{\langle ij \rangle} (c^{\dagger}_{i\sigma} c^{}_{j\sigma} + h.c.)
+ U \sum_i n_{i\uparrow} n_{i\downarrow} ,
\end{eqnarray}
where $c^{\dagger}_{i\sigma}$ ($c^{}_{i\sigma}$) creates 
(annihilates) the electron with the spin $\sigma$ at the lattice site $i$,
and the electron is at 1/2 filling such that there is one electron per site 
on average. When the Hubbard interaction is large enough, the electron 
will be Mott localized and form the spin-1/2 local moment ${\bf S}_i$ 
at each site. In the strong Mott regime, the nearest-neighbor Heisenberg model,
\begin{eqnarray}
\mc{H}_{\text{ex}} = J \sum_{\langle ij \rangle} {\bf S}_i \cdot {\bf S}_j, 
\end{eqnarray}
with ${J= 4t^2/U}$ from the second order perturbation theory of 
the Hubbard model would be sufficient to capture the interaction 
between the local spin moments, and the ground state
is the well-known 120-degree magnetic order. 
In the weak Mott regime when the Hubbard-$U$ interaction 
is still above the critical value for the Mott transition, 
the nearest-neighbor Heisenberg model is insufficient, and 
we need to consider the high order exchange interactions. 
It is generally accepted that, the presence of the sizable 
four-spin ring exchange interaction, 
\begin{eqnarray}
\mc{H}_{\text{ring}} &=& \frac{80t^4}{U^3} \sum_{\langle 
ijkl \rangle} \Big[
 ({\bf S}_i \cdot {\bf S}_j) ( {\bf S}_k \cdot {\bf S}_l) 
 + ({\bf S}_i \cdot {\bf S}_l) ( {\bf S}_k \cdot {\bf S}_j)  
 \nonumber \\
 &&  \quad\quad\quad\quad\quad\quad 
        - ({\bf S}_i \cdot {\bf S}_k) ( {\bf S}_j \cdot {\bf S}_l) \Big] ,
        \label{eqring}
\end{eqnarray}
together with other weaker further neighbor superexchange 
interactions~\cite{PhysRevB.99.054432}, 
immediately compete with the 
nearest-neighbor Heisenberg interaction and frustrate
the 120-degree magnetic order. 
Here, the ring exchange interaction is operating on 
the four spins around the elementary parallelogram 
of the triangular lattice. Indeed, various theoretical 
and numerical works have suggested the U(1) spin liquid 
with the spinon Fermi suface as the ground state 
for the triangular lattice Hubbard model in the weak Mott regime. 
A recent density matrix renormalization group (DMRG) 
calculation together with other supporting numerics, 
however, proposed the possibility of a fully-gapped
chiral spin liquid ground state instead of a gapless spinon
Fermi surface state~\cite{PhysRevX.10.021042,PhysRevLett.127.087201}. 
To resolve the discrepancy between the chiral spin liquid and 
the spinon Fermi surface spin liquid for the weak Mott regime 
is not the task of this review. 
In fact, the simple $J_1$-$J_2$ spin-1/2 Heisenberg model 
with $J_1$ ($J_2$) the first (second) neighbor exchange coupling 
on the triangular lattice was 
numerically shown to exhibit a U(1) Dirac spin liquid~\cite{PhysRevB.93.144411,PhysRevX.9.031026} 
that is suggested
as the mother state of many 2D magnetic states~\cite{DSLMotherState,Hermele2005}. 
It will be interesting if the chiral spin liquid could be induced 
from the Dirac mass generation of the Dirac spin liquid
by incorporating the four-spin ring exchange interaction. 
To clarify whether the spinon Fermi surface U(1) spin liquid 
or the chiral spin liquid is the ground state 
certainly requires further theoretical efforts. 
Since most theoretical works suggest the spinon Fermi surface U(1) 
spin liquid and the experimental system also seems to support
this state as the ground state or the parent state for further 
instabilities, we take this point of view and address the thermal
Hall effect for this state in the weak Mott insulating regime.

In this weak Mott insulating U(1) spin liquid, 
the electron experiences the spin-charge separation 
in which the charge degree of freedom develops a Mott gap and is 
Mott localized while the spin quantum number can still propagate
in the spin liquid phase.  
Due to the small charge gap in this regime, the localized electron is 
sometimes quoted as the ``quasi-itinerant'' electrons, 
and the fermionic spinons are actually not very far from 
the physical electrons. Because of this quasi-itinerant nature 
of the electrons and the strong charge fluctuations, 
it seems reasonable to expect the spin degrees of freedom 
to experience the external magnetic flux in some fashion~\cite{Motrunich2006}. 
From the high-order perturbation theory of the Hubbard model, 
the four-spin ring exchange interaction could naturally entrap 
the external magnetic flux that would modify the spinon behaviors. 
In fact, as the time reversal symmetry is explicitly broken by the magnetic field, 
the scalar spin chirality is allowed in the spin Hamiltonian in the 
third order perturbation of the Hubbard model. 
It was shown that, the external magnetic field couples 
linearly to the scalar spin chirality on the triangles
through a three-site ring-exchange process as,
\begin{equation}
\sim \frac{24t^3}{U^2}\sin \Phi_{\rm ext}\sum_{i,j,k\in \triangle} \big[ {\bf S}_i\cdot \big({\bf S}_j\times{\bf S}_k\big) \big],
\label{SSC_ring}
\end{equation}
where $\Phi_{\rm ext}$ is the external U(1) gauge flux 
on the elementary triangular plaquette
and ${i,j,k\in \triangle}$ 
are three sites on this elementary triangular plaquette.
As shown by Wen, 
Wilczek and Zee~\cite{Wen1989B}, this scalar spin chirality 
in the U(1) spin liquid is nothing but the internal U(1) gauge 
flux for the spinons with the relation, 
\begin{equation}
\sin \Phi= \frac{1}{2} \, {\bf S}_i\cdot \big({\bf S}_j\times{\bf S}_k\big),
\label{Phi_scalar_chi}
\end{equation}
where $\Phi$ refers to the internal U(1) gauge flux that is defined 
on the triangular plaquette formed by the sites ${i,j,k}$. 
Therefore, through the multiple-spin superexchange interaction,  
the external magnetic field could induce an internal U(1) gauge
flux that is experienced by the spinon hopping on the lattice.
As the external magnetic field is varied, the induced internal 
U(1) gauge flux could vary accordingly, which modifies the 
Landau level structures of the spinons. In analogy with to
the electron Landau-level population and the quantum oscillation 
in simple metals, one prominent consequence of the spinon-gauge 
coupling is a quantum oscillation in Mott insulators with the charge-neutral 
quasiparticles without an electron Fermi surface~\cite{Motrunich2005}.
As it is explained below, the thermal Hall effect of the spinons in this context 
turns out to have the same physical origin as the quantum oscillation. This
can probably be extended to other insulating systems where the quantum 
oscillation is observed, and a thermal Hall effect may naturally be 
expected.

In the presence of the induced internal U(1) gauge flux, 
the spinons can experience an emergent Lorentz force~\cite{Katsura2010}.
To see this, we turn to the spinon-gauge coupled model
and seek for the direct revelation by including the spinon 
degrees of freedom explicitly at the mean-field level.
The spinon-gauge coupled theory is reformulated by  
transforming the Hubbard model in the slave-rotor 
representation~\cite{Lee2005}. The electron is decomposed 
into the combination of the fermionic spinon and the bosonic 
charge represented by a U(1) charge rotor, namely 
${c_{i\sigma} \equiv f_{i\sigma} e^{-i\theta_i}}$,
where $f_{i\sigma}$ annihilates a fermionic spinon 
at the lattice site $i$ with the spin $\sigma$
and the charge rotor $e^{-i\theta_i}$ annihilates a charge boson 
with the electron charge $e$ at the site $i$. In this formulation,
 a U(1) gauge redundancy is introduced, and the spin-charge separation 
 of the weak Mott regime
is realized in the deconfined phase of the U(1) lattice gauge theory. 
Moreover, the fermionic spinon carries the internal U(1) gauge charge,
while the charge boson carries both the internal U(1) and the external 
U(1) gauge charges. In the metallic phase, the bosonic rotor is 
condensed, and the internal U(1) gauge field becomes massive via 
the Anderson-Higgs' mechanism such that the charge and the spin are
bound together to form the physical electrons. An equivalent description
of the internal gauge field is that, the condensed bosonic rotor locks the 
internal U(1) gauge field with the external U(1) gauge field.

In the weak Mott regime, the charge sector is gapped, and 
one then integrates out the charge sector to arrive with the 
low-energy effective theory for the spin sector. 
This generates a Maxwell term for the gauge fields
\begin{eqnarray}
{\cal L}_{g}= g^{-1} \int dr ({\cal F}_{\mu\nu}-F_{\mu\nu})^2 ,
\end{eqnarray}
where ${\cal F}_{\mu\nu}, F_{\mu\nu}$ are the electromagnetic tensor
for the internal and external gauge fields, respectively. The
low-energy effective Lagrangian for the U(1) spin liquid is 
given by~\cite{Katsura2010},
\begin{eqnarray}
&& {\cal L}[f,a_\mu] = {\cal L}_{m}[f,a_\mu] + {\cal L}_{g}[a_\mu-A_\mu], \\
&& {\cal L}_{m}[f,a_\mu]  =- t_s \sum_{\langle i,j\rangle} 
\big[ e^{i a_{ij}} f_{i\sigma}^\dagger f^{}_{j\sigma}  + h.c. \big]
 \nonumber \\
 && \quad\quad\quad\quad\quad + \sum_{j,\sigma} 
f_{j\sigma}^\dagger (\partial_0 - ia^0_j-\mu) f^{}_{j\sigma} ,
\label{L_mg_weak_Mott}
\end{eqnarray}
where ${\cal L}_m$ in the first line describes the spinon hopping on the
 triangular lattice and minimally coupled to the dynamical and internal 
 U(1) gauge field ${\boldsymbol a}$, and $t_s$ defines the spinon hopping.
In the presence of a finite external magnetic field with ${F_{xy} \ne 0}$,
the internal gauge flux will be induced as
dictated by the Maxwell gauge term ${\cal L}_g$. 
This field theoretical point of view is consistent 
with the microscopic result in Eq.~\eqref{SSC_ring}. 
Based on this argument, H. Katsura et al. suggested that
the spinons are subjected to an effective internal magnetic 
field that acts with an effective Lorentz force on the spinons, 
and the motion of the spinon is expected to be twisted 
under the temperature gradient, which would lead to 
a thermal Hall effect~\cite{Katsura2010}. They were able 
to make an estimate about the spinon thermal Hall conductivity  
${\kappa_{xy}\sim(\omega_c\tau)\kappa_{xx}}$, 
where ${\omega_c=eB/m_c}$ is the cyclotron frequency 
with the electrical charge $e$, the external magnetic field 
$B$ and the effective mass $m_c$ of the spinons. Since there 
are no spinon edge states from the modulated spinon band 
structure under the induced internal U(1) gauge flux,  
the thermal Hall effect for this case is generally not quantized.

\subsection{Strong Mott insulating U(1) spin liquids}
\label{sec32}

In the strong Mott insulating regime, the electrons are fully localized. 
The high-order ring exchange interactions in Eq.~\eqref{eqring} 
and Eq.~\eqref{SSC_ring} are strongly suppressed. 
The mechanism from the induction of the internal U(1) gauge flux 
by the external magnetic field in the previous subsection for the 
weak Mott regime becomes negligible.     
Since the large thermal Hall effects were observed in the strong Mott
insulating spin liquid candidates, it is then natural to expect fundamentally 
different microscopic mechanisms for the thermal Hall effects in these 
systems.

The microscopic degrees of freedom in the strong Mott regime are 
purely local spin moments rather than the quasi-itinerant electrons.  
Thus, the external magnetic field here mainly has a Zeeman coupling 
to the local spin moment. For the U(1) spin liquids in this regime, 
the matter fields are the charge-neutral spinons that are absent 
of the external U(1) gauge charge. The spinons carry the emergent 
U(1) gauge charges and are minimally coupled to the internal U(1) 
gauge field as the spinons hop on the lattice. 
To go beyond the simple and rigid spin splitting of the spinon bands by
the Zeeman coupling in the same way as what has been discussed for 
the $\mathbb{Z}_2$ spin liquids, 
we then consider the possibility of 
varying the internal U(1) gauge flux via the Zeeman coupling rather 
than the orbital effect of the external magnetic field in the previous 
subsection. Again, to twist the spinon motion and generate 
the thermal Hall effects, one would still hope to modify the 
internal U(1) gauge flux via the 
scalar spin chirality ${{\bf S}_i\cdot ({\bf S}_j\times{\bf S}_k)}$. 
Apparently, the magnetic field does not directly nor 
at the linear order induce the scalar spin chirality. 
It is observed that, the scalar spin chirality involves
the vector spin chirality ${{\bf S}_j\times{\bf S}_k}$,
and this object is directly related to the antisymmetric 
DM interaction. If the model Hamiltonian 
of the system allows a DM interaction, 
the corresponding ${\langle{  {\bf S}_j\times{\bf S}_k } \rangle}$
is naturally non-vanishing. When the external magnetic field is 
applied, the combination of the DM interaction
and the Zeeman coupling could generate a distribution of a 
finite scalar spin chirality with ${\langle{ {\bf S}_i\cdot ({\bf S}_j\times{\bf S}_k)} \rangle \neq 0}$. 
Unlike the uniform scalar spin chirality from the mechanism of 
the ring exchange for the weak Mott insulators, the induced    
scalar spin chirality is no longer \emph{uniform} throughout the lattice. 
This is because the DM interaction depends on 
the lattice symmetry and the DM vector could vary 
from bond to bond. While the spinons in this case still pick up the internal 
U(1) gauge flux from the scalar spin chirality, it is quite different from 
the uniform gauge flux case for the weak Mott regime in the previous
subsection.

\begin{figure}[t] 
	\centering
	\includegraphics[width=6.5cm]{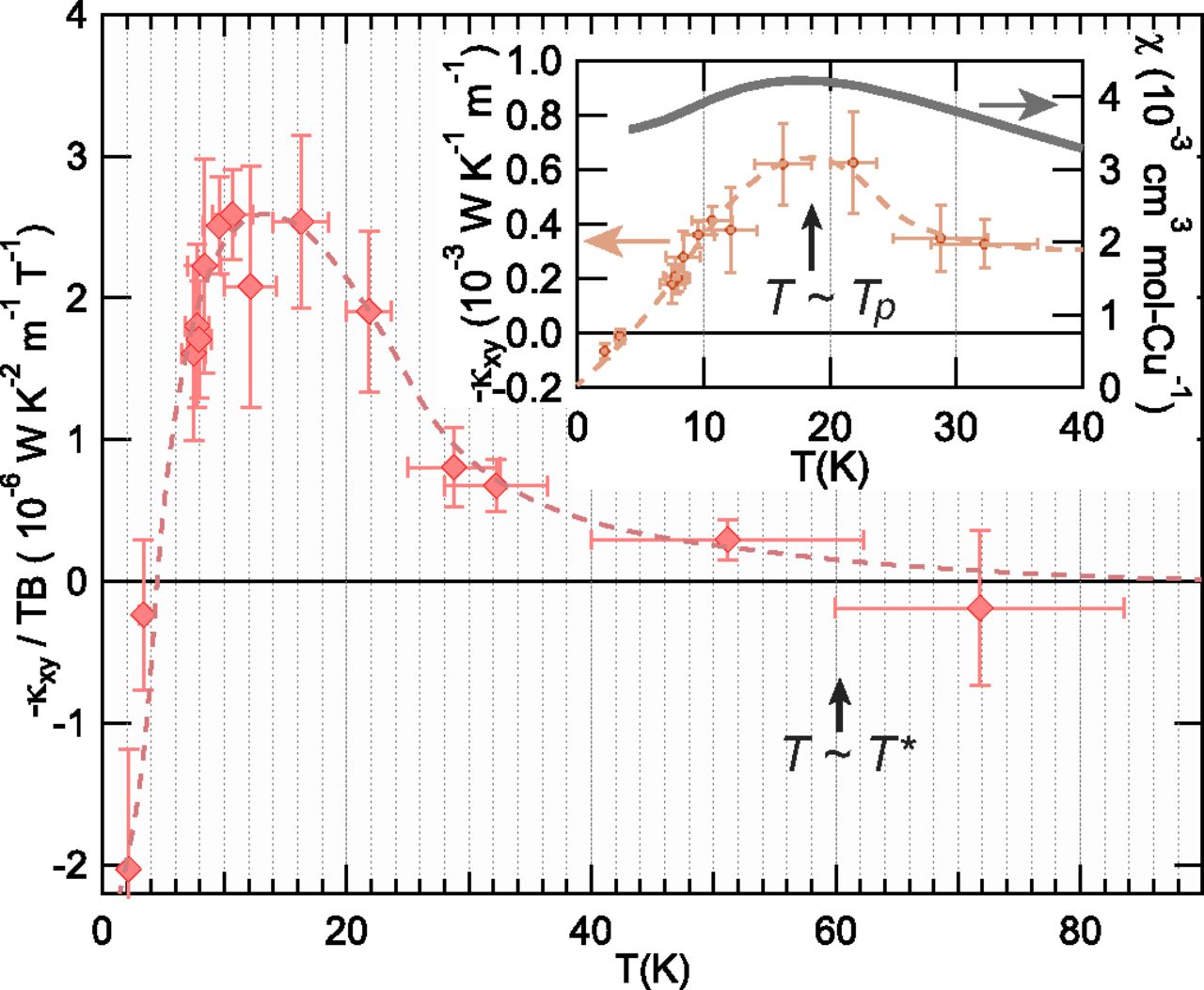}
	\caption{Temperature dependence of the thermal Hall coefficient $-\kappa_{xy}/TB$ at $15$ T. 
		(Inset) temperature dependence of the $-\kappa_{xy}$ at $15$ T and susceptibility $\chi$. Here $T_p$ represents the peak temperature of the magnetic susceptibility, and $T^*$ corresponds to the effective
		spin-interaction energy scale. 
		Figure is reprinted from Ref.~\onlinecite{Watanabe2016}. }
	\label{KHA_fig1}
\end{figure}

To illustrate the above idea of the DM interaction, 
we will demonstrate this mechanism for the strong Mott insulating 
U(1) spin liquids on a kagom\'{e} lattice below. For this purpose, 
we first introduce some of the early attempts about the 
thermal Hall effects for the strong Mott insulating 
kagom\'{e} magnets, and then introduce the theory 
with the gapless U(1) spin 
liquids.

\begin{figure*}[htbp] 
	\centering
	\includegraphics[width=14cm]{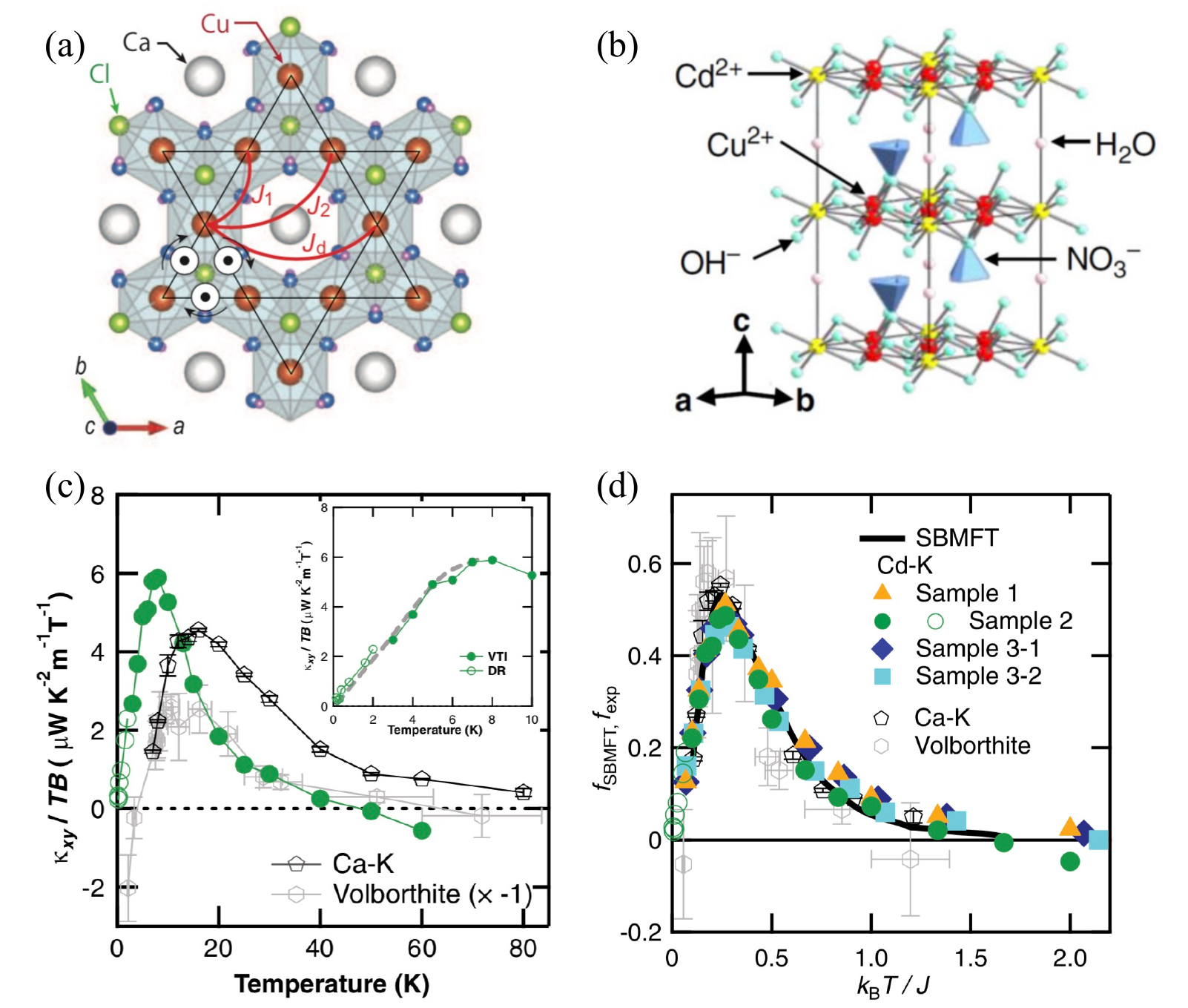}
	\caption{The thermal Hall experiments on the kagom\'{e} kapellasite and volborthite. 
		(a) Crystal structure of Ca kapellasite viewed along the c axis.
		(b) Crystal structure of Cd kapellasite in 3D where the 2D slice has similar structure shown in (a).
		(c) Comparison of $\kappa_{xy}/TB$ of 
		Cd-K~\cite{Yamashita2020} (green circles), 
		Ca-K~\cite{Yamashita2018} (open gray pentagon) and 
		volborthite\,\cite{Watanabe2016} (open gray hexagon). 
		For clarity, $\kappa_{xy}/TB$ of volborthite is multiplied by $-1$. 
		The inset shows an enlarged view of the low temperature data of $\kappa_{xy}/TB$ for Cd-K.
		(d) Normalized thermal Hall conductivity $f_\textrm{exp}$ of kagom\'e lattice antiferromagnet fitted by the parameters listed in Table\,\ref{KHA_table}. 
		The solid line shows a numerical calculation of $f_\textrm{SBMFT}$ at $D/J=0.1$ by the SBMFT\,\cite{Yamashita2020}.
		The data of Cd-K, Ca-K and volborthite are taken from 
		Ref.~\onlinecite{Yamashita2020}, 
		Ref.~\onlinecite{Yamashita2018} 
		and Ref.~\onlinecite{Watanabe2016}, respectively.
		Panel (a) is reprinted from Ref.~\onlinecite{Yamashita2018};
		Panel (b-d) are reprinted from Ref.~\onlinecite{Yamashita2020}.
	}
	\label{KHA_fig2}
\end{figure*}

\subsubsection{Early attempts with gapped Schwinger bosons}

It has been proposed that the kagom\'{e} magnet family could 
harbor a gapped $\mathbb{Z}_2$ spin liquid, U(1) Dirac spin liquid, 
spinon Fermi surface U(1) spin liquid and even chiral spin liquid. 
Experimentally, the thermal Hall signatures have been observed 
in the disordered regimes of the kagom\'{e} lattice magnets 
volborthite Cu$_3$V$_2$O$_7$(OH)$_2\cdot$2H$_2$O 
and Ca-based kapellasite CaCu$_3$(OH)$_6$Cl$_2\cdot$0.6H$_2$O.
These materials are clearly strong Mott insulators with large Mott gaps.  
Even though the volborthite develops a magnetic 
order at very low temperatures,
the spin liquid regimes were suggested to be relevant 
for or govern the low-temperature magnetic properties.
 Meanwhile, the inelastic neutron scattering 
experiment has reported the spinon continuum in the excitation 
spectrum for the more well-known kagom\'{e} magnet herbertsmithite 
ZnCu$_3$(OH)$_6$Cl$_2$~\cite{Han2012}, where the Cu$^{2+}$ ions 
form the ${S = 1/2}$ local moments on a 2D kagom\'{e} lattice. 

\begin{table}[thb]
	\caption{
		Values of $J$ and $\vert D/J\vert$ used to fit the thermal Hall conductivity $\kappa_{xy}$ 
		to the SBMFT simulation [Fig.\,\ref{KHA_fig2}(f)] for kagom\'{e} lattice antiferromagnet. 
		The data of Ca-K and that of volborthite is taken from Ref.~\onlinecite{Yamashita2018} 
		and Ref.~\onlinecite{Watanabe2016}, respectively.
		Table is reprinted from Ref.~\onlinecite{Yamashita2020}.
	}
	\label{KHA_table}
		\begin{tabular}{cccc}
		\hline\hline
			Material& Sample No. & $J/k_{\text{B}}$(K) & $D/J$ \\
			\hline
			Cd-Kapellasite\cite{Yamashita2020}&1&30&0.28 \\
				& 2&30&0.09 \\
				&3-1&29&0.65\\
				&3-2&28&0.6\\
			Ca-Kapellasite\cite{Yamashita2018}& &66&0.12\\
			Volborthite\cite{Watanabe2016}& &60&-0.07\\	
		\hline\hline
		\end{tabular}
\end{table}

For the volborthite, the susceptibility $\chi$ and the specific heat 
$C/T$ measurements strongly suggest the gapless nature of the spin excitation 
for the candidate spin liquid~\cite{Watanabe2016,Yoshida2012}.
The longitudinal thermal conductivity $\kappa_{xx}$ 
provides the evidence on the spin contribution
for the thermal transport and supports the itinerant nature of the elementary excitations.
$\kappa_{xx}/T$ is suppressed at a relative high magnetic field of $15$T,
below which the longitudinal thermal conductivity is mostly contributed by the spins.
The transverse thermal conductivity $\kappa_{xy}/T$ (plotted in Fig.~\ref{KHA_fig1})
shows more interesting structures and 
 displays a sign change upon approaching the N\'eel temperature.

The crystal structure of the Ca kapellasite is illustrated in
Fig.~\ref{KHA_fig2}(a). The thermodynamic~\cite{Yoshida2017} 
and spectroscopic~\cite{Ihara2017} studies imply the presence 
of gapless spin excitations in the spin liquid phase. 
The more inspiring thermal Hall signal, plotted in Fig.~\ref{KHA_fig2}(c),
shows a surprisingly similar temperature dependence 
as the volborthite~\cite{Yamashita2018}.  
In fact, the temperature dependence of $\kappa_{xy}$ 
in the Ca kapellasite and the volborthite converge to 
one single curve in Fig.~\ref{KHA_fig2}(d).
Moreover, this experimental curve was well-reproduced by the 
numerical simulation adopting the Schwinger boson mean-field (SBMF) 
approach by incorporating the influence of the DM
 interaction~\cite{Yamashita2018,Yamashita2020}. 

To illustrate the idea of the SBMF approach, one can simply consider the representative 
spin model in the strong Mott insulator on a kagom\'{e} lattice with the magnetic field 
normal to the kagom\'{e} plane,
\begin{equation}
\label{kagomeHam}
\mc{H}= J \sum_{\langle ij\rangle} {\bf S}_i \cdot {\bf S}_j 
+  \sum_{\langle ij\rangle} {\bf D}_{ij}\cdot ( {\bf S}_i \times {\bf S}_j)
- B \sum_{i}  S^z_i.
\end{equation}
In the SBMF theory, the underlying assumption for the 2D kagom\'{e} magnet is 
a \emph{gapped} $\mb{Z}_2$ spin liquid with the gapped bosonic spinon excitations, 
and the physical spin ${S=1/2}$ is expressed by the Schwinger bosons 
$b^{\dagger}_{\alpha}$ ($b_{\alpha}$) as~\cite{PhysRevB.45.12377} 
\begin{eqnarray}
\bs{S}_i=\frac{1}{2}\sum_{\alpha,\beta=
\uparrow,\downarrow}b^{\dagger}_{i\alpha}\bs{\sigma}_{\alpha\beta}b_{i\beta}^{}
\end{eqnarray} 
with $\bs{\sigma}$ being a vector of the Pauli matrices. 
Normally, one introduces two types of mean-field parameters 
$\langle b_{i\alpha}b_{j(-\alpha)} \rangle$ and $\langle b_{i\alpha}^{\dagger} b^{}_{j\alpha}\rangle$ 
to self-consistently solve for the mean-field ground state. It is necessary to stress that the choice 
of these two types of parameters only involves the spin singlet part, and a more general decoupling 
scheme requires the triplet terms. The latter is often adopted for more anisotropic spin models 
and/or competing ferromagnetic interactions.

According to the $\mb{Z}_2$ spin liquid classification in Ref.~\cite{PhysRevB.74.174423}, 
four different classes are obtained for the kagom\'{e} lattice geometry,
 and they can be generally categorized into the 0-flux and the $\pi$-flux states. 
 The 0-flux and the $\pi$-flux states are often adopted to study 
the kagom\'{e} spin liquid and the adjacent magnetic order~\cite{PhysRevB.45.12377}. 
Specifically, the zero-flux state is connected 
to the ${{\bs q}={\bs 0}}$ magnetic order by condensing the low-energy bosonic spinons, 
while the $\pi$-flux state is connected to the ${\sqrt{3} \times \sqrt{3}}$ order
with an enlarged unit cell. It is also believed that the ${{\bs q}={\bs 0}}$ configuration 
could take the energetic advantage in the presence of DM interaction~\cite{PhysRevB.81.064428,PhysRevB.81.144432}. Moreover, the recent numerical study~\cite{PhysRevLett.118.267201} suggests the ${\bs q}={\bf 0}$  
ordered state as a possible ground state of the Ca kapellasite. Therefore, it is reasonable to adopt 
the 0-flux ansatz instead of the $\pi$-flux state to study the thermal Hall effect of the Ca kapellasite.

In the numerical calculation, \citet{Yamashita2018} employed the 0-flux ansatz 
and dropped the spinon pairing $\langle b_{i\alpha}b_{j(-\alpha)} \rangle$ 
term by assuming that it is less effective at the high temperature region  
${k_B T > 0.1 J}$. One then considers the mean-field decoupling of Eq.~\eqref{kagomeHam}
and analyzes the spinon Hamiltonian for the spin liquid state with the mean-field parameter 
$\chi_{ij}^{\alpha}={\langle b^{\dagger}_{i\alpha}b_{j\alpha}^{} \rangle}$,
\begin{align}
\mathcal{H}_{\rm SBMF} = 
   \sum_{\langle i,j \rangle, \sigma} 
   t^{\sigma}_{ij} b_{i\sigma}^{\dagger}b_{j\sigma}^{} 
+ \sum_{i,\sigma} (\mu - \sigma B) b_{i\sigma}^{\dagger} b_{i\sigma}^{},
\end{align}
where 
\begin{eqnarray}
t^{\sigma}_{ij} & = & J \chi_{ji}^{\sigma} + J' \chi_{ji}^{-\sigma} e^{-i\sigma \phi_{ij}} ,  \\
J' &=&(J^2 + D_{ij}^2)^{1/2} ,
\end{eqnarray}
 and ${\tan \phi_{ij} = D_{ij}/J}$.  
 Remarkably, these relations are quite similar to the ones in Eq.~\eqref{eq16sec2}
 for the magnon hopping in the pyrochlore ferromagnets. Although the expressions look similar to the 
 magnon's case, the physics is fundamentally different. 
The sign of $D_{ij}$ is assumed to be positive if ${i \rightarrow j}$ is 
clockwise from the center of each triangle plaquette in the kagom\'e lattice,
and ${|D_{ij}|=D}$. 
The Lagrange multiplier $\mu$ is introduced to impose the Hilbert space 
constraint ${\sum_{\alpha} \langle b_{i\alpha}^{\dagger} b_{i\alpha}^{} \rangle = 2S}$. 
Due to the complex number $t^{\alpha}_{ij}$ from the DM interaction,  
the gapped spinon bands now carry the non-vanishing Berry curvatures, 
which are directly related to the thermal Hall conductivity through the 
relation Eq.~\eqref{k_xy_boson}. The fitting parameters to reconcile 
the numerical and experimental thermal Hall conductivities over there 
are the nearest-neighbor Heisenberg exchange coupling $J$ and
the DM interaction $D$, which are consistent with the estimation from 
thermodynamics. The reasonable agreement leads to the conclusion that
the relevant spin liquid in the Ca kapellasite is well captured by the SBMF theory 
and the observed thermal Hall effect arises from the spins.


The Cd kapellasite~\cite{Okuma2017,Okuma2019} has an identical 
crystal structure as the Ca kapellasite [see Fig.~\ref{KHA_fig2}(b)],   
but the thermal Hall measurement shows some distinct features~\cite{Yamashita2020}.
The temperature dependence of $\kappa_{xy}/T$ has a similar line-shape
as the previous two kagom\'{e} compounds, as shown in Fig.~\ref{KHA_fig2}(c).
The magnitude, however, deviates from the numerical result 
drastically~\cite{Yamashita2020}. Following the same strategy 
of fitting the thermal Hall data with $J$ and $D$,
one adopts the SBMF theory and still finds a nearly perfect convergence 
in Fig.~\ref{KHA_fig2}(d). As it is shown in Table.~\ref{KHA_table},
the required values of $J$ and $D$ are incompatible with 
the ones deduced from thermodynamics~\cite{Okuma2019}.
This mismatch led to the suggestion that the spins might not be the only 
source for the observed thermal Hall effect~\cite{Yamashita2018},
and the phonons might play some role in the Cd kapellasite
and its coupling to the spins remains to be explored.

From the viewpoint of the SBMF theory, the thermal Hall effect is 
intimately related to the DM interaction (see Table.~\ref{KHA_table}).
The opposite sign of the thermal Hall conductivity in volborthite and kapellasites,
as depicted in Fig.~\ref{KHA_fig2}(c), are then related to the sign of the DM 
interaction that determines the sign of the Berry curvatures for the bosonic spinons. 
Despite that the SBMF theory with the DM interaction seems to be a successful theory to 
understand the thermal Hall effect in the disordered phase of kagom\'e magnets 
in the strong Mott regime, there still remain some puzzles. In the SBMF theory 
for the spinon thermal Hall effect, the bosonic pairing terms are dropped out~\cite{Lee2015,Yamashita2018, Yamashita2020}. 
If the SBMF theory describes a spin liquid, it should correspond a gapped U(1) 
spin liquid in 2D. But as we mentioned above the gapped U(1) spin liquid cannot 
stably exist in 2D. Moreover, the gapped spinon bands in the SBMF 
theory can not capture the gapless nature of the spin excitations. 
The stability might be cured by incorporating the spinon pairing channel. 
Although the spinon bands are still gapped, the resulting state now is no longer a U(1) spin liquid
 but a $\mb{Z}_2$ spin liquid with bosonic spinon excitations. The involvement of the spinon pairing channel,
 however, may change the current values of the fitting parameters $J$ and $D$ in Table.~\ref{KHA_table}. 
More recently, the SBMF theory with the spinon pairing terms was actually employed to explore the thermal Hall effect
 in the square-lattice spin liquids, and gives the finite thermal Hall conductivities in a self-consistent fashion~\cite{Sachdev2019}.
 In addition, the SBMF theory can be adopted to describe the ordered magnets via the boson condensation,
 and was suggested to be better than linear spin-wave theory in evaluating the thermal Hall conductivity~\cite{PhysRevB.102.214421}.

\begin{figure}[t] 
	\centering
	\includegraphics[width=8.6cm]{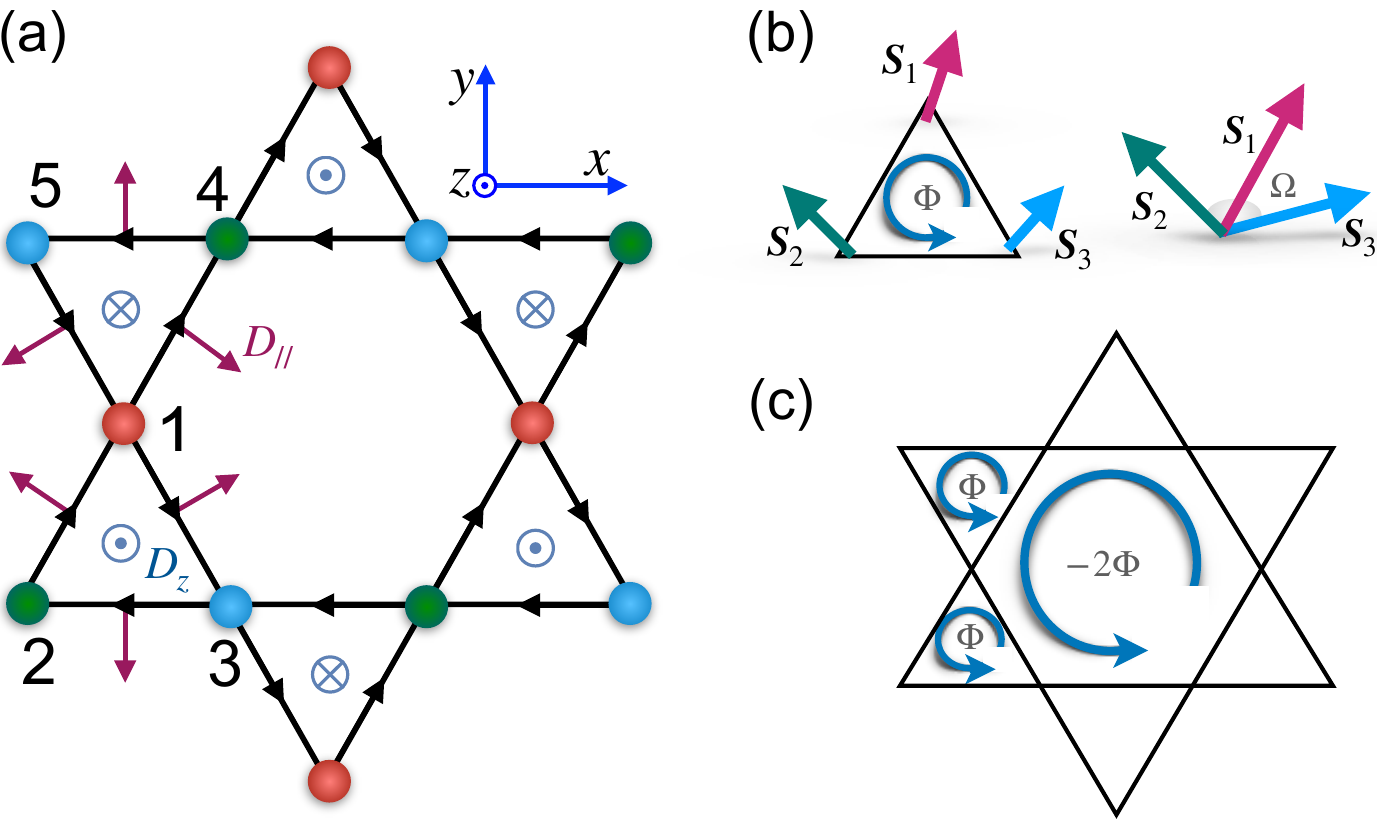}
	\caption{(a) Symmetry allowed DMIs between 
		first neighbors on the kagome lattice, where $D_z$ ($D_{\parallel}$) 
		is the $z$ (in-plane) component. The black arrows on the bonds specify 
		the order of the cross product $\boldsymbol{S_i}\times\boldsymbol{S_j}$. 
		The sublattices are labelled by colors.  (b) Schematic view 
		of scalar spin chirality for a non-collinear spin configuration, where $\Phi$  
		is the corresponding gauge flux through the plaquette and $\Omega$ is 
		the solid angle subtended by the three spins. (c) Internal U(1) flux 
		distribution induced on the kagom\'e lattice.
		Figures are reprinted from Ref.~\onlinecite{Gao2020}.}
	\label{DMI_fig1}
\end{figure}


\subsubsection{Internal flux generation via Dzyaloshinskii-Moriya interaction}
\label{sec422}

A different theory from the SBMF framework for the thermal Hall effects in the strong Mott 
insulating regime was developed in Ref.~\cite{Gao2020} for the 2D gapless U(1) spin liquids. 
In the theoretical formulation, the crucial role of the DM interaction for
the thermal Hall effect is exemplified in a bit more transparent manner. 
Under the external magnetic field, the combination of the linear Zeeman
coupling and the DM interaction induces a non-collinear canting of adjacent 
three spins on a triangular plaquette~\cite{LeeNagaosa2013,Gao2020}, 
which is shown in Fig.~\ref{DMI_fig1}(b). 
The solid angle subtended by the three spins is associated with the scalar 
spin chirality that was discussed for the magnons in Sec.~\ref{sec2}. 
For the U(1) spin liquids, however, the internal U(1) gauge flux 
is related to the scalar spin chirality~\cite{Zee1989,LeeNagaosa1992}.
These relations provide one roadmap for the controlling 
of the internal and emergent degrees of freedom
and a potential generation of thermal Hall currents 
in the strong Mott insulating spin liquids. The DM interaction
induces a ``vector spin chirality'' with~\cite{LeeNagaosa2013} 
\begin{equation}
\langle {\bf S}_i \times {\bf S}_j\rangle = \lambda {\bf D}_{ij},
\label{DM_vec_chi}
\end{equation}
where $\lambda$ is a proportionality constant with ${\lambda \sim {\mathcal O}(J^{-1})}$
and $J$ being the largest exchange coupling. The equality in 
Eq.~\eqref{DM_vec_chi} immediately gives birth to a linear relation 
between a single spin ${\bf S}_i$ and the scalar spin chirality 
${{\bf S}_i\cdot ({\bf S}_j \times {\bf S}_k) \simeq {\bf S}_i \cdot {\bf D}_{jk}}$.
For the kagom\'e lattice where the DM vectors on all bonds are related 
by symmetry, 
one then has the relation 
${ \langle\boldsymbol{S}_2\times\boldsymbol{S}_3\rangle
= \langle\boldsymbol{S}_4\times\boldsymbol{S}_5\rangle
= \lambda \boldsymbol{D}_{23}=\lambda \boldsymbol{D}_{45} }$.
This configuration is illustrated in Fig.~\ref{DMI_fig1}. Therefore,  
the $z$-component of the spin contains a piece proportional to the scalar spin chirality. 
Although this observation was made specifically 
for the neutron scattering experiments to detect the gauge field fluctuations, 
it establishes the microscopic link between the Zeeman coupling and 
the scalar spin chirality.

The Zeeman coupling generates a finite spin polarization with 
${\langle S_i^z\rangle =\chi B}$. 
Here $\chi$ is the uniform magnetic susceptibility and would take a constant value for 
the spinon Fermi surface spin liquid state. 
The internal U(1) gauge flux, proportional to the scalar spin 
charity [see Eq.~\eqref{Phi_scalar_chi}], is further decomposed into two pieces,
\begin{eqnarray}
&&
\langle {\bf S}_i \rangle \cdot \langle {\bf S}_j \times {\bf S}_k \rangle 
=   \lambda D_z\langle S^z_i\rangle  =  \lambda D_z \chi B. 
\label{Phi_B_DM}
\end{eqnarray}
In practice, one needs to consider the expectation of all possible decomposition on the 
plaquette. For the illustration purpose, this procedure explains the mechanism for the internal U(1) flux generation. 
From the orientations of the DM vector, 
one can conclude that the induced internal U(1) gauge fluxes by the external magnetic 
field on both the up triangle and the down triangle
of the kagom\'{e} lattice are equal and denoted
as $\Phi$. The direction of the flux loop is depicted in Fig.~\ref{DMI_fig1}(c).
Moreover, the flux through the hexagon is determined by fluxes in 
its six neighboring triangles and it equals 
$-2\Phi$ if adopting the anticlockwise loop convention 
in Fig.~\ref{DMI_fig1}(c). Here one can compare the flux pattern
with the fictitious U(1) gauge flux 
for the magnons in the pyrochlore ferromagnet where the spins remain
ferromagnetically aligned even with weak DM interactions.

The spinon Fermi surface spin liquid was suggested 
for some kagom\'e materials~\cite{Watanabe2016,Janson2010}. 
Given the possible presence of the gapless spinon Fermi surface 
U(1) spin liquid in the kagom\'e system, the spinon-gauge 
coupled model can be obtained by 
introducing the U(1) gauge field to the spinon mean-field 
theory~\cite{Lee2008,Motrunich2005,Gang2017}.
The corresponding Lagrangian reads,
\begin{eqnarray}
\mathcal{L} &=& {\cal L}_m+ 
\int {dr}  \sum_{\mu} \frac{1}{g} 
(  \epsilon_{\mu\nu\lambda} \partial_{\nu} a_{\lambda} )^2,
\label{lag }
\end{eqnarray}
where ${\cal L}_m$ is same as the one for the weak Mott 
 regime (see Eq.~\eqref{L_mg_weak_Mott}) and 
the second term describes the fluctuation of ${\boldsymbol a}$.
For the weak Mott regime, the internal U(1) gauge flux 
is directly coupled with the external one as shown in Eq.~\eqref{SSC_ring}
and is thus uniformly distributed. For a strong Mott insulating spin liquid, 
however, the induced internal U(1) gauge flux experienced 
by the spinons is attributed to the DM interaction (including its direction) 
and the lattice geometry.

\begin{figure}[t] 
	\centering
	\includegraphics[width=8.6cm]{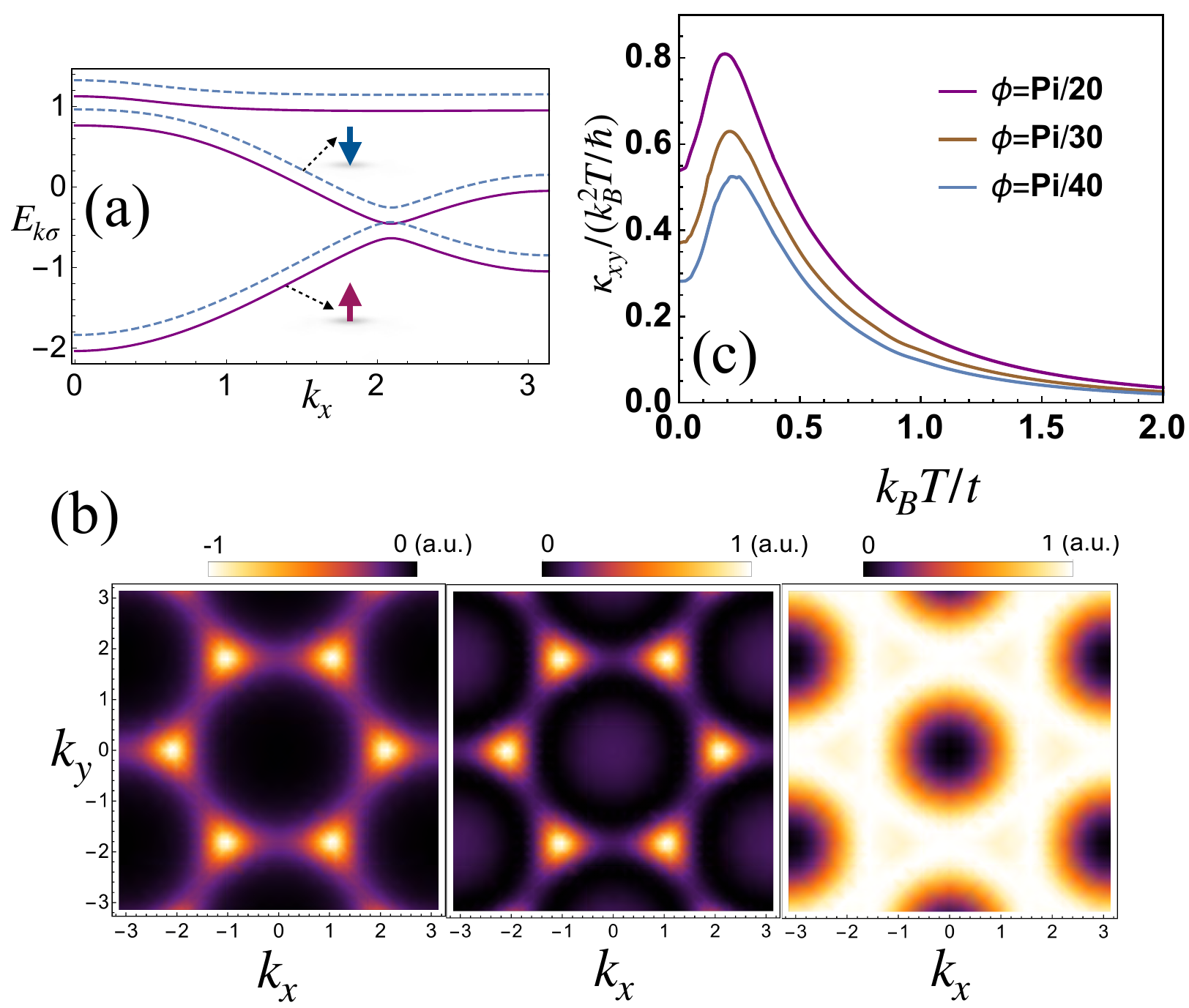} 
	\caption{
		(a) Spinon band structure with ${\phi=\pi/10}$. The solid (dashed) 
		lines are the bands for spin-$\uparrow$ (-$\downarrow$) spinons;
		(b) Density plot of the Berry curvature
		$\Omega_{n\boldsymbol{k}\sigma}$ of the lowest,  
		middle and highest bands for spin-$\uparrow$ 
		spinons with ${\phi=\pi/3}$;
		(c) The thermal Hall conductivity as a function of temperature.
		Figures are reprinted from Ref.~\onlinecite{Gao2020}.
	}
	\label{TTHE_fig1}
\end{figure}

This non-zero U(1) gauge flux renders finite Berry curvatures to the spinon bands 
and creates the distribution of the Berry curvature (see Fig.~\ref{TTHE_fig1}(b)) 
that is responsible for the thermal Hall conductivity $\kappa_{xy}$ 
from Eq.~\eqref{k_xy_fermi}. The temperature and flux dependence 
are plotted in Fig.~\ref{TTHE_fig1}(c)~\cite{Gao2020}. 
With a finite $\phi$, there exists a residual value for $\kappa_{xy}/T$ 
even in the zero temperature limit, indicating the existence of the spinon Fermi surface and gapless excitations.  
At finite temperatures, a non-monotonic line-shape is plotted in Fig.~\ref{TTHE_fig1}(c), 
which is consistent with the main feature of the experimental $\kappa_{xy}$ 
in the spin liquid regime. This result may be related to the clear thermal Hall 
signal observed in kagom\'e materials volborthite and kapellasite~\cite{Watanabe2016,Yamashita2018,Yamashita2020}, and could capture the gapless nature of the spin excitations. 
Again, the opposite signs of the thermal Hall conductivities in volborthite and kapellasite 
could arise from the opposite signs of the DM interaction 
that induces the internal U(1) gauge fluxes with the opposite signs.  
It is also necessary to stress that the U(1) gauge fluctuations 
in the gapless Fermi surface phase might result in significant 
corrections to the spinon thermal Hall conductivity~\cite{PhysRevB.101.195126}, 
which has not been considered in the formulation of Fig.~\ref{TTHE_fig1}(b).

\begin{figure*}[htbp]
	\centering
	\includegraphics[width=16cm]{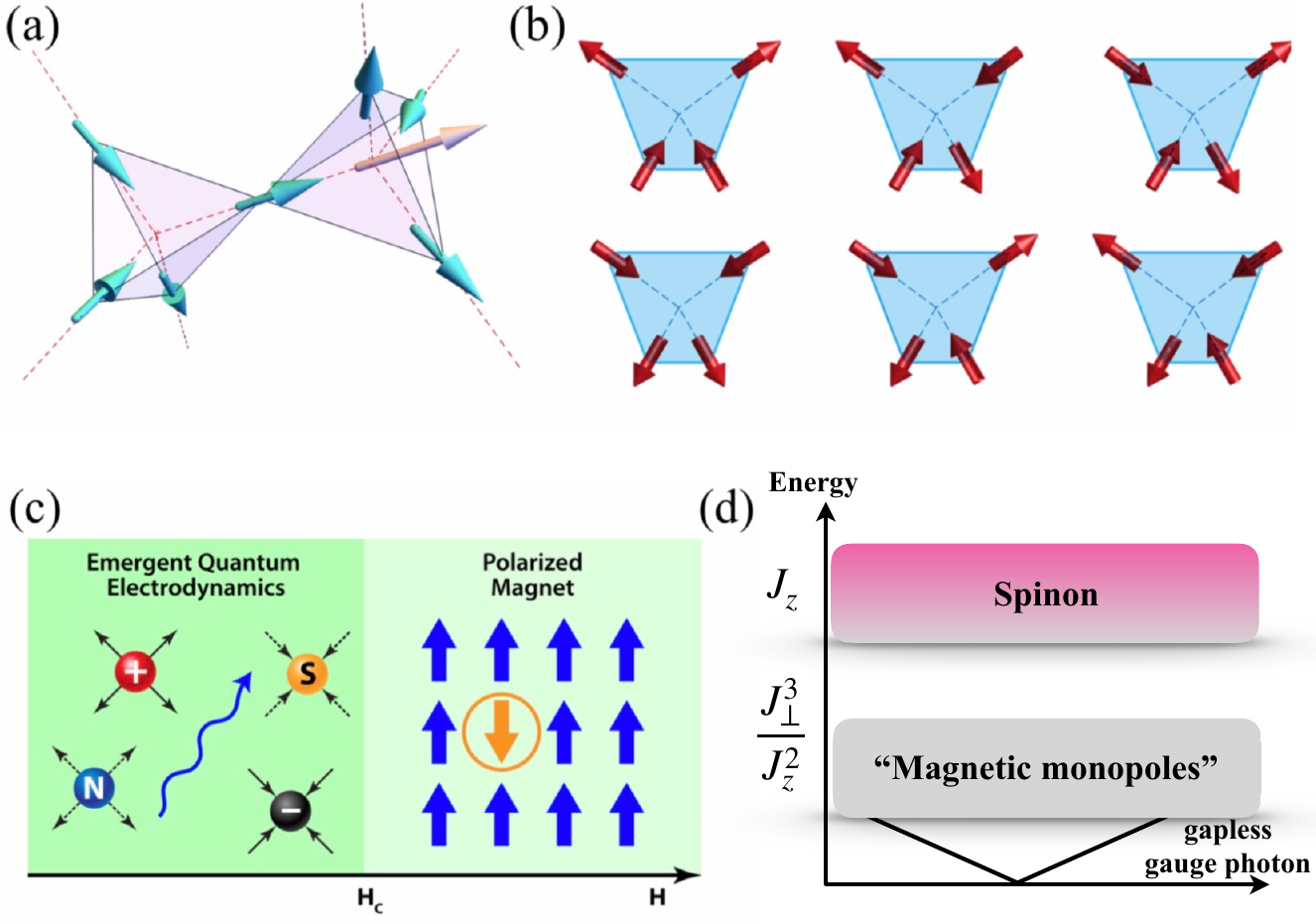}
	\caption{
	(a) Effective pseudospins on a corner-sharing tetrahedra, 
	which constitutes a particular configuration within the two-in-two-out spin ice manifold.
	(b) Spin configurations on a tetrahedra that obey the ``ice-rules''.
	(c) Proposed low-temperature phases of a quantum spin ice consist of 
	a magnetically polarized high-field phase with usual spin-flip excitations (shown at right) 
	and a low-field phase consisting of ``emergent'' dynamical photons (shown by a wavy line) 
	and electric and magnetic charges (shown schematically as plus and minus charges 
	with emanating electric fields and north (N) and south (S) poles with emanating magnetic fields).
	(d) Schematic of the spectrum of excitations in quantum spin ice including the approximate energy scales. 
	From low to high, the excitations are dubbed gapless photon, spinon, ``magnetic monopole'', respectively.
	Panel (b-c) are reprinted from Ref.~\onlinecite{Singh2011}.	}
	\label{QSI_fig1}
\end{figure*}

\section{Thermal Hall effects for 3D U(1) spin liquids in the pyrochlore magnets}
\label{sec5}


In this section, we move one spatial dimensional higher than Sec.~\ref{sec4}
and turn to the thermal Hall effects for the 3D U(1) spin liquids in the strong Mott regime. 
If one considers a 3D U(1) spin liquids with the gapless spinon matter such as the spinon Fermi surface, 
the underlying mechanisms for the spinon thermal Hall effects in the 3D bulk will most likely not bring 
more physical contents than the 2D cases, except the effects from the gauge fluctuations can be less
severe. In the special case of the 3D U(1) spin liquids with the 
topological band structures for the spinons~\cite{Pesin_2010,PhysRevX.6.011034}, 
the charge-neutral surface states could bring new ingredients 
for the surface thermal Hall transports that behave very much like the electric Hall transports of the 
surface Dirac electrons in the 3D topological band insulator. We will not give much discussion about 
the properties of these states. Instead, we devote this section to the thermal Hall effects for the 3D 
U(1) spin liquids in the context of the pyrochlore quantum spin ice that is probably more experimentally 
relevant at the current stage.


The rare-earth pyrochlore materials with the chemical formula $R_2$TM$_2$O$_7$ 
($R$: rare-earth elements, TM: transition metal elements) are promising candidate systems 
harboring the U(1) spin liquid of quantum spin ice~\cite{Canals2007,Tanaka2010,Savary2012,Balents2012}.
The interplay between the strong spin-orbit coupling and the crystal electric field  
on the rare-earth ions gives rise to the pseudospin-1/2 local moments that form    
a pyrochlore lattice~\cite{Tanaka2010,Huang2014,Balents2014,Gingras2014}. 
Because of the spin-orbit entangled nature of the relevant rare-earth ion, the 
relevant spin models to describe the interaction between the local moments are 
highly anisotropic~\cite{Gingras2014}. Depending on the microscopic origin and the symmetry properties 
of the local moments, there exist two types of realistic spin models~\cite{Tanaka2010,Ross2011,Huang2014}. 
One of the spin models applies to the conventional Kramers doublets~\cite{Ross2011,Savary2012}
and the non-Kramers doublets~\cite{Tanaka2010,Balents2012}.
For instance, the crystal field ground state of the Yb$^{3+}$ ion in 
Yb$_2$Ti$_2$O$_7$~\cite{Savary2012,Ross2009,Ross2011} 
belongs to the conventional Kramers doublets, and the Pr$^{3+}$ ion 
in Pr$_2$Zr$_2$O$_7$~\cite{Broholm2017}, Pr$_2$Ir$_2$O$_7$ 
and Tb$^{3+}$ ion in Tb$_2$Ti$_2$O$_7$~\cite{PhysRevB.83.094411,Canals2007,Gang2016,Gang2017D,PhysRevB.62.6496,PhysRevB.68.172407} 
belong to the non-Kramers doublets. 
The other model, known as the XYZ spin model~\cite{Gang2017C,Huang2014}, 
applies for the dipole-octuple doublets, such as 
the Ce$^{3+}$ ion in Ce$_2$Sn$_2$O$_7$ and 
Ce$_2$Zr$_2$O$_7$~\cite{Sibille2015,Gang2017C,Pengcheng2019,Yao2020,PhysRevX.12.021015,PhysRevLett.122.187201,PhysRevB.106.094425},
the Nd$^{3+}$ ion in Nd$_2$Zr$_2$O$_7$, Nd$_2$Sn$_2$O$_7$ and Nd$_2$Ir$_2$O$_7$~\cite{chen2022longrange,PhysRevLett.124.097203,PhysRevB.86.235129,PhysRevB.92.224430,Petit2016}, 
the Dy$^{3+}$ ion in Dy$_2$Ti$_2$O$_7$~\cite{Bertin_2012,Gang2016B}, 
the Er$^{3+}$ ion in the spinel compound MgEr$_2$Se$_4$
and the Yb$^{3+}$ ion in the spinel compound CdYb$_2$S$_4$~\cite{Gang2016B,PhysRevB.99.134438,PAWLAK1980195}
and even the $5d$ Os$^{4+}$ ion in Cd$_2$Os$_2$O$_7$~\cite{Gang2016B,PhysRevLett.108.247205}.
It is clear that both spin models can be reduced to the XXZ model in certain limits, 
and the XXZ model on the pyrochlore lattice could support the quantum spin ice    
U(1) spin liquid~\cite{Hermele2004B}. 
This 3D U(1) spin liquid is a stable phase and robust against local and weak perturbations. 
Although theoretical approaches are valid in the perturbative spin ice regime~\cite{Hermele2004B}, 
the stability of the pyrochlore U(1) spin liquid goes beyond the perturbative 
spin ice regime~\cite{Balents2012}. Therefore, we adopt the 
notion of the ``pyrochlore U(1) spin liquid''.

For the purpose to explain the origin of the thermal Hall transports in the pyrochlore U(1) spin liquid,
one needs to elucidate the emergent U(1) gauge structure in the pyrochlore U(1) spin liquid
and establish the relationship between the physical spin degrees of freedom and the emergent 
ones. In the following, we give a brief introduction of this physics. For a more thorough
discussion, one can refer to the discussion by \citet{Hermele2004B} and/or 
the reviews by \citet{Savary2016,Gingras2014}.
To capture the essential physics of the pyrochlore U(1) spin liquid, one can focus  
on the parent spin-1/2 XXZ model on the pyrochlore lattice with,   
\begin{equation}
{\cal H}=
\sum_{\langle i,j\rangle} [ J_{z}^{} S_i^z S_j^z-{J_{\perp}}(S_i^{+}S_j^{-}+S_i^{-}S_j^{+}) ], 
\label{QSI_Ham}
\end{equation}
where the $S^z$ moment is defined on the local $111$ axis 
(i.e. pointing in or out from the centers of each tetrahedron), 
and ${J_{z}>0}$ is needed to generate the frustration. 
In the Ising limit with ${J_{\perp} =0}$, 
the Ising moments are disordered due 
to the geometrical frustration, and an exponentially large number 
of classical ground state degeneracies were found~\cite{Anderson1956},
obeying the well-known ice rule with the ``two-in-two-out'' spin configuration 
on each tetrahedron~\cite{Ramirez1999,Castelnovo2008,Pauling1935,Bramwell2001}
(see the illustration in Fig.~\ref{QSI_fig1}). The transverse exchange  
in Eq.~\eqref{QSI_Ham} then generates the quantum fluctuation and lifts 
the massive degeneracy of the classical ground states. 
If the transverse exchange coupling is not strong enough to drive a quantum phase transition,
the system should be in the pyrochlore U(1) spin liquid.

The quantum effect introduced by the small $J_{\perp}$ can be readily 
taken into account by the standard third order degenerate perturbation 
theory. The effective Hamiltonian contains the ring exchange terms 
that live on the hexagonal loops on the pyrochlore lattice~\cite{Hermele2004B},
which is therefore named as the ring-exchange model. 
Such perturbative process is a bit more involved for the more realistic spin models. 
Hermele, Fisher, and Balents 
theoretically demonstrated the existence of a U(1) spin liquid phase adjacent to 
a soluble point from the quantum dimer model reasoning~\cite{Hermele2004B} 
and mapped the effective ring exchange model to a compact U(1) lattice gauge theory. 
Although the U(1) lattice gauge theory is constructed from the effective ring exchange model, 
it actually captures both the low-energy physics and the intermediate energy physics 
of the pyrochlore U(1) spin liquid. The U(1) lattice gauge theory is written 
on the diamond lattice formed by the tetrahedral centers of the pyrochlore lattice 
and takes the form
\begin{equation}
{\cal H}_{\rm LGT} =  -K  \sum_{{\hexagon_d}} 
\cos\big[ {\rm curl} A \big] 
+ \frac{U}{2}\sum_{ \langle {\boldsymbol r}{\boldsymbol r}^\prime \rangle} 
\big(E_{  {\boldsymbol r}{\boldsymbol r}^\prime  }-\frac{\epsilon_{\boldsymbol r}}{2} \big)^2  ,
\label{Zhang_LGT}
\end{equation}
where ${K\propto J_{\perp}^3/J_z^2}$, and a pair of conjugated gauge fields 
on the diamond lattice link ${\boldsymbol r}{\boldsymbol r}^\prime$ 
are introduced, {\sl i.e.} the electric field $E_{{\boldsymbol r}{\boldsymbol r}^\prime}$ 
and vector gauge potential $A_{{\boldsymbol r}{\boldsymbol r}^\prime}$
with ${[E_{{\boldsymbol r}{\boldsymbol r}^\prime},  A_{{\boldsymbol r}{\boldsymbol r}^\prime}] = -i}$,
 ${\epsilon_{\boldsymbol r} = \pm 1}$ for two different sublattices of the diamond lattice. 
Here ``$\hexagon_d$'' refers to the elementary 
hexagon on the diamond lattice formed by the tetrahedral centers of the pyrochlore lattice. 
The pyrochlore U(1) spin liquid is identified with an emergent U(1) gauge structure, that gives rise 
to several remarkable properties: there is the gapless emergent ``gauge photon'', the gapped spinons 
carrying the emergent ``electric'' charge, a gapped ``magnetic monopole'' carrying the emergent 
``magnetic'' charge, and an emergent $1/r$ ``Coulomb'' interaction between the pairs of spinons  
or magnetic monopoles (see Fig.~\ref{QSI_fig1}(c)). 
The two sets of commonly used terminology are clarified in Table.~\ref{notation}, 
where `notation 1' is used here. The gapless gauge photons, spinons and 
``magnetic monopoles'' are located at different energy scales as schematically 
depicted in Fig.~\ref{QSI_fig1}(d). Many interesting experimental signatures 
have been suggested, yet, the firm establishment of the pyrochlore U(1) spin liquid 
has not been settled. To resolve this task require 
a combination of many different experimental techniques on the candidate systems. 
In this section, we mainly explain the manifestation of the emergent U(1) gauge structure 
and the emergent quasiparticles from the thermal Hall transports.

\begin{table}[t]
\begin{tabular}{cc}
	\hline\hline
	Excitations (notation 1) & Excitations (notation 2) \\
	\hline
	Spinon & Magnetic monopole \\
	``Magnetic monopole'' & Electric monopole \\
	Gauge photon & Gauge photon \\
	\hline \hline 
\end{tabular}
\caption{Correspondence between two different notations for the elementary excitations
in the pyrochlore U(1) spin liquid. ``Magnetic monopole'' is sometimes referred as visons
in some literature. Usually ``vison'' refers to the $\mathbb{Z}_2$ magnetic flux~\cite{Fisher2000,Senthil2001,Senthil2001B} 
for the $\mathbb{Z}_2$ topological order in two dimensions~\cite{Kitaev2003}. }
\label{notation}
\end{table}

\begin{figure*}[t]
	\centering
	\includegraphics[width=17cm]{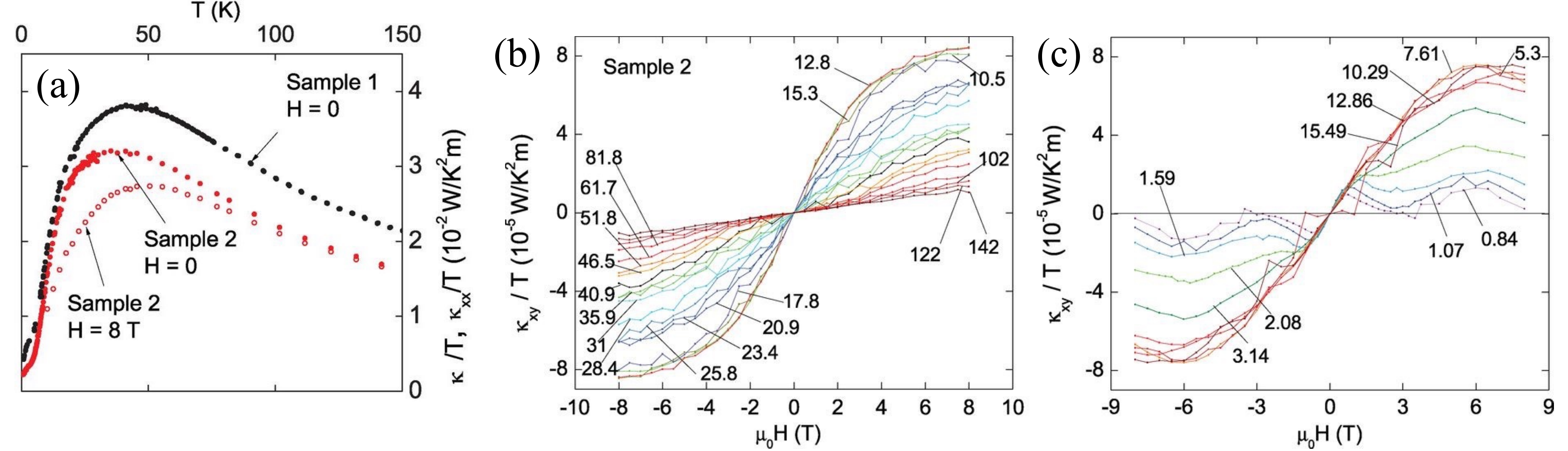}
	\caption{
	(a) The temperature dependence of the longitudinal thermal coefficient $\kappa/T$.
	(b,c) Thermal Hall conductivity $\kappa_{xy}/T$ versus magnetic field $H$;
	(b) From 140K to 50K, $\kappa_{xy}/T$ is $H$-linear. 
	Below 45 K, it develops pronounced curvature at large $H$, 
	reaching its largest value near 12 K. The sign is always holelike;
	(c) Below 15 K, the weak-field slope $[\kappa_{xy}/TB]_{B\rightarrow 0}$ is nearly temperature-independent. 
	Below 3 K, the field profile shows additional features that become prominent as $T\rightarrow 0$,
	namely, the sharp peak near 1T and the broad maximum at 6 T.
	Figures are reprinted from Ref.~\onlinecite{Hirschberger2015}.	}
	\label{QSI_fig3}
\end{figure*}

Although the thermal Hall transport in the pyrochlore magnets is not as popular 
as the Kitaev materials, there have already been some attempts~\cite{Hirschberger2015,PhysRevB.107.054429,Hirschberger2019}. 
Hirschberger et al. observed a large thermal Hall signal in the pyrochlore magnet 
Tb$_2$Ti$_2$O$_7$ below ${T=15}$K, which is clearly distinct from the conventional 
quasi-particle (such as magnons and phonons) dominated behavior~\cite{Hirschberger2015}.
In the high-$T$ regime (${T\ge 80}$K), the longitudinal thermal coefficient $\kappa_{xx}/T$
is independent to the magnetic field $H$. This is consistent with heat conduction dominated 
by phonons. In the low-temperature regime (below 1K), a fairly large $H$-dependence is observed, 
and the longitudinal thermal conductivity in Fig.~\ref{QSI_fig3}(a), is incompatible with the 
assumption that the magnon and/or the phonon are the main contributor. 
Moreover, the longitudinal thermal coefficient $\kappa_{xx}/T$ saturates to a constant value, indicating that the heat current 
is conveyed by the neutral, gapless fermionic excitations as suggested by
\citet{Hirschberger2015}. In Fig.~\ref{QSI_fig3}(b), the thermal Hall coefficient 
$\kappa_{xy}/T$ keeps growing with decreasing temperatures up to 15K, which coincides 
with the energy scale that the system is described by the effective pseudospin-1/2 doublets. 
With even lower temperatures in Fig.~\ref{QSI_fig3}(c), 
the thermal Hall conductivity $\kappa_{xy}/T$ shows a complicated 
nonlinear dependence on the magnetic field; the thermal Hall signal is suppressed by 
the magnetic field with a magnitude around 1T, then displays a broad peak around 6T.
This may lead to an impression that the low-temperature spin liquid state,
due to the strong quantum fluctuations, is readily suppressed by a large $H$.
This observation is further supported by the thermal Hall angle measurement
that excludes the possibility of the contribution from phonons or magnons. 
Motivated by this experiment and others, theorists started to investigate on the neutral, 
fractional excitation dominated thermal Hall effect in the pyrochlore U(1) spin liquid. 
In the following, we review the physics related to thermal Hall effects 
with the spinons and the ``magnetic monopoles'' by focusing on the context
of the pyrochlore U(1) spin liquid, respectively. 


\begin{figure*}[htbp]
	\centering
	\includegraphics[width=16cm]{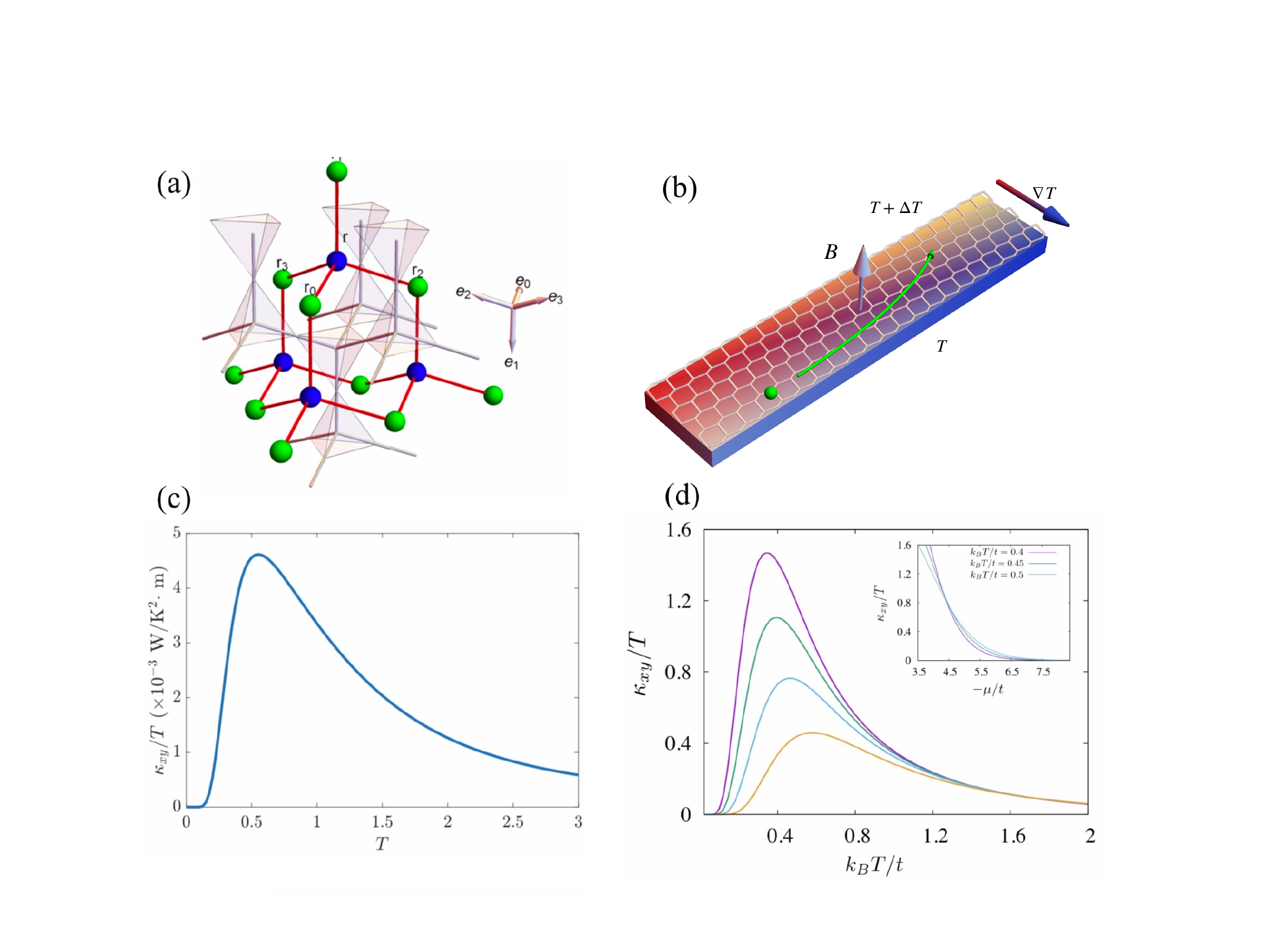}
	\caption{The ``magnetic monopole'' thermal Hall effects in the pyrochlore U(1) spin liquid. 
	(a) The diamond lattice (in gray line) formed by the tetrahedral centers of the original pyrochlore lattice, 
	and the dual diamond lattice 	(in red line). The physical spin is located in the middle of the link on the diamond 
	lattice.
	The dual diamond lattice is formed by the bond links that penetrate each hexagon plaquette of the original diamond lattice. 
	The spinons (``magnetic monopoles'') hop on the diamond (dual diamond) 
	lattice. The colored balls correspond to the position of  ``magnetic monopoles.''
	(b) Schematic picture of the thermal Hall effect from the ``magnetic monopoles'' 
	on the dual diamond lattice for the pyrochlore U(1) spin liquid. 
	We single out the ``magnetic monopoles'' (in green) that are suggested to contribute to 
	the thermal Hall effect in this work.
	(c) Numerical plot of the thermal Hall coefficient $\kappa_{xy}/T$ as the temperature $T$ changes.
	(d) The ``magnetic monopole'' thermal Hall coefficient $\kappa_{xy}/T$ 
	versus the temperature $k_B T/t$. Curves with different colors (from top to bottom) are plotted 
	with a decreasing sequence of chemical potential $\mu$.  
	The thermal Hall coefficient $\kappa_{xy}/T$ is 
	in a unit of $k_B^2/(2\pi \hbar a)\simeq 2.8\times10^{-4}$ W/(K$^2$m).
	Inset: The thermal hall coefficient $\kappa_{xy}/T$ is plotted 
	versus the chemical potential $-\mu/t$ for a set of temperatures.
	Panel (a,b,d) are reprinted from Ref.~\onlinecite{Zhang2020};
	Panel (c) is reprinted from Ref.~\onlinecite{Sungbin2020}.	}
	\label{QSI_fig2}
\end{figure*}

\subsection{Thermal Hall effect for spinons}
\label{secV1}

We start with the spinon thermal Hall effect. Before explaining the underlying physical mechanism, 
one can make some key understanding simply based on the energy scale and quantum ground state. 
The spinon is a much higher energy excitation than the gauge photons 
and the ``magnetic monopoles'', and is bosonic. 
To have the gapped bosonic excitations to contribute to the thermal Hall effect, one has to thermally 
excite these quasiparticles, and the required temperature would be around the energy scale of them. 
For the XXZ model in Eq.~\eqref{QSI_Ham}, the spinon energy scale is $\mathcal{O} (J_z)$ that
can be renormalized a bit by $J_\perp$. This is clearly a much larger energy scale.
At this temperature, the system is probably far from the spin liquid ground state, and 
the quantum coherence should be lost. Such a regime was dubbed ``thermal spin liquid"
that is similar to the low temperature
regime of classical spin ice~\cite{Savary2013} where the 
spinons are confined. How about the low temperature quantum coherent regime? 
Due to the large spinon gap, the thermally activated 
spinons are very few and are probably invisible in the thermal transports~\cite{doi:10.7566/JPSJ.87.064702}. 
From the above reasoning, the spinons should not contribute to the thermal 
Hall transport. Nevertheless, if one follows the spirit of calculating the thermal Hall
effects of spins in the finite-temperature paramagnet using the Schwinger boson construction~\cite{Lee2015}, 
one may still use the spinon formulation to calculate the thermal Hall response for the pyrochlore 
U(1) spin liquid materials at the finite temperatures. This task has been carried 
out by \citet{Sungbin2020}.

To generate the non-trivial Berry curvature for the spinon thermal Hall transport, 
as we have remarked in the previous section for 2D U(1) spin liquids, 
one could modify the U(1) gauge flux that is experienced by the spinons. 
For the pyrochlore U(1) spin liquids, however, the internal U(1) gauge flux 
for the spinons cannot be immediately or continuously tuned by weak 
external fields. This is different from the results of the previous section and also 
different from the ``magnetic monopole'' thermal Hall effect in the next subsection. 
One can actually see this from the XXZ model in Eq.~\eqref{QSI_Ham}. 
The background U(1) gauge flux on each hexagonal plaquette 
can either be 0 or $\pi$, and this depends on the ring exchange coupling or 
the flux coupling $K$ in Eq.~\eqref{Zhang_LGT}~\cite{Hermele2004B}. 
For the unfrustrated (${J_\perp >0}$) / frustrated (${J_\perp <0}$) case, 
the spinon experiences a background 0 / $\pi$ flux through 
the hexagonal palquette of the diamond lattice. The $\pi$-flux one can lead to the 
translation symmetry enrichment for the spinon spectrum~\cite{Balents2012,Gang2017E},
which has actually been suggested for the real materials, 
Ce-pyrochlore Ce$_2$Sn$_2$O$_7$ and 
Ce$_2$Zr$_2$O$_7$~\cite{Yao2020,PhysRevX.12.021015,Sibille_2020,Bhardwaj_2022,chen2023distinguishing,poree2023fractional}. 
Once the U(1) gauge flux pattern is determined by the microscopic spin interaction, it cannot 
be immediately modified by infinitesimal external magnetic fields. The more underlying reason
is that, the emergent magnetic field is {\sl even} under time reversal and does not couple linearly 
to the external magnetic field. This aspect is quite different 
from the case for ``magnetic monopoles'' in the next subsection.

Despite some level of the rigidity of the background U(1) gauge flux, 
a finite magnetic field would in principle modify the U(1) gauge flux and 
thereby twist the spinon motion. Ref.~\cite{Sungbin2020} was able to 
demonstrate that, the external magnetic field along the $[111]$-direction 
destabilizes the uniform $0$ flux (or $\pi$ flux) via the Zeeman coupling to the 
transverse spin components, 
and can continuously change the flux by generating 
additional ring exchange couplings when the field reaches a critical values. 
As a result, the spinons experience a complex U(1) gauge flux pattern. 
Around the temperature associated with the spinon gap,
this complex flux should lead to a spinon thermal Hall effect 
stimulated by the spinon band Berry curvature.
For the magnetic field along [110] direction, however, no fractional flux 
was obtained~\cite{Sungbin2020}.

To describe the spinon dynamics, one could directly insert the spinon
hopping on top of the lattice gauge theory in Eq.~\eqref{Zhang_LGT},
and the spinon hopping part is simply given as 
\begin{eqnarray}
{\mathcal H}_{\text{spinon}} = \sum_{{\boldsymbol r} ,{\boldsymbol r}'} \big[ 
 {-t_s} \Phi^\dagger_{\boldsymbol r} \Phi^{}_{ {\boldsymbol r}' } 
   e^{-iA_{{\boldsymbol r}{\boldsymbol r}'}} + h.c. \big],
\label{eqspinon3D}
\end{eqnarray}
where $\Phi^\dagger_{\boldsymbol r} $ ($\Phi^{}_{\boldsymbol r} $)
is the spinon creation (annihilation) operator at the diamond lattice site ${\boldsymbol r}$. 
In the Ising limit, the spinon is reduced to the defect tetrahedron that violates 
the spin ice rule. 
For the XXZ spin model in Eq.~\eqref{QSI_Ham}, the spinon hopping $t_s$ is 
approximately set by $J_{\perp}$. In addition to the minimal coupling
to the U(1) gauge field $A$, there is an energy penalty for changing the spinon density at the 
lattice site ${\boldsymbol r}$ due to the Ising coupling $J_z$.

The modified U(1) flux by the external magnetic field directly influences the vector gauge field $A$ in Eq.~\eqref{eqspinon3D}
and thereby generates the spinon Berry curvature.
Based on the spinon band structure, the thermal Hall coefficient $\kappa_{xy}(T)/T$ 
is evaluated~\cite{Sungbin2020}. 
Fig.~\ref{QSI_fig2}(c) shows the spinon thermal Hall conductivity $\kappa_{xy}/T$ as a function of temperature from the 
contribution of the
 thermally activated spinon bands,
and nonvanishing coefficient $\kappa_{xy}$ will signal the relevant evidence for the staggered flux phase as proposed 
by \citet{Sungbin2020} for the finite magnetic field. 
 As the temperature gradually increases, it grows from ${\kappa_{xy}/T=0}$ 
 towards the peak due to the thermal population of the non-trivial spinon bands.
The maximum thermal Hall coefficient is numerically evaluated to be 
$\kappa_{xy}/T\sim 4.6\times 10^{-3}\text{W}/(\text{K}^2\cdot\text{m})$ 
at low temperature $T<1\text{K}$, which is large and thus is expected to be accessible 
in experiments~\cite{Hirschberger2015,Ong2015,Kasahara2018,Shiyan2019,Hirschberger2019}.


\subsection{Thermal Hall effect for ``magnetic monopoles''}
\label{secV2}

The ``magnetic monopole" is the topological defect of the emergent vector gauge potential 
in the compact U(1) quantum electrodynamics.
Unlike the spinons that are connected to the defect tetrahedra
in the classical Ising limit, the ``magnetic monopoles'' are purely 
of quantum origin and have no classical analogue.  
The existence of the ``magnetic monopole" is one of the key properties 
of the compact U(1) lattice gauge theory in 3D~\cite{Fradkin2013} 
and the pyrochlore ice U(1) spin liquid~\cite{Hermele2004B}, 
and it is of great importance to demonstrate that this ``magnetic monopole" 
is a real physical entity rather than any artificial or fictitious excitation.
The existence of the ``magnetic monopole" continuum in the inelastic neutron 
scattering has been theoretically proposed in Ref.~\cite{Gang2017D}. 
Again, the wavefunction properties, however, are not revealed in the scattering experiments,
and thus it is demanding to think about the thermal Hall effect of the ``magnetic monopoles''.

As shown in Fig.~\ref{QSI_fig1}(d), the ``magnetic monopole" has a much lower
energy scale than the spinon excitations and overlaps partially with the gapless gauge photon. 
Thermal Hall effect for the ``magnetic monopoles'' in the pyrochlore U(1) spin liquid 
has been proposed by \citet{Zhang2020} and considered as 
a positive and direct evidence of the emergent U(1) gauge structure
and the coupling between the ``magnetic monopoles'' 
and the dual U(1) gauge field. 
The observation stems from the physical meaning of the microscopic spin variables 
in the pyrochlore U(1) spin liquid. 
It is observed that, the Ising component of the local moment, $S^z$, 
functions as an emergent electric field in the U(1) lattice gauge theory. 
The emergent electric field turns out to be the dual U(1) gauge flux for the 
``magnetic monopoles'', and thus, the ``magnetic monopoles'' could acquire
the Berry curvatures from the induced internal electric field. 
We have the following schematic flow of reasoning:
 \begin{eqnarray}
&&\text{Zeeman coupling to external magnetic field,}  
\\
&& \quad\quad\quad\quad \quad\quad\quad\quad {\Downarrow}  \nonumber 
\\
&&    \text{Finite magnetization } \langle S^z \rangle \neq 0 ,
 \\
&& \quad\quad\quad\quad \quad\quad\quad\quad {\Downarrow} \nonumber
 \\
&& \text{Modified emergent electric field distribution,} 
 \\
&& \quad\quad\quad\quad \quad\quad\quad\quad {\Downarrow} \nonumber
 \\
&&  \text{Induced dual U(1) flux distribution,}
 \\
&& \quad\quad\quad\quad \quad\quad\quad\quad {\Downarrow} \nonumber
 \\
&&   \text{Twisted ``magnetic monopole'' motion,} 
\\
&& \quad\quad\quad\quad \quad\quad\quad\quad {\Downarrow} \nonumber
\\
&&    \text{``Magnetic monopole'' Berry curvature,} 
\\
&& \quad\quad\quad\quad \quad\quad\quad\quad {\Downarrow} \nonumber
\\
&& \text{``Magnetic monopole'' thermal Hall effect.}
\end{eqnarray}

Unlike the spinon thermal Hall effect that requires a finite magnetic field 
to generate the internal U(1) gauge flux, an infinitesimal 
magnetic field is able to generate the thermal Hall effect of 
``magnetic monopoles'' by continuously tuning the internal dual U(1) gauge flux.  
From the above reasoning, one first writes down the Zeeman coupling to the external magnetic field, 
\begin{eqnarray}
{\cal H}_{\rm Zeeman}& = &  - h \sum_{i} ({\hat n} \cdot {\hat z}_i) S_i^z + \cdots \\
& \simeq & - h \sum_{\langle {\boldsymbol r}{\boldsymbol r}^\prime \rangle}  ({\hat n} \cdot {\hat z}_i)
({\rm curl}\ a -\bar{E}_{{\boldsymbol r}{\boldsymbol r}^\prime}) ,
\label{Zhang_Zeeman}
\end{eqnarray}
where $\hat{n}$ is the direction of the magnetic field, 
and the ``$\cdots$'' refers to the omitted coupling to the transverse components. 
For the non-Kramers doublets, as only the local $z$ component of the 
effective spin is odd under time reversal symmetry, the transverse coupling is absent automatically. 
The first line of Eq.~\eqref{Zhang_Zeeman} is written with the microscopic spin language
while the second line is expressed in terms of the emergent variables
for the pyrochlore U(1) spin liquid. 

Here, the $a$ field in Eq.~\eqref{Zhang_Zeeman} refers to the dual U(1) gauge field 
on the dual diamond lattice link as depicted in Fig.~\ref{QSI_fig2}(a).
The dual U(1) gauge theory can be obtained by performing 
an electromagnetic duality transformation on the original U(1) lattice gauge theory on the diamond lattice
of the tetrahedral centers. 
Similar to the spinon case, the dual U(1) charged particles, namely the ``magnetic monopoles'',
are implicit in the pure dual U(1) gauge theory and certain technical treatments are utilized to arrive at the desired 
full matter-gauge coupled theory~\cite{Hermele2004B,Gang2017D,Zhang2020,PhysRevB.73.134402,PhysRevB.71.125102}. 
The full dual theory describes ``magnetic monopoles'' hopping 
on the dual diamond lattice and coupling minimally with a dual U(1) gauge field,
which is written as 
\begin{eqnarray}
 {\cal H}_{\rm dual} &=
& - t\sum_{ \textsf{r} \textsf{r}^\prime} 
\Phi^\dagger_{\textsf{r}} \Phi^{}_{\textsf{r}^\prime} 
e^{-i 2\pi a_{\textsf{r}\textsf{r}^\prime}}
-\mu \sum_{\textsf{r}} 
\Phi^\dagger_{\textsf{r}} \Phi^{}_{\textsf{r}}
\nonumber \\
& + &  \sum_{{\boldsymbol r}{\boldsymbol r}^\prime} \frac{U}{2} 
({\rm curl}\ a-\bar{E}_{{\boldsymbol r}{\boldsymbol r}^\prime})^2 
- K \sum_{ \textsf{r} \textsf{r}^\prime}  
\cos B_{\textsf{r} \textsf{r}^\prime} ,
\label{Zhang_dual_ham}
\end{eqnarray}
where the first line describes the hopping of the ``magnetic monopoles'' on
the dual diamond lattice and minimally couples to the dual dynamical U(1)
gauge field, and the second line is the Maxwell term of the U(1) gauge field. 
Here the symbols ``$\textsf{r},\textsf{r}'$'' refer to the sites of the dual
diamond lattice and are distinguished from the diamond lattice sites ${\boldsymbol r},{\boldsymbol r}'$.
$t$ is the ``monopole'' hopping and is roughly set by the ring exchange energy scales. 
From this dual gauge theory, the emergent and internal electric 
field behaves as a dual U(1) gauge flux for the ``magnetic monopoles''.  
One can make a comparison with the spinon hopping and the coupling to the 
U(1) gauge field in Eq.~\eqref{eqspinon3D}.

The external magnetic field term in Eq.~\eqref{Zhang_Zeeman}
polarizes the internal and emergent electric field 
and modifies the dual U(1) gauge flux in Eq.~\eqref{Zhang_Zeeman}.
Internally, this corresponds to the induction of 
emergent dual U(1) gauge flux for the ``magnetic monopoles.'' 
As a result, the coupling between the internal variable and 
the external field effectively generates an {\sl emergent Lorentz force} on 
the ``magnetic monopoles'', and a thermal Hall effect 
is generated for the ``magnetic monopoles'' under the temperature gradient as illustrated in Fig.~\ref{QSI_fig2}(b).
Below the spinon gap, this is a direct manifestation and unbiased signature of 
the emergent ``monopole''-gauge coupling. 
This phenomenon serves as another interesting analog of the Lorentz force 
for the electron motion on the lattice, except that the Lorentz force 
here is emergent and arises from the induction of 
the internal dual U(1) gauge flux via the Zeeman coupling.

\begin{figure*}[t]
	\centering
	\includegraphics[width=14cm]{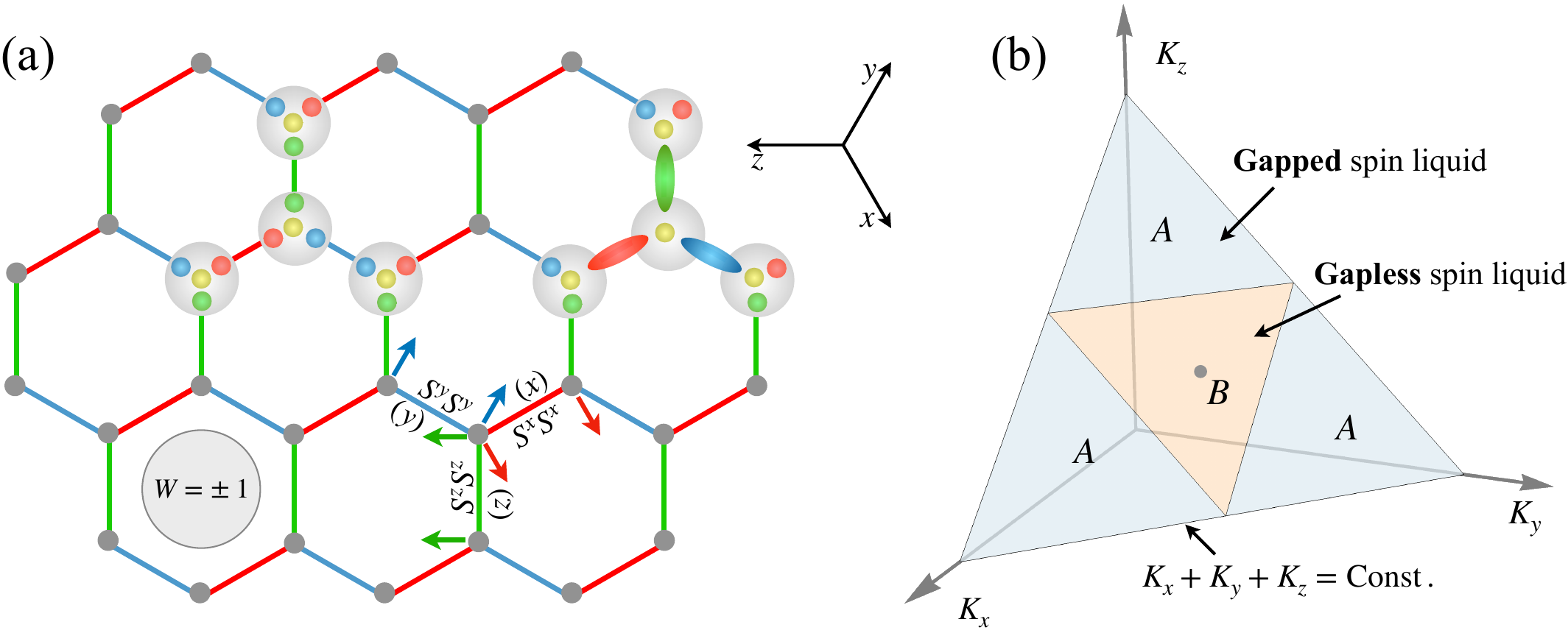}
	\caption{(a) The honeycomb lattice Kitaev model with the bond-directional couplings $K_x$ ($x$-link), $K_y$ ($y$-link) and $K_z$ ($z$-link). 
		The model can be analytically solved by introducing four flavors of Majorana fermions (indicated by the yellow, blue, green and red balls) 
		and recombining them into a static $\mb{Z}_2$ gauge field (indicated by the blue, green and red ovals) and a remaining gapless itinerant Majorana fermion 
		(yellow ball). $W$ represents a gauge flux of each hexagonal plaquette, and $W=-1$ is referred to as a $\pi$ flux excitation, while $W=+1$ indicates that there is no flux excitation. (b) Phase diagram of the Kitaev model plotted for a plane ${K_x + K_y + K_z = \rm{Const}}$. If one of the three couplings dominates,
		the system forms a gapped spin liquid (A Phase) indicated by the blue shading. Around the point of isotropic coupling strengths ${K_x = K_y = K_z}$ 
		(indicated by the gray dot) a gapless spin liquid (B Phase) emerges, which can be best characterized as a (semi-)metal of the Majorana fermions. Figures are reprinted from Ref.~\onlinecite{Trebst2022}.}
	\label{KitaevModel}
\end{figure*}

In the weak field regime, the U(1) spin liquid is assumed to be preserved and 
deconfinement is maintained, the ``magnetic monopole'' remains to be 
a good description of the low-energy magnetic excitation.  
The Zeeman coupling depends sensitively on the local crystal-field axis. 
Thus, the induced dual U(1) gauge flux depends 
on the lattice geometry and the field orientation.
For an arbitrary field strength, one can in principal
work out the dual gauge configuration experienced by the ``magnetic monopoles''~\cite{Zhang2020}
and with ignoring gauge fluctuation 
and treat the ``magnetic monopoles'' on the mean-field level.
The temperature dependence of the thermal Hall coefficient is shown in Fig.~\ref{QSI_fig2}(d) 
for a convenient choice of the external magnetic field.
With the increasing of temperature, 
the thermal Hall conductivity $\kappa_{xy}/T$ grows from zero and shows a non-monotonic behavior;
then, it drops to zero in high-temperature limit.

\subsection{Different sources in thermal Hall effects for different local moments}

We introduced the spinon and ``magnetic monopole'' thermal Hall effects from the more model and theoretical description.
The spinons would contribute to the thermal Hall conductivity $\kappa_{xy}$ 
when the temperature is relatively high to thermally activate the spinons. 
From the theoretical understanding, the ``magnetic monopole'' would be solely responsible for the thermal Hall effect 
at low temperatures below the spinon gap in most cases. 
We will explain the different cases below. 
As we have described in the beginning of this section, there exist three different types of local moments 
for these rare-earth pyrochlore magnets that are relevant for the pyrochlore U(1) spin liquids. 
It is a bit illuminating to clarify or summarize different sources in the thermal Hall effects for different local moments
in the pyrochlore U(1) spin liquids for more realistic systems. This will be useful for the experimental realization and the detection 
of different contributions in the actual materials.

We start from the conventional Kramers doublets. 
For the effective spin-1/2 local moment of the conventional Kramers doublets, all the three spin components are 
linearly coupled to the external magnetic fields. According to the results in Sec.~\ref{secV1} and Sec.~\ref{secV2}, 
both the spinons and the ``magnetic monopoles'' will contribute to the thermal Hall conductivity. 
There is, however, a large energy scale separation between the spinons and the ``magnetic monopoles''.
Thus, their contributions should appear at very different temperatures, assuming the spinon 
can still be a valid description of the magnetic excitations for the correlated paramagnetic regime. 
Moreover, they have rather different directional dependence on the external magnetic fields. 
These properties should be sufficient to differentiate their contributions. In fact, their contributions 
may also been revealed from or compared with the longitudinal thermal conductance measurements~\cite{doi:10.7566/JPSJ.87.064702} 
as they have different energy scales, 
except that the emergent gauge photons could complicate the interpretation of the low-temperature 
results due to the similar energy scale as the ``magnetic monopoles''.

For the effective spin-1/2 local moments of the non-Kramers doublets, 
only the Ising component $S^z$ is linearly coupled to the external magnetic field,
and the thermal Hall effect would be primarily contributed by the ``magnetic monopoles''. 
The spinon cannot be ``twisted'' by the magnetic field to generate the thermal Hall effect.
Both the spinons and the ``magnetic monopoles'' show up in the longitudinal thermal conductance,
but the signals should appear at rather separated temperature scales~\cite{doi:10.7566/JPSJ.87.064702}.
In addition, the gapless gauge photon should also contribute directly to 
the longitudinal thermal conductivity $\kappa_{xx}$~\cite{doi:10.7566/JPSJ.87.064702} 
around the same energy scale as the ``magnetic monopoles'' except that it remains active down to the lowest temperature. 
By comparing the thermal Hall conductance and the longitudinal thermal conductance, one can 
single out the visible contribution of the ``magnetic monopole'' in the thermal Hall conductance. 
In fact, the thermal Hall conductivity has been measured
for the pyrochlore U(1) spin liquid material Tb$_2$Ti$_2$O$_7$~\cite{PhysRevB.107.054429,Hirschberger2015}.
Although the Tb$^{3+}$ ion carries a non-Kramers doublet, the crystal field
gap to the excited crystal field states is relatively weak~\cite{PhysRevB.68.172407,Canals2007,PhysRevB.62.6496,PhysRevB.99.224407}, 
and this would lead to some complication
for the finite-temperature results.

 The effective spin-1/2 local moments of the dipole-octupole doublets bring 
 some complication for the understanding of the thermal Hall effect~\cite{Huang2014,Gang2017C}. 
 Here, there exists two symmetry enriched U(1) spin liquids. One is the dipolar
 U(1) spin liquid, and the other is the octupolar U(1) spin liquid. They are distinct 
 in the symmetry properties of the emergent fields and quasiparticles~\cite{Huang2014,Gang2017C}.
 For the dipolar U(1) spin liquid, the {\sl effective} Ising moment behaves as the 
 magnetic dipole moment and couples linearly with the external magnetic field. 
 By ``effective'' here, the Ising moment does not have to be the actual $S^z$ component. 
According to the explanation in Sec.~\ref{secV2}, 
 this Zeeman coupling generates the thermal Hall effect for the ``magnetic monopoles''. 
 For the effective transverse components of the local moment, there still exist 
 a linear coupling with the magnetic field. This coupling arises microscopically from the
 mixing term of ${J_{xz} (S_i^x S_j^z + S_i^z S_j^x)}$~\cite{Huang2014,Yao2020,Gang2017C}.
 Thus, the spinon thermal Hall effect should be expected, assuming the spinon 
can still be a valid description of the magnetic excitations for the correlated paramagnetic regime.
For the octupolar U(1) spin liquid, the effective Ising component is now the octupole moment 
and does not couple to the magnetic field linearly. 
As a result, the ``magnetic monopoles'' remarkably do not contribute to the thermal Hall effect.
This is the same reason that  ``magnetic monopoles'' and gauge photons do not directly show up in the 
inelastic neutron scattering (INS), only spinon continuum shows up in the INS and 
other magnetic probes~\cite{Yao2020,Gang2017C,chen2023distinguishing}. 
On the other hand, the magnetic field couples to the effective transverse spin component
and directly influences the spinon excitations and the spinon continuum. According to Ref.~\cite{Sungbin2020} and
the description in Sec.~\ref{secV2}, 
this transverse field coupling could in principle generate the spinon thermal Hall effect assuming
that one can still use the spinons to describe the magnetic excitations in the correlated paramagnetic 
regime.

\section{Thermal Hall effect for honeycomb Kitaev materials}
\label{sec6}

Kitaev spin liquids and Kitaev materials have emerged as a large field in recent years,
and the thermal Hall effect is particularly important in the study of these systems and the 
relevant quantum excitations. Very much like the rare-earth pyrochlore magnets in Sec.~\ref{sec5},
these systems have the rather anisotropic and bond-dependent spin interactions that arise from
the strong spin-orbit coupling of the heavy transition metal elements. 
Due to the nature of the extended $d$ electron orbitals from the transition metal ions, 
the energy scales of the couplings in these systems are usually much larger 
than the rare-earth ones. In the existing literature~\cite{Trebst2022,Hermanns2018}, 
these systems have been treated as the strong Mott insulators of Sec.~\ref{sec32}. 
Although we hold the point that many of these Kitaev magnets may be 
viewed as the weak Mott insulators~\cite{Gao2019,Chenunpub2023}, we will follow the conventional view 
of the strong Mott insulating regime for these systems in this review.

\subsection{Kitaev honeycomb model and Kitaev materials}

We begin with the Kitaev's exactly solvable model. 
Kitaev's honeycomb model appears to be an important spin model both for spin liquid theories 
and the experimental synthesis of spin liquid materials~\cite{Kitaev2006}. 
The original Kitaev model consists of the nearest-neighbor interactions 
between the usual SU(2) ${S=1/2}$ degrees of freedom.
The geometrical frustration is absent on the honeycomb lattice, instead, 
it is the presence of the bond-dependent Ising-like Kitaev interactions 
that induces strong quantum fluctuations and frustrates the spin orders, 
as shown in Fig.~\ref{KitaevModel}(a). The Hamiltonian is written as
\begin{equation}
\mc{H}_K = K_x\sum_{x~\rm{link}}S_i^x S_j^x 
                  +K_y\sum_{y~\rm{link}}S_i^y S_j^y +K_z\sum_{z~\rm{link}}S_i^z S_j^z,
\label{kitaevHam}
\end{equation}
which in fact belongs to the immense class of compass models. 
The Kitaev honeycomb model is \emph{exactly} solvable 
and its ground state supports both gapless and gapped $\mb{Z}_2$ spin liquids
depending on the relative strength of the Kitaev interactions along three different 
bonds [see Fig.~\ref{KitaevModel}(b)]. 
The emergent gauge sector, that glues the fractionalized particles together, 
is a deconfined $\mb{Z}_2$ gauge field~\cite{Fisher2001}
with gapped $\mathbb{Z}_2$ visons.  
The gapped phase of Kitaev model hosting the Abelian $\mb{Z}_2$ topological order is equivalent to 
the well-known toric code model~\cite{Kitaev2003} and gives the Abelian anyons, 
while the gapless one is a Majorana semi-metal with the Dirac-cone dispersion. 
In the presence of a weak out-of-plane 
magnetic field or other time-reversal symmetry-breaking perturbation,  
a gapped chiral spin liquid with gapped Ising anyons 
and non-Abelian statistics is obtained from the gapless state by opening up a bulk gap. 
For this remarkable state, the non-Abelian character is identical to 
that of the Moore-Read state related to the ${\nu = 5/2}$ fractional quantum Hall liquid, 
the ${p_x + ip_y}$ superconductor, 
and other intriguing physical systems proposed for realizing 
fault-tolerant topological quantum computation~\cite{Trebst2022}. 

It might be a bit illuminating to describe the parton construction for the honeycomb Kitaev model 
where the emergent and fractionalized spinons are understood. In the original exact treatment 
by A. Kitaev, the Majorana fermion bilinear representation for the spins was used and four flavors 
of Majorana fermions were adopted with the Hilbert space constraint~\cite{Kitaev2006}. 
The gapped and gapless $\mathbb{Z}_2$ spin liquids without the magnetic fields are 
understood from the spectral properties of one flavor of Majorana fermion. 
The gapped chiral spin liquid in the magnetic field is understood as an effective Haldane model 
for this flavor of Majorana fermion where a non-zero Chern number is obtained 
for the Majorana fermion bands. An equivalent understanding can be achieved 
with the conventional Abrikosov fermion representation of the spins~\cite{Nayak2011}. 
In fact, one can make a linear combination of the Kitaev's four Majorana fermions and 
relate to the two Abrikosov fermions. The spin liquid states in Kitaev's exact results
are re-interpreted as
the $p$-wave paired states of the Abrikosov fermions. In particular, 
the gapped chiral spin liquid in the magnetic field
is nothing but a $p+ip$ paired state, where all the understanding
of the $p+ip$ superconductivity can be applied.

Besides the prospect for topological quantum computation, 
 the gapped chiral spin liquid for the Kitaev model in the magnetic field supports 
a non-zero Chern number ${\cal C}$ for the filled Majorana fermions,
and thus, the gapless chiral Majorana edge mode. Like the Chern 
insulator of the electrons where there exists a quantized Hall conductance
due to the edge mode, here the Majorana mode is charge neutral,
and Kitaev further predicted the half-quantized thermal Hall effect with 
\begin{equation}
\frac{\kappa_{xy}}{T} = \frac{1}{2} {\cal C} K_0,
\label{KSLkxy}
\end{equation}
where ${K_0=\pi k_B^2/6\hbar}$ is a constant representing the thermal conductivity quanta 
and $ {{\cal C} =1}$. Although this result is not really the major result in Kitaev's work, 
it remarkably provides the smoking gun evidence of Kitaev spin liquid and is proved to 
be significant by the later experiments as we will discuss.

\begin{figure*}[htbp] 
	\centering
	\includegraphics[width=17cm]{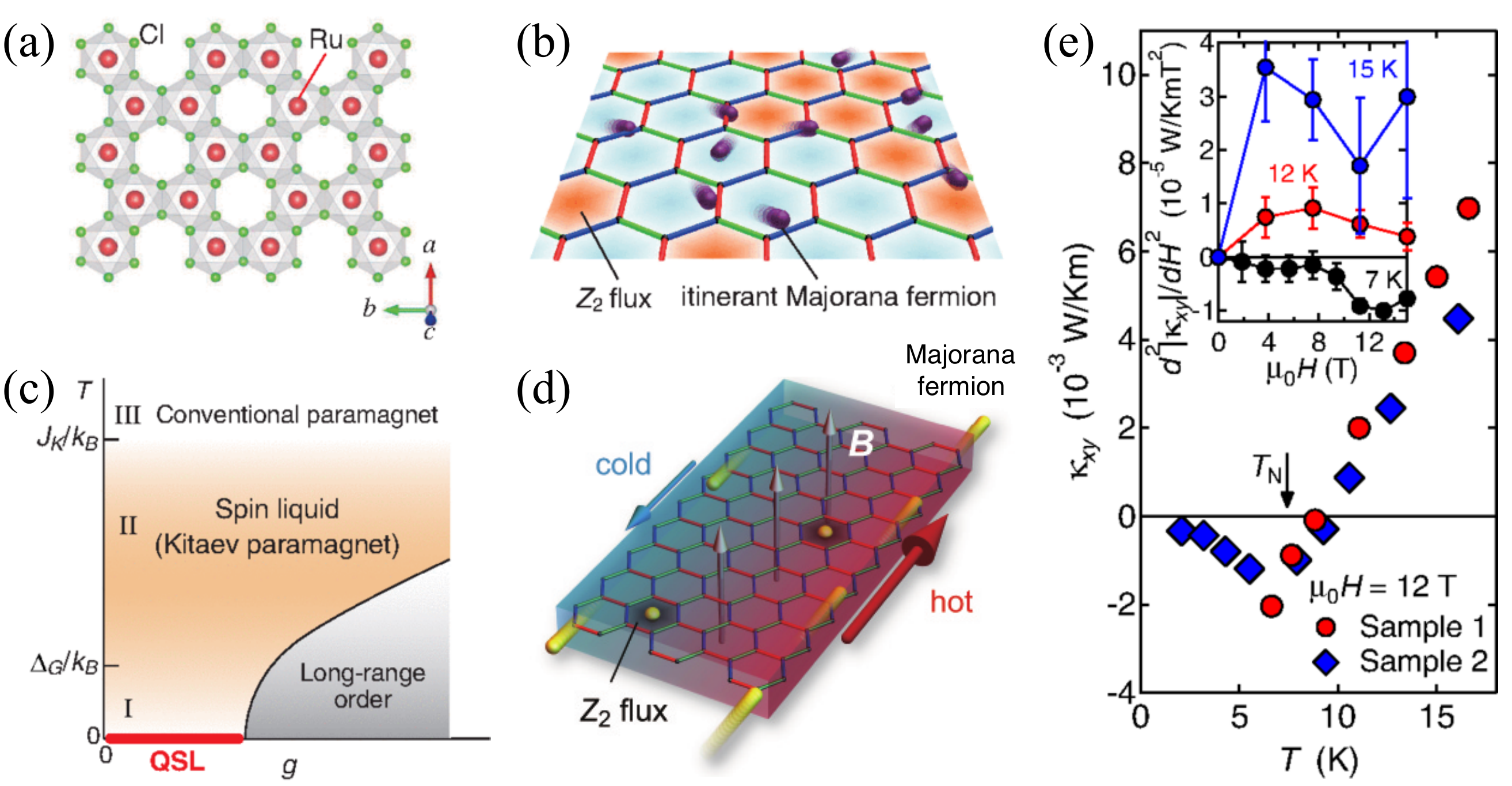} 
	\caption{	
	     (a) Crystal structure of $\alpha$-RuCl$_3$ in the $ab$ plane.
		(b) Schematic of Kitaev model on honeycomb lattice. Blue, green, and red bonds 
		      represent Ising-like bond-directional interactions at the apexes of the hexagons. 
		      The quantum spins are fractionalized into gapless itinerant Majorana fermions (purple spheres) 
		      and $\mathbb{Z}_2$ fluxes with $W_p=\pm 1$ (orange for $W_p=-1$).
		(c) Schematic phase diagram for the 2D Kitaev model as a function of $g$, 
		      where $g$ is the ratio of the Kitaev interaction $J_K$ and non-Kitaev interactions, 
		      such as Heisenberg and off-diagonal terms (in the pure Kitaev model, $g=0$).
		(d) Schematic illustrations of heat conductions in the Kitaev spin liquid state with magnetic field 
		      applied perpendicular to the planes (gray arrows). In the red (blue) regime, the temperature is higher (lower). 
		      The red and blue arrows represent thermal flow. The heat is carried by chiral edge currents of charge neutral Majorana fermions.
		(e) Temperature dependence of $\kappa_{xy}$ near $T_N$. 
		Inset shows the field dependence of $d^2|\kappa_{xy}(H)|/dH^2$ below and above $T_N$.
		Panels (a,b,c,e) reprinted from Ref.~\onlinecite{Matsuda2018B};
		Panel (d) reprinted from Ref.~\onlinecite{Matsuda2018}.
	}
	\label{KSL_fig2}
\end{figure*}

Unlike the Kitaev's toric code model that is also exactly solvable but involves the four-spin interactions~\cite{Kitaev2003}, 
the Kitaev's honeycomb model only involves the pairwise spin interactions that are more 
physical. Jackeli and Khaliullin~\cite{Jackeli2009} soon suggested its material relevance 
in a family of spin-orbit coupled Mott insulating iridates with the honeycomb geometry, and due to the
three-fold rotational symmetry demands three Kitaev couplings are equal with
\begin{eqnarray}
J_K \equiv K_x = K_y = K_z. 
\end{eqnarray}
  Their suggestion inspired the search for the Kitaev model in real materials, 
which is often referred as Kitaev materials~\cite{Trebst2022,Hermanns2018}.
Prominent examples are iridates family with spin-orbit entangled moments,
for instance, A$_2$IrO$_3$ with A = Na, Li~\cite{Singh2010,Singh2012},
and $\alpha$-RuCl$_3$~\cite{Sandilands2015,Nasu2016,Banerjee2017}. 
The theoretical and numerical studies laid out the essential ingredients 
for the realization of the Kitaev model, motivating the experimental search 
for Kitaev spin liquid and Majorana fermion.
The progress made along the line of Kitaev spin liquid and spin-orbit entangled physics are covered by many nice articles~\cite{Balents2014,Rau2016,Hermanns2018,Knolle2019,Takagi2019,Trebst2022}.

In fact, Kitaev materials go beyond iridates and ruthenates and have been extended
to $4f$ rare-earth magnets and $3d$ cobaltates~\cite{Feiye2017,Rau2018,Motome2019,PhysRevB.97.014407,PhysRevB.97.014408}. 
What gives the Kitaev interaction is the strong spin-orbit coupling, 
which is common to magnetic materials with heavy atoms. 
Other Mott insulators with 
spin-orbit-entangled effective spin-1/2 moments and a proper lattice geometry 
can also be Kitaev materials~\cite{Feiye2017}.  
Despite the growing list of Kitaev materials, all these systems face one crucial issue.
The real materials contain many competing interactions that can be as important as the Kitaev interaction, 
and other spin liquid states beyond Kitaev spin liquids can be relevant~\cite{Gang2019,song2022translationenriched}. 
Since vast experimental results especially that about the thermal Hall transport measurements
were conducted on the representative honeycomb Kitaev materials $\alpha$-RuCl$_3$, 
we will focus most discussion on $\alpha$-RuCl$_3$ and 
these discussion may be well extended to other systems and models.

\subsection{Prominent honeycomb Kitaev material $\alpha$-RuCl$_3$}
\label{RuCl3}

The $\alpha$-RuCl$_3$ compound is a $4d$ magnet with weakly coupled layers of 
the edge-sharing RuCl$_6$ octahedra, and the Ru$^{3+}$ ($4d^5$) ions 
form an almost ideal honeycomb lattice  
[see Fig.~\ref{KSL_fig2}(a)]. The study on $\alpha$-RuCl$_3$ traces   
back to the 70's, and it was considered to be a conventional semiconductor 
according to the transport measurement~\cite{Binotto1971}.
In the 90's, the spectroscopic measurement suggested the formation of 
a Mott insulator~\cite{PhysRevB.53.12769}. The angle-resolved photoemission spectroscopy
 confirmed that $\alpha$-RuCl$_3$ is indeed a spin-orbit-coupled 
Mott insulator with ${j_{\rm eff}=1/2}$~\cite{Plumb2014}.
The bond-dependent exchange qualifies $\alpha$-RuCl$_3$ for the Kitaev material
where a large Kitaev interaction (with ${J_K \simeq 80}$K) is 
present~\cite{Rau2014,Valenti2017}. Below the critical temperature ${T_N\simeq 7.5}$K, 
however, $\alpha$-RuCl$_3$ shows a zigzag antiferromagnetic order~\cite{Coldea2015}.

 Despite the low-temperature magnetic order, the spin liquid regime was still proposed for 
$\alpha$-RuCl$_3$ and was suggested to have the bounded characteristic temperatures 
${T_N<T< J_K}$ where this finite-temperature magnetic properties may be understood 
from the proximity to the gapless Kitaev spin liquid. The putative phase diagram subtended 
by the Kitaev and non-Kitaev interactions is depicted in Fig.~\ref{KSL_fig2}(c). 
In the spin liquid regime, the local spins fractionalize into itinerant and gapless Majorana 
fermions in the bulk and the gapped $\mb{Z}_2$ gauge flux. 
 Several experimental features 
may be interpreted as the consequences of the fractionalization.
These include, (i) a double peak shows up in the temperature dependence of the specific 
heat~\cite{Do2017,Loidl2019}, which is regarded as a two-step release of the magnetic 
entropy through the gapless Majorana fermions and the gapped fluxes~\cite{Yoshitake_2016}. 
 (ii) Raman scattering detects unusual magnetic scattering
 with a broad continuum~\cite{Sandilands2015}. 
 Instead of the monotonic decrease in temperature
 due to the thermal Bose factor for the conventional phonon-damped continuum,
the magnetic scattering here does not change appreciably within 
a large range of temperatures. 
Scattering with the fractionalized excitations is therefore an appealing interpretation
for the magnetic continuum. The fermionic nature of the Majorana spinons 
is suggested to be related to the temperature dependence 
of the Raman scattering intensity~\cite{Nasu2016}.
 Though quite striking, certain level of ambiguity is present in the data analysis
where the bosonic background contribution (from phonons) to the scattering intensity is subtracted.
(iii) The INS experiments~\cite{Banerjee2016,Banerjee2017,Do2017,PhysRevB.100.060405} 
seem to resolve the ambiguity in the Raman scattering. 
Above ${T_N}$, a broad magnetic continuum is found at high energy
and is interpreted as the consequence of fractionalization~\cite{Banerjee2016}.
The integrated neutron scattering data in certain energy window 
shows a star-like feature~\cite{Banerjee2017}
which is reproduced by the numerical study of Kitaev-Heisenberg model~\cite{Pollmann2017}.
It was also proposed that the observed continuum can represent the
incoherent excitations from strong magnetic anharmonicity 
that naturally occurs in such materials~\cite{Winter2017}.
This scenario seems to explain the observed INS spectrum of $\alpha$-RuCl$_3$.

 Thermal transport measurement is expected to be a sensitive way to detect
the itinerant Majorana spinons at low energies. The experiments were carried out 
with magnetic field slightly tilted out-of-plane direction~\cite{Minhyea2017,Hess2018}.
With a moderate field, the magnetic order is suppressed and
the system may be driven into the gapped Kitaev spin liquid with the non-Abelian character.
The longitudinal thermal conductivity $\kappa_{xx}$ shows a dramatic enhancement~\cite{Minhyea2017}
which onsets at the field-induced phase transition.
The temperature and field dependence of $\kappa_{xx}$
shows a peculiar and complex behavior,
which excludes the conventional contribution from phonons and magnons.
 The compelling scenario is then from the fractionalized excitations in the proximate Kitaev spin liquid.
A later study~\cite{Hess2018} observed similar behaviors, 
but attributed the unusual field dependence
to the phonon scattering off the exotic excitations proximate to Kitaev spin liquid.
Despite the difference,
the unusual features of $\kappa_{xx}$ and certain role of fractionalized excitations 
are in common. A recent study~\cite{Matsuda2018B} confirmed the $\kappa_{xx}$ 
results and proposed that both spins and phonons contribute to the heat current.
The field dependence comes from both, which are, however, difficult to decipher 
due to the lack of detailed spin-phonon coupling.

\begin{figure}[htbp] 
	\centering
	\includegraphics[width=8.6cm]{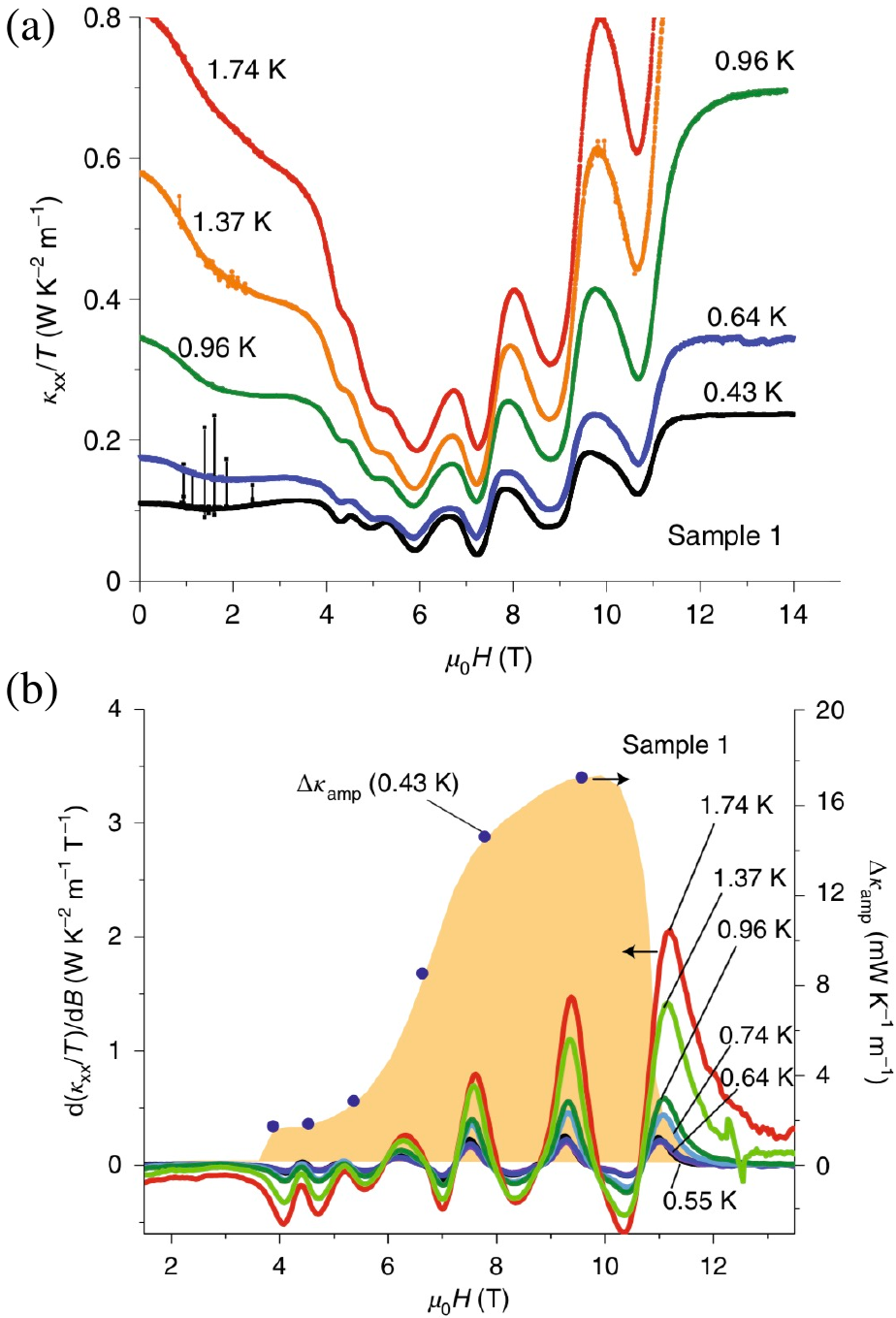} 
	\caption{Quantum oscillations in the spin liquid regime of $\alpha$-RuCl$_3$. 
	(a) The oscillations of the longitudinal thermal conductivity $\kappa_{xx}/T$ over the full field range at the selected temperatures (colored curves). 
	The data were recorded continuously, as well as with the stepped-field method. 
	At ${\sim 11}$T, the longitudinal thermal conductivity $\kappa_{xx}$ displays a step increase to a plateau-like profile 
	in the polarized state in which oscillations are strictly absent. 
	(b) The derivative curves $d(\kappa_{xx}/T)/dB$ for a range of temperatures (colored curves) 
	show that the oscillations onset abruptly at 4T. 
	The large derivative peak centered at  ${\sim 11.3}$T corresponds to the step increase 
	in $\kappa_{xx}$ mentioned, 
	and is not part of the oscillation sequence. Arrows indicate the relevant axes for the quantity plotted. 
	The amplitude $\Delta\kappa_{\rm amp}$ (solid circles) is strikingly prominent in the spin liquid regime. 
	Its profile (shaded orange) distinguishes the spin liquid from adjacent phases. 
	A weak remnant tail extends below 7T to 4T in the zigzag state. Figure adapted from Ref.~\cite{Ong2021}.}
	\label{kxxOscillation}
\end{figure}

In the even lower temperature regime ${T\simeq 0.4}$K, the longitudinal thermal 
conductivity of $\alpha$-RuCl$_3$ shows an interesting oscillation within the field 
range where the spin liquid is expected~\cite{Ong2021}. 
At larger fields, the oscillation amplitude displays a step increase to a flat plateau.
Eventually for the high-field polarized state, $\kappa_{xx}$ is dominated
by the phonon conductivity, and the oscillation feature disappears.
As the temperature is raised above $0.5$K, the oscillation amplitude decreases exponentially.
The field period in $\kappa_{xx}$ presents an intriguing analogy with the Shubnikov de Haas (SdH)
oscillation, despite the lack of free electrons in the layered Mott insulator $\alpha$-RuCl$_3$. 
This oscillation may be explained by a field-induced 
gapless spin liquid with spinon fermi surface~\cite{Motrunich2005,PhysRevB.97.045152}, 
which has been anticipated by theories and numerical calculations 
in $\alpha$-RuCl$_3$~\cite{Liu2018,Hickey2019,Patel2019,Jiang2018,Fujimoto2019}.
We will turn to the possibility of non-Kitaev spin liquids 
in this Kitaev material later and discuss its physical consequences.

\subsection{Thermal Hall effect of $\alpha$-RuCl$_3$}

Due to the complicated contributions in the longitudinal thermal conductance $\kappa_{xx}$, \citet{Matsuda2018B} further
conducted the thermal Hall measurement that {\sl may} effectively avoid the phonons
and unveil the nature of spin contribution.
The schematic picture of the heat conduction in 2D honeycomb layer
is illustrated in Fig.~\ref{KSL_fig2}(d). The finite thermal Hall conductivity $\kappa_{xy}$
is clearly resolved and the value is three orders of magnitude smaller
than the longitudinal thermal conductance $\kappa_{xx}$, and the sign of the
thermal Hall conductivity $\kappa_{xy}$ is 
changed from negative to positive upon crossing
the N\'eel temperature $T_N$ as shown in Fig.~\ref{KSL_fig2}(e). 
This peculiar temperature dependence is very different 
from the one for the phonons~\cite{PhysRevLett.95.155901,Yamashita2017}. 
Moreover, $d^2|\kappa_{xy}|/dH^2$ is negative (positive) 
below (above) $T_N$ and the temperature dependence
shows a distinct scaling behavior [see Fig.~\ref{KSL_fig2}(e)].
The temperature dependence of the thermal Hall coefficient
$\kappa_{xy}/T$ that was discovered in 
Ref.~\cite{Matsuda2018B} is mostly consistent with theoretical prediction
based on the Kitaev spin liquid~\cite{Motome2017}
except the quantization in the zero temperature limit.
As plotted in Fig.~\ref{KSL_fig2}(c), there are three distinct temperature regimes
in the phase diagram.
The quantization of 
the thermal Hall coefficient $\kappa_{xy}/T$ for the Majorana fermions is expected 
to occur for temperatures below the vison gap with ${T<\Delta_G}$ with $\Delta_G$
the vison gap.
Since in the intermediate temperature regime ${\Delta_G< T< J_K}$
the thermal excitation of the $\mathbb{Z}_2$ fluxes or visons would compromise
the quantization into a broad peak in the temperature domain.
In the high temperature, $\kappa_{xy}/T$ eventually goes to zero
in the uncorrelated paramagnetic regime.

\begin{figure}[htbp]
	\centering
	\includegraphics[width=8.6cm]{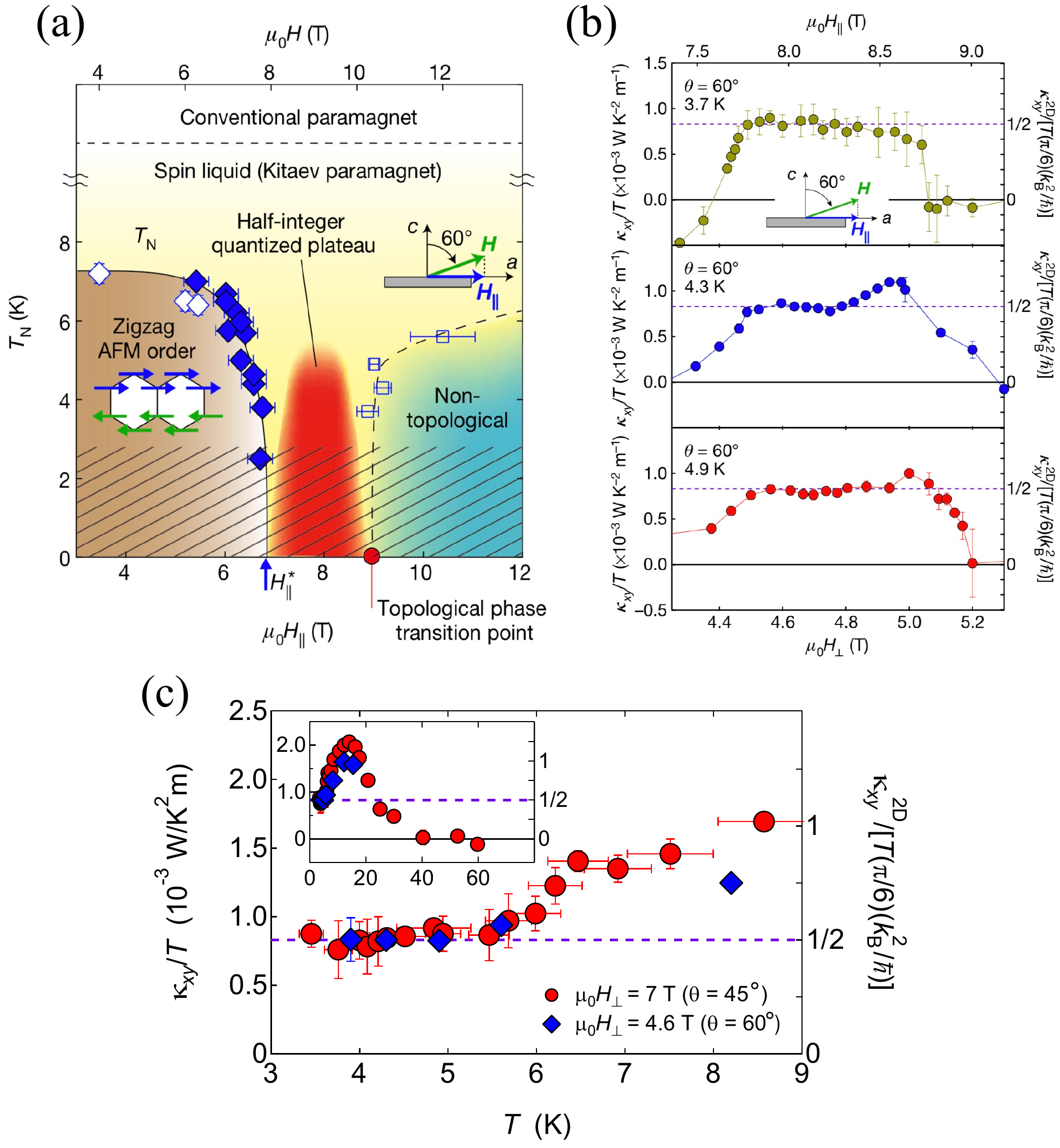}
	\caption{(a) Phase diagram of $\alpha$-RuCl$_3$ in tilted field of ${\theta=60^\circ}$. 
	Open and closed diamonds represent the onset temperature of zigzag order. 
	Below ${T\sim J_K}$, the spin liquid (Kitaev paramagnetic) regime appears. 
	At $\mu_0H_\parallel^\ast\sim7$\,T, $T_N$ vanishes (blue arrow). 
	A half-integer quantized plateau of 2D thermal Hall coefficient $\kappa_{xy}/T$ is observed in the red regime. 
	Open blue squares represent the fields at which the thermal Hall response disappears.  
	Red circle indicates a topological phase transition that separates  
	the non-trivial spin liquid state with  
	topologically protected chiral Majorana edge currents and a trivial state;
		(b) $\kappa_{xy}/T$ in tilted field of ${\theta=60^\circ}$ 
		plotted as a function of $H_\perp$ at low temperatures. 
		The right scales represent the 2D $\kappa_{xy}/T$ 
		in units of $\pi k_B^2/6\hbar$. 
		Violet dashed lines represent the half-integer thermal Hall conductance;
		(c) The temperature dependence of $\kappa_{xy}/T$. The main panel shows $\kappa_{xy}/T$ 
		in tilted fields of ${\theta=45^\circ}$ and $60^\circ$ at  ${\mu_0H_{\perp}=7}$T and 4.6T, respectively, 
		at which the quantized thermal Hall conductance plateau is observed at low temperatures. 
		The right scale and violet dashed lines are same as in (b). Inset shows the same data in a wider temperature regime.
		Figures are reprinted from Ref.~\cite{Matsuda2018}.}
	\label{KSL_fig1}
\end{figure}

\subsubsection{Half-integer quantized thermal Hall effect}

The breaking result was reported by Matsuda's group that a half-integer 
quantized thermal Hall conductivity $\kappa_{xy}/T$ is observed in the crystalline samples 
of $\alpha$-RuCl$_3$~\cite{Matsuda2018}. The temperature-magnetic field 
(perpendicular component) phase diagram is plotted in Fig.~\ref{KSL_fig1}(a).
In the low temperature regime, $\kappa_{xy}/T$ displays a half-integer quantized 
plateau within intermediate field regime where the gapped non-Abelian Kitaev spin liquid 
was expected, and this exotic state sets in the intermediate fields 
between the zigzag order in low field limit and a trivial polarized state at high fields.
As shown in Fig.~\ref{KSL_fig1}(b), $\kappa_{xy}/T$ is tiny in the zigzag ordered phase, 
and increases rapidly upon entering the field-induced Kitaev spin liquid.
The most striking feature is the presence of a near half-integer plateau 
with the intermediate field strength at low temperatures.
The plateau starts to disappear at a characteristic temperature [see Fig.~\ref{KSL_fig1}(c)] 
that approximately corresponds to the vison gap.
At higher temperatures, $\kappa_{xy}/T$ reaches a maximum then drops to zero,
(see the inset of Fig.~\ref{KSL_fig1}(c)). The presence of the quantized plateau 
is inertial to the field tilting angle 
and can be robustly reproduced in different samples.

The observation of half-integer thermal Hall conductance reveals 
that the topologically protected chiral Majorana edge current is likely 
to persist in $\alpha$-RuCl$_3$, and has sparked extensive theoretical 
studies~\cite{Rosch2018,Balents2018,Moore2018,Gao2019,Gao2020B,Hwang2022,Fuchun2020}. 
It was pointed out~\cite{Rosch2018,Balents2018} that the gapless acoustic phonons 
is unavoidably intertwined with the spins in $\alpha$-RuCl$_3$.
In contrast to the quantized electrical Hall conductivity, 
$\kappa_{xy}/T$ is never exactly quantized in real materials.
The coupling to the phonons destroys the ballistic thermal transport of the edge mode completely.
The heat can leak into the bulk, thus drastically modifying the edge picture of the thermal Hall effect.
The systematic investigation on the phonon effect concludes that the thermal Hall coefficient
$\kappa_{xy}/T$ remains approximately quantized, and counter-intuitively, 
an appropriate amount of phonon coupling to the edge mode is a necessary condition 
for the observation of the quantized thermal Hall effect~\cite{Rosch2018}. 
If the coupling is too strong (tuned by the magnetic field), the phonon dissipation 
channel broadens the field-induced gap.
This may close the Majorana fermion gap that protects the chiral edge current,
thus, the quantitated plateau of $\kappa_{xy}/T$ is sabotaged.
The coupling strength is reflected in the field dependence of the longitudinal thermal 
conductance $\kappa_{xx}$~\cite{Yamashita2020B}.
With the larger $\kappa_{xx}$, and thus the better suppressed magnetic 
scattering effects on the phonon thermal conduction, 
a clearer signal for the quantized thermal Hall effect shows up. 
The follow-up study~\cite{Yamashita2020B} reports a sample dependence of 
$\kappa_{xx}$ and $\kappa_{xy}$, as well as the relation between them;
thereby, it gives some confidence in the reproducibility of 
the half-integer quantized thermal Hall effect in samples with certain features.

\begin{figure}[htbp]
	\centering
	\includegraphics[width=8.6cm]{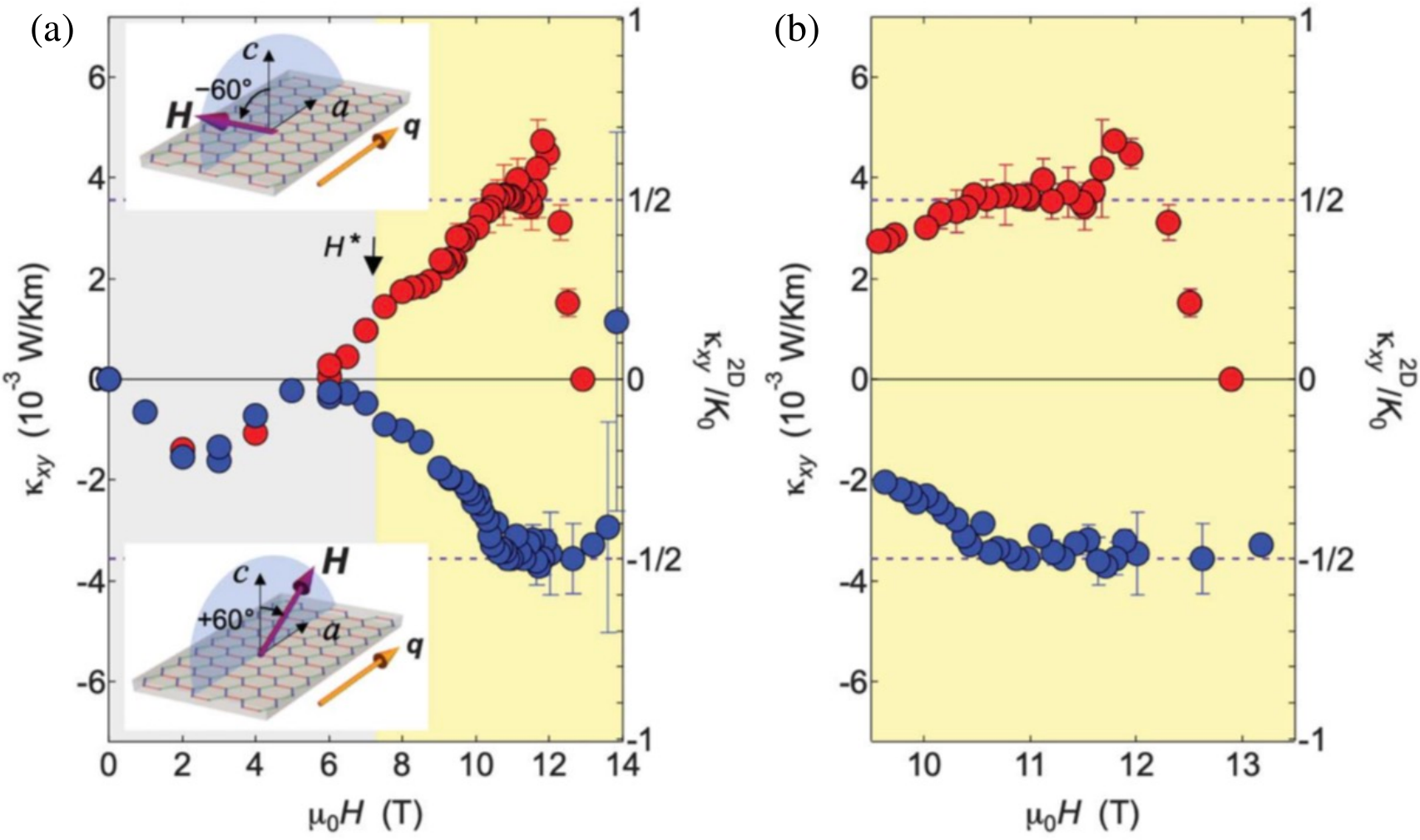}
	\caption{Thermal Hall effect in the tilted fields in $\alpha$-RuCl$_3$ sample. 
	(a) Field dependence of $\kappa_{xy}$ at 4.3K in tilted field of ${\theta=-60^\circ}$ (red circles) 
	and ${\theta=60^\circ}$ (blue circles) away from the $c$ axis in the $ac$ plane. 
	Gray and yellow shaded areas represent the AFM-ordered and spin-liquid states, respectively. 
	Thermal current $\bs{q}$ is applied along the $a$ axis. 
	The AFM transition field determined by the minimum of $\kappa_{xy}(H)$ is ${H^{\ast} = 7.0}$T. 
	(B) The same data in the high-field spin-liquid state. Figures are reprinted from Ref.~\onlinecite{Matsuda2021}.	
	}
	\label{signStructure}
\end{figure}

\subsubsection{Sign structure of thermal Hall conductivity}

Another important concern under the heated debate is whether
the experimental observation of half-quantized thermal Hall effect 
at intermediate magnetic fields is truly related to the non-Abelian Kitaev spin liquid 
of the pure Kitaev spin model, and what is the role of the non-Kitaev interactions 
whose presence has been demonstrated. 
Matsuda's group keeps pursuing along this direction 
by studying the field-angular variation of the thermal conductance plateau 
and its relation with the theoretical prediction based on pure Kitaev model~\cite{Matsuda2021}. 
A remarkable anomalous quantized thermal Hall effect 
is found even at zero out-of-plane magnetic field. 
Meanwhile, a non-trivial sign change feature in the spin-liquid regime
with azimuthal angle within the $ac$-plane is also observed, 
as depicted in Fig.~\ref{signStructure} where the thermal Hall conductivity 
$\kappa_{xy}$ reverses sign when the in-plane magnetic field direction 
switches from $-a$ to $a$. This sign structure is well-consistent 
with the sign of the topological Chern number of the pure Kitaev spin liquid  
through Eq.~\eqref{KSLkxy}. 
Together with the half-integer quantized thermal Hall conductivity, 
the authors concluded that the Kitaev interactions are crucially important 
for the quantized thermal Hall effect, 
while the contribution from the non-Kitaev ones are 
vastly outnumbered and negligible.

The angle dependent thermal Hall response of the system
under an applied magnetic field provokes theorists' interest 
and is shown to serve as an indicator for the non-Abelian 
Kitaev spin liquid~\cite{Hwang2022}. 
Recent theoretical calculation, however, studied a more realistic spin model of Kitaev materials 
with extended bond-dependent interactions, and found that the sign structure of the thermal Hall conductivity 
is a generic property of the polarized state in the presence of in-plane magnetic fields~\cite{PhysRevLett.126.147201,PhysRevB.103.174402}. 
Moreover, the thermal Hall conductivity can have a magnitude comparable to that observed 
in the experiments and can even have a plateau-like behavior~\cite{PhysRevB.103.174402}. 
In this case, the thermal Hall effect arises from the (topological) magnons with finite Berry curvatures. 
Therefore, a simple sign structure of the thermal Hall conductivity alone cannot serve as a  
decisive evidence for the non-Abelian Kitaev spin liquid. 
The ultimate testimony for this exotic state
is still the robust half-integer quantization of the thermal Hall conductivity in the zero temperature limit, 
where the (bosonic) magnon contribution always vanishes.

\subsection{Unquantized thermal Hall effect of field-induced gapless U(1) spin liquid}

\begin{figure}[t]
	\centering
	\includegraphics[width=8.6cm]{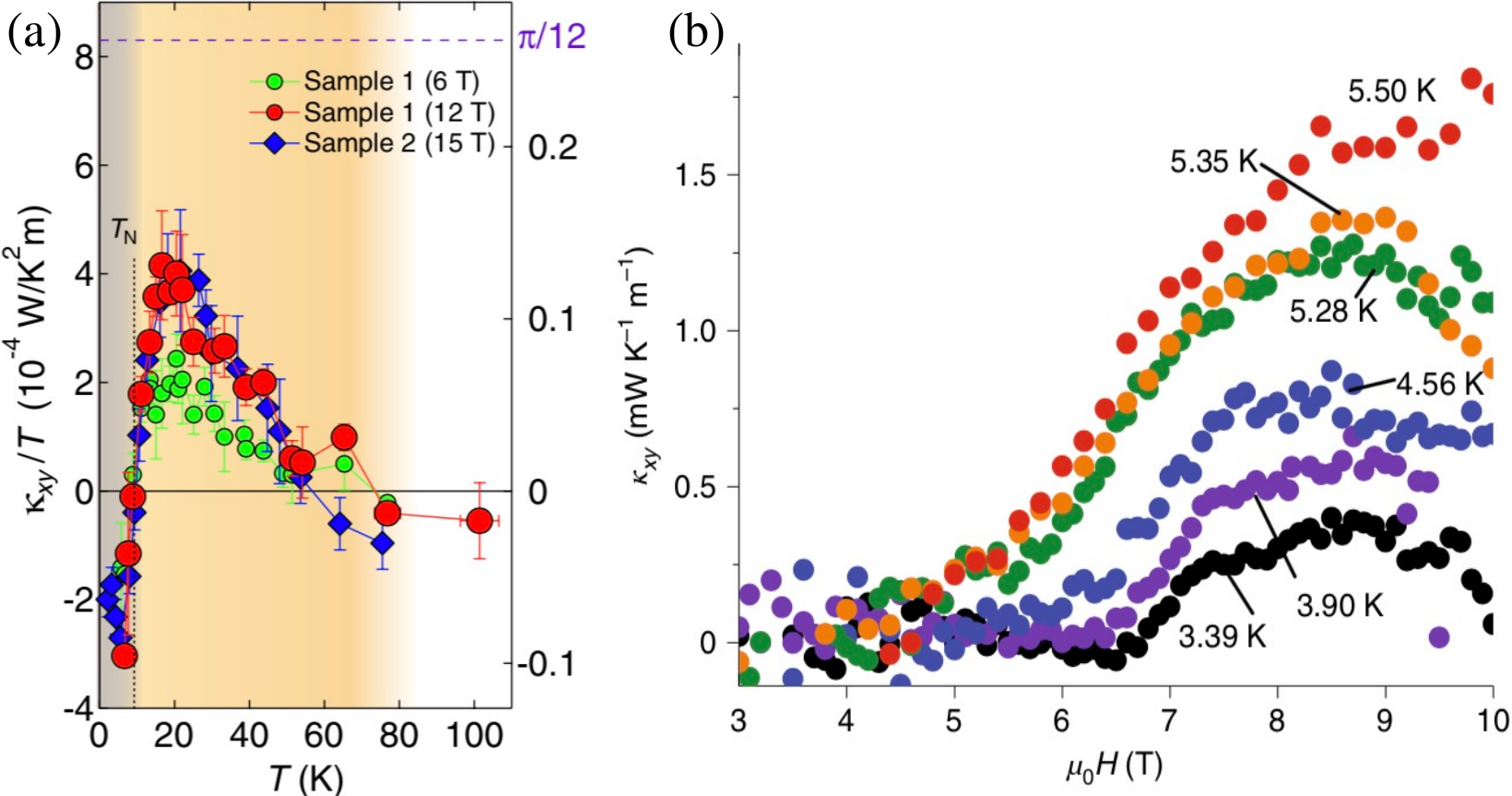}
	\caption{The measured unquantized planar thermal Hall response of $\alpha$-RuCl$_3$ sample. 
	(a) Temperature dependence of $\kappa_{xy}/T$ of $\alpha$-RuCl$_3$ in several magnetic fields 
	in the spin-liquid and magnetically ordered states. Reprinted from Ref.~\onlinecite{Matsuda2018B}.  
	(b)  Magnetic field dependence of thermal Hall conductivity $\kappa_{xy}$ of $\alpha$-RuCl$_3$ 
	 plotted for several temperatures from 3.4 K to 5.5 K. Reprinted from Ref.~\onlinecite{Ong2021}.	
	 Both experimental results are not quantized.
	}
	\label{nonQuantkxy}
\end{figure}

\begin{figure*}[htbp]
	\centering
	\includegraphics[width=15cm]{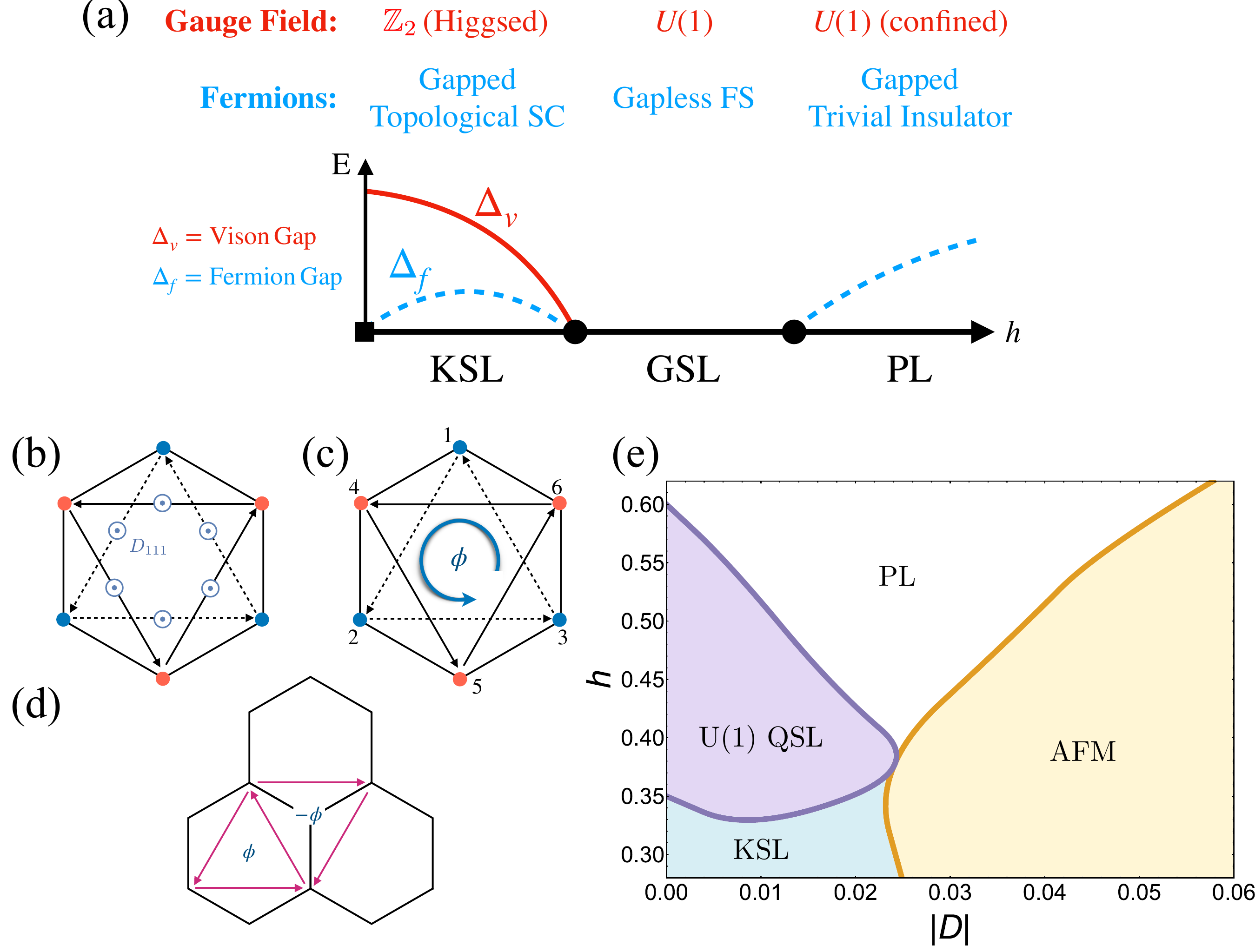}
	\caption{(a) Schematic phase diagram from the perspective of fermionic partons.
	 The behavior of the fermions and associated gauge field indicated for the KSL, GSL and PL phases. 
	 The flux (vison) gap $\Delta_v$ and the fermion gap $\Delta_f$ is also shown.
		(b) Symmetry allowed DM interactions between 
		second neighbors on the honeycomb lattice, where $D_{111}$ is the $[111]$ component. 
		The arrows specify the order of the cross product 
		$\boldsymbol{S_i}\times\boldsymbol{S_j}$. The two sublattices are labeled by colors.  
		(c) Schematic view of the gauge flux $\phi$ induced by the external magnetic field 
		in the presence of the next-nearest neighbor DM interaction. 
		(d) The net flux in one unit cell is zero and the space translation symmetry is well preserved.
		(e) Phase diagram for an extended Kitaev model
		in the combined presence of a finite DM interaction 
		and a finite magnetic field. In the figure, 
		``U(1) QSL'' specifically refers to the spinon Fermi surface U(1) spin liquid,
		``KSL'' refers to the Kitaev spin liquid, 
``GSL'' refers to the gapless spin liquid,		
		``AFM'' refers to the antiferromagnetic 
		ordered state, and ``PL'' refers to the polarized state.
		Panel (a) is reprinted from Ref.~\onlinecite{Hickey2019};
		Panel (b-e) are reprinted from Ref.~\onlinecite{Gao2019}.
	}
	\label{KSL_fig3}
\end{figure*}

Motivated by the observation of half-integer quantized plateau in $\alpha$-RuCl$_3$
with tilted magnetic fields, theorists have conducted systematically studies
on the original Kitaev model and explore its phase diagram 
in the presence of tilted magnetic fields~\cite{Hickey2019,Patel2019,Jiang2018,Zou2020}.
Meanwhile, the finite but non-quantized thermal Hall effect was also measured in the same material, 
over a broad range of temperatures and magnetic fields~\cite{Matsuda2018B,Hess2019,Ong2021}, 
as shown in Fig.~\ref{nonQuantkxy}. 
This points toward a scenario where the effect of the field yields an additional U(1) gapless 
spin liquid and the corresponding unquantized thermal Hall effect~\cite{Gao2020}.  
Specifically, as we have mentioned in Sec.~\ref{RuCl3}, the periodic oscillations of the longitudinal thermal conductivity
$\kappa_{xx}$ of $\alpha$-RuCl$_3$ is observed~\cite{Ong2021} 
in the moderate field strength range, much analogous to quantum oscillations in metals. 
The amplitude of the oscillations is very large within the intermediate field range, 
while strongly suppressed on either side, as shown in Fig~\ref{kxxOscillation}(b). 
Since $\alpha$-RuCl$_3$ is an excellent insulator with a charge gap,  
it seems that a field-induced spinon Fermi surface is a reasonable explanation 
for the oscillation~\cite{Motrunich2005,PhysRevB.97.045152,Ong2021}. 
The existence of a distinct gapless spin liquid at the intermediate field strength is 
also justified numerically.
The phase diagram is schematically depicted in Fig.~\ref{KSL_fig3}(a).
Various numerical and theoretical methods are employed, for instance
exact diagonalization (ED) techniques~\cite{Hickey2019}, 
density matrix renormalization group (DMRG)~\cite{Patel2019},
and recent dualities of gauge theories~\cite{Zou2020}.

From the spin susceptibility, magnetization, specific heat and thermodynamic entropy, 
there are clear signals for the existence of an intermediate phase~\cite{Patel2019}.
The static spin correlation function shows a power-law decay in the momentum space. 
All evidences indicates that the intermediate state is a gapless spin liquid.
Hickey and Trebst perform an ED calculation on small clusters, which also claims
the stable existence of an intermediate spin liquid 
in between the Kitaev spin liquid and the trivial paramagnetic insulator~\cite{Hickey2019}.
They find an increase of density of states at low energies with the field strength,
which indicates that the intermediate spin liquid is gapless.
Detailed information of the Fermi surface can be extracted from the spin structure factor~\cite{Patel2019}.
This gapless degree of freedom is attributed to the closing of $\mathbb{Z}_2$ vison gap,
and the flux is heavily fluctuating around the transition 
while the Majorana fermions are relatively unaffected [see Fig.~\ref{KSL_fig3}(a)].
The key role of the gauge sector is further supported 
by the thermodynamics at finite temperatures. 
An intuitive theoretical scenario for this transition is proposed by considering
the Abrikosov fermionic spinon decomposition,
where an emergent U(1) gauge field naturally occurs with the fractionalization.
The gapless spin liquid is composed of gapless spinon Fermi surface coupled 
with the U(1) gauge field. The transition from the gapless spin liquid to Kitaev spin liquid
can be understood from the spinon pairing instability and 
the superconducting condensate of the spinons is formed which Higgses 
the U(1) gauge structure down to $\mathbb{Z}_2$. 
The filled spinon bands are topologically non-trivial and correspond to the gapped Kitaev spin liquid. 
The transition from gapless spin liquid to the high-field polarized phase can also be incorporated in this framework.

On the materials' side, the actual Kitaev materials may not necessarily support Kitaev spin liquid, 
and it is well-known that having a Kitaev term in the spin interaction is not the sufficient 
condition for the Kitaev spin liquid ground state~\cite{Gang2019}. The
$ab$-initio modeling for the honeycomb Kitaev materials 
RuCl$_3$, Na$_2$IrO$_3$, and $\alpha$-Li$_2$IrO$_3$
shows that a variety of additional interactions 
beyond the bond-dependent Kitaev exchange
is present~\cite{Khaliullin2010,Rau2014,Rau2016,Liu2018}, 
particularly, in the layered spin-orbit Mott insulator RuCl$_3$~\cite{Valenti2016,Valenti2017}. 
The non-Kitaev interactions, such as the nearest-neighbor Heisenberg exchange interaction 
and the off-diagonal pseudo-dipole interactions, also play an important role~\cite{Rau2014,Liu2018,Valenti2016}. 
Meanwhile, the stability of the field-induced gapless spin liquid is tested against non-Kitaev interactions~\cite{Hickey2019}.

The field-driven U(1) gapless spin liquid~\cite{Hickey2019,Patel2019,Jiang2018,Fu2018,Zou2020} 
in Kitaev magnets and the inevitable presence of non-Kitaev interactions 
provide an alternative theoretical explanation for the non-quantized thermal Hall effect.  
Gao et al. considered a particular off-diagonal interaction,
the second-neighbor antisymmetric DM interaction with the form~\cite{Gao2019},
\begin{eqnarray}
{\mc{H}_{\rm DM}=\sum_{ \langle \langle i,j\rangle\rangle }
\boldsymbol{D}_{ij}\cdot(\boldsymbol{S_i}\times\boldsymbol{S_j})} ,
\end{eqnarray}
which is also symmetry-allowed for the honeycomb Kitaev materials. 
According to the Moriya's rules~\cite{Moriya1960}, 
there exist components of $\boldsymbol{D_{ij}}$ 
perpendicular to the planes with the strength $D_{111}$, 
as schematically depicted in Fig.~\ref{KSL_fig3}(b), and all the in-plane components 
vanish when the honeycomb plane is a mirror plane of the crystal structure. 
It has been estimated~\cite{Valenti2016} that a large second neighbor DM term 
${|\boldsymbol{D}_{ij}|>4}$meV is present for the Kitaev material 
$\alpha$-Li$_2$IrO$_3$, which however is often omitted in the literature. 
The field-driven U(1) spin liquid in Kitaev magnets survives to a finite DM interaction, 
which is checked using ED~\cite{Gao2019}.
The phase diagram in Fig.~\ref{KSL_fig3}(e) shows that
the U(1) spin liquid region stable up to a finite DM interaction. 
It should be noted that additional interactions relevant for real Kitaev materials, 
could further increase or decrease the stability of the U(1) spin liquid 
against the effects of the finite DM term. 
In any case, the U(1) spin liquid is stable to adding finite, 
though small, DM interactions. 

Like what we have described in Sec.~\ref{sec422}, 
the combination of the microscopic DM interaction
and Zeeman coupling further induces an internal 
U(1) gauge flux distribution on the honeycomb plane.
One can then establish ${\sin \phi\propto \lambda D_{111}\chi B}$
under an external magnetic field $B$ (with $\chi$ being the magnetic susceptibility). 
Detailed configuration of the gauge flux on honeycomb lattice 
is schematically illustrated in Fig.~\ref{KSL_fig3}(b). 
The net flux in one unit cell is zero and the space translation 
symmetry is preserved, as shown in Fig.~\ref{KSL_fig3}(c) and (d).

The field-induced U(1) spin liquid harbors an internal U(1) gauge flux facilitated 
by the DM interaction. The spinons carry the emergent U(1) gauge charges and 
are minimally coupled to the U(1) gauge field, thus the spinons feel 
the U(1) gauge flux as the spinons hop between the second-neighbor 
sites on the lattice. The first-neighbor spinon hopping 
does not pick up any phase since the 
net flux in a unit cell is zero, much like the Haldane model for spinless fermions. 
By performing a relevant projective symmetry 
group analysis~\cite{Wen2002}, 
three kinds of U(1) spin liquids were obtained~\cite{Jiang2018} 
that are connected to the Kitaev $\mathbb{Z}_2$ spin liquid state 
through a continuous phase transition without symmetry breaking. 
Moreover, only one of them, labeled as $U_1A_{k=0}$ in Ref.~\cite{Jiang2018},
is shown to support robust spinon Fermi surfaces. 
A representative mean-field Hamiltonian can be obtained for this U(1) spin liquid
with a neutral spinon Fermi surface on the honeycomb lattice~\cite{Gao2019}.
The internal U(1) gauge flux reconstructs the spinon bands 
and the Fermi pockets are survived up to some critical magnetic field.
For the stable deconfined spin liquid avoiding the confinement, 
the induced internal gauge flux is responsible for the spinon 
thermal Hall effect under a temperature gradient. 
The mechanism of spinon thermal Hall effect
is generalized for a field-induced U(1) spin liquid 
with a spinon Fermi surface in honeycomb Kitaev magnets~\cite{Gao2019},
where $\kappa_{xy}$ is evaluated by Eq.~\eqref{k_xy_fermi}. 
In contrast to the Kitaev spin liquid, 
the thermal Hall coefficient $\kappa_{xy}/T$ trends to a 
finite but non-quantized value in the zero-temperature limit, 
which is the prominent feature of a gapless spin liquid. 
In the intermediate temperature region, $\kappa_{xy}/T$ 
decreases monotonically or exhibits a broad peak, depending on the detailed 
mean-field spinon Hamiltonian, and eventually vanishes at high temperatures~\cite{Gao2019}. 
The vanishing $\kappa_{xy}$ in the high temperature region originates 
from the almost equally populated spinon bands and the corresponding 
Berry curvature cancellation.

An independent study~\cite{Sachdev2020} examines the thermal Hall response
in the pure Kitaev model for a wide variety of field strengths and orientations 
using a parton mean-field theory~\cite{Nayak2011,YongBaek2012,Okamoto2013}. 
The quantization ceases to hold as the system 
undergoes a phase transition to a field-induced U(1) gapless spin liquid.
The temperature and field dependence of the thermal Hall effect in the gapless spin liquid is investigated, 
which is shown to be consistent with the experiments, particularly, the quantization is field-angle-dependent~\cite{Matsuda2021}. 
This finding demonstrates that an unquantized response can be obtained from pure Kitaev model under magnetic field, 
namely without requiring any of the auxiliary DM interactions as suggested previously in Ref.~\cite{Gao2019}.
These many possibility towards the destruction of the original quantization hints that unquantized thermal Hall effect 
associated with the onset of Fermi surfaces is a more general phenomenon. 
\citet{Sachdev2020} reinforced this understanding by considering a different class of spin models, {\sl i.e.}
third neighbor Heisenberg antiferromagnets on a triangular lattice. 
Recent numerical evidence~\cite{Sheng2019} suggests that triangular lattice Heisenberg model 
with competing interactions is another example of a spin liquid with the emergent Fermi surfaces.
Similar temperature dependence of 
the thermal Hall conductance $\kappa_{xy}/T$ is found compared with the Kitaev honeycomb case.

\section{Thermal Hall signatures of topological and magnetic phase transitions}
\label{sec11}

To inspire the content of this section, we first provide a different perspective 
for the thermal Hall effects in the Kitaev spin liquids. 
Without of the magnetic field, the Kitaev model with three identical couplings
has a gapless $\mathbb{Z}_2$ spin liquid with the Dirac-cone band structure
for the gapless Majorana fermions. This state can be viewed as a gapless critical 
state. A generic magnetic field drives a topological phase transition into a gapped
Kitaev spin liquid with the topologically protected chiral Majorana edge mode. 
The half-quantized thermal Hall effect is thus an indication of topological phase 
transition. 
This view may be a bit insightful when one turns to the three-dimensional Kitaev models 
and considers different forms of $\mathbb{Z}_2$ spin liquids with 
different realizations of the Majorana bands, and the time reversal breaking leads to 
rich structures of topological transitions	~\cite{PhysRevB.93.085101}.
Leaving the context of the Kitaev materials, one can rely on the thermal
Hall signatures in order to identify the topological and magnetic phase transitions
in more broader contexts. 
This would provide some characteristic information about the topological features 
 beyond the usual thermodynamic characterization of the 
quantum phase transitions.
We will use this section to demonstrate the general application of thermal Hall transport to the diagnosis of 
topological phase transition beyond the Kitaev systems
and elaborate this point through the concrete contexts of the spin models on the honeycomb lattice.

\begin{figure}[t]
	\centering
	\includegraphics[width=8.6cm]{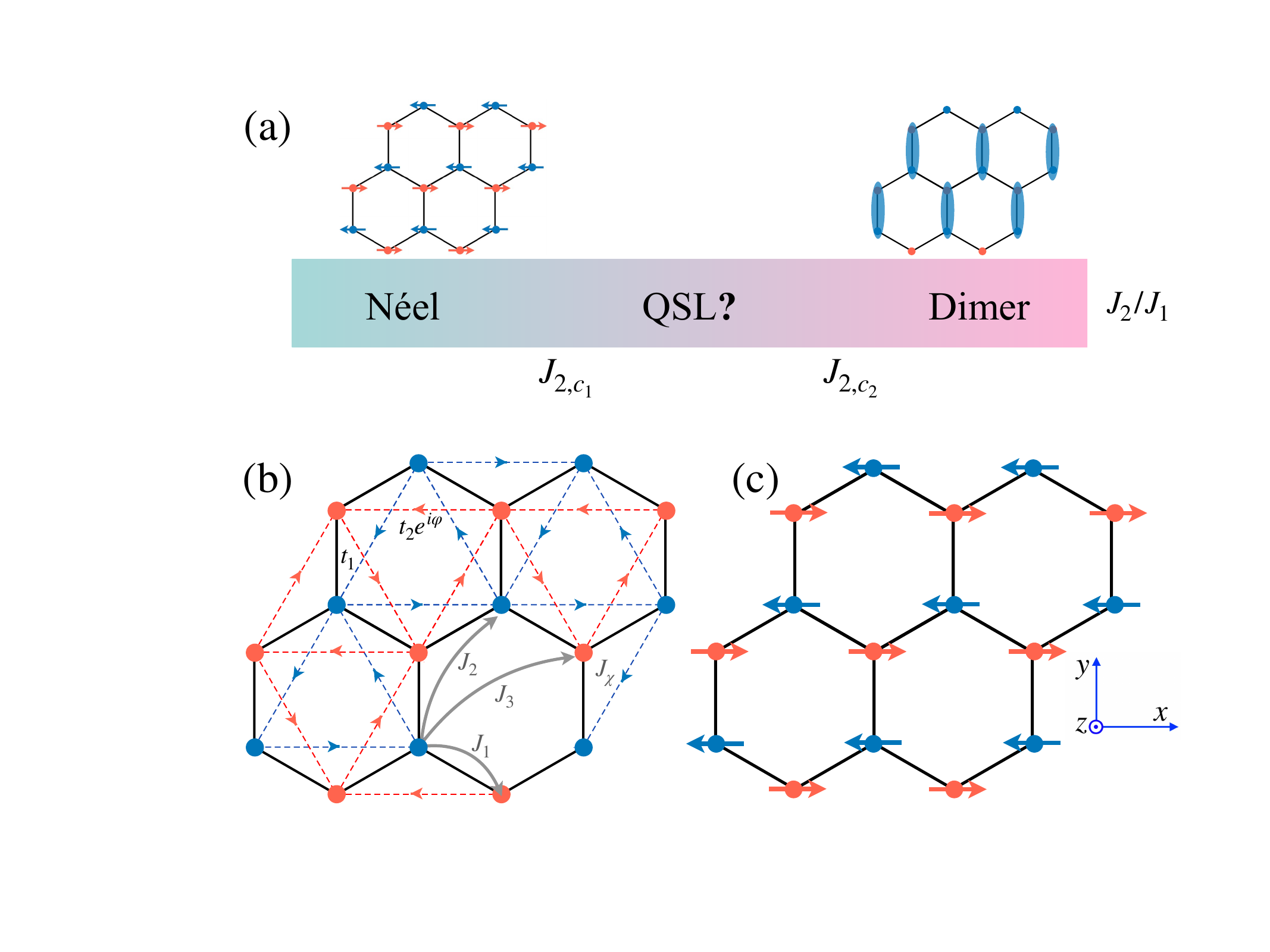}
	\caption{ (a) General phase diagram from the numerical studies for a pure $J_1$-$J_2$ 
	Heisenberg model on the honeycomb lattice. 	For the small $J_2$ region, 
	the ground state is generally believed to be a long-range N\'eel order, 
	while $J_2$ becomes comparable to $J_1$, 
	a dimer state or a stripe order could be stabilized, 
	and the intermediate regime is proposed as a spin liquid, both gapped and gapless. 
	(b) Schematic illustration of the hopping matrix up to second neighbors on a honeycomb lattice, 
	where for the nearest-neighbor hopping ${t_{1,ij}=t_{1,ji}=t_1}$; the second-neighbor hopping 
	acquires a positive phase ${t_{2,ij}=t_2e^{i\varphi}}$ when the spinon hops along 
	the (dashed) arrows. The (light) gray curve arrows represent Heisenberg exchanges 
	up to third neighbor, while $J_{\chi}$ refers to the scalar spin chirality term related to three neighbor sites.
	(c) The N\'eel state. Here we choose the order along $x$-direction to 
	minimize the  energy under a $z$-direction external magnetic field. 
	}
	\label{J1J2_fig1}
\end{figure}

 Unlike the Kitaev model, the general spin models on a honeycomb lattice are usually not exactly solvable, 
but the physics emergent from them can also be rather rich and appealing. One such example is the 
spin-1/2 honeycomb lattice antiferromagnetic Heisenberg model. 
Although the ground state of the nearest-neighbor Heisenberg model on the honeycomb lattice is 
a conventional antiferromagnetic N\'{e}el order (see Fig.~\ref{J1J2_fig1}), 
switching on the second-neighbor interaction would melt this long-range order and drive the system 
into a quantum disordered phase. 
The numerical studies~\cite{Sondhi2011,Fisher2013,Zhu2013,Massimo2012,Bishop2012,Liu2020,Albuquerque2011,Satoshi2013,Becca2017} 
have suggested that the spin liquid could emerge 
from the spin-1/2 antiferromagnetic $J_1$-$J_2$ Heisenberg model 
on the honeycomb lattice for the intermediate $J_2/J_1$, 
while the specific parameter range of it has been debated 
and the detailed properties of the candidate spin liquids have not yet 
reached a consensus. 
Fig.~\ref{J1J2_fig1}(a) plots a generic phase diagram of the $J_1$-$J_2$ 
Heisenberg model on a honeycomb lattice.

Beyond the simple $J_1$-$J_2$ model in Fig.~\ref{J1J2_fig1}, 
a series of recent studies have shown that the phase diagrams of 
the honeycomb lattice magnets with various other models 
could even harbor topological nontrivial phases, 
including both the magnetically orders with topological magnon 
bands~\cite{PhysRevLett.117.227201,PhysRevLett.127.217202,PhysRevLett.128.117201,Kim2022,Morimoto2022} and the spin liquids 
with fractionalized excitations and semion topological order~\cite{Hickey2016,Liu2020,Gao2020B,PhysRevB.105.024401}. 
These nontrivial phases and the corresponding phase transitions can in principle
manifest themselves in the thermal Hall effect with distinct signatures. 

For the rest of this section, we mainly discuss two categories of thermal Hall signatures 
that could be predicated for certain quantum magnets harboring topological phases or features. 
The first is a non-trivial thermal Hall signature arises from the proximity to a quantum 
critical point between the conventional magnetic order and the coexisting state 
with the semion topological order. The other one is a thermal Hall effect 
for the more conventional magnons in magnetically ordered states, 
where the thermal Hall conductivity changes signs or in several orders of 
magnitude near two types of phase transitions. 
Both examples indicate that the thermal Hall measurement provides a powerful 
and sensitive tool to experimentally discern the topological and magnetic transitions.

\subsection{Non-trivial thermal Hall signatures proximate to quantum critical point}
\label{sec111}

\subsubsection{Parton mean-field theory for the spinon topology}
\label{sec1111}

As we have stated, the intermediate phase of the spin-1/2 antiferromagnetic
 $J_1$-$J_2$ honeycomb Heisenberg model was found to be disordered.  
 A recent work~\cite{Liu2020} found two topologically different phases 
in the intermediate disordered regime, and one of them is the $\pi/2$-flux chiral spin liquid (CSL) 
with the semion topological order. 
In this chiral spin liquid, the second-neighbor exchange $J_2$ behaves 
with similar properties as the flux term in the Haldane model~\cite{Haldane1988,Liu2020}, 
and a large $J_2$ term renders the spinons with a topological phase similar 
to the spin-orbital coupling in the Kane-Mele model~\cite{Kane2005}. 
Beyond the pure $J_1$-$J_2$ Heisenberg model, 
\citet{Hickey2016} studied the honeycomb lattice Haldane-Hubbard Mott insulator 
of the spin-$1/2$ fermions, where the third-neighbor exchange $J_3$ 
and the scalar spin chirality $J_\chi$,
\begin{eqnarray}
H_{\chi} = J_{\chi}\sum_{i,j,k\in\triangle}{\bf S}_i\cdot({\bf S}_j\times{\bf S}_k),
\end{eqnarray}
are taken into consideration and a parameter window is singled out 
 for the chiral spin liquid proximate to the conventional N\'{e}el order. 
As the chiral spin liquid is described by a topological field theory,
the N\'{e}el-CSL transition is an exotic one and is facilitated by fractionalizing the spins.
The low energy excitations are gapped spinons that carry the semionic statistics 
due to the Chern-Simons term. On the magnetically ordered side, 
the spinon (or anyon) condensation destroys the topological order 
and generates the magnetic order. 
Thus, the relevant quantum critical point for the above honeycomb lattice models
is the N\'{e}el-CSL transition.

To explore the thermal Hall effect of the N\'{e}el-CSL transition, 
one needs to think about the field-induced response of the system.  
The potential coexistence of multiple competing phases~\cite{PhysRevB.105.155104}, 
namely the coexistence of the chiral spin liquid and long-range magnetic order, 
brings the complexity to the field-induced phenomenon. 
As the chiral spin liquid carries the topological order, it can coexist with the conventional orders such as the N\'{e}el order.
In the coexisting regime, the physical ingredients at play are the conventionally ordered spins, the
fractionalized and deconfined spinon excitations from the spin liquid and the Zeeman coupling. 
The topological phase transition could be driven by the external magnetic field,
and the spinon confinement-deconfinement is affected by the coexisting magnetic order.
Besides the unusual thermodynamic appearance of the quantum critical points, 
the topological features that are related to the many-body wavefunctions near the 
N\'{e}el-CSL criticality 
can be experimentally revealed by the nontrivial thermal Hall signatures~\cite{Samajdar2019,Gao2020B}.

\begin{figure*}[htbp]
	\centering
	\includegraphics[width=\textwidth]{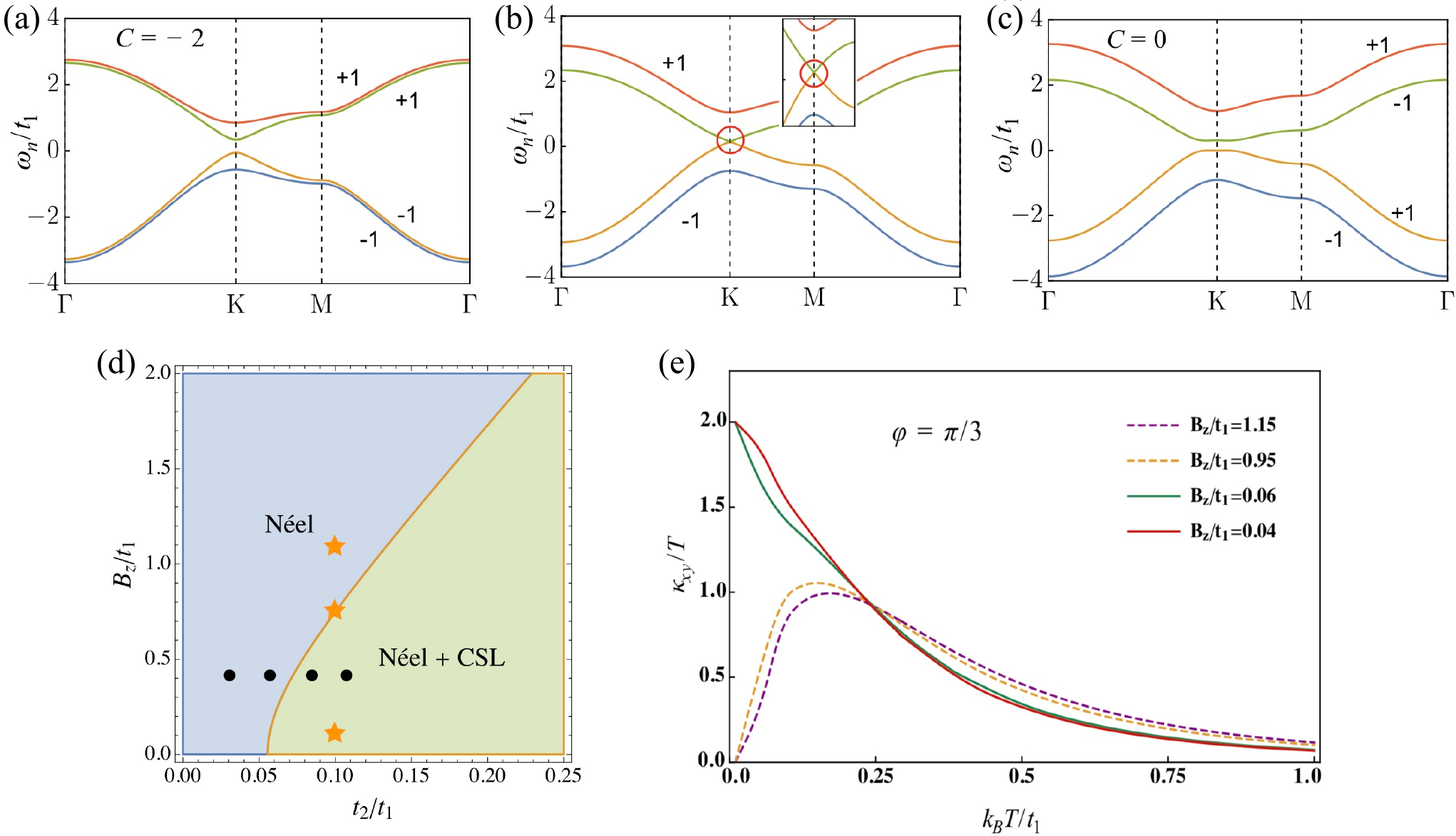}
	\caption{(a-c) Representative spinon bands along the high symmetry points in the Brillouin zone. 
	The numbers $\pm 1$ near the bands stand for the corresponding Chern numbers, 
	and the number $C$ represents the total Chern number of the fully occupied spinon bands. 
	The calculation is performed with ${t_2/t_1=0.1}$ while varying the magnetic fields for 
	(a) ${B_z/t_1=0.1}$ (b) ${B_z/t_1=0.75}$ and (c) ${B_z/t_1=1.1}$ for the mean-field model of 
	Eq.~\eqref{eq65}. With the increasing of fields, 
	the spinon bands experience a gap closing and reopening.
		(d) Mean-field phase diagram with varying $t_2$ and magnetic field $B_z$. 
		The colored phase boundary represents for a phase transition from the coexisting 
		phase of magnetic order and CSL to the conventional antiferromagnetic N\'eel state, 
		well compatible with the fact that the second neighbor exchange brings the competing interaction. 
		The orange stars correspond to the parameters used in the spinon band structures (a-c).
		(e) The temperature dependence of thermal Hall conductivity calculated with fixed second-neighbor 
		hopping amplitude $t_2/t_1=0.09$, while varying the temperature and magnetic field $B_z$. 
		The unit of $\kappa_{xy}/T$ here is also $\pi k_B^2/6\hbar$.
		Results shown in all panels adopt a parameter convention ${\varphi=\pi/3}$ and ${m/t_1=0.5}$.
		Figures are reprinted from Ref.~\cite{Gao2020B}.
	}
	\label{J1J2_fig2}
\end{figure*}

In the coexisting phase of the chiral spin liquid and the long-range magnetic order, 
the fractional spinons couple not only to the external Zeeman field, 
but also to the conventional magnetic order. 
To describe the chiral spin liquid with the fractionalized excitations,
one can adopt the Abrikosov fermion construction for the physical spin operator
and reformulate the interaction by a quadratic decoupling.
This is the procedure that was adopted for the strong Mott insulating regime in Sec.~\ref{sec422}. 
For this purpose, one expresses the spin operator as ${\bf S}_i = 1/2 \sum_{\alpha\beta} f^{\dagger}_{i\alpha} {\boldsymbol{\sigma}}_{\alpha\beta} f^{}_{i\beta}$
with the Hilbert space constraint $\sum_{\alpha} f^{\dagger}_{i\alpha} f^{}_{i\alpha} =1$.
The fractionalized spinon constitutes a mean-field Hamiltonian
by neglecting the gauge fluctuation and hopping parameters are constrained by the 
projective symmetry group,  
\begin{eqnarray}
H_{\text{MF}} &=& -\sum_{ij,\alpha} \, [\, t_{1,{ij}} f^{\dagger}_{i\alpha} f^{}_{j\alpha} + t_{2,ij} f^{\dagger}_{i\alpha} f^{}_{j\alpha} + h.c.  ] 
\nonumber \\
&&  + \frac{m}{2} \sum_{i,\alpha\beta} \nu_i f^{\dagger}_{i\alpha} \sigma^x_{\alpha\beta} f^{}_{i\beta}  
       - \frac{B_z}{2} \sum_{i,\alpha\beta}  f^{\dagger}_{i\alpha} \sigma^z_{\alpha\beta} f^{}_{i\beta}   
       \nonumber \\
       && -\mu \sum_{i\alpha} f^{\dagger}_{i\alpha} f^{}_{i\alpha} ,
       \label{eq65}
\end{eqnarray}
where the first line describes on the spinon hopping in the chiral spin liquid [see Fig.~\ref{J1J2_fig1}(b)],
the second line introduces the coupling to the N\'{e}el order and the external magnetic field in $z$,
and the third line is a chemical potential term that enforces the Hilbert space constraint on average. 
Since the field is applied in the $z$ direction, the N\'{e}el order is arranged in the perpendicular direction, e.g. 
the $x$ direction to gain energy. Here $\nu_i = \pm 1$ for two different sublattices and $m$ characterizes
the coupling strength from the N\'{e}el order.

$H_{\text{MF}}$ provides a mean-field description for the N\'{e}el order, the spinons in the chiral spin liquid,
and also the coexisting state of the N\'{e}el order and the chiral spin liquid, as well as the quantum phase transition
between the N\'{e}el state and the coexisting state. 
Based on the mean-field analysis of the spinons, 
the representative spinon band evolution is depicted 
in Figs.~\ref{J1J2_fig2}(a-c) with various magnetic fields.
In the small field limit, the influence of the long-range N\'{e}el order 
is transmitted into the fractionalized spinon degree of freedom  
and splits both the occupied and unoccupied spinon bands [see Fig.~\ref{J1J2_fig2}(a)]. 
The occupied spinon bands acquire a nonzero net Chern number, and then
the low-energy effective field theory involves the Chern-Simons action
\begin{equation}
\mathcal{S}_{\rm CS} = \frac{K}{4\pi} \int dt d{\bf r}\ \epsilon_{\mu\nu\lambda}a_{\mu}\partial_{\nu} a_{\lambda},
\end{equation}
with $K$ being the Chern number of the filled spinon bands,
and $a_{\mu}$ is the emergent U(1) gauge field.  
With the increasing of magnetic fields, 
the spinon bands experience a gap closing and reopening
[see Figs.~\ref{J1J2_fig2}(b-c)].
The net Chern number of the occupied bands vanishes, corresponding to a compact U(1) gauge theory in 2D. 
The gapped spinons can be integrated out safely
generating a pure compact U(1) gauge field that 
is always confined in 2D due to the proliferation of instantons~\cite{Polyakov1977}.
The change in the net Chern number signals a topological quantum phase transition
from the topologically ordered state to a phase with a trivial topology. 
Both the external magnetic field and the second-neighbor hopping coefficient $t_2$ could drive the system 
from a nontrivial coexisting phase into a conventional N\'{e}el state.

As we show in Fig.~\ref{J1J2_fig2}, the
thermal Hall effect is proved to be a powerful experimental probe to examine the topological quantum phase transition 
and its critical behavior. 
The finite thermal Hall conductivity is found in the coexisting phase of the long-range magnetic order and the 
chiral spin liquid, as well as in the proximate confined ordered phase. 
To observe the field-driven transition in Fig.~\ref{J1J2_fig2}(d),
the thermal Hall coefficient is plotted with varying temperatures 
at different magnetic fields in Fig.~\ref{J1J2_fig2}(e).
The colored solid curves represent the thermal Hall conductivity 
in the coexisting phase with the semion topological order. 
The curve is quantized to 2 in the zero-temperature limit due to the chiral edge modes,
then decreases monotonically with increasing temperature. 
On the other hand, the dashed curves represent the thermal Hall conductivity 
in the proximate confined phase, which is exactly 0 in the zero-temperature limit.
At finite temperatures, however, it increases rapidly with temperatures and then decreases gradually 
after reaching a maximum in the finite-temperature regime. 
The non-quantized and finite thermal Hall conductivity of 
the proximate confined phase with the same order of magnitude 
in the finite-temperature region is rather nontrivial 
since the magnon picture from the N\'eel ordered phase 
only gives rise to a much smaller thermal Hall conductivity~\cite{PhysRevB.99.205157}. 
This implies that the proximity effect of the quantum critical point for topological phase transition 
could result in an unusual thermal Hall effect, 
and the thermal Hall measurement is an effective approach to single out this topological phase transition. 
It should also be kept in mind that the above result is completely a mean-field analysis at this stage, 
and the gauge theories of thermal Hall effect beyond mean-field level will be discussed in the following subsection.

\begin{figure}[htbp]
	\centering
	\includegraphics[width=8.6cm]{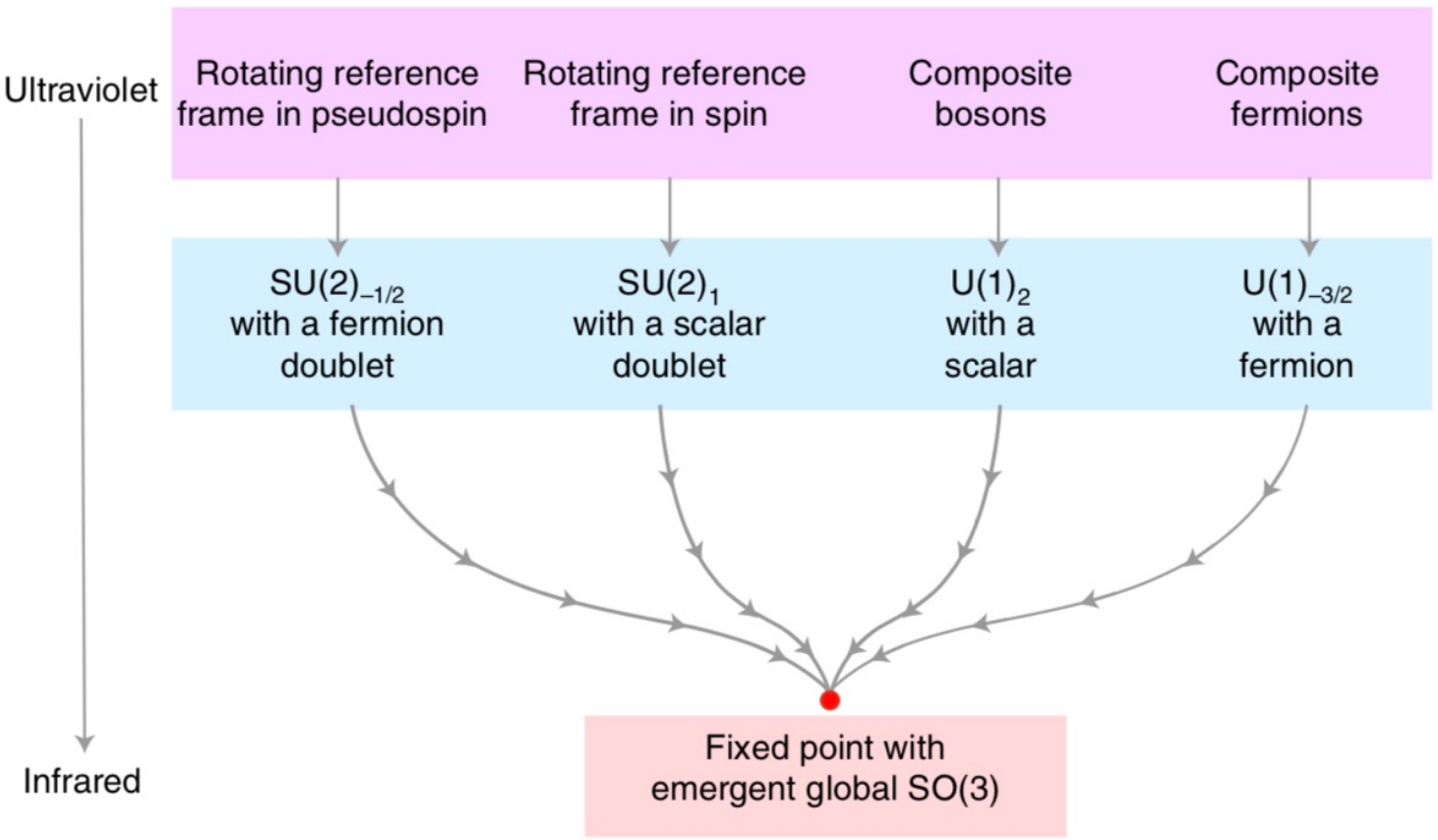}
	\caption{Four dual-field theories for the antiferromagnet flow to the same fixed point. 
	Distinct approaches to the lattice antiferromagnet (violet) lead to four different continuum 
	field theories (blue) for the transition from the N\'eel  to the N\'eel + CSL state [Fig.~\ref{J1J2_fig2}(d)]. 
	Universality then implies that these describe the same renormalization group fixed point (red) 
	with an emergent global SO(3) symmetry. Figure reprinted from Ref.~\cite{Samajdar2019}.}
	\label{dualFieldTheory}
\end{figure}

\subsubsection{Thermal Hall effect with gauge fluctuations}

Within the parton mean-field framework, the key nontrivial thermal Hall signatures 
proximate to the topological phase transition are obtained by ignoring the gauge fluctuations 
that could give out significant corrections to the resulting thermal Hall conductivity. 
The complete theory of the above thermal Hall effect corresponds to fermionic matters 
coupled to emergent gauge fields in $2+1$ dimensions. 
In fact, to explain the experimental observation of the giant thermal Hall conductivity 
in the pseudogap phase of the cuprate superconductors~\cite{Grissonnanche2019}, 
Samajdar et al. 
theoretically explored the N\'{e}el-VBS criticality on the square lattice and 
considered the instability to the chiral spin liquid 
under the perturbation from the uniform scalar spin chirality~\cite{Samajdar2019}.  
The chiral spin liquid is expected to support a large thermal Hall signal due to the
chiral edge mode. 
From this setting, they started from a $\pi$-flux spin liquid on a square lattice 
whose gauge confinement
leads to the N\'{e}el state,
and computed the enhanced 
thermal Hall conductivity $\kappa_{xy}$ using the gauge theory for the critical point 
between the N\'eel state and a coexisting phase with the N\'eel order and 
the chiral spin liquid with the semion topological order~\cite{Samajdar2019}, 
which has four different formulations in 
terms of the SU(2) or U(1) gauge theories, 
all dual to each other, as shown in Fig.~\ref{dualFieldTheory}. 
They prove that, in the topological phase with
the Chern-Simons term, the gauge-fluctuation corrections 
renormalize the thermal Hall conductivity
such that,
\begin{eqnarray}
\frac{\kappa_{xy}}{T} = 2\times\pi k_B^2/6\hbar \rightarrow \pi k_B^2/6\hbar
\end{eqnarray}
 as the temperature approaches zero, where a factor of 2 reduction 
 was obtained after the including the gauge fluctuations.  
Although persisting the quantized character of 
the zero-temperature thermal Hall conductivity, 
the gauge fluctuations indeed significantly change the parton 
mean-field result in Sec.~\ref{sec1111}.

In the finite temperature regime, an expansion in the inverse number of matter  
flavors is employed to compute the universal non-zero temperature thermal Hall effect 
with using the gauge theories near the quantum critical point~\cite{PhysRevB.101.195126}. 
Specifically, in the vicinity of the onset of the chiral spin liquid with the semion topological order in the N\'eel state, 
four duality-equivalent gauge theories emerge (see Fig.~\ref{dualFieldTheory}), 
one of which is the SU(2) gauge theory at Chern-Simons level ${k=-1/2}$ coupled 
to a single flavor of a two-component Dirac fermion $\Psi$ with the mass $M$~\cite{Samajdar2019}.  
The ${M > 0}$ phase is topologically trivial and corresponds to the conventional N\'eel state, 
while the ${M < 0}$ phase harbors the chiral spin liquid with the semion topological order and the chiral edge mode.
This square lattice result is essentially similar to the mean-field phase diagram for the honeycomb lattice 
case in Fig.~\ref{J1J2_fig2}(d). Near this topological quantum critical point ${M=0}$, 
the gauge theory is strongly coupled and the calculations for the finite temperature regime 
should rely on an expansion in $1/N_f$, with $N_f$ being the number of flavors of spinon matter fields. 
By generalizing the fermions to $\Psi_l$ with $l =1, ..., N_f$ flavors, 
the resulting Lagrangian becomes 
\begin{equation}
\mathcal{L} = \bar{\Psi}_l[i\gamma^{\mu}(\partial_{\mu}-iA_{\mu})]{\Psi}_l+M \bar{\Psi}_l {\Psi}_l+ k \rm{CS}[A_{\mu}],
\label{eq48}
\end{equation}
where $\Psi_l$ (${\bar{\Psi}_l=\Psi_l^{\dagger}\gamma^0}$) is a continuum spinon matter field 
of $l$-th flavor, $A_{\mu}$ is the emergent SU(2) gauge field, $\rm{CS}[A_{\mu}]$ represents 
the SU(2) Chern–Simons term and $k$ is the corresponding Chern-Simons level. 

With this generalization in Eq.~\eqref{eq48}, Guo et al. 
showed in detail that an important component of the $1/N_f$ corrections 
to $\kappa_{xy}$ can be interpreted as the contribution of the collective mode associated with the emergent 
gauge fields~\cite{PhysRevB.101.195126}.  The leading contribution can be viewed as that of the fermionic spinon matters, 
while the next-to-leading order term arises from the quantum fluctuations of the emergent gauge fields, 
which owns a sign opposite to that of the matter contribution. 
For the doped antiferromagnet, Guo et al. further developed an effective theory containing the fermionic matter that forms 
the pocket Fermi surfaces. 
They deduced the thermal Hall contribution of the gauge fluctuation from the Maxwell-Chern-Simons effective action, 
and find that this contribution also has the opposite sign from the Wiedemann-Franz contribution expected for the Fermi pockets, 
well consistent with the observed trends in the thermal Hall measurement of cuprate superconductors~\cite{Grissonnanche2019}. 
Although the giant thermal Hall conductivity in cuprates has been attributed to the phonons 
in some later literature~\cite{PhysRevB.105.115101,PhysRevX.12.041031,Grissonnanche2020,lyu2023phonons},
the theoretical results by including gauge fluctuations into the thermal Hall effect remain to be a 
great advance in the field.

\subsection{Sign and orders of magnitude switches of thermal Hall signatures near phase transitions}
\label{sec112}

In the last subsection,  
we have discussed a non-trivial thermal Hall signature 
 due to the proximity to a quantum critical point between the conventional magnetic order and the coexisting phase hosting 
 both N\'eel order and semion topological order with fractionalized excitations. 
 The corresponding gauge theories of the thermal Hall effect near the quantum critical point are also reviewed. 
 Here we turn to the fully ordered states and show that, in the framework of the linear spin-wave theory, 
 the magnetic and topological phase transitions and even the change of the magnon band topology 
 could also result in rich structures of the magnon thermal Hall signatures, 
  including the sign change and/or the several orders of magnitude change in 
  the thermal Hall conductance.

To study the thermal Hall effect of magnons in collinear antiferromagnetic insulators, 
Neumann et al. considered a spin Hamiltonian on the honeycomb lattice 
with~\cite{PhysRevLett.128.117201}
\begin{equation}
\label{eqHamNNB}
\mathcal{H}= \mathcal{H}_{\rm NN} +\mathcal{H}_{\rm B},
\end{equation}
where the nearest-neighbor term $\mathcal{H}_{\rm NN}$ was first proposed for 
the manganese thiophosphate MnPS$_3$, and its explicit form is given by
\begin{align}
\mathcal{H}_{\rm NN}= \frac{1}{2\hbar^2}\sum_{\langle ij\rangle} \bs{S}_i^T 
\left(\begin{array}{ccc}
J+J_a c_{ij}& -J_a s_{ij}& 0
\vspace{2mm}
\\
-J_a s_{ij} &  J-J_a c_{ij} &0 
\vspace{2mm}
\\
0& 0& J_z\\  
\end{array}
\right)
\bs{S}_j,
\label{rot1}
\end{align}
with the antiferromagnetic couplings ${J_z >J > 0}$
and $c_{ij} = \cos \theta_{ij}, s_{ij} = \sin \theta_{ij}$. 
The $J_a$ related exchange interaction, originating from spin-orbit coupling,  
is traceless and symmetric. It is related to the nearest-neighbor bonds ${\langle ij\rangle}$ 
by the bond-dependent phases $\theta_{ij}=0, 2\pi/3$, and $-2\pi/3$ (from the three-fold rotation).  
In the ground state of $\mathcal{H}_{\rm NN}$ with sufficiently small $J_a$, the spins of the sublattice A (B) 
point in the $+z$ $(-z)$ direction that is favored by $J$ and $J_z$. 
This term alone could produce a nonreciprocal magnon spectrum with the
asymmetric spin-wave dispersion.

\begin{figure}[htbp]
	\centering
	\includegraphics[width=8.6cm]{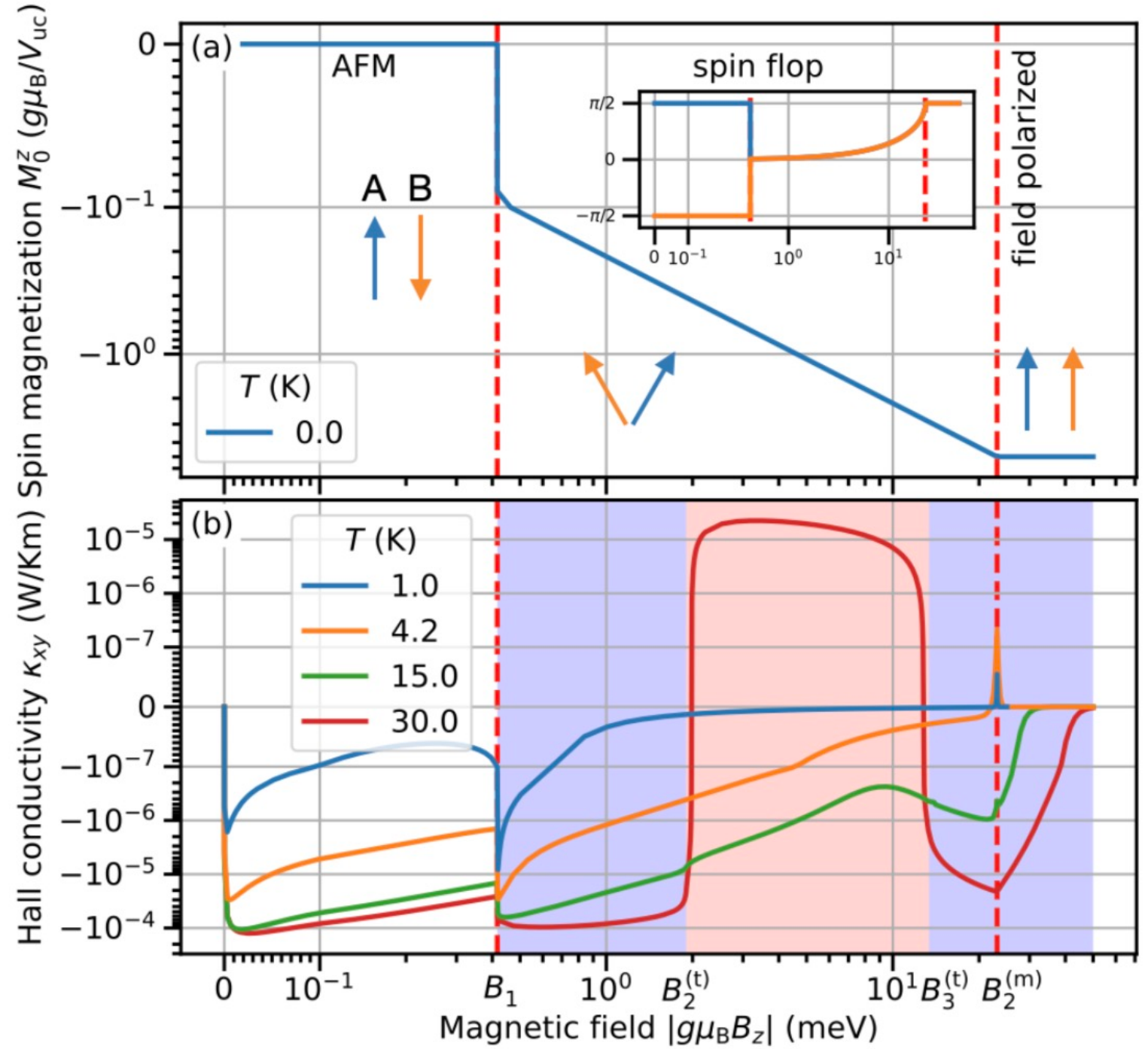}
	\caption{Magnetic, topological, and transport properties of (bulk)
		MnPS$_3$ (A = 0). (a) Classical ground state magnetization versus the
		magnetic field. Inset: angles $\theta_A$ and $\theta_B$ of the sublattice A (blue) and B
		(orange) spins with the $xy$ plane. (b) Thermal Hall conductivity $\kappa_{xy}$ for
		four selected temperatures ($T$ = 1.0 K, 4.2 K, 15 K, and 30 K). 
		The white/blue/red background color indicates topological phases with Chern numbers $\mathcal{C} = 0, -1, +1$ 
		of the lowest magnon band. 
		Dashed red lines mark the magnetic phase transitions at the critical fields $B_1^{(m)}$ and $B_2^{(m)}$. 
		All four panels have logarithmic ordinates and abscissae with linear-scale segments around 0, 
		which are identified by equally spaced minor ticks. 
		Figures are reprinted from Ref.~\cite{PhysRevLett.128.117201}.}
	\label{signSwitch}
\end{figure}

By applying an external out-of-plane magnetic field, 
the Zeeman coupling $\mathcal{H}_{\rm B}$ can be introduced as
\begin{align}
\mathcal{H}_{\rm B}= \frac{g\mu_B B_z}{\hbar}\sum_{i} S_i^z,
\end{align}
which can destabilize the antiferromagnetic order and induce the corresponding magnetic phase transitions. 
There are two critical magnetic fields ${B_1^{(m)}<B_2^{(m)}}$ that separate the phase diagram of the above 
spin Hamiltonian Eq.~\eqref{eqHamNNB}. The classical ground state, below the first critical field $B_1^{(m)}$, 
is a collinear antiferromagnet with a N\'eel vector pointing in the $z$ direction, while the field-polarized state,  
with all spins pointing along $+z$, appears when the magnetic field is larger than $B_2^{(m)}$. 
In the intermediate field range between $B_1^{(m)}$ and $B_2^{(m)}$, 
the system enters a coplanar spin flop phase, as depicted in Fig.~\ref{signSwitch}(a). 

Remarkably, the external field not only destabilizes the antiferromagnetic order, 
but also breaks an effective time-reversal symmetry that was discussed in Sec.~\ref{sec222}. 
It is expected that the magnon thermal Hall signatures could emerge to 
manifest the related phase transitions. Based on the linear spin-wave theory, 
one can follow the standard approach to calculate the field-dependent 
thermal Hall conductivity $\kappa_{xy}(B_z)$ for fields starting from zero, 
and the result is shown in Fig.~\ref{signSwitch}(b). 
It is clear that with the increasing magnetic field, there are four magnetic 
and topological phase transitions occur at ${B_1<B_2^{(t)}< B_3^{(t)}< B_2^{(m)}}$. 
Here the notations $B^{(m)}$ and $B^{(t)}$ represent the magnetic phase transition 
and ``topological transition'' points, respectively. $B_1$ is the critical field where the topological 
and magnetic phase transitions coincide. 
Moreover, we clarify that the ``topological transition'' is used simply for the convenience of the communication. 
The change of the magnon band topology is usually not a real phase transition, which is quite different
from the magnetic structure transition and the topological transition of the previous subsection. 

The changes of the thermal Hall conductivity $\kappa_{xy}$ in Fig.~\ref{signSwitch}(b) 
are traced back to the evolution of the magnon spectrum and the corresponding Berry curvatures. 
At the antiferromagnetic-spin flop phase transition point $B_1$,  
the Chern number of the lowest magnon band jumps from $0$ to $-1$, therefore, 
the magnetic phase transition is accompanied by a simultaneous topological change of the magnon band structure. 
The resulting thermal Hall conductivity $\kappa_{xy}$ increases abruptly and manifests as an obvious peak around $B_1$ field, especially at lower temperatures.

 The second and third ``topological transitions'' are both located in the intermediate spin flop phase 
 and attributed to the band inversions of the magnon bands. 
 At $B_2^{(t)}$, two magnon bands intersect and their Chern numbers are interchanged, 
 \emph{i.e.}, the Chern number of the lowest magnon band changes from $-1$ to $+1$. As a consequence, 
 the thermal Hall signature manifests the prominent sign changes at the elevated temperatures. 
 Similarly, at $B_3^{(t)}$, two magnon bands intersect again with the Chern number of the lowest magnon band 
 changing from $+1$ to $-1$. 
 The thermal Hall signature shows another sign switches at high temperatures around $B_3^{(t)}$. 
 Finally, the second-order phase transition from the spin flop to the field polarized states also manifests 
  in $\kappa_{xy}$ as a clear peak at $B_2^{(m)}$ for lower temperatures.

The temperature and field dependence of the magnon thermal Hall signatures 
proves that the magnonic thermal Hall conductivity $\kappa_{xy}$ is very sensitive 
to the magnetic structures at low temperatures. 
More specifically, the magnonic thermal Hall conductivity $\kappa_{xy}$ exhibits 
the pronounced peaks at the magnetic phase transitions, 
which is rather unaffected by the ``topological phase transitions''. 
On the other hand, the magnonic thermal Hall conductivity $\kappa_{xy}$ traces 
the ``topological phase transitions'' at high temperatures, 
which is insensitive to the magnetic transitions. 
The thermal Hall conductivity may change by several orders near a magnetic phase transition, 
and it may also change sign near the ``topological phase transition''. 
The spin Hamiltonian Eq.~\eqref{eqHamNNB} is not just a toy model, 
and the evolution of the thermal Hall signature with temperature and magnetic field
 is intimately related to the bulk MnPS$_3$. 
 The theoretical results can be directly compared to the future thermal Hall measurements.

 Moreover, for the van der Waals magnet MnPS$_3$,  
 the sample on a substrate or in a heterostructure would produce the local environments 
 to break the sublattice symmetry. Usually, it can be  mimicked 
 by introducing an extra on-site anisotropy term $\mathcal{H}_{\rm On}$ 
 to the spin Hamiltonian Eq.~\eqref{eqHamNNB}. 
 For the spins on the sublattice A, it is considered as
\begin{align}
\mathcal{H}_{\rm On}= -\frac{A}{\hbar^2}\sum_{i\in A}(S_i^z)^2.
\end{align}
The thermal Hall signatures that reveal the phase transitions for the van der Waals magnet MnPS$_3$ 
under magnetic fields are also analyzed by Neumann et al., 
and some key differences to bulk MnPS$_3$ are obtained ~\cite{PhysRevLett.128.117201}.

The sign switches of magnon thermal Hall signatures are not limited to the Hamiltonian 
Eq.~\eqref{eqHamNNB} and/or the material MnPS$_3$, 
instead, it is a general phenomenon for the topological phase transitions of the magnonic systems~\cite{PhysRevLett.127.217202,PhysRevB.104.144422,Zhuo2022,Kim2022,Morimoto2022}. 
As another example, Lu et al. considered the magnons in a honeycomb ferromagnet 
containing both ferromagnetic $J_1$-$J_2$ Heisenberg Hamiltonian and out-of-plane DM interaction, 
as well as the Zeeman coupling under magnetic fields~\cite{PhysRevLett.127.217202}, 
which is related to the honeycomb lattice magnet CrI$_3$~\cite{PhysRevX.8.041028}. 
By incorporating the Hartree-type self-energy associated 
with the magnon-magnon interactions into the quadratic magnon Hamiltonian, 
they demonstrated that the magnonic system exhibits the temperature driven topological phase transitions. 
Specifically, when the temperature increases, the magnon band gap at the Dirac points experiences 
a closing and reopening near a critical temperature $T_c$.  
The Chern numbers of the magnon bands are distinct below and above $T_c$, 
and the corresponding Dirac mass changes sign. 
 Lu et al. then confirmed that the gap-closing phenomenon is indeed a signature 
for the topological phase transition of the magnonic system and their analysis indicates that the magnon thermal Hall conductivity 
exhibits a sign reversal at $T_c$, which can serve as an experimental probe of this topological character~\cite{PhysRevLett.127.217202}.  
This sign switches of thermal Hall signature could be potentially realized in the honeycomb lattice magnet CrI$_3$ 
and is experimentally accessible.
 In summary, the magnetic phase transition and the magnon ``topological phase transition'' cause the distinct signatures in $\kappa_{xy}$, 
 and the measurement of thermal Hall conductivity can be used to identify these transitions.

\section{Conclusion}
\label{sec7}

In this review, we have provided a comprehensive introduction to the recent development of thermal Hall effects 
for various charge-neutral excitations in different quantum magnets and the related quantum phases. 
The finite thermal Hall conductivity in different regimes of quantum magnets could be attributed to the Berry curvature
properties of the corresponding elementary excitations. 
For the magnetic ordered states, the topological magnon bands owns non-vanishing 
Berry curvatures and contribute to the magnon thermal Hall effects. 
We explained the rich mechanisms for the magnons to acquire the non-trivial Berry curvatures in different magnets, 
and the underlying principle for the magnon thermal Hall effect is the effective time-reversal symmetry breaking 
and the magnetic point group compatible with ferromagnetism.  
For the disordered valence bond singlet states and the plaquette singlet states whose 
 elementary excitations are the triplon quasiparticles, the DM interactions could render the triplon with a non-zero Berry phase, 
  which can explain the topological character in the triplon spectrum of 
  the Sutherland-Shastry material SrCu$_2$(BO$_3$)$_2$, and result in triplon thermal Hall effect.

In the spin liquids, the related magnetic excitations are usually fractionalized, 
and the gauge coupling are generally strong. It is often hard to obtain an accurate quadratic Hamiltonian 
for these fractional magnetic excitations. Therefore, unlike the cases for the magnons and triplons, 
one should not completely rely on the quadratic mean-field Hamiltonian of fractionalized excitations to 
understand the corresponding thermal Hall effects. It is better to think about the thermal Hall effect in 
spin liquid from the underlying gauge structures first, as the fractionalized excitations with the emergent gauge 
charges are directly coupled to the emergent gauge fields. Thermal Hall effects were suggested for 
the spinon Fermi surface U(1) spin liquid in the weak Mott regime. 
This effect is actually quite natural in the weak Mott regime. Over there, 
the concept of spinons are not so distinct from the physical 
electrons due to the weak Mott gap and strong charge fluctuations. 
Physically, this can be understood from the fact that the 
external gauge flux enters into the four-spin ring exchange 
interaction. 
From the gauge theory description, the internal U(1) gauge 
flux is locked to the external U(1) gauge flux through the strong 
charge fluctuations, such that the spinon motion is twisted by 
the induced internal U(1) gauge flux. For the strong Mott insulators, 
the charge gap is large and the charge fluctuation is strongly suppressed. 
This induction of the internal U(1) gauge flux via the strong charge fluctuations 
does not apply to the strong Mott regimes. 
In the U(1) spin liquids of the strong Mott regime, 
different physical mechanisms are needed to understand the large thermal Hall effect. 
For the U(1) spin liquids whose gauge flux is
related to the scalar spin chirality ${\bs S_i}\cdot({\bs S}_j \times {\bs S}_k)$, 
the combination of the DM interaction and a simple Zeeman coupling could generate an internal U(1) gauge flux,
and thus twists the motion of the spinons. This mechanism broadly applies to the U(1) spin liquids with the fermionic spinons 
and explain the corresponding spinon thermal Hall effect. 
 Overall, for the spin liquid with a continuous gauge theory description, one key to 
resolve the mechanism for the thermal Hall effect is to understand the physical 
manifestation of the internal gauge flux and then the role of the external probes. 
This is related to the relation between the microscopic degrees of freedom and 
the emergent degrees of freedom in the lattice gauge theory formulation.

For the 3D pyrochlore U(1) spin liquid that is also in the strong Mott regime, 
the relation of the internal emergent variables and the physical spin variables is much simpler than the one 
described in the previous paragraph, 
and this relation immediately points to the underlying mechanism for thermal Hall transports. 
Here, the linear Zeeman coupling already induces 
an internal dual U(1) gauge flux and twists the motion of the ``magnetic monopoles'' (lower energy scale) 
for the ``magnetic monopoles'' thermal Hall effects,
 or an internal U(1) gauge flux and twists the motion of the gapped spinons (higher energy scale) 
 for the spinon thermal Hall effects under certain conditions.  
 
 For the Kitaev materials, the thermal Hall effect could arise from the Kitaev spin liquid 
 and various non-Kitaev spin liquids, 
 in particular the gapless U(1) spin liquids with a spinon Fermi surface, Dirac spin liquids, and variants of Abelian chiral spin liquids. 
 Moreover, the sign structure of thermal Hall effect observed in the experiments alone cannot serve as smoking-gun signature 
 for the Kitaev spin liquid, and only the half-integer quantization at the low temperatures would be a decisive probe 
 as the magnon contribution vanishes in the zero-temperature limit, and other types of spin liquids can never give a half-integer 
 quantized thermal Hall conductivity. 
 
 Beyond the explanation of different physical mechanisms for thermal Hall effects in different contexts, 
  we further review the non-trivial thermal Hall signatures that can be used to diagnose 
   quantum phase transitions and/or topological transition/changes of the different quantum systems. 
   The thermal Hall conductivity develops interesting proximity signatures near the quantum critical points. 
   It was even shown that, thermal Hall conductivity 
   associated with these transitions can change signs or change by several orders of magnitude. 
   All these understanding and applications can only be achieved based on the microscopic physical mechanisms
   for thermal Hall effects.

Finally, we briefly discuss the phonon thermal Hall effect in quantum magnets. 
The spin-lattice coupling is inevitable and can be rather strong in many materials.  
Although we have stated in the introduction section that, to avoid the complication 
in the interpretation of the longitudinal thermal conductivity $\kappa_{xx}$, 
it can be more selective to measure the transverse thermal Hall coefficient, 
the phonon thermal Hall effect can also emerge and 
contribute to the total transverse thermal conductivity $\kappa_{xy}$. 
Thermal Hall effects of the phonons have been reported in various 
compounds~\cite{PhysRevLett.95.155901,PhysRevLett.118.145902,PhysRevB.99.134419,PhysRevLett.124.105901,Yamashita2020,Grissonnanche2020,nmat4905,s41467,PhysRevLett.127.247202,PhysRevB.107.L060404,Ataei_2024,Cu3TeO6}. 
So far, to our knowledge, there exist two major mechanisms 
for the generation of the phonon thermal Hall 
effects. 
The first one is based on the scattering of phonons with the local magnetic defects or with the collective magnetic fluctuations,
where such phonon scattering produces the thermal Hall effects~\cite{PhysRevX.12.041031,PhysRevResearch.5.033197,pnas.2215141119}. 
The second one is based on the coupling and/or the hybridization of the phonons or lattices with
the magnetic degrees of freedom in magnetic insulators or the electric polarization of multiferroic materials,
such that the phonons can acquire the Berry curvature distribution from such coupling or hybridization~\cite{PhysRevLett.117.217205,PhysRevLett.123.167202,Jingyuan2020,PhysRevLett.123.237207,ma2023chiral,PhysRevB.105.L100402}.  
In particular, in the nonmagnetic insulator SrTiO$_3$ that is believed to be nearly ferroelectric, 
only phonons are responsible for the thermal Hall transport~\cite{Jingyuan2020}.
The significant phonon thermal Hall contribution, accompanied with the thermal Hall effect of magnetic excitations, 
is also observed in the kagom\'e antiferromagnet Cd-Kapellasite, 
in which the spin-phonon coupling gives rise to the phonon thermal Hall effect~\cite{Yamashita2020}. 
Moreover, it is argued that inside the pseudogap phase of cuprate, 
the phonons should be chiral to generate the Hall response and
 explain the observed large negative thermal Hall conductivity~\cite{Grissonnanche2019,Grissonnanche2020}. 
 The mechanism of the chiral phonons remains to be identified, but it must be intrinsic. 
 The phonon thermal Hall effect can be an important subject on its own and should require
 a different treatment from what has been included in this review.

Due to the limited scope and knowledge of the current authors, we are probably far from being 
complete in reviewing the rapid development of this topic. 
One obvious missing point, that is likely quite important, 
is about the apparent thermal Hall transport in the correlated paramagnetic 
regime where the weakly-interacting quasiparticle-like excitations or description do not seem to apply. 
The transports of these unparticles  
would require a different formulation~\cite{PhysRevB.95.155131,RevModPhys.82.1743} from what has been introduced in this review. 
We here apologize for all the missing points and other important results that have been ignored.
 In this review, we mainly explain the thermal Hall effects and the microscopic origins in different quantum
 magnets. As we insert the physical explanation by connecting the experimental results with 
 the understanding of the exotic quantum phases and interesting quantum systems, this review contains 
 extra contents about topological magnons, topological triplons, excitonic magnetism, Kitaev materials, 
 quantum spin liquid theories, 
 quantum oscillations, quantum spin ice, spin liquid materials, and 
many frustrated magnetic systems. We hope our review can be of some usefulness to experimentalists 
and theorists who are interested in the thermal Hall effects and quantum magnets in general.

\section*{Credit authorship contribution statement}
{\bf Gang Chen:} Initiated the whole project, Developed the scope and focus of the review, Writing the whole manuscript, Figure creation. 
{\bf Xiao-Tian Zhang:} Developed the focus of the review, Writing parts of the manuscript, Figure creation. {\bf Yong Hao Gao:}
Developed the focus of the review, Writing parts of the manuscript, Figure creation.

\section*{Declaration of competing interest}
The authors declare that they have no known competing financial interests or personal relationships that could have
appeared to influence the work reported in this paper.


\section*{Acknowledgments}

This work is supported by the National Science Foundation of 
China with Grant No.~92065203, the Ministry of Science and 
Technology of China with Grants No.2021YFA1400300 
and by the Research Grants Council of Hong Kong with 
Collaborative Research Fund C7012-21GF.

\newpage



  \bibliography{Ref.bib}

\begin{thebibliography}{352}%
\makeatletter
\providecommand \@ifxundefined [1]{%
 \@ifx{#1\undefined}
}%
\providecommand \@ifnum [1]{%
 \ifnum #1\expandafter \@firstoftwo
 \else \expandafter \@secondoftwo
 \fi
}%
\providecommand \@ifx [1]{%
 \ifx #1\expandafter \@firstoftwo
 \else \expandafter \@secondoftwo
 \fi
}%
\providecommand \natexlab [1]{#1}%
\providecommand \enquote  [1]{``#1''}%
\providecommand \bibnamefont  [1]{#1}%
\providecommand \bibfnamefont [1]{#1}%
\providecommand \citenamefont [1]{#1}%
\providecommand \href@noop [0]{\@secondoftwo}%
\providecommand \href [0]{\begingroup \@sanitize@url \@href}%
\providecommand \@href[1]{\@@startlink{#1}\@@href}%
\providecommand \@@href[1]{\endgroup#1\@@endlink}%
\providecommand \@sanitize@url [0]{\catcode `\\12\catcode `\$12\catcode
  `\&12\catcode `\#12\catcode `\^12\catcode `\_12\catcode `\%12\relax}%
\providecommand \@@startlink[1]{}%
\providecommand \@@endlink[0]{}%
\providecommand \url  [0]{\begingroup\@sanitize@url \@url }%
\providecommand \@url [1]{\endgroup\@href {#1}{\urlprefix }}%
\providecommand \urlprefix  [0]{URL }%
\providecommand \Eprint [0]{\href }%
\providecommand \doibase [0]{http://dx.doi.org/}%
\providecommand \selectlanguage [0]{\@gobble}%
\providecommand \bibinfo  [0]{\@secondoftwo}%
\providecommand \bibfield  [0]{\@secondoftwo}%
\providecommand \translation [1]{[#1]}%
\providecommand \BibitemOpen [0]{}%
\providecommand \bibitemStop [0]{}%
\providecommand \bibitemNoStop [0]{.\EOS\space}%
\providecommand \EOS [0]{\spacefactor3000\relax}%
\providecommand \BibitemShut  [1]{\csname bibitem#1\endcsname}%
\let\auto@bib@innerbib\@empty
\bibitem [{\citenamefont {Abd-Elmeguid}\ \emph {et~al.}(2004)\citenamefont
  {Abd-Elmeguid}, \citenamefont {Ni}, \citenamefont {Khomskii}, \citenamefont
  {Pocha}, \citenamefont {Johrendt}, \citenamefont {Wang},\ and\ \citenamefont
  {Syassen}}]{PhysRevLett.93.126403}%
  \BibitemOpen
  \bibfield  {author} {\bibinfo {author} {\bibnamefont {Abd-Elmeguid},
  \bibfnamefont {M~M}}, \bibinfo {author} {\bibfnamefont {B.}~\bibnamefont
  {Ni}}, \bibinfo {author} {\bibfnamefont {D.~I.}\ \bibnamefont {Khomskii}},
  \bibinfo {author} {\bibfnamefont {R.}~\bibnamefont {Pocha}}, \bibinfo
  {author} {\bibfnamefont {D.}~\bibnamefont {Johrendt}}, \bibinfo {author}
  {\bibfnamefont {X.}~\bibnamefont {Wang}}, \ and\ \bibinfo {author}
  {\bibfnamefont {K.}~\bibnamefont {Syassen}}} (\bibinfo {year} {2004}),\
  \bibfield  {title} {\enquote {\bibinfo {title} {{Transition from Mott
  Insulator to Superconductor in
  ${\mathrm{G}\mathrm{a}\mathrm{N}\mathrm{b}}_{4}{\mathrm{S}\mathrm{e}}_{8}$
  and
  ${\mathrm{G}\mathrm{a}\mathrm{T}\mathrm{a}}_{4}{\mathrm{S}\mathrm{e}}_{8}$
  under High Pressure}},}\ }\href {\doibase 10.1103/PhysRevLett.93.126403}
  {\bibfield  {journal} {\bibinfo  {journal} {Phys. Rev. Lett.}\ }\textbf
  {\bibinfo {volume} {93}},\ \bibinfo {pages} {126403}}\BibitemShut {NoStop}%
\bibitem [{\citenamefont {Akazawa}\ \emph {et~al.}(2022)\citenamefont
  {Akazawa}, \citenamefont {Lee}, \citenamefont {Takeda}, \citenamefont
  {Fujima}, \citenamefont {Tokunaga}, \citenamefont {Arima}, \citenamefont
  {Han},\ and\ \citenamefont {Yamashita}}]{PhysRevResearch.4.043085}%
  \BibitemOpen
  \bibfield  {author} {\bibinfo {author} {\bibnamefont {Akazawa}, \bibfnamefont
  {Masatoshi}}, \bibinfo {author} {\bibfnamefont {Hyun-Yong}\ \bibnamefont
  {Lee}}, \bibinfo {author} {\bibfnamefont {Hikaru}\ \bibnamefont {Takeda}},
  \bibinfo {author} {\bibfnamefont {Yuri}\ \bibnamefont {Fujima}}, \bibinfo
  {author} {\bibfnamefont {Yusuke}\ \bibnamefont {Tokunaga}}, \bibinfo {author}
  {\bibfnamefont {Taka-hisa}\ \bibnamefont {Arima}}, \bibinfo {author}
  {\bibfnamefont {Jung~Hoon}\ \bibnamefont {Han}}, \ and\ \bibinfo {author}
  {\bibfnamefont {Minoru}\ \bibnamefont {Yamashita}}} (\bibinfo {year}
  {2022}),\ \bibfield  {title} {\enquote {\bibinfo {title} {{Topological
  thermal Hall effect of magnons in magnetic skyrmion lattice}},}\ }\href
  {\doibase 10.1103/PhysRevResearch.4.043085} {\bibfield  {journal} {\bibinfo
  {journal} {Phys. Rev. Res.}\ }\textbf {\bibinfo {volume} {4}},\ \bibinfo
  {pages} {043085}}\BibitemShut {NoStop}%
\bibitem [{\citenamefont {Akazawa}\ \emph {et~al.}(2020)\citenamefont
  {Akazawa}, \citenamefont {Shimozawa}, \citenamefont {Kittaka}, \citenamefont
  {Sakakibara}, \citenamefont {Okuma}, \citenamefont {Hiroi}, \citenamefont
  {Lee}, \citenamefont {Kawashima}, \citenamefont {Han},\ and\ \citenamefont
  {Yamashita}}]{Yamashita2020}%
  \BibitemOpen
  \bibfield  {author} {\bibinfo {author} {\bibnamefont {Akazawa}, \bibfnamefont
  {Masatoshi}}, \bibinfo {author} {\bibfnamefont {Masaaki}\ \bibnamefont
  {Shimozawa}}, \bibinfo {author} {\bibfnamefont {Shunichiro}\ \bibnamefont
  {Kittaka}}, \bibinfo {author} {\bibfnamefont {Toshiro}\ \bibnamefont
  {Sakakibara}}, \bibinfo {author} {\bibfnamefont {Ryutaro}\ \bibnamefont
  {Okuma}}, \bibinfo {author} {\bibfnamefont {Zenji}\ \bibnamefont {Hiroi}},
  \bibinfo {author} {\bibfnamefont {Hyun-Yong}\ \bibnamefont {Lee}}, \bibinfo
  {author} {\bibfnamefont {Naoki}\ \bibnamefont {Kawashima}}, \bibinfo {author}
  {\bibfnamefont {Jung~Hoon}\ \bibnamefont {Han}}, \ and\ \bibinfo {author}
  {\bibfnamefont {Minoru}\ \bibnamefont {Yamashita}}} (\bibinfo {year}
  {2020}),\ \bibfield  {title} {\enquote {\bibinfo {title} {{Thermal Hall
  Effects of Spins and Phonons in Kagome Antiferromagnet Cd-Kapellasite}},}\
  }\href {\doibase 10.1103/PhysRevX.10.041059} {\bibfield  {journal} {\bibinfo
  {journal} {Phys. Rev. X}\ }\textbf {\bibinfo {volume} {10}},\ \bibinfo
  {pages} {041059}}\BibitemShut {NoStop}%
\bibitem [{\citenamefont {Alahmed}\ \emph {et~al.}(2022)\citenamefont
  {Alahmed}, \citenamefont {Zhang}, \citenamefont {Wen}, \citenamefont {Xiong},
  \citenamefont {Li}, \citenamefont {chuan Zhang}, \citenamefont {Lux},
  \citenamefont {Freimuth}, \citenamefont {Mahdi}, \citenamefont {Mokrousov},
  \citenamefont {Novosad}, \citenamefont {Kwok}, \citenamefont {Yu},
  \citenamefont {Zhang}, \citenamefont {Lee},\ and\ \citenamefont
  {Li}}]{Alahmed_2022}%
  \BibitemOpen
  \bibfield  {author} {\bibinfo {author} {\bibnamefont {Alahmed}, \bibfnamefont
  {Laith}}, \bibinfo {author} {\bibfnamefont {Xiaoqian}\ \bibnamefont {Zhang}},
  \bibinfo {author} {\bibfnamefont {Jiajia}\ \bibnamefont {Wen}}, \bibinfo
  {author} {\bibfnamefont {Yuzan}\ \bibnamefont {Xiong}}, \bibinfo {author}
  {\bibfnamefont {Yi}~\bibnamefont {Li}}, \bibinfo {author} {\bibfnamefont
  {Li}~\bibnamefont {chuan Zhang}}, \bibinfo {author} {\bibfnamefont {Fabian}\
  \bibnamefont {Lux}}, \bibinfo {author} {\bibfnamefont {Frank}\ \bibnamefont
  {Freimuth}}, \bibinfo {author} {\bibfnamefont {Muntasir}\ \bibnamefont
  {Mahdi}}, \bibinfo {author} {\bibfnamefont {Yuriy}\ \bibnamefont
  {Mokrousov}}, \bibinfo {author} {\bibfnamefont {Valentine}\ \bibnamefont
  {Novosad}}, \bibinfo {author} {\bibfnamefont {Wai-Kwong}\ \bibnamefont
  {Kwok}}, \bibinfo {author} {\bibfnamefont {Dapeng}\ \bibnamefont {Yu}},
  \bibinfo {author} {\bibfnamefont {Wei}\ \bibnamefont {Zhang}}, \bibinfo
  {author} {\bibfnamefont {Young~S.}\ \bibnamefont {Lee}}, \ and\ \bibinfo
  {author} {\bibfnamefont {Peng}\ \bibnamefont {Li}}} (\bibinfo {year}
  {2022}),\ \bibfield  {title} {\enquote {\bibinfo {title} {{Evidence of
  Magnon-Mediated Orbital Magnetism in a Quasi-2D Topological Magnon
  Insulator}},}\ }\href {\doibase 10.1021/acs.nanolett.2c00562} {\bibfield
  {journal} {\bibinfo  {journal} {Nano Letters}\ }\textbf {\bibinfo {volume}
  {22}}~(\bibinfo {number} {13}),\ \bibinfo {pages} {5114--5119}}\BibitemShut
  {NoStop}%
\bibitem [{\citenamefont {Albuquerque}\ \emph {et~al.}(2011)\citenamefont
  {Albuquerque}, \citenamefont {Schwandt}, \citenamefont {Het\'enyi},
  \citenamefont {Capponi}, \citenamefont {Mambrini},\ and\ \citenamefont
  {L\"auchli}}]{Albuquerque2011}%
  \BibitemOpen
  \bibfield  {author} {\bibinfo {author} {\bibnamefont {Albuquerque},
  \bibfnamefont {A~F}}, \bibinfo {author} {\bibfnamefont {D.}~\bibnamefont
  {Schwandt}}, \bibinfo {author} {\bibfnamefont {B.}~\bibnamefont {Het\'enyi}},
  \bibinfo {author} {\bibfnamefont {S.}~\bibnamefont {Capponi}}, \bibinfo
  {author} {\bibfnamefont {M.}~\bibnamefont {Mambrini}}, \ and\ \bibinfo
  {author} {\bibfnamefont {A.~M.}\ \bibnamefont {L\"auchli}}} (\bibinfo {year}
  {2011}),\ \bibfield  {title} {\enquote {\bibinfo {title} {{Phase diagram of a
  frustrated quantum antiferromagnet on the honeycomb lattice: Magnetic order
  versus valence-bond crystal formation}},}\ }\href {\doibase
  10.1103/PhysRevB.84.024406} {\bibfield  {journal} {\bibinfo  {journal} {Phys.
  Rev. B}\ }\textbf {\bibinfo {volume} {84}},\ \bibinfo {pages}
  {024406}}\BibitemShut {NoStop}%
\bibitem [{\citenamefont {Anderson}(1956)}]{Anderson1956}%
  \BibitemOpen
  \bibfield  {author} {\bibinfo {author} {\bibnamefont {Anderson},
  \bibfnamefont {P~W}}} (\bibinfo {year} {1956}),\ \bibfield  {title} {\enquote
  {\bibinfo {title} {{Ordering and antiferromagnetism in ferrites}},}\ }\href
  {\doibase 10.1103/PhysRev.102.1008} {\bibfield  {journal} {\bibinfo
  {journal} {Phys. Rev.}\ }\textbf {\bibinfo {volume} {102}},\ \bibinfo {pages}
  {1008--1013}}\BibitemShut {NoStop}%
\bibitem [{\citenamefont {Anderson}(1987)}]{Anderson1987}%
  \BibitemOpen
  \bibfield  {author} {\bibinfo {author} {\bibnamefont {Anderson},
  \bibfnamefont {P~W}}} (\bibinfo {year} {1987}),\ \bibfield  {title} {\enquote
  {\bibinfo {title} {{The resonating valence bond state in La$_2$CuO$_4$ and
  superconductivity}},}\ }\href {\doibase 10.1126/science.235.4793.1196}
  {\bibfield  {journal} {\bibinfo  {journal} {Science}\ }\textbf {\bibinfo
  {volume} {235}}~(\bibinfo {number} {4793}),\ \bibinfo {pages}
  {1196--1198}}\BibitemShut {NoStop}%
\bibitem [{\citenamefont {Ando}(2013)}]{Ando_2013}%
  \BibitemOpen
  \bibfield  {author} {\bibinfo {author} {\bibnamefont {Ando}, \bibfnamefont
  {Yoichi}}} (\bibinfo {year} {2013}),\ \bibfield  {title} {\enquote {\bibinfo
  {title} {Topological insulator materials},}\ }\href {\doibase
  10.7566/jpsj.82.102001} {\bibfield  {journal} {\bibinfo  {journal} {Journal
  of the Physical Society of Japan}\ }\textbf {\bibinfo {volume}
  {82}}~(\bibinfo {number} {10}),\ \bibinfo {pages} {102001}}\BibitemShut
  {NoStop}%
\bibitem [{\citenamefont {Anisimov}\ \emph {et~al.}(2019)\citenamefont
  {Anisimov}, \citenamefont {Aust}, \citenamefont {Khaliullin},\ and\
  \citenamefont {Daghofer}}]{PhysRevLett.122.177201}%
  \BibitemOpen
  \bibfield  {author} {\bibinfo {author} {\bibnamefont {Anisimov},
  \bibfnamefont {Pavel~S}}, \bibinfo {author} {\bibfnamefont {Friedemann}\
  \bibnamefont {Aust}}, \bibinfo {author} {\bibfnamefont {Giniyat}\
  \bibnamefont {Khaliullin}}, \ and\ \bibinfo {author} {\bibfnamefont {Maria}\
  \bibnamefont {Daghofer}}} (\bibinfo {year} {2019}),\ \bibfield  {title}
  {\enquote {\bibinfo {title} {{Nontrivial Triplon Topology and Triplon Liquid
  in Kitaev-Heisenberg-type Excitonic Magnets}},}\ }\href {\doibase
  10.1103/PhysRevLett.122.177201} {\bibfield  {journal} {\bibinfo  {journal}
  {Phys. Rev. Lett.}\ }\textbf {\bibinfo {volume} {122}},\ \bibinfo {pages}
  {177201}}\BibitemShut {NoStop}%
\bibitem [{\citenamefont {Ataei}\ \emph {et~al.}(2024)\citenamefont {Ataei},
  \citenamefont {Grissonnanche}, \citenamefont {Boulanger}, \citenamefont
  {Chen}, \citenamefont {Lefrancois}, \citenamefont {Brouet},\ and\
  \citenamefont {Taillefer}}]{Ataei_2024}%
  \BibitemOpen
  \bibfield  {author} {\bibinfo {author} {\bibnamefont {Ataei}, \bibfnamefont
  {A}}, \bibinfo {author} {\bibfnamefont {G.}~\bibnamefont {Grissonnanche}},
  \bibinfo {author} {\bibfnamefont {M.-E.}\ \bibnamefont {Boulanger}}, \bibinfo
  {author} {\bibfnamefont {L.}~\bibnamefont {Chen}}, \bibinfo {author}
  {\bibfnamefont {E.}~\bibnamefont {Lefrancois}}, \bibinfo {author}
  {\bibfnamefont {V.}~\bibnamefont {Brouet}}, \ and\ \bibinfo {author}
  {\bibfnamefont {L.}~\bibnamefont {Taillefer}}} (\bibinfo {year} {2024}),\
  \bibfield  {title} {\enquote {\bibinfo {title} {{Phonon chirality from
  impurity scattering in the antiferromagnetic phase of Sr$_2$IrO$_4$}},}\
  }\href {\doibase 10.1038/s41567-024-02384-5} {\bibfield  {journal} {\bibinfo
  {journal} {Nature Physics}\ }10.1038/s41567-024-02384-5}\BibitemShut
  {NoStop}%
\bibitem [{\citenamefont {Ballou}\ \emph {et~al.}(2003)\citenamefont {Ballou},
  \citenamefont {Canals}, \citenamefont {Elhajal}, \citenamefont {Lacroix},\
  and\ \citenamefont {Wills}}]{BALLOU2003465}%
  \BibitemOpen
  \bibfield  {author} {\bibinfo {author} {\bibnamefont {Ballou}, \bibfnamefont
  {R}}, \bibinfo {author} {\bibfnamefont {B.}~\bibnamefont {Canals}}, \bibinfo
  {author} {\bibfnamefont {M.}~\bibnamefont {Elhajal}}, \bibinfo {author}
  {\bibfnamefont {C.}~\bibnamefont {Lacroix}}, \ and\ \bibinfo {author}
  {\bibfnamefont {A.S.}\ \bibnamefont {Wills}}} (\bibinfo {year} {2003}),\
  \bibfield  {title} {\enquote {\bibinfo {title} {{Models for ordering in the
  jarosites Kagom\'{e} systems}},}\ }\href {\doibase
  https://doi.org/10.1016/S0304-8853(03)00079-9} {\bibfield  {journal}
  {\bibinfo  {journal} {Journal of Magnetism and Magnetic Materials}\ }\textbf
  {\bibinfo {volume} {262}}~(\bibinfo {number} {3}),\ \bibinfo {pages}
  {465--471}}\BibitemShut {NoStop}%
\bibitem [{\citenamefont {Balz}\ \emph {et~al.}(2019)\citenamefont {Balz},
  \citenamefont {Lampen-Kelley}, \citenamefont {Banerjee}, \citenamefont {Yan},
  \citenamefont {Lu}, \citenamefont {Hu}, \citenamefont {Yadav}, \citenamefont
  {Takano}, \citenamefont {Liu}, \citenamefont {Tennant}, \citenamefont
  {Lumsden}, \citenamefont {Mandrus},\ and\ \citenamefont
  {Nagler}}]{PhysRevB.100.060405}%
  \BibitemOpen
  \bibfield  {author} {\bibinfo {author} {\bibnamefont {Balz}, \bibfnamefont
  {Christian}}, \bibinfo {author} {\bibfnamefont {Paula}\ \bibnamefont
  {Lampen-Kelley}}, \bibinfo {author} {\bibfnamefont {Arnab}\ \bibnamefont
  {Banerjee}}, \bibinfo {author} {\bibfnamefont {Jiaqiang}\ \bibnamefont
  {Yan}}, \bibinfo {author} {\bibfnamefont {Zhilun}\ \bibnamefont {Lu}},
  \bibinfo {author} {\bibfnamefont {Xinzhe}\ \bibnamefont {Hu}}, \bibinfo
  {author} {\bibfnamefont {Swapnil~M.}\ \bibnamefont {Yadav}}, \bibinfo
  {author} {\bibfnamefont {Yasu}\ \bibnamefont {Takano}}, \bibinfo {author}
  {\bibfnamefont {Yaohua}\ \bibnamefont {Liu}}, \bibinfo {author}
  {\bibfnamefont {D.~Alan}\ \bibnamefont {Tennant}}, \bibinfo {author}
  {\bibfnamefont {Mark~D.}\ \bibnamefont {Lumsden}}, \bibinfo {author}
  {\bibfnamefont {David}\ \bibnamefont {Mandrus}}, \ and\ \bibinfo {author}
  {\bibfnamefont {Stephen~E.}\ \bibnamefont {Nagler}}} (\bibinfo {year}
  {2019}),\ \bibfield  {title} {\enquote {\bibinfo {title} {{Finite field
  regime for a quantum spin liquid in
  $\ensuremath{\alpha}\text{\ensuremath{-}}{\mathrm{RuCl}}_{3}$}},}\ }\href
  {\doibase 10.1103/PhysRevB.100.060405} {\bibfield  {journal} {\bibinfo
  {journal} {Phys. Rev. B}\ }\textbf {\bibinfo {volume} {100}},\ \bibinfo
  {pages} {060405}}\BibitemShut {NoStop}%
\bibitem [{\citenamefont {Banerjee}\ \emph {et~al.}(2016)\citenamefont
  {Banerjee}, \citenamefont {Bridges}, \citenamefont {Yan}, \citenamefont
  {Aczel}, \citenamefont {Li}, \citenamefont {Stone}, \citenamefont {Granroth},
  \citenamefont {Lumsden}, \citenamefont {Yiu}, \citenamefont {Knolle},
  \citenamefont {Bhattacharjee}, \citenamefont {Kovrizhin}, \citenamefont
  {Moessner}, \citenamefont {Tennant}, \citenamefont {Mandrus},\ and\
  \citenamefont {Nagler}}]{Banerjee2016}%
  \BibitemOpen
  \bibfield  {author} {\bibinfo {author} {\bibnamefont {Banerjee},
  \bibfnamefont {A}}, \bibinfo {author} {\bibfnamefont {C.~A.}\ \bibnamefont
  {Bridges}}, \bibinfo {author} {\bibfnamefont {J.-Q.}\ \bibnamefont {Yan}},
  \bibinfo {author} {\bibfnamefont {A.~A.}\ \bibnamefont {Aczel}}, \bibinfo
  {author} {\bibfnamefont {L.}~\bibnamefont {Li}}, \bibinfo {author}
  {\bibfnamefont {M.~B.}\ \bibnamefont {Stone}}, \bibinfo {author}
  {\bibfnamefont {G.~E.}\ \bibnamefont {Granroth}}, \bibinfo {author}
  {\bibfnamefont {M.~D.}\ \bibnamefont {Lumsden}}, \bibinfo {author}
  {\bibfnamefont {Y.}~\bibnamefont {Yiu}}, \bibinfo {author} {\bibfnamefont
  {J.}~\bibnamefont {Knolle}}, \bibinfo {author} {\bibfnamefont
  {S.}~\bibnamefont {Bhattacharjee}}, \bibinfo {author} {\bibfnamefont {D.~L.}\
  \bibnamefont {Kovrizhin}}, \bibinfo {author} {\bibfnamefont {R.}~\bibnamefont
  {Moessner}}, \bibinfo {author} {\bibfnamefont {D.~A.}\ \bibnamefont
  {Tennant}}, \bibinfo {author} {\bibfnamefont {D.~G.}\ \bibnamefont
  {Mandrus}}, \ and\ \bibinfo {author} {\bibfnamefont {S.~E.}\ \bibnamefont
  {Nagler}}} (\bibinfo {year} {2016}),\ \bibfield  {title} {\enquote {\bibinfo
  {title} {{Proximate Kitaev quantum spin liquid behaviour in a honeycomb
  magnet}},}\ }\href {\doibase 10.1038/nmat4604} {\bibfield  {journal}
  {\bibinfo  {journal} {Nature Materials}\ }\textbf {\bibinfo {volume}
  {15}}~(\bibinfo {number} {7}),\ \bibinfo {pages} {733--740}}\BibitemShut
  {NoStop}%
\bibitem [{\citenamefont {Banerjee}\ \emph {et~al.}(2017)\citenamefont
  {Banerjee}, \citenamefont {Yan}, \citenamefont {Knolle}, \citenamefont
  {Bridges}, \citenamefont {Stone}, \citenamefont {Lumsden}, \citenamefont
  {Mandrus}, \citenamefont {Tennant}, \citenamefont {Moessner},\ and\
  \citenamefont {Nagler}}]{Banerjee2017}%
  \BibitemOpen
  \bibfield  {author} {\bibinfo {author} {\bibnamefont {Banerjee},
  \bibfnamefont {Arnab}}, \bibinfo {author} {\bibfnamefont {Jiaqiang}\
  \bibnamefont {Yan}}, \bibinfo {author} {\bibfnamefont {Johannes}\
  \bibnamefont {Knolle}}, \bibinfo {author} {\bibfnamefont {Craig~A.}\
  \bibnamefont {Bridges}}, \bibinfo {author} {\bibfnamefont {Matthew~B.}\
  \bibnamefont {Stone}}, \bibinfo {author} {\bibfnamefont {Mark~D.}\
  \bibnamefont {Lumsden}}, \bibinfo {author} {\bibfnamefont {David~G.}\
  \bibnamefont {Mandrus}}, \bibinfo {author} {\bibfnamefont {David~A.}\
  \bibnamefont {Tennant}}, \bibinfo {author} {\bibfnamefont {Roderich}\
  \bibnamefont {Moessner}}, \ and\ \bibinfo {author} {\bibfnamefont
  {Stephen~E.}\ \bibnamefont {Nagler}}} (\bibinfo {year} {2017}),\ \bibfield
  {title} {\enquote {\bibinfo {title} {{Neutron scattering in the proximate
  quantum spin liquid
  $\ensuremath{\alpha}\ensuremath{-}{\mathrm{RuCl}}_{3}$}},}\ }\href {\doibase
  10.1126/science.aah6015} {\bibfield  {journal} {\bibinfo  {journal}
  {Science}\ }\textbf {\bibinfo {volume} {356}}~(\bibinfo {number} {6342}),\
  \bibinfo {pages} {1055--1059}}\BibitemShut {NoStop}%
\bibitem [{\citenamefont {Banerjee}\ \emph {et~al.}(2018)\citenamefont
  {Banerjee}, \citenamefont {Heiblum}, \citenamefont {Umansky}, \citenamefont
  {Feldman}, \citenamefont {Oreg},\ and\ \citenamefont
  {Stern}}]{Banerjee_2018}%
  \BibitemOpen
  \bibfield  {author} {\bibinfo {author} {\bibnamefont {Banerjee},
  \bibfnamefont {Mitali}}, \bibinfo {author} {\bibfnamefont {Moty}\
  \bibnamefont {Heiblum}}, \bibinfo {author} {\bibfnamefont {Vladimir}\
  \bibnamefont {Umansky}}, \bibinfo {author} {\bibfnamefont {Dima~E.}\
  \bibnamefont {Feldman}}, \bibinfo {author} {\bibfnamefont {Yuval}\
  \bibnamefont {Oreg}}, \ and\ \bibinfo {author} {\bibfnamefont {Ady}\
  \bibnamefont {Stern}}} (\bibinfo {year} {2018}),\ \bibfield  {title}
  {\enquote {\bibinfo {title} {{Observation of half-integer thermal Hall
  conductance}},}\ }\href {\doibase 10.1038/s41586-018-0184-1} {\bibfield
  {journal} {\bibinfo  {journal} {Nature}\ }\textbf {\bibinfo {volume}
  {559}}~(\bibinfo {number} {7713}),\ \bibinfo {pages} {205--210}}\BibitemShut
  {NoStop}%
\bibitem [{\citenamefont {Bergman}\ \emph {et~al.}(2006)\citenamefont
  {Bergman}, \citenamefont {Fiete},\ and\ \citenamefont
  {Balents}}]{PhysRevB.73.134402}%
  \BibitemOpen
  \bibfield  {author} {\bibinfo {author} {\bibnamefont {Bergman}, \bibfnamefont
  {Doron~L}}, \bibinfo {author} {\bibfnamefont {Gregory~A.}\ \bibnamefont
  {Fiete}}, \ and\ \bibinfo {author} {\bibfnamefont {Leon}\ \bibnamefont
  {Balents}}} (\bibinfo {year} {2006}),\ \bibfield  {title} {\enquote {\bibinfo
  {title} {Ordering in a frustrated pyrochlore antiferromagnet proximate to a
  spin liquid},}\ }\href {\doibase 10.1103/PhysRevB.73.134402} {\bibfield
  {journal} {\bibinfo  {journal} {Phys. Rev. B}\ }\textbf {\bibinfo {volume}
  {73}},\ \bibinfo {pages} {134402}}\BibitemShut {NoStop}%
\bibitem [{\citenamefont {Bertin}\ \emph {et~al.}(2012)\citenamefont {Bertin},
  \citenamefont {Chapuis}, \citenamefont {de~Réotier},\ and\ \citenamefont
  {Yaouanc}}]{Bertin_2012}%
  \BibitemOpen
  \bibfield  {author} {\bibinfo {author} {\bibnamefont {Bertin}, \bibfnamefont
  {A}}, \bibinfo {author} {\bibfnamefont {Y}~\bibnamefont {Chapuis}}, \bibinfo
  {author} {\bibfnamefont {P~Dalmas}\ \bibnamefont {de~Réotier}}, \ and\
  \bibinfo {author} {\bibfnamefont {A}~\bibnamefont {Yaouanc}}} (\bibinfo
  {year} {2012}),\ \bibfield  {title} {\enquote {\bibinfo {title} {{Crystal
  electric field in the R$_2$Ti$_2$O$_7$ pyrochlore compounds}},}\ }\href
  {\doibase 10.1088/0953-8984/24/25/256003} {\bibfield  {journal} {\bibinfo
  {journal} {Journal of Physics: Condensed Matter}\ }\textbf {\bibinfo {volume}
  {24}}~(\bibinfo {number} {25}),\ \bibinfo {pages} {256003}}\BibitemShut
  {NoStop}%
\bibitem [{\citenamefont {Bhardwaj}\ \emph {et~al.}(2022)\citenamefont
  {Bhardwaj}, \citenamefont {Zhang}, \citenamefont {Yan}, \citenamefont
  {Moessner}, \citenamefont {Nevidomskyy},\ and\ \citenamefont
  {Changlani}}]{Bhardwaj_2022}%
  \BibitemOpen
  \bibfield  {author} {\bibinfo {author} {\bibnamefont {Bhardwaj},
  \bibfnamefont {Anish}}, \bibinfo {author} {\bibfnamefont {Shu}\ \bibnamefont
  {Zhang}}, \bibinfo {author} {\bibfnamefont {Han}\ \bibnamefont {Yan}},
  \bibinfo {author} {\bibfnamefont {Roderich}\ \bibnamefont {Moessner}},
  \bibinfo {author} {\bibfnamefont {Andriy~H.}\ \bibnamefont {Nevidomskyy}}, \
  and\ \bibinfo {author} {\bibfnamefont {Hitesh~J.}\ \bibnamefont {Changlani}}}
  (\bibinfo {year} {2022}),\ \bibfield  {title} {\enquote {\bibinfo {title}
  {{Sleuthing out exotic quantum spin liquidity in the pyrochlore magnet
  Ce$_2$Zr$_2$O$_7$}},}\ }\href {\doibase 10.1038/s41535-022-00458-2}
  {\bibfield  {journal} {\bibinfo  {journal} {npj Quantum Materials}\ }\textbf
  {\bibinfo {volume} {7}}~(\bibinfo {number} {1}),\
  10.1038/s41535-022-00458-2}\BibitemShut {NoStop}%
\bibitem [{\citenamefont {Binotto}\ \emph {et~al.}(1971)\citenamefont
  {Binotto}, \citenamefont {Pollini},\ and\ \citenamefont
  {Spinolo}}]{Binotto1971}%
  \BibitemOpen
  \bibfield  {author} {\bibinfo {author} {\bibnamefont {Binotto}, \bibfnamefont
  {L}}, \bibinfo {author} {\bibfnamefont {I.}~\bibnamefont {Pollini}}, \ and\
  \bibinfo {author} {\bibfnamefont {G.}~\bibnamefont {Spinolo}}} (\bibinfo
  {year} {1971}),\ \bibfield  {title} {\enquote {\bibinfo {title} {{Optical and
  transport properties of the magnetic semiconductor
  $\ensuremath{\alpha}\ensuremath{-}{\mathrm{RuCl}}_{3}$}},}\ }\href {\doibase
  10.1002/pssb.2220440126} {\bibfield  {journal} {\bibinfo  {journal} {Physica
  Status Solidi (b)}\ }\textbf {\bibinfo {volume} {44}}~(\bibinfo {number}
  {1}),\ \bibinfo {pages} {245--252}}\BibitemShut {NoStop}%
\bibitem [{\citenamefont {Bishop}\ \emph {et~al.}(2012)\citenamefont {Bishop},
  \citenamefont {Li}, \citenamefont {Farnell},\ and\ \citenamefont
  {Campbell}}]{Bishop2012}%
  \BibitemOpen
  \bibfield  {author} {\bibinfo {author} {\bibnamefont {Bishop}, \bibfnamefont
  {R~F}}, \bibinfo {author} {\bibfnamefont {P~H~Y}\ \bibnamefont {Li}},
  \bibinfo {author} {\bibfnamefont {D~J~J}\ \bibnamefont {Farnell}}, \ and\
  \bibinfo {author} {\bibfnamefont {C~E}\ \bibnamefont {Campbell}}} (\bibinfo
  {year} {2012}),\ \bibfield  {title} {\enquote {\bibinfo {title} {The
  frustrated heisenberg antiferromagnet on the honeycomb lattice:
  ${J}_1$-${J}_2$ model},}\ }\href {\doibase 10.1088/0953-8984/24/23/236002}
  {\bibfield  {journal} {\bibinfo  {journal} {Journal of Physics: Condensed
  Matter}\ }\textbf {\bibinfo {volume} {24}}~(\bibinfo {number} {23}),\
  \bibinfo {pages} {236002}}\BibitemShut {NoStop}%
\bibitem [{\citenamefont {Boulanger}\ \emph {et~al.}(2022)\citenamefont
  {Boulanger}, \citenamefont {Grissonnanche}, \citenamefont
  {Lefran\ifmmode~\mbox{\c{c}}\else \c{c}\fi{}ois}, \citenamefont {Gourgout},
  \citenamefont {Xu}, \citenamefont {Shen}, \citenamefont {Greene},\ and\
  \citenamefont {Taillefer}}]{PhysRevB.105.115101}%
  \BibitemOpen
  \bibfield  {author} {\bibinfo {author} {\bibnamefont {Boulanger},
  \bibfnamefont {Marie-Eve}}, \bibinfo {author} {\bibfnamefont {Ga\"el}\
  \bibnamefont {Grissonnanche}}, \bibinfo {author} {\bibfnamefont {\'Etienne}\
  \bibnamefont {Lefran\ifmmode~\mbox{\c{c}}\else \c{c}\fi{}ois}}, \bibinfo
  {author} {\bibfnamefont {Adrien}\ \bibnamefont {Gourgout}}, \bibinfo {author}
  {\bibfnamefont {Ke-Jun}\ \bibnamefont {Xu}}, \bibinfo {author} {\bibfnamefont
  {Zhi-Xun}\ \bibnamefont {Shen}}, \bibinfo {author} {\bibfnamefont
  {Richard~L.}\ \bibnamefont {Greene}}, \ and\ \bibinfo {author} {\bibfnamefont
  {Louis}\ \bibnamefont {Taillefer}}} (\bibinfo {year} {2022}),\ \bibfield
  {title} {\enquote {\bibinfo {title} {{Thermal Hall conductivity of
  electron-doped cuprates}},}\ }\href {\doibase 10.1103/PhysRevB.105.115101}
  {\bibfield  {journal} {\bibinfo  {journal} {Phys. Rev. B}\ }\textbf {\bibinfo
  {volume} {105}},\ \bibinfo {pages} {115101}}\BibitemShut {NoStop}%
\bibitem [{\citenamefont {Bourgeois-Hope}\ \emph {et~al.}(2019)\citenamefont
  {Bourgeois-Hope}, \citenamefont {Lalibert\'e}, \citenamefont
  {Lefran\ifmmode~\mbox{\c{c}}\else \c{c}\fi{}ois}, \citenamefont
  {Grissonnanche}, \citenamefont {de~Cotret}, \citenamefont {Gordon},
  \citenamefont {Kitou}, \citenamefont {Sawa}, \citenamefont {Cui},
  \citenamefont {Kato}, \citenamefont {Taillefer},\ and\ \citenamefont
  {Doiron-Leyraud}}]{PhysRevX.9.041051}%
  \BibitemOpen
  \bibfield  {author} {\bibinfo {author} {\bibnamefont {Bourgeois-Hope},
  \bibfnamefont {P}}, \bibinfo {author} {\bibfnamefont {F.}~\bibnamefont
  {Lalibert\'e}}, \bibinfo {author} {\bibfnamefont {E.}~\bibnamefont
  {Lefran\ifmmode~\mbox{\c{c}}\else \c{c}\fi{}ois}}, \bibinfo {author}
  {\bibfnamefont {G.}~\bibnamefont {Grissonnanche}}, \bibinfo {author}
  {\bibfnamefont {S.~Ren\'e}\ \bibnamefont {de~Cotret}}, \bibinfo {author}
  {\bibfnamefont {R.}~\bibnamefont {Gordon}}, \bibinfo {author} {\bibfnamefont
  {S.}~\bibnamefont {Kitou}}, \bibinfo {author} {\bibfnamefont
  {H.}~\bibnamefont {Sawa}}, \bibinfo {author} {\bibfnamefont {H.}~\bibnamefont
  {Cui}}, \bibinfo {author} {\bibfnamefont {R.}~\bibnamefont {Kato}}, \bibinfo
  {author} {\bibfnamefont {L.}~\bibnamefont {Taillefer}}, \ and\ \bibinfo
  {author} {\bibfnamefont {N.}~\bibnamefont {Doiron-Leyraud}}} (\bibinfo {year}
  {2019}),\ \bibfield  {title} {\enquote {\bibinfo {title} {{Thermal
  Conductivity of the Quantum Spin Liquid Candidate
  ${\mathrm{EtMe}}_{3}\mathrm{Sb}\mathbf{[}\mathrm{Pd}\mathbf{(}\mathrm{dmit}{\mathbf{)}}_{2}{\mathbf{]}}_{2}$:
  No Evidence of Mobile Gapless Excitations}},}\ }\href {\doibase
  10.1103/PhysRevX.9.041051} {\bibfield  {journal} {\bibinfo  {journal} {Phys.
  Rev. X}\ }\textbf {\bibinfo {volume} {9}},\ \bibinfo {pages}
  {041051}}\BibitemShut {NoStop}%
\bibitem [{\citenamefont {Bramwell}(2001)}]{Bramwell2001}%
  \BibitemOpen
  \bibfield  {author} {\bibinfo {author} {\bibnamefont {Bramwell},
  \bibfnamefont {S~T}}} (\bibinfo {year} {2001}),\ \bibfield  {title} {\enquote
  {\bibinfo {title} {Spin ice state in frustrated magnetic pyrochlore
  materials},}\ }\href {\doibase 10.1126/science.1064761} {\bibfield  {journal}
  {\bibinfo  {journal} {Science}\ }\textbf {\bibinfo {volume} {294}}~(\bibinfo
  {number} {5546}),\ \bibinfo {pages} {1495--1501}}\BibitemShut {NoStop}%
\bibitem [{\citenamefont {Bruno}\ \emph {et~al.}(2004)\citenamefont {Bruno},
  \citenamefont {Dugaev},\ and\ \citenamefont
  {Taillefumier}}]{PhysRevLett.93.096806}%
  \BibitemOpen
  \bibfield  {author} {\bibinfo {author} {\bibnamefont {Bruno}, \bibfnamefont
  {P}}, \bibinfo {author} {\bibfnamefont {V.~K.}\ \bibnamefont {Dugaev}}, \
  and\ \bibinfo {author} {\bibfnamefont {M.}~\bibnamefont {Taillefumier}}}
  (\bibinfo {year} {2004}),\ \bibfield  {title} {\enquote {\bibinfo {title}
  {{Topological Hall Effect and Berry Phase in Magnetic Nanostructures}},}\
  }\href {\doibase 10.1103/PhysRevLett.93.096806} {\bibfield  {journal}
  {\bibinfo  {journal} {Phys. Rev. Lett.}\ }\textbf {\bibinfo {volume} {93}},\
  \bibinfo {pages} {096806}}\BibitemShut {NoStop}%
\bibitem [{\citenamefont {Burnell}\ and\ \citenamefont
  {Nayak}(2011)}]{Nayak2011}%
  \BibitemOpen
  \bibfield  {author} {\bibinfo {author} {\bibnamefont {Burnell}, \bibfnamefont
  {F~J}}, \ and\ \bibinfo {author} {\bibfnamefont {Chetan}\ \bibnamefont
  {Nayak}}} (\bibinfo {year} {2011}),\ \bibfield  {title} {\enquote {\bibinfo
  {title} {{SU(2) slave fermion solution of the Kitaev honeycomb lattice
  model}},}\ }\href {\doibase 10.1103/PhysRevB.84.125125} {\bibfield  {journal}
  {\bibinfo  {journal} {Phys. Rev. B}\ }\textbf {\bibinfo {volume} {84}},\
  \bibinfo {pages} {125125}}\BibitemShut {NoStop}%
\bibitem [{\citenamefont {Cairns}\ \emph {et~al.}(2020)\citenamefont {Cairns},
  \citenamefont {Reid}, \citenamefont {Perry}, \citenamefont {Prabhakaran},\
  and\ \citenamefont {Huxley}}]{Cairns2020}%
  \BibitemOpen
  \bibfield  {author} {\bibinfo {author} {\bibnamefont {Cairns}, \bibfnamefont
  {Luke~Pritchard}}, \bibinfo {author} {\bibfnamefont {J.-Ph.}\ \bibnamefont
  {Reid}}, \bibinfo {author} {\bibfnamefont {Robin}\ \bibnamefont {Perry}},
  \bibinfo {author} {\bibfnamefont {Dharmalingam}\ \bibnamefont {Prabhakaran}},
  \ and\ \bibinfo {author} {\bibfnamefont {Andrew}\ \bibnamefont {Huxley}}}
  (\bibinfo {year} {2020}),\ \bibfield  {title} {\enquote {\bibinfo {title}
  {{Thermal Hall Effect of Topological Triplons in
  SrCu${}_{2}$(BO${}_{3}$)${}_{2}$}},}\ }in\ \href {\doibase
  10.7566/jpscp.30.011089} {\emph {\bibinfo {booktitle} {Proceedings of the
  International Conference on Strongly Correlated Electron Systems
  ({SCES}2019)}}}\ (\bibinfo  {publisher} {Journal of the Physical Society of
  Japan})\BibitemShut {NoStop}%
\bibitem [{\citenamefont {Castelnovo}\ \emph {et~al.}(2008)\citenamefont
  {Castelnovo}, \citenamefont {Moessner},\ and\ \citenamefont
  {Sondhi}}]{Castelnovo2008}%
  \BibitemOpen
  \bibfield  {author} {\bibinfo {author} {\bibnamefont {Castelnovo},
  \bibfnamefont {C}}, \bibinfo {author} {\bibfnamefont {R.}~\bibnamefont
  {Moessner}}, \ and\ \bibinfo {author} {\bibfnamefont {S.~L.}\ \bibnamefont
  {Sondhi}}} (\bibinfo {year} {2008}),\ \bibfield  {title} {\enquote {\bibinfo
  {title} {Magnetic monopoles in spin ice},}\ }\href {\doibase
  10.1038/nature06433} {\bibfield  {journal} {\bibinfo  {journal} {Nature}\
  }\textbf {\bibinfo {volume} {451}}~(\bibinfo {number} {7174}),\ \bibinfo
  {pages} {42--45}}\BibitemShut {NoStop}%
\bibitem [{\citenamefont {C\'epas}\ \emph {et~al.}(2001)\citenamefont
  {C\'epas}, \citenamefont {Kakurai}, \citenamefont {Regnault}, \citenamefont
  {Ziman}, \citenamefont {Boucher}, \citenamefont {Aso}, \citenamefont {Nishi},
  \citenamefont {Kageyama},\ and\ \citenamefont {Ueda}}]{Cepas2001}%
  \BibitemOpen
  \bibfield  {author} {\bibinfo {author} {\bibnamefont {C\'epas}, \bibfnamefont
  {O}}, \bibinfo {author} {\bibfnamefont {K.}~\bibnamefont {Kakurai}}, \bibinfo
  {author} {\bibfnamefont {L.~P.}\ \bibnamefont {Regnault}}, \bibinfo {author}
  {\bibfnamefont {T.}~\bibnamefont {Ziman}}, \bibinfo {author} {\bibfnamefont
  {J.~P.}\ \bibnamefont {Boucher}}, \bibinfo {author} {\bibfnamefont
  {N.}~\bibnamefont {Aso}}, \bibinfo {author} {\bibfnamefont {M.}~\bibnamefont
  {Nishi}}, \bibinfo {author} {\bibfnamefont {H.}~\bibnamefont {Kageyama}}, \
  and\ \bibinfo {author} {\bibfnamefont {Y.}~\bibnamefont {Ueda}}} (\bibinfo
  {year} {2001}),\ \bibfield  {title} {\enquote {\bibinfo {title}
  {{Dzyaloshinski-Moriya Interaction in the 2D Spin Gap System
  ${\mathrm{SrCu}}_{2}({\mathrm{BO}}_{3}{)}_{2}$}},}\ }\href {\doibase
  10.1103/PhysRevLett.87.167205} {\bibfield  {journal} {\bibinfo  {journal}
  {Phys. Rev. Lett.}\ }\textbf {\bibinfo {volume} {87}},\ \bibinfo {pages}
  {167205}}\BibitemShut {NoStop}%
\bibitem [{\citenamefont {Chaloupka}\ \emph {et~al.}(2010)\citenamefont
  {Chaloupka}, \citenamefont {Jackeli},\ and\ \citenamefont
  {Khaliullin}}]{Khaliullin2010}%
  \BibitemOpen
  \bibfield  {author} {\bibinfo {author} {\bibnamefont {Chaloupka},
  \bibfnamefont {Ji\ifmmode \check{r}\else~\v{r}\fi{}\'{\i}}}, \bibinfo
  {author} {\bibfnamefont {George}\ \bibnamefont {Jackeli}}, \ and\ \bibinfo
  {author} {\bibfnamefont {Giniyat}\ \bibnamefont {Khaliullin}}} (\bibinfo
  {year} {2010}),\ \bibfield  {title} {\enquote {\bibinfo {title}
  {{Kitaev-Heisenberg model on a honeycomb lattice: Possible exotic phases in
  Iridium oxides ${A}_{2}{\mathrm{IrO}}_{3}$}},}\ }\href {\doibase
  10.1103/PhysRevLett.105.027204} {\bibfield  {journal} {\bibinfo  {journal}
  {Phys. Rev. Lett.}\ }\textbf {\bibinfo {volume} {105}},\ \bibinfo {pages}
  {027204}}\BibitemShut {NoStop}%
\bibitem [{\citenamefont {Che}\ \emph {et~al.}(2023)\citenamefont {Che},
  \citenamefont {Li}, \citenamefont {Wu}, \citenamefont {Li}, \citenamefont
  {Guang}, \citenamefont {Xia}, \citenamefont {Yue}, \citenamefont {Wang},
  \citenamefont {Zhao}, \citenamefont {Li},\ and\ \citenamefont
  {Sun}}]{PhysRevB.107.054429}%
  \BibitemOpen
  \bibfield  {author} {\bibinfo {author} {\bibnamefont {Che}, \bibfnamefont
  {H~L}}, \bibinfo {author} {\bibfnamefont {S.~J.}\ \bibnamefont {Li}},
  \bibinfo {author} {\bibfnamefont {J.~C.}\ \bibnamefont {Wu}}, \bibinfo
  {author} {\bibfnamefont {N.}~\bibnamefont {Li}}, \bibinfo {author}
  {\bibfnamefont {S.~K.}\ \bibnamefont {Guang}}, \bibinfo {author}
  {\bibfnamefont {K.}~\bibnamefont {Xia}}, \bibinfo {author} {\bibfnamefont
  {X.~Y.}\ \bibnamefont {Yue}}, \bibinfo {author} {\bibfnamefont {Y.~Y.}\
  \bibnamefont {Wang}}, \bibinfo {author} {\bibfnamefont {X.}~\bibnamefont
  {Zhao}}, \bibinfo {author} {\bibfnamefont {Q.~J.}\ \bibnamefont {Li}}, \ and\
  \bibinfo {author} {\bibfnamefont {X.~F.}\ \bibnamefont {Sun}}} (\bibinfo
  {year} {2023}),\ \bibfield  {title} {\enquote {\bibinfo {title}
  {{Low-temperature specific heat and heat transport of
  ${\mathrm{Tb}}_{2}{\mathrm{Ti}}_{2\ensuremath{-}x}{\mathrm{Zr}}_{x}{\mathrm{O}}_{7}$
  single crystals}},}\ }\href {\doibase 10.1103/PhysRevB.107.054429} {\bibfield
   {journal} {\bibinfo  {journal} {Phys. Rev. B}\ }\textbf {\bibinfo {volume}
  {107}},\ \bibinfo {pages} {054429}}\BibitemShut {NoStop}%
\bibitem [{\citenamefont {Chen}\ and\ \citenamefont
  {Villadiego}(2023)}]{PhysRevB.107.045114}%
  \BibitemOpen
  \bibfield  {author} {\bibinfo {author} {\bibnamefont {Chen}, \bibfnamefont
  {Chuan}}, \ and\ \bibinfo {author} {\bibfnamefont {Inti~Sodemann}\
  \bibnamefont {Villadiego}}} (\bibinfo {year} {2023}),\ \bibfield  {title}
  {\enquote {\bibinfo {title} {{Nature of visons in the perturbed ferromagnetic
  and antiferromagnetic Kitaev honeycomb models}},}\ }\href {\doibase
  10.1103/PhysRevB.107.045114} {\bibfield  {journal} {\bibinfo  {journal}
  {Phys. Rev. B}\ }\textbf {\bibinfo {volume} {107}},\ \bibinfo {pages}
  {045114}}\BibitemShut {NoStop}%
\bibitem [{\citenamefont {Chen}(2016)}]{Gang2016}%
  \BibitemOpen
  \bibfield  {author} {\bibinfo {author} {\bibnamefont {Chen}, \bibfnamefont
  {Gang}}} (\bibinfo {year} {2016}),\ \bibfield  {title} {\enquote {\bibinfo
  {title} {{``Magnetic monopole'' condensation of the pyrochlore ice {U(1)}
  quantum spin liquid: Application to
  ${\mathrm{Pr}}_{2}{\mathrm{Ir}}_{2}{\mathrm{O}}_{7}$ and
  ${\mathrm{Yb}}_{2}{\mathrm{Ti}}_{2}{\mathrm{O}}_{7}$}},}\ }\href {\doibase
  10.1103/PhysRevB.94.205107} {\bibfield  {journal} {\bibinfo  {journal} {Phys.
  Rev. B}\ }\textbf {\bibinfo {volume} {94}},\ \bibinfo {pages}
  {205107}}\BibitemShut {NoStop}%
\bibitem [{\citenamefont {Chen}(2017{\natexlab{a}})}]{Gang2017D}%
  \BibitemOpen
  \bibfield  {author} {\bibinfo {author} {\bibnamefont {Chen}, \bibfnamefont
  {Gang}}} (\bibinfo {year} {2017}{\natexlab{a}}),\ \bibfield  {title}
  {\enquote {\bibinfo {title} {{Dirac's ``magnetic monopoles'' in pyrochlore
  ice U(1) spin liquids: Spectrum and classification}},}\ }\href {\doibase
  10.1103/PhysRevB.96.195127} {\bibfield  {journal} {\bibinfo  {journal} {Phys.
  Rev. B}\ }\textbf {\bibinfo {volume} {96}},\ \bibinfo {pages}
  {195127}}\BibitemShut {NoStop}%
\bibitem [{\citenamefont {Chen}(2017{\natexlab{b}})}]{PhysRevB.96.020412}%
  \BibitemOpen
  \bibfield  {author} {\bibinfo {author} {\bibnamefont {Chen}, \bibfnamefont
  {Gang}}} (\bibinfo {year} {2017}{\natexlab{b}}),\ \bibfield  {title}
  {\enquote {\bibinfo {title} {Quantum paramagnet and frustrated quantum
  criticality in a spin-one diamond lattice antiferromagnet},}\ }\href
  {\doibase 10.1103/PhysRevB.96.020412} {\bibfield  {journal} {\bibinfo
  {journal} {Phys. Rev. B}\ }\textbf {\bibinfo {volume} {96}},\ \bibinfo
  {pages} {020412}}\BibitemShut {NoStop}%
\bibitem [{\citenamefont {Chen}(2017{\natexlab{c}})}]{Gang2017E}%
  \BibitemOpen
  \bibfield  {author} {\bibinfo {author} {\bibnamefont {Chen}, \bibfnamefont
  {Gang}}} (\bibinfo {year} {2017}{\natexlab{c}}),\ \bibfield  {title}
  {\enquote {\bibinfo {title} {Spectral periodicity of the spinon continuum in
  quantum spin ice},}\ }\href {\doibase 10.1103/PhysRevB.96.085136} {\bibfield
  {journal} {\bibinfo  {journal} {Phys. Rev. B}\ }\textbf {\bibinfo {volume}
  {96}},\ \bibinfo {pages} {085136}}\BibitemShut {NoStop}%
\bibitem [{\citenamefont {Chen}(2019)}]{PhysRevResearch.1.033141}%
  \BibitemOpen
  \bibfield  {author} {\bibinfo {author} {\bibnamefont {Chen}, \bibfnamefont
  {Gang}}} (\bibinfo {year} {2019}),\ \bibfield  {title} {\enquote {\bibinfo
  {title} {Intrinsic transverse field in frustrated quantum ising magnets:
  Physical origin and quantum effects},}\ }\href {\doibase
  10.1103/PhysRevResearch.1.033141} {\bibfield  {journal} {\bibinfo  {journal}
  {Phys. Rev. Res.}\ }\textbf {\bibinfo {volume} {1}},\ \bibinfo {pages}
  {033141}}\BibitemShut {NoStop}%
\bibitem [{\citenamefont {Chen}(2022)}]{chen2022longrange}%
  \BibitemOpen
  \bibfield  {author} {\bibinfo {author} {\bibnamefont {Chen}, \bibfnamefont
  {Gang}}} (\bibinfo {year} {2022}),\ \href@noop {} {\enquote {\bibinfo {title}
  {{Long-range order coexisting with long-range entanglement: Is
  Nd$_2$Sn$_2$O$_7$ the first Coulombic antiferromagnet with a visible emergent
  gauge photon?}}}\ }\Eprint {http://arxiv.org/abs/2208.07939}
  {arXiv:2208.07939 [cond-mat.str-el]} \BibitemShut {NoStop}%
\bibitem [{\citenamefont {Chen}(2023{\natexlab{a}})}]{chen2023distinguishing}%
  \BibitemOpen
  \bibfield  {author} {\bibinfo {author} {\bibnamefont {Chen}, \bibfnamefont
  {Gang}}} (\bibinfo {year} {2023}{\natexlab{a}}),\ \href@noop {} {\enquote
  {\bibinfo {title} {{Distinguishing thermodynamics and spectroscopy for
  octupolar U(1) spin liquid of Ce-pyrochlores}},}\ }\Eprint
  {http://arxiv.org/abs/2304.01892} {arXiv:2304.01892 [cond-mat.str-el]}
  \BibitemShut {NoStop}%
\bibitem [{\citenamefont {Chen}(2023{\natexlab{b}})}]{Chenunpub2023}%
  \BibitemOpen
  \bibfield  {author} {\bibinfo {author} {\bibnamefont {Chen}, \bibfnamefont
  {Gang}}} (\bibinfo {year} {2023}{\natexlab{b}}),\ \href@noop {} {\enquote
  {\bibinfo {title} {{Kitaev materials as weak Mott insulators}},}\ }\bibinfo
  {note} {In preparation}\BibitemShut {NoStop}%
\bibitem [{\citenamefont {Chen}\ and\ \citenamefont
  {Balents}(2008)}]{Gang2008}%
  \BibitemOpen
  \bibfield  {author} {\bibinfo {author} {\bibnamefont {Chen}, \bibfnamefont
  {Gang}}, \ and\ \bibinfo {author} {\bibfnamefont {Leon}\ \bibnamefont
  {Balents}}} (\bibinfo {year} {2008}),\ \bibfield  {title} {\enquote {\bibinfo
  {title} {{Spin-orbit effects in
  ${\text{Na}}_{4}{\text{Ir}}_{3}{\text{O}}_{8}$: A hyper-kagom{\'{e}} lattice
  antiferromagnet}},}\ }\href {\doibase 10.1103/PhysRevB.78.094403} {\bibfield
  {journal} {\bibinfo  {journal} {Phys. Rev. B}\ }\textbf {\bibinfo {volume}
  {78}},\ \bibinfo {pages} {094403}}\BibitemShut {NoStop}%
\bibitem [{\citenamefont {Chen}\ and\ \citenamefont
  {Balents}(2011)}]{PhysRevB.84.094420}%
  \BibitemOpen
  \bibfield  {author} {\bibinfo {author} {\bibnamefont {Chen}, \bibfnamefont
  {Gang}}, \ and\ \bibinfo {author} {\bibfnamefont {Leon}\ \bibnamefont
  {Balents}}} (\bibinfo {year} {2011}),\ \bibfield  {title} {\enquote {\bibinfo
  {title} {Spin-orbit coupling in ${d}^{2}$ ordered double perovskites},}\
  }\href {\doibase 10.1103/PhysRevB.84.094420} {\bibfield  {journal} {\bibinfo
  {journal} {Phys. Rev. B}\ }\textbf {\bibinfo {volume} {84}},\ \bibinfo
  {pages} {094420}}\BibitemShut {NoStop}%
\bibitem [{\citenamefont {Chen}\ \emph
  {et~al.}(2009{\natexlab{a}})\citenamefont {Chen}, \citenamefont {Balents},\
  and\ \citenamefont {Schnyder}}]{PhysRevLett.102.096406}%
  \BibitemOpen
  \bibfield  {author} {\bibinfo {author} {\bibnamefont {Chen}, \bibfnamefont
  {Gang}}, \bibinfo {author} {\bibfnamefont {Leon}\ \bibnamefont {Balents}}, \
  and\ \bibinfo {author} {\bibfnamefont {Andreas~P.}\ \bibnamefont {Schnyder}}}
  (\bibinfo {year} {2009}{\natexlab{a}}),\ \bibfield  {title} {\enquote
  {\bibinfo {title} {{Spin-Orbital Singlet and Quantum Critical Point on the
  Diamond Lattice: ${\mathrm{FeSc}}_{2}{\mathbf{S}}_{4}$}},}\ }\href {\doibase
  10.1103/PhysRevLett.102.096406} {\bibfield  {journal} {\bibinfo  {journal}
  {Phys. Rev. Lett.}\ }\textbf {\bibinfo {volume} {102}},\ \bibinfo {pages}
  {096406}}\BibitemShut {NoStop}%
\bibitem [{\citenamefont {Chen}\ and\ \citenamefont
  {Hermele}(2012)}]{PhysRevB.86.235129}%
  \BibitemOpen
  \bibfield  {author} {\bibinfo {author} {\bibnamefont {Chen}, \bibfnamefont
  {Gang}}, \ and\ \bibinfo {author} {\bibfnamefont {Michael}\ \bibnamefont
  {Hermele}}} (\bibinfo {year} {2012}),\ \bibfield  {title} {\enquote {\bibinfo
  {title} {Magnetic orders and topological phases from $f$-$d$ exchange in
  pyrochlore iridates},}\ }\href {\doibase 10.1103/PhysRevB.86.235129}
  {\bibfield  {journal} {\bibinfo  {journal} {Phys. Rev. B}\ }\textbf {\bibinfo
  {volume} {86}},\ \bibinfo {pages} {235129}}\BibitemShut {NoStop}%
\bibitem [{\citenamefont {Chen}\ \emph {et~al.}(2012)\citenamefont {Chen},
  \citenamefont {Hermele},\ and\ \citenamefont
  {Radzihovsky}}]{PhysRevLett.109.016402}%
  \BibitemOpen
  \bibfield  {author} {\bibinfo {author} {\bibnamefont {Chen}, \bibfnamefont
  {Gang}}, \bibinfo {author} {\bibfnamefont {Michael}\ \bibnamefont {Hermele}},
  \ and\ \bibinfo {author} {\bibfnamefont {Leo}\ \bibnamefont {Radzihovsky}}}
  (\bibinfo {year} {2012}),\ \bibfield  {title} {\enquote {\bibinfo {title}
  {{Frustrated Quantum Critical Theory of Putative Spin-Liquid Phenomenology in
  $6H\mathrm{\text{\ensuremath{-}}}\mathrm{B}\mathrm{\text{\ensuremath{-}}}{\mathrm{Ba}}_{3}{\mathrm{NiSb}}_{2}{\mathrm{O}}_{9}$}},}\
  }\href {\doibase 10.1103/PhysRevLett.109.016402} {\bibfield  {journal}
  {\bibinfo  {journal} {Phys. Rev. Lett.}\ }\textbf {\bibinfo {volume} {109}},\
  \bibinfo {pages} {016402}}\BibitemShut {NoStop}%
\bibitem [{\citenamefont {Chen}\ \emph {et~al.}(2014)\citenamefont {Chen},
  \citenamefont {Kee},\ and\ \citenamefont {Kim}}]{PhysRevLett.113.197202}%
  \BibitemOpen
  \bibfield  {author} {\bibinfo {author} {\bibnamefont {Chen}, \bibfnamefont
  {Gang}}, \bibinfo {author} {\bibfnamefont {Hae-Young}\ \bibnamefont {Kee}}, \
  and\ \bibinfo {author} {\bibfnamefont {Yong~Baek}\ \bibnamefont {Kim}}}
  (\bibinfo {year} {2014}),\ \bibfield  {title} {\enquote {\bibinfo {title}
  {Fractionalized charge excitations in a spin liquid on partially filled
  pyrochlore lattices},}\ }\href {\doibase 10.1103/PhysRevLett.113.197202}
  {\bibfield  {journal} {\bibinfo  {journal} {Phys. Rev. Lett.}\ }\textbf
  {\bibinfo {volume} {113}},\ \bibinfo {pages} {197202}}\BibitemShut {NoStop}%
\bibitem [{\citenamefont {Chen}\ \emph {et~al.}(2016)\citenamefont {Chen},
  \citenamefont {Kee},\ and\ \citenamefont {Kim}}]{PhysRevB.93.245134}%
  \BibitemOpen
  \bibfield  {author} {\bibinfo {author} {\bibnamefont {Chen}, \bibfnamefont
  {Gang}}, \bibinfo {author} {\bibfnamefont {Hae-Young}\ \bibnamefont {Kee}}, \
  and\ \bibinfo {author} {\bibfnamefont {Yong~Baek}\ \bibnamefont {Kim}}}
  (\bibinfo {year} {2016}),\ \bibfield  {title} {\enquote {\bibinfo {title}
  {{Cluster Mott insulators and two Curie-Weiss regimes on an anisotropic
  kagome lattice}},}\ }\href {\doibase 10.1103/PhysRevB.93.245134} {\bibfield
  {journal} {\bibinfo  {journal} {Phys. Rev. B}\ }\textbf {\bibinfo {volume}
  {93}},\ \bibinfo {pages} {245134}}\BibitemShut {NoStop}%
\bibitem [{\citenamefont {Chen}\ and\ \citenamefont
  {Lee}(2018)}]{PhysRevB.97.035124}%
  \BibitemOpen
  \bibfield  {author} {\bibinfo {author} {\bibnamefont {Chen}, \bibfnamefont
  {Gang}}, \ and\ \bibinfo {author} {\bibfnamefont {Patrick~A.}\ \bibnamefont
  {Lee}}} (\bibinfo {year} {2018}),\ \bibfield  {title} {\enquote {\bibinfo
  {title} {{Emergent orbitals in the cluster Mott insulator on a breathing
  kagome lattice}},}\ }\href {\doibase 10.1103/PhysRevB.97.035124} {\bibfield
  {journal} {\bibinfo  {journal} {Phys. Rev. B}\ }\textbf {\bibinfo {volume}
  {97}},\ \bibinfo {pages} {035124}}\BibitemShut {NoStop}%
\bibitem [{\citenamefont {Chen}\ \emph {et~al.}(2010)\citenamefont {Chen},
  \citenamefont {Pereira},\ and\ \citenamefont {Balents}}]{PhysRevB.82.174440}%
  \BibitemOpen
  \bibfield  {author} {\bibinfo {author} {\bibnamefont {Chen}, \bibfnamefont
  {Gang}}, \bibinfo {author} {\bibfnamefont {Rodrigo}\ \bibnamefont {Pereira}},
  \ and\ \bibinfo {author} {\bibfnamefont {Leon}\ \bibnamefont {Balents}}}
  (\bibinfo {year} {2010}),\ \bibfield  {title} {\enquote {\bibinfo {title}
  {Exotic phases induced by strong spin-orbit coupling in ordered double
  perovskites},}\ }\href {\doibase 10.1103/PhysRevB.82.174440} {\bibfield
  {journal} {\bibinfo  {journal} {Phys. Rev. B}\ }\textbf {\bibinfo {volume}
  {82}},\ \bibinfo {pages} {174440}}\BibitemShut {NoStop}%
\bibitem [{\citenamefont {Chen}\ \emph
  {et~al.}(2009{\natexlab{b}})\citenamefont {Chen}, \citenamefont {Schnyder},\
  and\ \citenamefont {Balents}}]{PhysRevB.80.224409}%
  \BibitemOpen
  \bibfield  {author} {\bibinfo {author} {\bibnamefont {Chen}, \bibfnamefont
  {Gang}}, \bibinfo {author} {\bibfnamefont {Andreas~P.}\ \bibnamefont
  {Schnyder}}, \ and\ \bibinfo {author} {\bibfnamefont {Leon}\ \bibnamefont
  {Balents}}} (\bibinfo {year} {2009}{\natexlab{b}}),\ \bibfield  {title}
  {\enquote {\bibinfo {title} {{Excitation spectrum and magnetic field effects
  in a quantum critical spin-orbital system: The case of
  ${\text{FeSc}}_{2}{\text{S}}_{4}$}},}\ }\href {\doibase
  10.1103/PhysRevB.80.224409} {\bibfield  {journal} {\bibinfo  {journal} {Phys.
  Rev. B}\ }\textbf {\bibinfo {volume} {80}},\ \bibinfo {pages}
  {224409}}\BibitemShut {NoStop}%
\bibitem [{\citenamefont {Chen}\ and\ \citenamefont {Wu}(2021)}]{chen2021mott}%
  \BibitemOpen
  \bibfield  {author} {\bibinfo {author} {\bibnamefont {Chen}, \bibfnamefont
  {Gang}}, \ and\ \bibinfo {author} {\bibfnamefont {Congjun}\ \bibnamefont
  {Wu}}} (\bibinfo {year} {2021}),\ \href@noop {} {\enquote {\bibinfo {title}
  {{Mott insulators with large local Hilbert spaces in quantum materials and
  ultracold atoms}},}\ }\Eprint {http://arxiv.org/abs/2112.02630}
  {arXiv:2112.02630 [cond-mat.str-el]} \BibitemShut {NoStop}%
\bibitem [{\citenamefont {Chen}\ \emph {et~al.}(2020)\citenamefont {Chen},
  \citenamefont {Kivelson},\ and\ \citenamefont {Sun}}]{Jingyuan2020}%
  \BibitemOpen
  \bibfield  {author} {\bibinfo {author} {\bibnamefont {Chen}, \bibfnamefont
  {Jing-Yuan}}, \bibinfo {author} {\bibfnamefont {Steven~A.}\ \bibnamefont
  {Kivelson}}, \ and\ \bibinfo {author} {\bibfnamefont {Xiao-Qi}\ \bibnamefont
  {Sun}}} (\bibinfo {year} {2020}),\ \bibfield  {title} {\enquote {\bibinfo
  {title} {Enhanced thermal hall effect in nearly ferroelectric insulators},}\
  }\href {\doibase 10.1103/PhysRevLett.124.167601} {\bibfield  {journal}
  {\bibinfo  {journal} {Phys. Rev. Lett.}\ }\textbf {\bibinfo {volume} {124}},\
  \bibinfo {pages} {167601}}\BibitemShut {NoStop}%
\bibitem [{\citenamefont {Chen}\ \emph {et~al.}(2018)\citenamefont {Chen},
  \citenamefont {Chung}, \citenamefont {Gao}, \citenamefont {Chen},
  \citenamefont {Stone}, \citenamefont {Kolesnikov}, \citenamefont {Huang},\
  and\ \citenamefont {Dai}}]{PhysRevX.8.041028}%
  \BibitemOpen
  \bibfield  {author} {\bibinfo {author} {\bibnamefont {Chen}, \bibfnamefont
  {Lebing}}, \bibinfo {author} {\bibfnamefont {Jae-Ho}\ \bibnamefont {Chung}},
  \bibinfo {author} {\bibfnamefont {Bin}\ \bibnamefont {Gao}}, \bibinfo
  {author} {\bibfnamefont {Tong}\ \bibnamefont {Chen}}, \bibinfo {author}
  {\bibfnamefont {Matthew~B.}\ \bibnamefont {Stone}}, \bibinfo {author}
  {\bibfnamefont {Alexander~I.}\ \bibnamefont {Kolesnikov}}, \bibinfo {author}
  {\bibfnamefont {Qingzhen}\ \bibnamefont {Huang}}, \ and\ \bibinfo {author}
  {\bibfnamefont {Pengcheng}\ \bibnamefont {Dai}}} (\bibinfo {year} {2018}),\
  \bibfield  {title} {\enquote {\bibinfo {title} {{Topological Spin Excitations
  in Honeycomb Ferromagnet ${\mathrm{CrI}}_{3}$}},}\ }\href {\doibase
  10.1103/PhysRevX.8.041028} {\bibfield  {journal} {\bibinfo  {journal} {Phys.
  Rev. X}\ }\textbf {\bibinfo {volume} {8}},\ \bibinfo {pages}
  {041028}}\BibitemShut {NoStop}%
\bibitem [{\citenamefont {Chen}\ \emph {et~al.}(2022)\citenamefont {Chen},
  \citenamefont {Boulanger}, \citenamefont {Wang}, \citenamefont {Tafti},\ and\
  \citenamefont {Taillefer}}]{Cu3TeO6}%
  \BibitemOpen
  \bibfield  {author} {\bibinfo {author} {\bibnamefont {Chen}, \bibfnamefont
  {Lu}}, \bibinfo {author} {\bibfnamefont {Marie-Eve}\ \bibnamefont
  {Boulanger}}, \bibinfo {author} {\bibfnamefont {Zhi-Cheng}\ \bibnamefont
  {Wang}}, \bibinfo {author} {\bibfnamefont {Fazel}\ \bibnamefont {Tafti}}, \
  and\ \bibinfo {author} {\bibfnamefont {Louis}\ \bibnamefont {Taillefer}}}
  (\bibinfo {year} {2022}),\ \bibfield  {title} {\enquote {\bibinfo {title}
  {{Large phonon thermal Hall conductivity in the antiferromagnetic insulator
  Cu$_3$TeO$_6$}},}\ }\href {\doibase 10.1073/pnas.2208016119} {\bibfield
  {journal} {\bibinfo  {journal} {Proceedings of the National Academy of
  Sciences}\ }\textbf {\bibinfo {volume} {119}},\ \bibinfo {pages}
  {e2208016119}}\BibitemShut {NoStop}%
\bibitem [{\citenamefont {Cheng}\ \emph {et~al.}(2007)\citenamefont {Cheng},
  \citenamefont {C\'epas}, \citenamefont {Leung},\ and\ \citenamefont
  {Ziman}}]{Cheng2007}%
  \BibitemOpen
  \bibfield  {author} {\bibinfo {author} {\bibnamefont {Cheng}, \bibfnamefont
  {Y~F}}, \bibinfo {author} {\bibfnamefont {O.}~\bibnamefont {C\'epas}},
  \bibinfo {author} {\bibfnamefont {P.~W.}\ \bibnamefont {Leung}}, \ and\
  \bibinfo {author} {\bibfnamefont {T.}~\bibnamefont {Ziman}}} (\bibinfo {year}
  {2007}),\ \bibfield  {title} {\enquote {\bibinfo {title} {{Magnon dispersion
  and anisotropies in
  $\mathrm{Sr}{\mathrm{Cu}}_{2}{(\mathrm{B}{\mathrm{O}}_{3})}_{2}$}},}\ }\href
  {\doibase 10.1103/PhysRevB.75.144422} {\bibfield  {journal} {\bibinfo
  {journal} {Phys. Rev. B}\ }\textbf {\bibinfo {volume} {75}},\ \bibinfo
  {pages} {144422}}\BibitemShut {NoStop}%
\bibitem [{\citenamefont {Chern}\ \emph {et~al.}(2021)\citenamefont {Chern},
  \citenamefont {Zhang},\ and\ \citenamefont {Kim}}]{PhysRevLett.126.147201}%
  \BibitemOpen
  \bibfield  {author} {\bibinfo {author} {\bibnamefont {Chern}, \bibfnamefont
  {Li~Ern}}, \bibinfo {author} {\bibfnamefont {Emily~Z.}\ \bibnamefont
  {Zhang}}, \ and\ \bibinfo {author} {\bibfnamefont {Yong~Baek}\ \bibnamefont
  {Kim}}} (\bibinfo {year} {2021}),\ \bibfield  {title} {\enquote {\bibinfo
  {title} {{Sign structure of thermal Hall conductivity and topological magnons
  for in-plane field polarized Kitaev magnets}},}\ }\href {\doibase
  10.1103/PhysRevLett.126.147201} {\bibfield  {journal} {\bibinfo  {journal}
  {Phys. Rev. Lett.}\ }\textbf {\bibinfo {volume} {126}},\ \bibinfo {pages}
  {147201}}\BibitemShut {NoStop}%
\bibitem [{\citenamefont {Chisnell}\ \emph {et~al.}(2015)\citenamefont
  {Chisnell}, \citenamefont {Helton}, \citenamefont {Freedman}, \citenamefont
  {Singh}, \citenamefont {Bewley}, \citenamefont {Nocera},\ and\ \citenamefont
  {Lee}}]{Chisnell2015}%
  \BibitemOpen
  \bibfield  {author} {\bibinfo {author} {\bibnamefont {Chisnell},
  \bibfnamefont {R}}, \bibinfo {author} {\bibfnamefont {J.~S.}\ \bibnamefont
  {Helton}}, \bibinfo {author} {\bibfnamefont {D.~E.}\ \bibnamefont
  {Freedman}}, \bibinfo {author} {\bibfnamefont {D.~K.}\ \bibnamefont {Singh}},
  \bibinfo {author} {\bibfnamefont {R.~I.}\ \bibnamefont {Bewley}}, \bibinfo
  {author} {\bibfnamefont {D.~G.}\ \bibnamefont {Nocera}}, \ and\ \bibinfo
  {author} {\bibfnamefont {Y.~S.}\ \bibnamefont {Lee}}} (\bibinfo {year}
  {2015}),\ \bibfield  {title} {\enquote {\bibinfo {title} {{Topological magnon
  bands in a kagom{\'{e}} lattice ferromagnet}},}\ }\href {\doibase
  10.1103/PhysRevLett.115.147201} {\bibfield  {journal} {\bibinfo  {journal}
  {Phys. Rev. Lett.}\ }\textbf {\bibinfo {volume} {115}},\ \bibinfo {pages}
  {147201}}\BibitemShut {NoStop}%
\bibitem [{\citenamefont {Clark}\ \emph {et~al.}(2011)\citenamefont {Clark},
  \citenamefont {Abanin},\ and\ \citenamefont {Sondhi}}]{Sondhi2011}%
  \BibitemOpen
  \bibfield  {author} {\bibinfo {author} {\bibnamefont {Clark}, \bibfnamefont
  {B~K}}, \bibinfo {author} {\bibfnamefont {D.~A.}\ \bibnamefont {Abanin}}, \
  and\ \bibinfo {author} {\bibfnamefont {S.~L.}\ \bibnamefont {Sondhi}}}
  (\bibinfo {year} {2011}),\ \bibfield  {title} {\enquote {\bibinfo {title}
  {{Nature of the spin liquid state of the Hubbard model on a honeycomb
  lattice}},}\ }\href {\doibase 10.1103/PhysRevLett.107.087204} {\bibfield
  {journal} {\bibinfo  {journal} {Phys. Rev. Lett.}\ }\textbf {\bibinfo
  {volume} {107}},\ \bibinfo {pages} {087204}}\BibitemShut {NoStop}%
\bibitem [{\citenamefont {Cookmeyer}\ and\ \citenamefont
  {Moore}(2018)}]{Moore2018}%
  \BibitemOpen
  \bibfield  {author} {\bibinfo {author} {\bibnamefont {Cookmeyer},
  \bibfnamefont {Jonathan}}, \ and\ \bibinfo {author} {\bibfnamefont {Joel~E.}\
  \bibnamefont {Moore}}} (\bibinfo {year} {2018}),\ \bibfield  {title}
  {\enquote {\bibinfo {title} {{Spin-wave analysis of the low-temperature
  thermal Hall effect in the candidate Kitaev spin liquid
  $\ensuremath{\alpha}\ensuremath{-}{\mathrm{RuCl}}_{3}$}},}\ }\href {\doibase
  10.1103/PhysRevB.98.060412} {\bibfield  {journal} {\bibinfo  {journal} {Phys.
  Rev. B}\ }\textbf {\bibinfo {volume} {98}},\ \bibinfo {pages}
  {060412}}\BibitemShut {NoStop}%
\bibitem [{\citenamefont {Cookmeyer}\ \emph {et~al.}(2021)\citenamefont
  {Cookmeyer}, \citenamefont {Motruk},\ and\ \citenamefont
  {Moore}}]{PhysRevLett.127.087201}%
  \BibitemOpen
  \bibfield  {author} {\bibinfo {author} {\bibnamefont {Cookmeyer},
  \bibfnamefont {Tessa}}, \bibinfo {author} {\bibfnamefont {Johannes}\
  \bibnamefont {Motruk}}, \ and\ \bibinfo {author} {\bibfnamefont {Joel~E.}\
  \bibnamefont {Moore}}} (\bibinfo {year} {2021}),\ \bibfield  {title}
  {\enquote {\bibinfo {title} {{Four-Spin Terms and the Origin of the Chiral
  Spin Liquid in Mott Insulators on the Triangular Lattice}},}\ }\href
  {\doibase 10.1103/PhysRevLett.127.087201} {\bibfield  {journal} {\bibinfo
  {journal} {Phys. Rev. Lett.}\ }\textbf {\bibinfo {volume} {127}},\ \bibinfo
  {pages} {087201}}\BibitemShut {NoStop}%
\bibitem [{\citenamefont {Czajka}\ \emph {et~al.}(2021)\citenamefont {Czajka},
  \citenamefont {Gao}, \citenamefont {Hirschberger}, \citenamefont
  {Lampen-Kelley}, \citenamefont {Banerjee}, \citenamefont {Yan}, \citenamefont
  {Mandrus}, \citenamefont {Nagler},\ and\ \citenamefont {Ong}}]{Ong2021}%
  \BibitemOpen
  \bibfield  {author} {\bibinfo {author} {\bibnamefont {Czajka}, \bibfnamefont
  {Peter}}, \bibinfo {author} {\bibfnamefont {Tong}\ \bibnamefont {Gao}},
  \bibinfo {author} {\bibfnamefont {Max}\ \bibnamefont {Hirschberger}},
  \bibinfo {author} {\bibfnamefont {Paula}\ \bibnamefont {Lampen-Kelley}},
  \bibinfo {author} {\bibfnamefont {Arnab}\ \bibnamefont {Banerjee}}, \bibinfo
  {author} {\bibfnamefont {Jiaqiang}\ \bibnamefont {Yan}}, \bibinfo {author}
  {\bibfnamefont {David~G.}\ \bibnamefont {Mandrus}}, \bibinfo {author}
  {\bibfnamefont {Stephen~E.}\ \bibnamefont {Nagler}}, \ and\ \bibinfo {author}
  {\bibfnamefont {N.~P.}\ \bibnamefont {Ong}}} (\bibinfo {year} {2021}),\
  \bibfield  {title} {\enquote {\bibinfo {title} {{Oscillations of the thermal
  conductivity in the spin-liquid state of
  $\ensuremath{\alpha}\ensuremath{-}{\mathrm{RuCl}}_{3}$}},}\ }\href {\doibase
  10.1038/s41567-021-01243-x} {\bibfield  {journal} {\bibinfo  {journal}
  {Nature Physics}\ }\textbf {\bibinfo {volume} {17}},\ \bibinfo {pages}
  {915--919}}\BibitemShut {NoStop}%
\bibitem [{\citenamefont {Davison}\ \emph {et~al.}(2017)\citenamefont
  {Davison}, \citenamefont {Fu}, \citenamefont {Georges}, \citenamefont {Gu},
  \citenamefont {Jensen},\ and\ \citenamefont {Sachdev}}]{PhysRevB.95.155131}%
  \BibitemOpen
  \bibfield  {author} {\bibinfo {author} {\bibnamefont {Davison}, \bibfnamefont
  {Richard~A}}, \bibinfo {author} {\bibfnamefont {Wenbo}\ \bibnamefont {Fu}},
  \bibinfo {author} {\bibfnamefont {Antoine}\ \bibnamefont {Georges}}, \bibinfo
  {author} {\bibfnamefont {Yingfei}\ \bibnamefont {Gu}}, \bibinfo {author}
  {\bibfnamefont {Kristan}\ \bibnamefont {Jensen}}, \ and\ \bibinfo {author}
  {\bibfnamefont {Subir}\ \bibnamefont {Sachdev}}} (\bibinfo {year} {2017}),\
  \bibfield  {title} {\enquote {\bibinfo {title} {{Thermoelectric transport in
  disordered metals without quasiparticles: The Sachdev-Ye-Kitaev models and
  holography}},}\ }\href {\doibase 10.1103/PhysRevB.95.155131} {\bibfield
  {journal} {\bibinfo  {journal} {Phys. Rev. B}\ }\textbf {\bibinfo {volume}
  {95}},\ \bibinfo {pages} {155131}}\BibitemShut {NoStop}%
\bibitem [{\citenamefont {Do}\ \emph {et~al.}(2017)\citenamefont {Do},
  \citenamefont {Park}, \citenamefont {Yoshitake}, \citenamefont {Nasu},
  \citenamefont {Motome}, \citenamefont {Kwon}, \citenamefont {Adroja},
  \citenamefont {Voneshen}, \citenamefont {Kim}, \citenamefont {Jang},
  \citenamefont {Park}, \citenamefont {Choi},\ and\ \citenamefont
  {Ji}}]{Do2017}%
  \BibitemOpen
  \bibfield  {author} {\bibinfo {author} {\bibnamefont {Do}, \bibfnamefont
  {Seung-Hwan}}, \bibinfo {author} {\bibfnamefont {Sang-Youn}\ \bibnamefont
  {Park}}, \bibinfo {author} {\bibfnamefont {Junki}\ \bibnamefont {Yoshitake}},
  \bibinfo {author} {\bibfnamefont {Joji}\ \bibnamefont {Nasu}}, \bibinfo
  {author} {\bibfnamefont {Yukitoshi}\ \bibnamefont {Motome}}, \bibinfo
  {author} {\bibfnamefont {Yong~Seung}\ \bibnamefont {Kwon}}, \bibinfo {author}
  {\bibfnamefont {D.~T.}\ \bibnamefont {Adroja}}, \bibinfo {author}
  {\bibfnamefont {D.~J.}\ \bibnamefont {Voneshen}}, \bibinfo {author}
  {\bibfnamefont {Kyoo}\ \bibnamefont {Kim}}, \bibinfo {author} {\bibfnamefont
  {T.-H.}\ \bibnamefont {Jang}}, \bibinfo {author} {\bibfnamefont {J.-H.}\
  \bibnamefont {Park}}, \bibinfo {author} {\bibfnamefont {Kwang-Yong}\
  \bibnamefont {Choi}}, \ and\ \bibinfo {author} {\bibfnamefont {Sungdae}\
  \bibnamefont {Ji}}} (\bibinfo {year} {2017}),\ \bibfield  {title} {\enquote
  {\bibinfo {title} {{Majorana fermions in the Kitaev quantum spin system
  $\alpha$-{RuCl}$_3$}},}\ }\href {\doibase 10.1038/nphys4264} {\bibfield
  {journal} {\bibinfo  {journal} {Nature Physics}\ }\textbf {\bibinfo {volume}
  {13}}~(\bibinfo {number} {11}),\ \bibinfo {pages} {1079--1084}}\BibitemShut
  {NoStop}%
\bibitem [{\citenamefont {Doki}\ \emph {et~al.}(2018)\citenamefont {Doki},
  \citenamefont {Akazawa}, \citenamefont {Lee}, \citenamefont {Han},
  \citenamefont {Sugii}, \citenamefont {Shimozawa}, \citenamefont {Kawashima},
  \citenamefont {Oda}, \citenamefont {Yoshida},\ and\ \citenamefont
  {Yamashita}}]{Yamashita2018}%
  \BibitemOpen
  \bibfield  {author} {\bibinfo {author} {\bibnamefont {Doki}, \bibfnamefont
  {Hayato}}, \bibinfo {author} {\bibfnamefont {Masatoshi}\ \bibnamefont
  {Akazawa}}, \bibinfo {author} {\bibfnamefont {Hyun-Yong}\ \bibnamefont
  {Lee}}, \bibinfo {author} {\bibfnamefont {Jung~Hoon}\ \bibnamefont {Han}},
  \bibinfo {author} {\bibfnamefont {Kaori}\ \bibnamefont {Sugii}}, \bibinfo
  {author} {\bibfnamefont {Masaaki}\ \bibnamefont {Shimozawa}}, \bibinfo
  {author} {\bibfnamefont {Naoki}\ \bibnamefont {Kawashima}}, \bibinfo {author}
  {\bibfnamefont {Migaku}\ \bibnamefont {Oda}}, \bibinfo {author}
  {\bibfnamefont {Hiroyuki}\ \bibnamefont {Yoshida}}, \ and\ \bibinfo {author}
  {\bibfnamefont {Minoru}\ \bibnamefont {Yamashita}}} (\bibinfo {year}
  {2018}),\ \bibfield  {title} {\enquote {\bibinfo {title} {{Spin Thermal Hall
  Conductivity of a Kagome Antiferromagnet}},}\ }\href {\doibase
  10.1103/PhysRevLett.121.097203} {\bibfield  {journal} {\bibinfo  {journal}
  {Phys. Rev. Lett.}\ }\textbf {\bibinfo {volume} {121}},\ \bibinfo {pages}
  {097203}}\BibitemShut {NoStop}%
\bibitem [{\citenamefont {Elhajal}\ \emph {et~al.}(2002)\citenamefont
  {Elhajal}, \citenamefont {Canals},\ and\ \citenamefont
  {Lacroix}}]{Lacroix2002}%
  \BibitemOpen
  \bibfield  {author} {\bibinfo {author} {\bibnamefont {Elhajal}, \bibfnamefont
  {M}}, \bibinfo {author} {\bibfnamefont {B.}~\bibnamefont {Canals}}, \ and\
  \bibinfo {author} {\bibfnamefont {C.}~\bibnamefont {Lacroix}}} (\bibinfo
  {year} {2002}),\ \bibfield  {title} {\enquote {\bibinfo {title} {{Symmetry
  breaking due to Dzyaloshinsky-Moriya interactions in the kagom\'e
  lattice}},}\ }\href {\doibase 10.1103/PhysRevB.66.014422} {\bibfield
  {journal} {\bibinfo  {journal} {Phys. Rev. B}\ }\textbf {\bibinfo {volume}
  {66}},\ \bibinfo {pages} {014422}}\BibitemShut {NoStop}%
\bibitem [{\citenamefont {Ferrari}\ and\ \citenamefont
  {Becca}(2019)}]{PhysRevX.9.031026}%
  \BibitemOpen
  \bibfield  {author} {\bibinfo {author} {\bibnamefont {Ferrari}, \bibfnamefont
  {Francesco}}, \ and\ \bibinfo {author} {\bibfnamefont {Federico}\
  \bibnamefont {Becca}}} (\bibinfo {year} {2019}),\ \bibfield  {title}
  {\enquote {\bibinfo {title} {{Dynamical Structure Factor of the
  ${J}_{1}\ensuremath{-}{J}_{2}$ Heisenberg Model on the Triangular Lattice:
  Magnons, Spinons, and Gauge Fields}},}\ }\href {\doibase
  10.1103/PhysRevX.9.031026} {\bibfield  {journal} {\bibinfo  {journal} {Phys.
  Rev. X}\ }\textbf {\bibinfo {volume} {9}},\ \bibinfo {pages}
  {031026}}\BibitemShut {NoStop}%
\bibitem [{\citenamefont {Ferrari}\ \emph {et~al.}(2017)\citenamefont
  {Ferrari}, \citenamefont {Bieri},\ and\ \citenamefont {Becca}}]{Becca2017}%
  \BibitemOpen
  \bibfield  {author} {\bibinfo {author} {\bibnamefont {Ferrari}, \bibfnamefont
  {Francesco}}, \bibinfo {author} {\bibfnamefont {Samuel}\ \bibnamefont
  {Bieri}}, \ and\ \bibinfo {author} {\bibfnamefont {Federico}\ \bibnamefont
  {Becca}}} (\bibinfo {year} {2017}),\ \bibfield  {title} {\enquote {\bibinfo
  {title} {{Competition between spin liquids and valence-bond order in the
  frustrated spin-$\frac{1}{2}$ Heisenberg model on the honeycomb lattice}},}\
  }\href {\doibase 10.1103/PhysRevB.96.104401} {\bibfield  {journal} {\bibinfo
  {journal} {Phys. Rev. B}\ }\textbf {\bibinfo {volume} {96}},\ \bibinfo
  {pages} {104401}}\BibitemShut {NoStop}%
\bibitem [{\citenamefont {Fradkin}(2013)}]{Fradkin2013}%
  \BibitemOpen
  \bibfield  {author} {\bibinfo {author} {\bibnamefont {Fradkin}, \bibfnamefont
  {Eduardo}}} (\bibinfo {year} {2013}),\ \href@noop {} {\emph {\bibinfo {title}
  {Field theories of condensed matter physics}}}\ (\bibinfo  {publisher}
  {Cambridge University Press})\BibitemShut {NoStop}%
\bibitem [{\citenamefont {Fradkin}\ and\ \citenamefont
  {Shenker}(1979)}]{Fradkin1979}%
  \BibitemOpen
  \bibfield  {author} {\bibinfo {author} {\bibnamefont {Fradkin}, \bibfnamefont
  {Eduardo}}, \ and\ \bibinfo {author} {\bibfnamefont {Stephen~H.}\
  \bibnamefont {Shenker}}} (\bibinfo {year} {1979}),\ \bibfield  {title}
  {\enquote {\bibinfo {title} {{Phase diagrams of lattice gauge theories with
  Higgs fields}},}\ }\href {\doibase 10.1103/PhysRevD.19.3682} {\bibfield
  {journal} {\bibinfo  {journal} {Phys. Rev. D}\ }\textbf {\bibinfo {volume}
  {19}},\ \bibinfo {pages} {3682--3697}}\BibitemShut {NoStop}%
\bibitem [{\citenamefont {Ganesh}\ \emph {et~al.}(2013)\citenamefont {Ganesh},
  \citenamefont {van~den Brink},\ and\ \citenamefont
  {Nishimoto}}]{Satoshi2013}%
  \BibitemOpen
  \bibfield  {author} {\bibinfo {author} {\bibnamefont {Ganesh}, \bibfnamefont
  {R}}, \bibinfo {author} {\bibfnamefont {Jeroen}\ \bibnamefont {van~den
  Brink}}, \ and\ \bibinfo {author} {\bibfnamefont {Satoshi}\ \bibnamefont
  {Nishimoto}}} (\bibinfo {year} {2013}),\ \bibfield  {title} {\enquote
  {\bibinfo {title} {{Deconfined criticality in the frustrated Heisenberg
  honeycomb antiferromagnet}},}\ }\href {\doibase
  10.1103/PhysRevLett.110.127203} {\bibfield  {journal} {\bibinfo  {journal}
  {Phys. Rev. Lett.}\ }\textbf {\bibinfo {volume} {110}},\ \bibinfo {pages}
  {127203}}\BibitemShut {NoStop}%
\bibitem [{\citenamefont {Gao}\ \emph {et~al.}(2019{\natexlab{a}})\citenamefont
  {Gao}, \citenamefont {Chen}, \citenamefont {Tam}, \citenamefont {Huang},
  \citenamefont {Sasmal}, \citenamefont {Adroja}, \citenamefont {Ye},
  \citenamefont {Cao}, \citenamefont {Sala}, \citenamefont {Stone},
  \citenamefont {Baines}, \citenamefont {Verezhak}, \citenamefont {Hu},
  \citenamefont {Chung}, \citenamefont {Xu}, \citenamefont {Cheong},
  \citenamefont {Nallaiyan}, \citenamefont {Spagna}, \citenamefont {Maple},
  \citenamefont {Nevidomskyy}, \citenamefont {Morosan}, \citenamefont {Chen},\
  and\ \citenamefont {Dai}}]{Pengcheng2019}%
  \BibitemOpen
  \bibfield  {author} {\bibinfo {author} {\bibnamefont {Gao}, \bibfnamefont
  {Bin}}, \bibinfo {author} {\bibfnamefont {Tong}\ \bibnamefont {Chen}},
  \bibinfo {author} {\bibfnamefont {David~W.}\ \bibnamefont {Tam}}, \bibinfo
  {author} {\bibfnamefont {Chien-Lung}\ \bibnamefont {Huang}}, \bibinfo
  {author} {\bibfnamefont {Kalyan}\ \bibnamefont {Sasmal}}, \bibinfo {author}
  {\bibfnamefont {Devashibhai~T.}\ \bibnamefont {Adroja}}, \bibinfo {author}
  {\bibfnamefont {Feng}\ \bibnamefont {Ye}}, \bibinfo {author} {\bibfnamefont
  {Huibo}\ \bibnamefont {Cao}}, \bibinfo {author} {\bibfnamefont {Gabriele}\
  \bibnamefont {Sala}}, \bibinfo {author} {\bibfnamefont {Matthew~B.}\
  \bibnamefont {Stone}}, \bibinfo {author} {\bibfnamefont {Christopher}\
  \bibnamefont {Baines}}, \bibinfo {author} {\bibfnamefont {Joel A.~T.}\
  \bibnamefont {Verezhak}}, \bibinfo {author} {\bibfnamefont {Haoyu}\
  \bibnamefont {Hu}}, \bibinfo {author} {\bibfnamefont {Jae-Ho}\ \bibnamefont
  {Chung}}, \bibinfo {author} {\bibfnamefont {Xianghan}\ \bibnamefont {Xu}},
  \bibinfo {author} {\bibfnamefont {Sang-Wook}\ \bibnamefont {Cheong}},
  \bibinfo {author} {\bibfnamefont {Manivannan}\ \bibnamefont {Nallaiyan}},
  \bibinfo {author} {\bibfnamefont {Stefano}\ \bibnamefont {Spagna}}, \bibinfo
  {author} {\bibfnamefont {M.~Brian}\ \bibnamefont {Maple}}, \bibinfo {author}
  {\bibfnamefont {Andriy~H.}\ \bibnamefont {Nevidomskyy}}, \bibinfo {author}
  {\bibfnamefont {Emilia}\ \bibnamefont {Morosan}}, \bibinfo {author}
  {\bibfnamefont {Gang}\ \bibnamefont {Chen}}, \ and\ \bibinfo {author}
  {\bibfnamefont {Pengcheng}\ \bibnamefont {Dai}}} (\bibinfo {year}
  {2019}{\natexlab{a}}),\ \bibfield  {title} {\enquote {\bibinfo {title}
  {{Experimental signatures of a three-dimensional quantum spin liquid in
  effective spin-1/2 {{Ce}$_2${Zr}$_2${O}$_7$ pyrochlore}}},}\ }\href {\doibase
  10.1038/s41567-019-0577-6} {\bibfield  {journal} {\bibinfo  {journal} {Nature
  Physics}\ }\textbf {\bibinfo {volume} {15}}~(\bibinfo {number} {10}),\
  \bibinfo {pages} {1052--1057}}\BibitemShut {NoStop}%
\bibitem [{\citenamefont {Gao}\ \emph {et~al.}(2022)\citenamefont {Gao},
  \citenamefont {Chen}, \citenamefont {Yan}, \citenamefont {Duan},
  \citenamefont {Huang}, \citenamefont {Yao}, \citenamefont {Ye}, \citenamefont
  {Balz}, \citenamefont {Stewart}, \citenamefont {Nakajima}, \citenamefont
  {Ohira-Kawamura}, \citenamefont {Xu}, \citenamefont {Xu}, \citenamefont
  {Cheong}, \citenamefont {Morosan}, \citenamefont {Nevidomskyy}, \citenamefont
  {Chen},\ and\ \citenamefont {Dai}}]{PhysRevB.106.094425}%
  \BibitemOpen
  \bibfield  {author} {\bibinfo {author} {\bibnamefont {Gao}, \bibfnamefont
  {Bin}}, \bibinfo {author} {\bibfnamefont {Tong}\ \bibnamefont {Chen}},
  \bibinfo {author} {\bibfnamefont {Han}\ \bibnamefont {Yan}}, \bibinfo
  {author} {\bibfnamefont {Chunruo}\ \bibnamefont {Duan}}, \bibinfo {author}
  {\bibfnamefont {Chien-Lung}\ \bibnamefont {Huang}}, \bibinfo {author}
  {\bibfnamefont {Xu~Ping}\ \bibnamefont {Yao}}, \bibinfo {author}
  {\bibfnamefont {Feng}\ \bibnamefont {Ye}}, \bibinfo {author} {\bibfnamefont
  {Christian}\ \bibnamefont {Balz}}, \bibinfo {author} {\bibfnamefont
  {J.~Ross}\ \bibnamefont {Stewart}}, \bibinfo {author} {\bibfnamefont {Kenji}\
  \bibnamefont {Nakajima}}, \bibinfo {author} {\bibfnamefont {Seiko}\
  \bibnamefont {Ohira-Kawamura}}, \bibinfo {author} {\bibfnamefont {Guangyong}\
  \bibnamefont {Xu}}, \bibinfo {author} {\bibfnamefont {Xianghan}\ \bibnamefont
  {Xu}}, \bibinfo {author} {\bibfnamefont {Sang-Wook}\ \bibnamefont {Cheong}},
  \bibinfo {author} {\bibfnamefont {Emilia}\ \bibnamefont {Morosan}}, \bibinfo
  {author} {\bibfnamefont {Andriy~H.}\ \bibnamefont {Nevidomskyy}}, \bibinfo
  {author} {\bibfnamefont {Gang}\ \bibnamefont {Chen}}, \ and\ \bibinfo
  {author} {\bibfnamefont {Pengcheng}\ \bibnamefont {Dai}}} (\bibinfo {year}
  {2022}),\ \bibfield  {title} {\enquote {\bibinfo {title} {Magnetic field
  effects in an octupolar quantum spin liquid candidate},}\ }\href {\doibase
  10.1103/PhysRevB.106.094425} {\bibfield  {journal} {\bibinfo  {journal}
  {Phys. Rev. B}\ }\textbf {\bibinfo {volume} {106}},\ \bibinfo {pages}
  {094425}}\BibitemShut {NoStop}%
\bibitem [{\citenamefont {Gao}\ \emph {et~al.}(2020{\natexlab{a}})\citenamefont
  {Gao}, \citenamefont {Rosales}, \citenamefont {Albarrac{\'{\i}}n},
  \citenamefont {Tsurkan}, \citenamefont {Kaur}, \citenamefont {Fennell},
  \citenamefont {Steffens}, \citenamefont {Boehm}, \citenamefont
  {{\v{C}}erm{\'{a}}k}, \citenamefont {Schneidewind}, \citenamefont
  {Ressouche}, \citenamefont {Cabra}, \citenamefont {Rüegg},\ and\
  \citenamefont {Zaharko}}]{Gao_2020}%
  \BibitemOpen
  \bibfield  {author} {\bibinfo {author} {\bibnamefont {Gao}, \bibfnamefont
  {Shang}}, \bibinfo {author} {\bibfnamefont {H.~Diego}\ \bibnamefont
  {Rosales}}, \bibinfo {author} {\bibfnamefont {Flavia A.~G{\'{o}}mez}\
  \bibnamefont {Albarrac{\'{\i}}n}}, \bibinfo {author} {\bibfnamefont
  {Vladimir}\ \bibnamefont {Tsurkan}}, \bibinfo {author} {\bibfnamefont
  {Guratinder}\ \bibnamefont {Kaur}}, \bibinfo {author} {\bibfnamefont {Tom}\
  \bibnamefont {Fennell}}, \bibinfo {author} {\bibfnamefont {Paul}\
  \bibnamefont {Steffens}}, \bibinfo {author} {\bibfnamefont {Martin}\
  \bibnamefont {Boehm}}, \bibinfo {author} {\bibfnamefont {Petr}\ \bibnamefont
  {{\v{C}}erm{\'{a}}k}}, \bibinfo {author} {\bibfnamefont {Astrid}\
  \bibnamefont {Schneidewind}}, \bibinfo {author} {\bibfnamefont {Eric}\
  \bibnamefont {Ressouche}}, \bibinfo {author} {\bibfnamefont {Daniel~C.}\
  \bibnamefont {Cabra}}, \bibinfo {author} {\bibfnamefont {Christian}\
  \bibnamefont {Rüegg}}, \ and\ \bibinfo {author} {\bibfnamefont {Oksana}\
  \bibnamefont {Zaharko}}} (\bibinfo {year} {2020}{\natexlab{a}}),\ \bibfield
  {title} {\enquote {\bibinfo {title} {Fractional antiferromagnetic skyrmion
  lattice induced by anisotropic couplings},}\ }\href {\doibase
  10.1038/s41586-020-2716-8} {\bibfield  {journal} {\bibinfo  {journal}
  {Nature}\ }\textbf {\bibinfo {volume} {586}}~(\bibinfo {number} {7827}),\
  \bibinfo {pages} {37--41}}\BibitemShut {NoStop}%
\bibitem [{\citenamefont {Gao}\ and\ \citenamefont {Chen}(2020)}]{Gao2020}%
  \BibitemOpen
  \bibfield  {author} {\bibinfo {author} {\bibnamefont {Gao}, \bibfnamefont
  {Yong~Hao}}, \ and\ \bibinfo {author} {\bibfnamefont {Gang}\ \bibnamefont
  {Chen}}} (\bibinfo {year} {2020}),\ \bibfield  {title} {\enquote {\bibinfo
  {title} {{Topological thermal Hall effect for topological excitations in spin
  liquid: Emergent Lorentz force on the spinons}},}\ }\href {\doibase
  10.21468/scipostphyscore.2.2.004} {\bibfield  {journal} {\bibinfo  {journal}
  {{SciPost} Physics Core}\ }\textbf {\bibinfo {volume} {2}}~(\bibinfo {number}
  {2}),\ \bibinfo {pages} {004}}\BibitemShut {NoStop}%
\bibitem [{\citenamefont {Gao}\ \emph {et~al.}(2019{\natexlab{b}})\citenamefont
  {Gao}, \citenamefont {Hickey}, \citenamefont {Xiang}, \citenamefont
  {Trebst},\ and\ \citenamefont {Chen}}]{Gao2019}%
  \BibitemOpen
  \bibfield  {author} {\bibinfo {author} {\bibnamefont {Gao}, \bibfnamefont
  {Yong~Hao}}, \bibinfo {author} {\bibfnamefont {Ciar\'an}\ \bibnamefont
  {Hickey}}, \bibinfo {author} {\bibfnamefont {Tao}\ \bibnamefont {Xiang}},
  \bibinfo {author} {\bibfnamefont {Simon}\ \bibnamefont {Trebst}}, \ and\
  \bibinfo {author} {\bibfnamefont {Gang}\ \bibnamefont {Chen}}} (\bibinfo
  {year} {2019}{\natexlab{b}}),\ \bibfield  {title} {\enquote {\bibinfo {title}
  {{Thermal Hall signatures of non-Kitaev spin liquids in honeycomb Kitaev
  materials}},}\ }\href {\doibase 10.1103/PhysRevResearch.1.013014} {\bibfield
  {journal} {\bibinfo  {journal} {Phys. Rev. Research}\ }\textbf {\bibinfo
  {volume} {1}},\ \bibinfo {pages} {013014}}\BibitemShut {NoStop}%
\bibitem [{\citenamefont {Gao}\ \emph {et~al.}(2020{\natexlab{b}})\citenamefont
  {Gao}, \citenamefont {Yao},\ and\ \citenamefont {Chen}}]{Gao2020B}%
  \BibitemOpen
  \bibfield  {author} {\bibinfo {author} {\bibnamefont {Gao}, \bibfnamefont
  {Yonghao}}, \bibinfo {author} {\bibfnamefont {Xu-Ping}\ \bibnamefont {Yao}},
  \ and\ \bibinfo {author} {\bibfnamefont {Gang}\ \bibnamefont {Chen}}}
  (\bibinfo {year} {2020}{\natexlab{b}}),\ \bibfield  {title} {\enquote
  {\bibinfo {title} {{Topological phase transition and nontrivial thermal Hall
  signatures in honeycomb lattice magnets}},}\ }\href {\doibase
  10.1103/PhysRevResearch.2.043071} {\bibfield  {journal} {\bibinfo  {journal}
  {Phys. Rev. Research}\ }\textbf {\bibinfo {volume} {2}},\ \bibinfo {pages}
  {043071}}\BibitemShut {NoStop}%
\bibitem [{\citenamefont {Gaudet}\ \emph {et~al.}(2019)\citenamefont {Gaudet},
  \citenamefont {Smith}, \citenamefont {Dudemaine}, \citenamefont {Beare},
  \citenamefont {Buhariwalla}, \citenamefont {Butch}, \citenamefont {Stone},
  \citenamefont {Kolesnikov}, \citenamefont {Xu}, \citenamefont {Yahne},
  \citenamefont {Ross}, \citenamefont {Marjerrison}, \citenamefont {Garrett},
  \citenamefont {Luke}, \citenamefont {Bianchi},\ and\ \citenamefont
  {Gaulin}}]{PhysRevLett.122.187201}%
  \BibitemOpen
  \bibfield  {author} {\bibinfo {author} {\bibnamefont {Gaudet}, \bibfnamefont
  {J}}, \bibinfo {author} {\bibfnamefont {E.~M.}\ \bibnamefont {Smith}},
  \bibinfo {author} {\bibfnamefont {J.}~\bibnamefont {Dudemaine}}, \bibinfo
  {author} {\bibfnamefont {J.}~\bibnamefont {Beare}}, \bibinfo {author}
  {\bibfnamefont {C.~R.~C.}\ \bibnamefont {Buhariwalla}}, \bibinfo {author}
  {\bibfnamefont {N.~P.}\ \bibnamefont {Butch}}, \bibinfo {author}
  {\bibfnamefont {M.~B.}\ \bibnamefont {Stone}}, \bibinfo {author}
  {\bibfnamefont {A.~I.}\ \bibnamefont {Kolesnikov}}, \bibinfo {author}
  {\bibfnamefont {Guangyong}\ \bibnamefont {Xu}}, \bibinfo {author}
  {\bibfnamefont {D.~R.}\ \bibnamefont {Yahne}}, \bibinfo {author}
  {\bibfnamefont {K.~A.}\ \bibnamefont {Ross}}, \bibinfo {author}
  {\bibfnamefont {C.~A.}\ \bibnamefont {Marjerrison}}, \bibinfo {author}
  {\bibfnamefont {J.~D.}\ \bibnamefont {Garrett}}, \bibinfo {author}
  {\bibfnamefont {G.~M.}\ \bibnamefont {Luke}}, \bibinfo {author}
  {\bibfnamefont {A.~D.}\ \bibnamefont {Bianchi}}, \ and\ \bibinfo {author}
  {\bibfnamefont {B.~D.}\ \bibnamefont {Gaulin}}} (\bibinfo {year} {2019}),\
  \bibfield  {title} {\enquote {\bibinfo {title} {{Quantum Spin Ice Dynamics in
  the Dipole-Octupole Pyrochlore Magnet
  ${\mathrm{Ce}}_{2}{\mathrm{Zr}}_{2}{\mathrm{O}}_{7}$}},}\ }\href {\doibase
  10.1103/PhysRevLett.122.187201} {\bibfield  {journal} {\bibinfo  {journal}
  {Phys. Rev. Lett.}\ }\textbf {\bibinfo {volume} {122}},\ \bibinfo {pages}
  {187201}}\BibitemShut {NoStop}%
\bibitem [{\citenamefont {Gaulin}\ \emph {et~al.}(2004)\citenamefont {Gaulin},
  \citenamefont {Lee}, \citenamefont {Haravifard}, \citenamefont {Castellan},
  \citenamefont {Berlinsky}, \citenamefont {Dabkowska}, \citenamefont {Qiu},\
  and\ \citenamefont {Copley}}]{Gaulin2004}%
  \BibitemOpen
  \bibfield  {author} {\bibinfo {author} {\bibnamefont {Gaulin}, \bibfnamefont
  {B~D}}, \bibinfo {author} {\bibfnamefont {S.~H.}\ \bibnamefont {Lee}},
  \bibinfo {author} {\bibfnamefont {S.}~\bibnamefont {Haravifard}}, \bibinfo
  {author} {\bibfnamefont {J.~P.}\ \bibnamefont {Castellan}}, \bibinfo {author}
  {\bibfnamefont {A.~J.}\ \bibnamefont {Berlinsky}}, \bibinfo {author}
  {\bibfnamefont {H.~A.}\ \bibnamefont {Dabkowska}}, \bibinfo {author}
  {\bibfnamefont {Y.}~\bibnamefont {Qiu}}, \ and\ \bibinfo {author}
  {\bibfnamefont {J.~R.~D.}\ \bibnamefont {Copley}}} (\bibinfo {year} {2004}),\
  \bibfield  {title} {\enquote {\bibinfo {title} {{High-Resolution Study of
  Spin Excitations in the Singlet Ground State of
  ${\mathrm{S}\mathrm{r}\mathrm{C}\mathrm{u}}_{2}({\mathrm{B}\mathrm{O}}_{3}{)}_{2}$}},}\
  }\href {\doibase 10.1103/PhysRevLett.93.267202} {\bibfield  {journal}
  {\bibinfo  {journal} {Phys. Rev. Lett.}\ }\textbf {\bibinfo {volume} {93}},\
  \bibinfo {pages} {267202}}\BibitemShut {NoStop}%
\bibitem [{\citenamefont {Giamarchi}\ \emph {et~al.}(2008)\citenamefont
  {Giamarchi}, \citenamefont {Rüegg},\ and\ \citenamefont
  {Tchernyshyov}}]{Giamarchi_2008}%
  \BibitemOpen
  \bibfield  {author} {\bibinfo {author} {\bibnamefont {Giamarchi},
  \bibfnamefont {Thierry}}, \bibinfo {author} {\bibfnamefont {Christian}\
  \bibnamefont {Rüegg}}, \ and\ \bibinfo {author} {\bibfnamefont {Oleg}\
  \bibnamefont {Tchernyshyov}}} (\bibinfo {year} {2008}),\ \bibfield  {title}
  {\enquote {\bibinfo {title} {Bose{\textendash}einstein condensation in
  magnetic~insulators},}\ }\href {\doibase 10.1038/nphys893} {\bibfield
  {journal} {\bibinfo  {journal} {Nature Physics}\ }\textbf {\bibinfo {volume}
  {4}}~(\bibinfo {number} {3}),\ \bibinfo {pages} {198--204}}\BibitemShut
  {NoStop}%
\bibitem [{\citenamefont {Gingras}\ \emph {et~al.}(2000)\citenamefont
  {Gingras}, \citenamefont {den Hertog}, \citenamefont {Faucher}, \citenamefont
  {Gardner}, \citenamefont {Dunsiger}, \citenamefont {Chang}, \citenamefont
  {Gaulin}, \citenamefont {Raju},\ and\ \citenamefont
  {Greedan}}]{PhysRevB.62.6496}%
  \BibitemOpen
  \bibfield  {author} {\bibinfo {author} {\bibnamefont {Gingras}, \bibfnamefont
  {M~J~P}}, \bibinfo {author} {\bibfnamefont {B.~C.}\ \bibnamefont {den
  Hertog}}, \bibinfo {author} {\bibfnamefont {M.}~\bibnamefont {Faucher}},
  \bibinfo {author} {\bibfnamefont {J.~S.}\ \bibnamefont {Gardner}}, \bibinfo
  {author} {\bibfnamefont {S.~R.}\ \bibnamefont {Dunsiger}}, \bibinfo {author}
  {\bibfnamefont {L.~J.}\ \bibnamefont {Chang}}, \bibinfo {author}
  {\bibfnamefont {B.~D.}\ \bibnamefont {Gaulin}}, \bibinfo {author}
  {\bibfnamefont {N.~P.}\ \bibnamefont {Raju}}, \ and\ \bibinfo {author}
  {\bibfnamefont {J.~E.}\ \bibnamefont {Greedan}}} (\bibinfo {year} {2000}),\
  \bibfield  {title} {\enquote {\bibinfo {title} {{Thermodynamic and single-ion
  properties of ${\mathrm{Tb}}^{3+}$ within the collective paramagnetic-spin
  liquid state of the frustrated pyrochlore antiferromagnet
  ${\mathrm{Tb}}_{2}{\mathrm{Ti}}_{2}{\mathrm{O}}_{7}$}},}\ }\href {\doibase
  10.1103/PhysRevB.62.6496} {\bibfield  {journal} {\bibinfo  {journal} {Phys.
  Rev. B}\ }\textbf {\bibinfo {volume} {62}},\ \bibinfo {pages}
  {6496--6511}}\BibitemShut {NoStop}%
\bibitem [{\citenamefont {Gingras}\ and\ \citenamefont
  {McClarty}(2014)}]{Gingras2014}%
  \BibitemOpen
  \bibfield  {author} {\bibinfo {author} {\bibnamefont {Gingras}, \bibfnamefont
  {M~J~P}}, \ and\ \bibinfo {author} {\bibfnamefont {P.~A.}\ \bibnamefont
  {McClarty}}} (\bibinfo {year} {2014}),\ \bibfield  {title} {\enquote
  {\bibinfo {title} {Quantum spin ice: a search for gapless quantum spin
  liquids in pyrochlore magnets},}\ }\href {\doibase
  10.1088/0034-4885/77/5/056501} {\bibfield  {journal} {\bibinfo  {journal}
  {Reports on Progress in Physics}\ }\textbf {\bibinfo {volume} {77}}~(\bibinfo
  {number} {5}),\ \bibinfo {pages} {056501}}\BibitemShut {NoStop}%
\bibitem [{\citenamefont {Go}\ \emph {et~al.}(2019)\citenamefont {Go},
  \citenamefont {Kim},\ and\ \citenamefont {Lee}}]{PhysRevLett.123.237207}%
  \BibitemOpen
  \bibfield  {author} {\bibinfo {author} {\bibnamefont {Go}, \bibfnamefont
  {Gyungchoon}}, \bibinfo {author} {\bibfnamefont {Se~Kwon}\ \bibnamefont
  {Kim}}, \ and\ \bibinfo {author} {\bibfnamefont {Kyung-Jin}\ \bibnamefont
  {Lee}}} (\bibinfo {year} {2019}),\ \bibfield  {title} {\enquote {\bibinfo
  {title} {{Topological Magnon-Phonon Hybrid Excitations in Two-Dimensional
  Ferromagnets with Tunable Chern Numbers}},}\ }\href {\doibase
  10.1103/PhysRevLett.123.237207} {\bibfield  {journal} {\bibinfo  {journal}
  {Phys. Rev. Lett.}\ }\textbf {\bibinfo {volume} {123}},\ \bibinfo {pages}
  {237207}}\BibitemShut {NoStop}%
\bibitem [{\citenamefont {Gohlke}\ \emph {et~al.}(2017)\citenamefont {Gohlke},
  \citenamefont {Verresen}, \citenamefont {Moessner},\ and\ \citenamefont
  {Pollmann}}]{Pollmann2017}%
  \BibitemOpen
  \bibfield  {author} {\bibinfo {author} {\bibnamefont {Gohlke}, \bibfnamefont
  {Matthias}}, \bibinfo {author} {\bibfnamefont {Ruben}\ \bibnamefont
  {Verresen}}, \bibinfo {author} {\bibfnamefont {Roderich}\ \bibnamefont
  {Moessner}}, \ and\ \bibinfo {author} {\bibfnamefont {Frank}\ \bibnamefont
  {Pollmann}}} (\bibinfo {year} {2017}),\ \bibfield  {title} {\enquote
  {\bibinfo {title} {{Dynamics of the Kitaev-Heisenberg model}},}\ }\href
  {\doibase 10.1103/PhysRevLett.119.157203} {\bibfield  {journal} {\bibinfo
  {journal} {Phys. Rev. Lett.}\ }\textbf {\bibinfo {volume} {119}},\ \bibinfo
  {pages} {157203}}\BibitemShut {NoStop}%
\bibitem [{\citenamefont {Gong}\ \emph {et~al.}(2013)\citenamefont {Gong},
  \citenamefont {Sheng}, \citenamefont {Motrunich},\ and\ \citenamefont
  {Fisher}}]{Fisher2013}%
  \BibitemOpen
  \bibfield  {author} {\bibinfo {author} {\bibnamefont {Gong}, \bibfnamefont
  {Shou-Shu}}, \bibinfo {author} {\bibfnamefont {D.~N.}\ \bibnamefont {Sheng}},
  \bibinfo {author} {\bibfnamefont {Olexei~I.}\ \bibnamefont {Motrunich}}, \
  and\ \bibinfo {author} {\bibfnamefont {Matthew P.~A.}\ \bibnamefont
  {Fisher}}} (\bibinfo {year} {2013}),\ \bibfield  {title} {\enquote {\bibinfo
  {title} {{Phase diagram of the spin-$\frac{1}{2}$ ${J}_{1}$-${J}_{2}$
  Heisenberg model on a honeycomb lattice}},}\ }\href {\doibase
  10.1103/PhysRevB.88.165138} {\bibfield  {journal} {\bibinfo  {journal} {Phys.
  Rev. B}\ }\textbf {\bibinfo {volume} {88}},\ \bibinfo {pages}
  {165138}}\BibitemShut {NoStop}%
\bibitem [{\citenamefont {Gong}\ \emph {et~al.}(2019)\citenamefont {Gong},
  \citenamefont {Zheng}, \citenamefont {Lee}, \citenamefont {Lu},\ and\
  \citenamefont {Sheng}}]{Sheng2019}%
  \BibitemOpen
  \bibfield  {author} {\bibinfo {author} {\bibnamefont {Gong}, \bibfnamefont
  {Shou-Shu}}, \bibinfo {author} {\bibfnamefont {Wayne}\ \bibnamefont {Zheng}},
  \bibinfo {author} {\bibfnamefont {Mac}\ \bibnamefont {Lee}}, \bibinfo
  {author} {\bibfnamefont {Yuan-Ming}\ \bibnamefont {Lu}}, \ and\ \bibinfo
  {author} {\bibfnamefont {D.~N.}\ \bibnamefont {Sheng}}} (\bibinfo {year}
  {2019}),\ \bibfield  {title} {\enquote {\bibinfo {title} {{Chiral spin liquid
  with spinon Fermi surfaces in the spin-$\frac{1}{2}$ triangular Heisenberg
  model}},}\ }\href {\doibase 10.1103/PhysRevB.100.241111} {\bibfield
  {journal} {\bibinfo  {journal} {Phys. Rev. B}\ }\textbf {\bibinfo {volume}
  {100}},\ \bibinfo {pages} {241111}}\BibitemShut {NoStop}%
\bibitem [{\citenamefont {Grissonnanche}\ \emph {et~al.}(2019)\citenamefont
  {Grissonnanche}, \citenamefont {Legros}, \citenamefont {Badoux},
  \citenamefont {Lefran{\c{c}}ois}, \citenamefont {Zatko}, \citenamefont
  {Lizaire}, \citenamefont {Lalibert{\'{e}}}, \citenamefont {Gourgout},
  \citenamefont {Zhou}, \citenamefont {Pyon}, \citenamefont {Takayama},
  \citenamefont {Takagi}, \citenamefont {Ono}, \citenamefont {Doiron-Leyraud},\
  and\ \citenamefont {Taillefer}}]{Grissonnanche2019}%
  \BibitemOpen
  \bibfield  {author} {\bibinfo {author} {\bibnamefont {Grissonnanche},
  \bibfnamefont {G}}, \bibinfo {author} {\bibfnamefont {A.}~\bibnamefont
  {Legros}}, \bibinfo {author} {\bibfnamefont {S.}~\bibnamefont {Badoux}},
  \bibinfo {author} {\bibfnamefont {E.}~\bibnamefont {Lefran{\c{c}}ois}},
  \bibinfo {author} {\bibfnamefont {V.}~\bibnamefont {Zatko}}, \bibinfo
  {author} {\bibfnamefont {M.}~\bibnamefont {Lizaire}}, \bibinfo {author}
  {\bibfnamefont {F.}~\bibnamefont {Lalibert{\'{e}}}}, \bibinfo {author}
  {\bibfnamefont {A.}~\bibnamefont {Gourgout}}, \bibinfo {author}
  {\bibfnamefont {J.-S.}\ \bibnamefont {Zhou}}, \bibinfo {author}
  {\bibfnamefont {S.}~\bibnamefont {Pyon}}, \bibinfo {author} {\bibfnamefont
  {T.}~\bibnamefont {Takayama}}, \bibinfo {author} {\bibfnamefont
  {H.}~\bibnamefont {Takagi}}, \bibinfo {author} {\bibfnamefont
  {S.}~\bibnamefont {Ono}}, \bibinfo {author} {\bibfnamefont {N.}~\bibnamefont
  {Doiron-Leyraud}}, \ and\ \bibinfo {author} {\bibfnamefont {L.}~\bibnamefont
  {Taillefer}}} (\bibinfo {year} {2019}),\ \bibfield  {title} {\enquote
  {\bibinfo {title} {{Giant thermal Hall conductivity in the pseudogap phase of
  cuprate superconductors}},}\ }\href {\doibase 10.1038/s41586-019-1375-0}
  {\bibfield  {journal} {\bibinfo  {journal} {Nature}\ }\textbf {\bibinfo
  {volume} {571}}~(\bibinfo {number} {7765}),\ \bibinfo {pages}
  {376--380}}\BibitemShut {NoStop}%
\bibitem [{\citenamefont {Grissonnanche}\ \emph {et~al.}(2020)\citenamefont
  {Grissonnanche}, \citenamefont {Theriault}, \citenamefont {Gourgout},
  \citenamefont {Boulanger}, \citenamefont {Lefrançois}, \citenamefont
  {Ataei}, \citenamefont {Laliberte}, \citenamefont {Dion}, \citenamefont
  {Zhou}, \citenamefont {Pyon}, \citenamefont {Takayama}, \citenamefont
  {Takagi}, \citenamefont {Doiron-Leyraud},\ and\ \citenamefont
  {Taillefer}}]{Grissonnanche2020}%
  \BibitemOpen
  \bibfield  {author} {\bibinfo {author} {\bibnamefont {Grissonnanche},
  \bibfnamefont {G}}, \bibinfo {author} {\bibfnamefont {S.}~\bibnamefont
  {Theriault}}, \bibinfo {author} {\bibfnamefont {A.}~\bibnamefont {Gourgout}},
  \bibinfo {author} {\bibfnamefont {M.-E.}\ \bibnamefont {Boulanger}}, \bibinfo
  {author} {\bibfnamefont {E.}~\bibnamefont {Lefrançois}}, \bibinfo {author}
  {\bibfnamefont {A.}~\bibnamefont {Ataei}}, \bibinfo {author} {\bibfnamefont
  {F.}~\bibnamefont {Laliberte}}, \bibinfo {author} {\bibfnamefont
  {M.}~\bibnamefont {Dion}}, \bibinfo {author} {\bibfnamefont {J.-S.}\
  \bibnamefont {Zhou}}, \bibinfo {author} {\bibfnamefont {S.}~\bibnamefont
  {Pyon}}, \bibinfo {author} {\bibfnamefont {T.}~\bibnamefont {Takayama}},
  \bibinfo {author} {\bibfnamefont {H.}~\bibnamefont {Takagi}}, \bibinfo
  {author} {\bibfnamefont {N.}~\bibnamefont {Doiron-Leyraud}}, \ and\ \bibinfo
  {author} {\bibfnamefont {L.}~\bibnamefont {Taillefer}}} (\bibinfo {year}
  {2020}),\ \bibfield  {title} {\enquote {\bibinfo {title} {Chiral phonons in
  the pseudogap phase of cuprates},}\ }\href {\doibase
  10.1038/s41567-020-0965-y} {\bibfield  {journal} {\bibinfo  {journal} {Nature
  Physics}\ }\textbf {\bibinfo {volume} {16}}~(\bibinfo {number} {12}),\
  \bibinfo {pages} {1108–1111}}\BibitemShut {NoStop}%
\bibitem [{\citenamefont {Grohol}\ \emph
  {et~al.}(2005{\natexlab{a}})\citenamefont {Grohol}, \citenamefont {Matan},
  \citenamefont {Cho}, \citenamefont {Lee}, \citenamefont {Lynn}, \citenamefont
  {Nocera},\ and\ \citenamefont {Lee}}]{Grohol2005}%
  \BibitemOpen
  \bibfield  {author} {\bibinfo {author} {\bibnamefont {Grohol}, \bibfnamefont
  {Daniel}}, \bibinfo {author} {\bibfnamefont {Kittiwit}\ \bibnamefont
  {Matan}}, \bibinfo {author} {\bibfnamefont {Jin-Hyung}\ \bibnamefont {Cho}},
  \bibinfo {author} {\bibfnamefont {Seung-Hun}\ \bibnamefont {Lee}}, \bibinfo
  {author} {\bibfnamefont {Jeffrey~W.}\ \bibnamefont {Lynn}}, \bibinfo {author}
  {\bibfnamefont {Daniel~G.}\ \bibnamefont {Nocera}}, \ and\ \bibinfo {author}
  {\bibfnamefont {Young~S.}\ \bibnamefont {Lee}}} (\bibinfo {year}
  {2005}{\natexlab{a}}),\ \bibfield  {title} {\enquote {\bibinfo {title} {Spin
  chirality on a two-dimensional frustrated lattice},}\ }\href {\doibase
  10.1038/nmat1353} {\bibfield  {journal} {\bibinfo  {journal} {Nature
  Materials}\ }\textbf {\bibinfo {volume} {4}}~(\bibinfo {number} {4}),\
  \bibinfo {pages} {323--328}}\BibitemShut {NoStop}%
\bibitem [{\citenamefont {Grohol}\ \emph
  {et~al.}(2005{\natexlab{b}})\citenamefont {Grohol}, \citenamefont {Matan},
  \citenamefont {Cho}, \citenamefont {Lee}, \citenamefont {Lynn}, \citenamefont
  {Nocera},\ and\ \citenamefont {Lee}}]{Grohol_2005}%
  \BibitemOpen
  \bibfield  {author} {\bibinfo {author} {\bibnamefont {Grohol}, \bibfnamefont
  {Daniel}}, \bibinfo {author} {\bibfnamefont {Kittiwit}\ \bibnamefont
  {Matan}}, \bibinfo {author} {\bibfnamefont {Jin-Hyung}\ \bibnamefont {Cho}},
  \bibinfo {author} {\bibfnamefont {Seung-Hun}\ \bibnamefont {Lee}}, \bibinfo
  {author} {\bibfnamefont {Jeffrey~W.}\ \bibnamefont {Lynn}}, \bibinfo {author}
  {\bibfnamefont {Daniel~G.}\ \bibnamefont {Nocera}}, \ and\ \bibinfo {author}
  {\bibfnamefont {Young~S.}\ \bibnamefont {Lee}}} (\bibinfo {year}
  {2005}{\natexlab{b}}),\ \bibfield  {title} {\enquote {\bibinfo {title} {Spin
  chirality on a two-dimensional frustrated lattice},}\ }\href {\doibase
  10.1038/nmat1353} {\bibfield  {journal} {\bibinfo  {journal} {Nature
  Materials}\ }\textbf {\bibinfo {volume} {4}}~(\bibinfo {number} {4}),\
  \bibinfo {pages} {323--328}}\BibitemShut {NoStop}%
\bibitem [{\citenamefont {Gromov}\ \emph {et~al.}(2015)\citenamefont {Gromov},
  \citenamefont {Cho}, \citenamefont {You}, \citenamefont {Abanov},\ and\
  \citenamefont {Fradkin}}]{PhysRevLett.114.016805}%
  \BibitemOpen
  \bibfield  {author} {\bibinfo {author} {\bibnamefont {Gromov}, \bibfnamefont
  {Andrey}}, \bibinfo {author} {\bibfnamefont {Gil~Young}\ \bibnamefont {Cho}},
  \bibinfo {author} {\bibfnamefont {Yizhi}\ \bibnamefont {You}}, \bibinfo
  {author} {\bibfnamefont {Alexander~G.}\ \bibnamefont {Abanov}}, \ and\
  \bibinfo {author} {\bibfnamefont {Eduardo}\ \bibnamefont {Fradkin}}}
  (\bibinfo {year} {2015}),\ \bibfield  {title} {\enquote {\bibinfo {title}
  {{Framing Anomaly in the Effective Theory of the Fractional Quantum Hall
  Effect}},}\ }\href {\doibase 10.1103/PhysRevLett.114.016805} {\bibfield
  {journal} {\bibinfo  {journal} {Phys. Rev. Lett.}\ }\textbf {\bibinfo
  {volume} {114}},\ \bibinfo {pages} {016805}}\BibitemShut {NoStop}%
\bibitem [{\citenamefont {Guo}(2023)}]{PhysRevResearch.5.033197}%
  \BibitemOpen
  \bibfield  {author} {\bibinfo {author} {\bibnamefont {Guo}, \bibfnamefont
  {Haoyu}}} (\bibinfo {year} {2023}),\ \bibfield  {title} {\enquote {\bibinfo
  {title} {{Phonon thermal Hall effect in a non-Kramers paramagnet}},}\ }\href
  {\doibase 10.1103/PhysRevResearch.5.033197} {\bibfield  {journal} {\bibinfo
  {journal} {Phys. Rev. Res.}\ }\textbf {\bibinfo {volume} {5}},\ \bibinfo
  {pages} {033197}}\BibitemShut {NoStop}%
\bibitem [{\citenamefont {Guo}\ \emph {et~al.}(2022)\citenamefont {Guo},
  \citenamefont {Joshi},\ and\ \citenamefont {Sachdev}}]{pnas.2215141119}%
  \BibitemOpen
  \bibfield  {author} {\bibinfo {author} {\bibnamefont {Guo}, \bibfnamefont
  {Haoyu}}, \bibinfo {author} {\bibfnamefont {Darshan}\ \bibnamefont {Joshi}},
  \ and\ \bibinfo {author} {\bibfnamefont {Subir}\ \bibnamefont {Sachdev}}}
  (\bibinfo {year} {2022}),\ \bibfield  {title} {\enquote {\bibinfo {title}
  {Resonant thermal hall effect of phonons coupled to dynamical defects},}\
  }\href {\doibase 10.1073/pnas.2215141119} {\bibfield  {journal} {\bibinfo
  {journal} {Proceedings of the National Academy of Sciences}\ }\textbf
  {\bibinfo {volume} {119}},\ \bibinfo {pages} {e2215141119}}\BibitemShut
  {NoStop}%
\bibitem [{\citenamefont {Guo}\ \emph {et~al.}(2020)\citenamefont {Guo},
  \citenamefont {Samajdar}, \citenamefont {Scheurer},\ and\ \citenamefont
  {Sachdev}}]{PhysRevB.101.195126}%
  \BibitemOpen
  \bibfield  {author} {\bibinfo {author} {\bibnamefont {Guo}, \bibfnamefont
  {Haoyu}}, \bibinfo {author} {\bibfnamefont {Rhine}\ \bibnamefont {Samajdar}},
  \bibinfo {author} {\bibfnamefont {Mathias~S.}\ \bibnamefont {Scheurer}}, \
  and\ \bibinfo {author} {\bibfnamefont {Subir}\ \bibnamefont {Sachdev}}}
  (\bibinfo {year} {2020}),\ \bibfield  {title} {\enquote {\bibinfo {title}
  {{Gauge theories for the thermal Hall effect}},}\ }\href {\doibase
  10.1103/PhysRevB.101.195126} {\bibfield  {journal} {\bibinfo  {journal}
  {Phys. Rev. B}\ }\textbf {\bibinfo {volume} {101}},\ \bibinfo {pages}
  {195126}}\BibitemShut {NoStop}%
\bibitem [{\citenamefont {{H.-C. Jiang, C.-Y. Wang, B. Huang, Y.-M
  Lu}}(2018)}]{Jiang2018}%
  \BibitemOpen
  \bibfield  {author} {\bibinfo {author} {\bibnamefont {{H.-C. Jiang, C.-Y.
  Wang, B. Huang, Y.-M Lu}},}} (\bibinfo {year} {2018}),\ \href@noop {}
  {\enquote {\bibinfo {title} {{{Field induced quantum spin liquid with spinon
  Fermi surfaces in the Kitaev model}}},}\ }\Eprint
  {http://arxiv.org/abs/1809.08247} {arXiv:1809.08247} \BibitemShut {NoStop}%
\bibitem [{\citenamefont {Haldane}(1988)}]{Haldane1988}%
  \BibitemOpen
  \bibfield  {author} {\bibinfo {author} {\bibnamefont {Haldane}, \bibfnamefont
  {F~D~M}}} (\bibinfo {year} {1988}),\ \bibfield  {title} {\enquote {\bibinfo
  {title} {{Model for a quantum Hall effect without Landau levels:
  Condensed-matter realization of the ``parity anomaly"}},}\ }\href {\doibase
  10.1103/physrevlett.61.2015} {\bibfield  {journal} {\bibinfo  {journal}
  {Physical Review Letters}\ }\textbf {\bibinfo {volume} {61}}~(\bibinfo
  {number} {18}),\ \bibinfo {pages} {2015--2018}}\BibitemShut {NoStop}%
\bibitem [{\citenamefont {Han}\ \emph {et~al.}(2019)\citenamefont {Han},
  \citenamefont {Park},\ and\ \citenamefont {Lee}}]{PhysRevB.99.205157}%
  \BibitemOpen
  \bibfield  {author} {\bibinfo {author} {\bibnamefont {Han}, \bibfnamefont
  {Jung~Hoon}}, \bibinfo {author} {\bibfnamefont {Jin-Hong}\ \bibnamefont
  {Park}}, \ and\ \bibinfo {author} {\bibfnamefont {Patrick~A.}\ \bibnamefont
  {Lee}}} (\bibinfo {year} {2019}),\ \bibfield  {title} {\enquote {\bibinfo
  {title} {{Consideration of thermal Hall effect in undoped cuprates}},}\
  }\href {\doibase 10.1103/PhysRevB.99.205157} {\bibfield  {journal} {\bibinfo
  {journal} {Phys. Rev. B}\ }\textbf {\bibinfo {volume} {99}},\ \bibinfo
  {pages} {205157}}\BibitemShut {NoStop}%
\bibitem [{\citenamefont {Han}\ \emph {et~al.}(2012)\citenamefont {Han},
  \citenamefont {Helton}, \citenamefont {Chu}, \citenamefont {Nocera},
  \citenamefont {Rodriguez-Rivera}, \citenamefont {Broholm},\ and\
  \citenamefont {Lee}}]{Han2012}%
  \BibitemOpen
  \bibfield  {author} {\bibinfo {author} {\bibnamefont {Han}, \bibfnamefont
  {Tian-Heng}}, \bibinfo {author} {\bibfnamefont {Joel~S.}\ \bibnamefont
  {Helton}}, \bibinfo {author} {\bibfnamefont {Shaoyan}\ \bibnamefont {Chu}},
  \bibinfo {author} {\bibfnamefont {Daniel~G.}\ \bibnamefont {Nocera}},
  \bibinfo {author} {\bibfnamefont {Jose~A.}\ \bibnamefont {Rodriguez-Rivera}},
  \bibinfo {author} {\bibfnamefont {Collin}\ \bibnamefont {Broholm}}, \ and\
  \bibinfo {author} {\bibfnamefont {Young~S.}\ \bibnamefont {Lee}}} (\bibinfo
  {year} {2012}),\ \bibfield  {title} {\enquote {\bibinfo {title}
  {Fractionalized excitations in the spin-liquid state of a kagome-lattice
  antiferromagnet},}\ }\href {\doibase 10.1038/nature11659} {\bibfield
  {journal} {\bibinfo  {journal} {Nature}\ }\textbf {\bibinfo {volume}
  {492}}~(\bibinfo {number} {7429}),\ \bibinfo {pages} {406--410}}\BibitemShut
  {NoStop}%
\bibitem [{\citenamefont {He}\ \emph {et~al.}(2018)\citenamefont {He},
  \citenamefont {Xu}, \citenamefont {Chen}, \citenamefont {Law},\ and\
  \citenamefont {Lee}}]{Lee2018}%
  \BibitemOpen
  \bibfield  {author} {\bibinfo {author} {\bibnamefont {He}, \bibfnamefont
  {Wen-Yu}}, \bibinfo {author} {\bibfnamefont {Xiao~Yan}\ \bibnamefont {Xu}},
  \bibinfo {author} {\bibfnamefont {Gang}\ \bibnamefont {Chen}}, \bibinfo
  {author} {\bibfnamefont {K.~T.}\ \bibnamefont {Law}}, \ and\ \bibinfo
  {author} {\bibfnamefont {Patrick~A.}\ \bibnamefont {Lee}}} (\bibinfo {year}
  {2018}),\ \bibfield  {title} {\enquote {\bibinfo {title} {{Spinon Fermi
  surface in a cluster Mott insulator model on a triangular lattice and
  possible application to
  $1T\text{\ensuremath{-}}{\mathrm{Ta}\mathrm{S}}_{2}$}},}\ }\href {\doibase
  10.1103/PhysRevLett.121.046401} {\bibfield  {journal} {\bibinfo  {journal}
  {Phys. Rev. Lett.}\ }\textbf {\bibinfo {volume} {121}},\ \bibinfo {pages}
  {046401}}\BibitemShut {NoStop}%
\bibitem [{\citenamefont {Hentrich}\ \emph {et~al.}(2019)\citenamefont
  {Hentrich}, \citenamefont {Roslova}, \citenamefont {Isaeva}, \citenamefont
  {Doert}, \citenamefont {Brenig}, \citenamefont {B\"uchner},\ and\
  \citenamefont {Hess}}]{Hess2019}%
  \BibitemOpen
  \bibfield  {author} {\bibinfo {author} {\bibnamefont {Hentrich},
  \bibfnamefont {Richard}}, \bibinfo {author} {\bibfnamefont {Maria}\
  \bibnamefont {Roslova}}, \bibinfo {author} {\bibfnamefont {Anna}\
  \bibnamefont {Isaeva}}, \bibinfo {author} {\bibfnamefont {Thomas}\
  \bibnamefont {Doert}}, \bibinfo {author} {\bibfnamefont {Wolfram}\
  \bibnamefont {Brenig}}, \bibinfo {author} {\bibfnamefont {Bernd}\
  \bibnamefont {B\"uchner}}, \ and\ \bibinfo {author} {\bibfnamefont
  {Christian}\ \bibnamefont {Hess}}} (\bibinfo {year} {2019}),\ \bibfield
  {title} {\enquote {\bibinfo {title} {{Large thermal Hall effect in
  $\ensuremath{\alpha}\text{\ensuremath{-}}{\mathrm{RuCl}}_{3}$: Evidence for
  heat transport by Kitaev-Heisenberg paramagnons}},}\ }\href {\doibase
  10.1103/PhysRevB.99.085136} {\bibfield  {journal} {\bibinfo  {journal} {Phys.
  Rev. B}\ }\textbf {\bibinfo {volume} {99}},\ \bibinfo {pages}
  {085136}}\BibitemShut {NoStop}%
\bibitem [{\citenamefont {Hentrich}\ \emph {et~al.}(2018)\citenamefont
  {Hentrich}, \citenamefont {Wolter}, \citenamefont {Zotos}, \citenamefont
  {Brenig}, \citenamefont {Nowak}, \citenamefont {Isaeva}, \citenamefont
  {Doert}, \citenamefont {Banerjee}, \citenamefont {Lampen-Kelley},
  \citenamefont {Mandrus}, \citenamefont {Nagler}, \citenamefont {Sears},
  \citenamefont {Kim}, \citenamefont {B\"uchner},\ and\ \citenamefont
  {Hess}}]{Hess2018}%
  \BibitemOpen
  \bibfield  {author} {\bibinfo {author} {\bibnamefont {Hentrich},
  \bibfnamefont {Richard}}, \bibinfo {author} {\bibfnamefont {Anja U.~B.}\
  \bibnamefont {Wolter}}, \bibinfo {author} {\bibfnamefont {Xenophon}\
  \bibnamefont {Zotos}}, \bibinfo {author} {\bibfnamefont {Wolfram}\
  \bibnamefont {Brenig}}, \bibinfo {author} {\bibfnamefont {Domenic}\
  \bibnamefont {Nowak}}, \bibinfo {author} {\bibfnamefont {Anna}\ \bibnamefont
  {Isaeva}}, \bibinfo {author} {\bibfnamefont {Thomas}\ \bibnamefont {Doert}},
  \bibinfo {author} {\bibfnamefont {Arnab}\ \bibnamefont {Banerjee}}, \bibinfo
  {author} {\bibfnamefont {Paula}\ \bibnamefont {Lampen-Kelley}}, \bibinfo
  {author} {\bibfnamefont {David~G.}\ \bibnamefont {Mandrus}}, \bibinfo
  {author} {\bibfnamefont {Stephen~E.}\ \bibnamefont {Nagler}}, \bibinfo
  {author} {\bibfnamefont {Jennifer}\ \bibnamefont {Sears}}, \bibinfo {author}
  {\bibfnamefont {Young-June}\ \bibnamefont {Kim}}, \bibinfo {author}
  {\bibfnamefont {Bernd}\ \bibnamefont {B\"uchner}}, \ and\ \bibinfo {author}
  {\bibfnamefont {Christian}\ \bibnamefont {Hess}}} (\bibinfo {year} {2018}),\
  \bibfield  {title} {\enquote {\bibinfo {title} {{Unusual Phonon Heat
  Transport in $\ensuremath{\alpha}\text{\ensuremath{-}}{\mathrm{RuCl}}_{3}$:
  Strong Spin-Phonon Scattering and Field-Induced Spin Gap}},}\ }\href
  {\doibase 10.1103/PhysRevLett.120.117204} {\bibfield  {journal} {\bibinfo
  {journal} {Phys. Rev. Lett.}\ }\textbf {\bibinfo {volume} {120}},\ \bibinfo
  {pages} {117204}}\BibitemShut {NoStop}%
\bibitem [{\citenamefont {Herbut}\ and\ \citenamefont
  {Seradjeh}(2003)}]{Herbut2003}%
  \BibitemOpen
  \bibfield  {author} {\bibinfo {author} {\bibnamefont {Herbut}, \bibfnamefont
  {Igor~F}}, \ and\ \bibinfo {author} {\bibfnamefont {Babak~H.}\ \bibnamefont
  {Seradjeh}}} (\bibinfo {year} {2003}),\ \bibfield  {title} {\enquote
  {\bibinfo {title} {{Permanent confinement in the compact
  ${\mathrm{Q}\mathrm{E}\mathrm{D}}_{3}$ with fermionic matter}},}\ }\href
  {\doibase 10.1103/PhysRevLett.91.171601} {\bibfield  {journal} {\bibinfo
  {journal} {Phys. Rev. Lett.}\ }\textbf {\bibinfo {volume} {91}},\ \bibinfo
  {pages} {171601}}\BibitemShut {NoStop}%
\bibitem [{\citenamefont {Herbut}\ \emph {et~al.}(2003)\citenamefont {Herbut},
  \citenamefont {Seradjeh}, \citenamefont {Sachdev},\ and\ \citenamefont
  {Murthy}}]{Herbut2003B}%
  \BibitemOpen
  \bibfield  {author} {\bibinfo {author} {\bibnamefont {Herbut}, \bibfnamefont
  {Igor~F}}, \bibinfo {author} {\bibfnamefont {Babak~H.}\ \bibnamefont
  {Seradjeh}}, \bibinfo {author} {\bibfnamefont {Subir}\ \bibnamefont
  {Sachdev}}, \ and\ \bibinfo {author} {\bibfnamefont {Ganpathy}\ \bibnamefont
  {Murthy}}} (\bibinfo {year} {2003}),\ \bibfield  {title} {\enquote {\bibinfo
  {title} {Absence of {U(1)} spin liquids in two dimensions},}\ }\href
  {\doibase 10.1103/PhysRevB.68.195110} {\bibfield  {journal} {\bibinfo
  {journal} {Phys. Rev. B}\ }\textbf {\bibinfo {volume} {68}},\ \bibinfo
  {pages} {195110}}\BibitemShut {NoStop}%
\bibitem [{\citenamefont {Hermanns}\ \emph {et~al.}(2018)\citenamefont
  {Hermanns}, \citenamefont {Kimchi},\ and\ \citenamefont
  {Knolle}}]{Hermanns2018}%
  \BibitemOpen
  \bibfield  {author} {\bibinfo {author} {\bibnamefont {Hermanns},
  \bibfnamefont {M}}, \bibinfo {author} {\bibfnamefont {I.}~\bibnamefont
  {Kimchi}}, \ and\ \bibinfo {author} {\bibfnamefont {J.}~\bibnamefont
  {Knolle}}} (\bibinfo {year} {2018}),\ \bibfield  {title} {\enquote {\bibinfo
  {title} {{Physics of the Kitaev model: Fractionalization, dynamic
  correlations, and material connections}},}\ }\href {\doibase
  10.1146/annurev-conmatphys-033117-053934} {\bibfield  {journal} {\bibinfo
  {journal} {Annual Review of Condensed Matter Physics}\ }\textbf {\bibinfo
  {volume} {9}}~(\bibinfo {number} {1}),\ \bibinfo {pages}
  {17--33}}\BibitemShut {NoStop}%
\bibitem [{\citenamefont {Hermele}\ \emph
  {et~al.}(2004{\natexlab{a}})\citenamefont {Hermele}, \citenamefont {Fisher},\
  and\ \citenamefont {Balents}}]{Hermele2004B}%
  \BibitemOpen
  \bibfield  {author} {\bibinfo {author} {\bibnamefont {Hermele}, \bibfnamefont
  {Michael}}, \bibinfo {author} {\bibfnamefont {Matthew P.~A.}\ \bibnamefont
  {Fisher}}, \ and\ \bibinfo {author} {\bibfnamefont {Leon}\ \bibnamefont
  {Balents}}} (\bibinfo {year} {2004}{\natexlab{a}}),\ \bibfield  {title}
  {\enquote {\bibinfo {title} {{Pyrochlore photons: The U(1) spin liquid in a
  $S=\frac{1}{2}$ three-dimensional frustrated magnet}},}\ }\href {\doibase
  10.1103/PhysRevB.69.064404} {\bibfield  {journal} {\bibinfo  {journal} {Phys.
  Rev. B}\ }\textbf {\bibinfo {volume} {69}},\ \bibinfo {pages}
  {064404}}\BibitemShut {NoStop}%
\bibitem [{\citenamefont {Hermele}\ \emph {et~al.}(2008)\citenamefont
  {Hermele}, \citenamefont {Ran}, \citenamefont {Lee},\ and\ \citenamefont
  {Wen}}]{Hermele2008}%
  \BibitemOpen
  \bibfield  {author} {\bibinfo {author} {\bibnamefont {Hermele}, \bibfnamefont
  {Michael}}, \bibinfo {author} {\bibfnamefont {Ying}\ \bibnamefont {Ran}},
  \bibinfo {author} {\bibfnamefont {Patrick~A.}\ \bibnamefont {Lee}}, \ and\
  \bibinfo {author} {\bibfnamefont {Xiao-Gang}\ \bibnamefont {Wen}}} (\bibinfo
  {year} {2008}),\ \bibfield  {title} {\enquote {\bibinfo {title} {{Properties
  of an algebraic spin liquid on the kagom{\'{e}} lattice}},}\ }\href {\doibase
  10.1103/PhysRevB.77.224413} {\bibfield  {journal} {\bibinfo  {journal} {Phys.
  Rev. B}\ }\textbf {\bibinfo {volume} {77}},\ \bibinfo {pages}
  {224413}}\BibitemShut {NoStop}%
\bibitem [{\citenamefont {Hermele}\ \emph {et~al.}(2005)\citenamefont
  {Hermele}, \citenamefont {Senthil},\ and\ \citenamefont
  {Fisher}}]{Hermele2005}%
  \BibitemOpen
  \bibfield  {author} {\bibinfo {author} {\bibnamefont {Hermele}, \bibfnamefont
  {Michael}}, \bibinfo {author} {\bibfnamefont {T.}~\bibnamefont {Senthil}}, \
  and\ \bibinfo {author} {\bibfnamefont {Matthew P.~A.}\ \bibnamefont
  {Fisher}}} (\bibinfo {year} {2005}),\ \bibfield  {title} {\enquote {\bibinfo
  {title} {Algebraic spin liquid as the mother of many competing orders},}\
  }\href {\doibase 10.1103/PhysRevB.72.104404} {\bibfield  {journal} {\bibinfo
  {journal} {Phys. Rev. B}\ }\textbf {\bibinfo {volume} {72}},\ \bibinfo
  {pages} {104404}}\BibitemShut {NoStop}%
\bibitem [{\citenamefont {Hermele}\ \emph
  {et~al.}(2004{\natexlab{b}})\citenamefont {Hermele}, \citenamefont {Senthil},
  \citenamefont {Fisher}, \citenamefont {Lee}, \citenamefont {Nagaosa},\ and\
  \citenamefont {Wen}}]{Hermele2004}%
  \BibitemOpen
  \bibfield  {author} {\bibinfo {author} {\bibnamefont {Hermele}, \bibfnamefont
  {Michael}}, \bibinfo {author} {\bibfnamefont {T.}~\bibnamefont {Senthil}},
  \bibinfo {author} {\bibfnamefont {Matthew P.~A.}\ \bibnamefont {Fisher}},
  \bibinfo {author} {\bibfnamefont {Patrick~A.}\ \bibnamefont {Lee}}, \bibinfo
  {author} {\bibfnamefont {Naoto}\ \bibnamefont {Nagaosa}}, \ and\ \bibinfo
  {author} {\bibfnamefont {Xiao-Gang}\ \bibnamefont {Wen}}} (\bibinfo {year}
  {2004}{\natexlab{b}}),\ \bibfield  {title} {\enquote {\bibinfo {title}
  {{Stability of U(1) spin liquids in two dimensions}},}\ }\href {\doibase
  10.1103/PhysRevB.70.214437} {\bibfield  {journal} {\bibinfo  {journal} {Phys.
  Rev. B}\ }\textbf {\bibinfo {volume} {70}},\ \bibinfo {pages}
  {214437}}\BibitemShut {NoStop}%
\bibitem [{\citenamefont {Hickey}\ \emph {et~al.}(2016)\citenamefont {Hickey},
  \citenamefont {Cincio}, \citenamefont {Papi{\'{c}}},\ and\ \citenamefont
  {Paramekanti}}]{Hickey2016}%
  \BibitemOpen
  \bibfield  {author} {\bibinfo {author} {\bibnamefont {Hickey}, \bibfnamefont
  {Ciar{\'{a}}n}}, \bibinfo {author} {\bibfnamefont {Lukasz}\ \bibnamefont
  {Cincio}}, \bibinfo {author} {\bibfnamefont {Zlatko}\ \bibnamefont
  {Papi{\'{c}}}}, \ and\ \bibinfo {author} {\bibfnamefont {Arun}\ \bibnamefont
  {Paramekanti}}} (\bibinfo {year} {2016}),\ \bibfield  {title} {\enquote
  {\bibinfo {title} {{Haldane-Hubbard Mott insulator: From tetrahedral spin
  crystal to chiral spin liquid}},}\ }\href {\doibase
  10.1103/physrevlett.116.137202} {\bibfield  {journal} {\bibinfo  {journal}
  {Physical Review Letters}\ }\textbf {\bibinfo {volume} {116}}~(\bibinfo
  {number} {13}),\ \bibinfo {pages} {137202}}\BibitemShut {NoStop}%
\bibitem [{\citenamefont {Hickey}\ and\ \citenamefont
  {Trebst}(2019)}]{Hickey2019}%
  \BibitemOpen
  \bibfield  {author} {\bibinfo {author} {\bibnamefont {Hickey}, \bibfnamefont
  {Ciar{\'{a}}n}}, \ and\ \bibinfo {author} {\bibfnamefont {Simon}\
  \bibnamefont {Trebst}}} (\bibinfo {year} {2019}),\ \bibfield  {title}
  {\enquote {\bibinfo {title} {{Emergence of a field-driven U(1) spin liquid in
  the Kitaev honeycomb model}},}\ }\href {\doibase 10.1038/s41467-019-08459-9}
  {\bibfield  {journal} {\bibinfo  {journal} {Nature Communications}\ }\textbf
  {\bibinfo {volume} {10}}~(\bibinfo {number} {1}),\ \bibinfo {pages}
  {530}}\BibitemShut {NoStop}%
\bibitem [{\citenamefont {Hirokane}\ \emph {et~al.}(2019)\citenamefont
  {Hirokane}, \citenamefont {Nii}, \citenamefont {Tomioka},\ and\ \citenamefont
  {Onose}}]{PhysRevB.99.134419}%
  \BibitemOpen
  \bibfield  {author} {\bibinfo {author} {\bibnamefont {Hirokane},
  \bibfnamefont {Yuji}}, \bibinfo {author} {\bibfnamefont {Yoichi}\
  \bibnamefont {Nii}}, \bibinfo {author} {\bibfnamefont {Yasuhide}\
  \bibnamefont {Tomioka}}, \ and\ \bibinfo {author} {\bibfnamefont {Yoshinori}\
  \bibnamefont {Onose}}} (\bibinfo {year} {2019}),\ \bibfield  {title}
  {\enquote {\bibinfo {title} {{Phononic thermal Hall effect in diluted terbium
  oxides}},}\ }\href {\doibase 10.1103/PhysRevB.99.134419} {\bibfield
  {journal} {\bibinfo  {journal} {Phys. Rev. B}\ }\textbf {\bibinfo {volume}
  {99}},\ \bibinfo {pages} {134419}}\BibitemShut {NoStop}%
\bibitem [{\citenamefont {Hirschberger}\ \emph
  {et~al.}(2015{\natexlab{a}})\citenamefont {Hirschberger}, \citenamefont
  {Chisnell}, \citenamefont {Lee},\ and\ \citenamefont {Ong}}]{Ong2015}%
  \BibitemOpen
  \bibfield  {author} {\bibinfo {author} {\bibnamefont {Hirschberger},
  \bibfnamefont {Max}}, \bibinfo {author} {\bibfnamefont {Robin}\ \bibnamefont
  {Chisnell}}, \bibinfo {author} {\bibfnamefont {Young~S.}\ \bibnamefont
  {Lee}}, \ and\ \bibinfo {author} {\bibfnamefont {N.~P.}\ \bibnamefont {Ong}}}
  (\bibinfo {year} {2015}{\natexlab{a}}),\ \bibfield  {title} {\enquote
  {\bibinfo {title} {{Thermal Hall effect of spin excitations in a kagom{\'{e}}
  magnet}},}\ }\href {\doibase 10.1103/PhysRevLett.115.106603} {\bibfield
  {journal} {\bibinfo  {journal} {Phys. Rev. Lett.}\ }\textbf {\bibinfo
  {volume} {115}},\ \bibinfo {pages} {106603}}\BibitemShut {NoStop}%
\bibitem [{\citenamefont {Hirschberger}\ \emph
  {et~al.}(2015{\natexlab{b}})\citenamefont {Hirschberger}, \citenamefont
  {Krizan}, \citenamefont {Cava},\ and\ \citenamefont
  {Ong}}]{Hirschberger2015}%
  \BibitemOpen
  \bibfield  {author} {\bibinfo {author} {\bibnamefont {Hirschberger},
  \bibfnamefont {Max}}, \bibinfo {author} {\bibfnamefont {Jason~W.}\
  \bibnamefont {Krizan}}, \bibinfo {author} {\bibfnamefont {R.~J.}\
  \bibnamefont {Cava}}, \ and\ \bibinfo {author} {\bibfnamefont {N.~P.}\
  \bibnamefont {Ong}}} (\bibinfo {year} {2015}{\natexlab{b}}),\ \bibfield
  {title} {\enquote {\bibinfo {title} {{Large thermal Hall conductivity of
  neutral spin excitations in a frustrated quantum magnet}},}\ }\href {\doibase
  10.1126/science.1257340} {\bibfield  {journal} {\bibinfo  {journal}
  {Science}\ }\textbf {\bibinfo {volume} {348}}~(\bibinfo {number} {6230}),\
  \bibinfo {pages} {106--109}}\BibitemShut {NoStop}%
\bibitem [{\citenamefont {Holstein}\ and\ \citenamefont
  {Primakoff}(1940)}]{PhysRev.58.1098}%
  \BibitemOpen
  \bibfield  {author} {\bibinfo {author} {\bibnamefont {Holstein},
  \bibfnamefont {T}}, \ and\ \bibinfo {author} {\bibfnamefont {H.}~\bibnamefont
  {Primakoff}}} (\bibinfo {year} {1940}),\ \bibfield  {title} {\enquote
  {\bibinfo {title} {Field dependence of the intrinsic domain magnetization of
  a ferromagnet},}\ }\href {\doibase 10.1103/PhysRev.58.1098} {\bibfield
  {journal} {\bibinfo  {journal} {Phys. Rev.}\ }\textbf {\bibinfo {volume}
  {58}},\ \bibinfo {pages} {1098--1113}}\BibitemShut {NoStop}%
\bibitem [{\citenamefont {{Hongsoek Kim, Se Kwon Kim}}(2022)}]{Kim2022}%
  \BibitemOpen
  \bibfield  {author} {\bibinfo {author} {\bibnamefont {{Hongsoek Kim, Se Kwon
  Kim}},}} (\bibinfo {year} {2022}),\ \href@noop {} {\enquote {\bibinfo {title}
  {{{Topological phase transition in magnon bands in a honeycomb ferromagnet
  driven by sublattice symmetry breaking}}},}\ }\Eprint
  {http://arxiv.org/abs/2203.03845} {arXiv:2203.03845} \BibitemShut {NoStop}%
\bibitem [{\citenamefont {van Hoogdalem}\ \emph {et~al.}(2013)\citenamefont
  {van Hoogdalem}, \citenamefont {Tserkovnyak},\ and\ \citenamefont
  {Loss}}]{PhysRevB.87.024402}%
  \BibitemOpen
  \bibfield  {author} {\bibinfo {author} {\bibnamefont {van Hoogdalem},
  \bibfnamefont {Kevin~A}}, \bibinfo {author} {\bibfnamefont {Yaroslav}\
  \bibnamefont {Tserkovnyak}}, \ and\ \bibinfo {author} {\bibfnamefont
  {Daniel}\ \bibnamefont {Loss}}} (\bibinfo {year} {2013}),\ \bibfield  {title}
  {\enquote {\bibinfo {title} {{Magnetic texture-induced thermal Hall
  effects}},}\ }\href {\doibase 10.1103/PhysRevB.87.024402} {\bibfield
  {journal} {\bibinfo  {journal} {Phys. Rev. B}\ }\textbf {\bibinfo {volume}
  {87}},\ \bibinfo {pages} {024402}}\BibitemShut {NoStop}%
\bibitem [{\citenamefont {Huang}\ \emph {et~al.}(0)\citenamefont {Huang},
  \citenamefont {Wang}, \citenamefont {Wang},\ and\ \citenamefont
  {Chen}}]{s0217979224500401}%
  \BibitemOpen
  \bibfield  {author} {\bibinfo {author} {\bibnamefont {Huang}, \bibfnamefont
  {Chun-Jiong}}, \bibinfo {author} {\bibfnamefont {Xiaoqun}\ \bibnamefont
  {Wang}}, \bibinfo {author} {\bibfnamefont {Ziqiang}\ \bibnamefont {Wang}}, \
  and\ \bibinfo {author} {\bibfnamefont {Gang}\ \bibnamefont {Chen}}} (\bibinfo
  {year} {0}),\ \bibfield  {title} {\enquote {\bibinfo {title} {{Emergent
  Halperin–Saslow mode, gauge glass and quenched disorders in quantum Ising
  magnet TmMgGaO$_4$}},}\ }\href {\doibase 10.1142/S0217979224500401}
  {\bibfield  {journal} {\bibinfo  {journal} {International Journal of Modern
  Physics B}\ }\textbf {\bibinfo {volume} {0}}~(\bibinfo {number} {0}),\
  \bibinfo {pages} {2450040}}\BibitemShut {NoStop}%
\bibitem [{\citenamefont {Huang}\ \emph {et~al.}(2014)\citenamefont {Huang},
  \citenamefont {Chen},\ and\ \citenamefont {Hermele}}]{Huang2014}%
  \BibitemOpen
  \bibfield  {author} {\bibinfo {author} {\bibnamefont {Huang}, \bibfnamefont
  {Yi-Ping}}, \bibinfo {author} {\bibfnamefont {Gang}\ \bibnamefont {Chen}}, \
  and\ \bibinfo {author} {\bibfnamefont {Michael}\ \bibnamefont {Hermele}}}
  (\bibinfo {year} {2014}),\ \bibfield  {title} {\enquote {\bibinfo {title}
  {Quantum spin ices and topological phases from dipolar-octupolar doublets on
  the pyrochlore lattice},}\ }\href {\doibase 10.1103/PhysRevLett.112.167203}
  {\bibfield  {journal} {\bibinfo  {journal} {Phys. Rev. Lett.}\ }\textbf
  {\bibinfo {volume} {112}},\ \bibinfo {pages} {167203}}\BibitemShut {NoStop}%
\bibitem [{\citenamefont {Huang}\ \emph {et~al.}(2022)\citenamefont {Huang},
  \citenamefont {Zhu}, \citenamefont {Gong}, \citenamefont {Jiang},\ and\
  \citenamefont {Sheng}}]{PhysRevB.105.155104}%
  \BibitemOpen
  \bibfield  {author} {\bibinfo {author} {\bibnamefont {Huang}, \bibfnamefont
  {Yixuan}}, \bibinfo {author} {\bibfnamefont {W.}~\bibnamefont {Zhu}},
  \bibinfo {author} {\bibfnamefont {Shou-Shu}\ \bibnamefont {Gong}}, \bibinfo
  {author} {\bibfnamefont {Hong-Chen}\ \bibnamefont {Jiang}}, \ and\ \bibinfo
  {author} {\bibfnamefont {D.~N.}\ \bibnamefont {Sheng}}} (\bibinfo {year}
  {2022}),\ \bibfield  {title} {\enquote {\bibinfo {title} {Coexistence of
  non-abelian chiral spin liquid and magnetic order in a spin-1
  antiferromagnet},}\ }\href {\doibase 10.1103/PhysRevB.105.155104} {\bibfield
  {journal} {\bibinfo  {journal} {Phys. Rev. B}\ }\textbf {\bibinfo {volume}
  {105}},\ \bibinfo {pages} {155104}}\BibitemShut {NoStop}%
\bibitem [{\citenamefont {Huh}\ \emph {et~al.}(2010)\citenamefont {Huh},
  \citenamefont {Fritz},\ and\ \citenamefont {Sachdev}}]{PhysRevB.81.144432}%
  \BibitemOpen
  \bibfield  {author} {\bibinfo {author} {\bibnamefont {Huh}, \bibfnamefont
  {Yejin}}, \bibinfo {author} {\bibfnamefont {Lars}\ \bibnamefont {Fritz}}, \
  and\ \bibinfo {author} {\bibfnamefont {Subir}\ \bibnamefont {Sachdev}}}
  (\bibinfo {year} {2010}),\ \bibfield  {title} {\enquote {\bibinfo {title}
  {{Quantum criticality of the kagome antiferromagnet with
  Dzyaloshinskii-Moriya interactions}},}\ }\href {\doibase
  10.1103/PhysRevB.81.144432} {\bibfield  {journal} {\bibinfo  {journal} {Phys.
  Rev. B}\ }\textbf {\bibinfo {volume} {81}},\ \bibinfo {pages}
  {144432}}\BibitemShut {NoStop}%
\bibitem [{\citenamefont {Hwang}\ \emph {et~al.}(2022)\citenamefont {Hwang},
  \citenamefont {Go}, \citenamefont {Seong}, \citenamefont {Shibauchi},\ and\
  \citenamefont {Moon}}]{Hwang2022}%
  \BibitemOpen
  \bibfield  {author} {\bibinfo {author} {\bibnamefont {Hwang}, \bibfnamefont
  {Kyusung}}, \bibinfo {author} {\bibfnamefont {Ara}\ \bibnamefont {Go}},
  \bibinfo {author} {\bibfnamefont {Ji~Heon}\ \bibnamefont {Seong}}, \bibinfo
  {author} {\bibfnamefont {Takasada}\ \bibnamefont {Shibauchi}}, \ and\
  \bibinfo {author} {\bibfnamefont {Eun-Gook}\ \bibnamefont {Moon}}} (\bibinfo
  {year} {2022}),\ \bibfield  {title} {\enquote {\bibinfo {title}
  {{Identification of a Kitaev quantum spin liquid by magnetic field angle
  dependence}},}\ }\href {\doibase 10.1038/s41467-021-27943-9} {\bibfield
  {journal} {\bibinfo  {journal} {Nature Communications}\ }\textbf {\bibinfo
  {volume} {13}}~(\bibinfo {number} {1}),\
  10.1038/s41467-021-27943-9}\BibitemShut {NoStop}%
\bibitem [{\citenamefont {Ideue}\ \emph {et~al.}(2012)\citenamefont {Ideue},
  \citenamefont {Onose}, \citenamefont {Katsura}, \citenamefont {Shiomi},
  \citenamefont {Ishiwata}, \citenamefont {Nagaosa},\ and\ \citenamefont
  {Tokura}}]{PhysRevB.85.134411}%
  \BibitemOpen
  \bibfield  {author} {\bibinfo {author} {\bibnamefont {Ideue}, \bibfnamefont
  {T}}, \bibinfo {author} {\bibfnamefont {Y.}~\bibnamefont {Onose}}, \bibinfo
  {author} {\bibfnamefont {H.}~\bibnamefont {Katsura}}, \bibinfo {author}
  {\bibfnamefont {Y.}~\bibnamefont {Shiomi}}, \bibinfo {author} {\bibfnamefont
  {S.}~\bibnamefont {Ishiwata}}, \bibinfo {author} {\bibfnamefont
  {N.}~\bibnamefont {Nagaosa}}, \ and\ \bibinfo {author} {\bibfnamefont
  {Y.}~\bibnamefont {Tokura}}} (\bibinfo {year} {2012}),\ \bibfield  {title}
  {\enquote {\bibinfo {title} {{Effect of lattice geometry on magnon Hall
  effect in ferromagnetic insulators}},}\ }\href {\doibase
  10.1103/PhysRevB.85.134411} {\bibfield  {journal} {\bibinfo  {journal} {Phys.
  Rev. B}\ }\textbf {\bibinfo {volume} {85}},\ \bibinfo {pages}
  {134411}}\BibitemShut {NoStop}%
\bibitem [{\citenamefont {Ideue}\ \emph {et~al.}(2017)\citenamefont {Ideue},
  \citenamefont {Kurumaji}, \citenamefont {Ishiwata},\ and\ \citenamefont
  {Tokura}}]{nmat4905}%
  \BibitemOpen
  \bibfield  {author} {\bibinfo {author} {\bibnamefont {Ideue}, \bibfnamefont
  {Toshiya}}, \bibinfo {author} {\bibfnamefont {Takashi}\ \bibnamefont
  {Kurumaji}}, \bibinfo {author} {\bibfnamefont {Shintaro}\ \bibnamefont
  {Ishiwata}}, \ and\ \bibinfo {author} {\bibfnamefont {Yoshinori}\
  \bibnamefont {Tokura}}} (\bibinfo {year} {2017}),\ \bibfield  {title}
  {\enquote {\bibinfo {title} {Giant thermal hall effect in multiferroics},}\
  }\href {\doibase 10.1038/nmat4905} {\bibfield  {journal} {\bibinfo  {journal}
  {Nature materials}\ }\textbf {\bibinfo {volume} {16}},\ \bibinfo {pages}
  {797}}\BibitemShut {NoStop}%
\bibitem [{\citenamefont {Ihara}\ \emph {et~al.}(2017)\citenamefont {Ihara},
  \citenamefont {Sasaki}, \citenamefont {Noguchi}, \citenamefont {Ishii},
  \citenamefont {Oda},\ and\ \citenamefont {Yoshida}}]{Ihara2017}%
  \BibitemOpen
  \bibfield  {author} {\bibinfo {author} {\bibnamefont {Ihara}, \bibfnamefont
  {Y}}, \bibinfo {author} {\bibfnamefont {T.}~\bibnamefont {Sasaki}}, \bibinfo
  {author} {\bibfnamefont {N.}~\bibnamefont {Noguchi}}, \bibinfo {author}
  {\bibfnamefont {Y.}~\bibnamefont {Ishii}}, \bibinfo {author} {\bibfnamefont
  {M.}~\bibnamefont {Oda}}, \ and\ \bibinfo {author} {\bibfnamefont
  {H.}~\bibnamefont {Yoshida}}} (\bibinfo {year} {2017}),\ \bibfield  {title}
  {\enquote {\bibinfo {title} {{Gapless magnetic excitations in the kagome
  antiferromagnet Ca-kapellasite probed by $^{35}\mathrm{Cl}$ NMR
  spectroscopy}},}\ }\href {\doibase 10.1103/PhysRevB.96.180409} {\bibfield
  {journal} {\bibinfo  {journal} {Phys. Rev. B}\ }\textbf {\bibinfo {volume}
  {96}},\ \bibinfo {pages} {180409}}\BibitemShut {NoStop}%
\bibitem [{\citenamefont {Ioffe}\ and\ \citenamefont
  {Larkin}(1989)}]{Ioffe1989}%
  \BibitemOpen
  \bibfield  {author} {\bibinfo {author} {\bibnamefont {Ioffe}, \bibfnamefont
  {L~B}}, \ and\ \bibinfo {author} {\bibfnamefont {A.~I.}\ \bibnamefont
  {Larkin}}} (\bibinfo {year} {1989}),\ \bibfield  {title} {\enquote {\bibinfo
  {title} {Gapless fermions and gauge fields in dielectrics},}\ }\href
  {\doibase 10.1103/PhysRevB.39.8988} {\bibfield  {journal} {\bibinfo
  {journal} {Phys. Rev. B}\ }\textbf {\bibinfo {volume} {39}},\ \bibinfo
  {pages} {8988--8999}}\BibitemShut {NoStop}%
\bibitem [{\citenamefont {Iqbal}\ \emph {et~al.}(2016)\citenamefont {Iqbal},
  \citenamefont {Hu}, \citenamefont {Thomale}, \citenamefont {Poilblanc},\ and\
  \citenamefont {Becca}}]{PhysRevB.93.144411}%
  \BibitemOpen
  \bibfield  {author} {\bibinfo {author} {\bibnamefont {Iqbal}, \bibfnamefont
  {Yasir}}, \bibinfo {author} {\bibfnamefont {Wen-Jun}\ \bibnamefont {Hu}},
  \bibinfo {author} {\bibfnamefont {Ronny}\ \bibnamefont {Thomale}}, \bibinfo
  {author} {\bibfnamefont {Didier}\ \bibnamefont {Poilblanc}}, \ and\ \bibinfo
  {author} {\bibfnamefont {Federico}\ \bibnamefont {Becca}}} (\bibinfo {year}
  {2016}),\ \bibfield  {title} {\enquote {\bibinfo {title} {{Spin liquid nature
  in the Heisenberg ${J}_{1}\ensuremath{-}{J}_{2}$ triangular
  antiferromagnet}},}\ }\href {\doibase 10.1103/PhysRevB.93.144411} {\bibfield
  {journal} {\bibinfo  {journal} {Phys. Rev. B}\ }\textbf {\bibinfo {volume}
  {93}},\ \bibinfo {pages} {144411}}\BibitemShut {NoStop}%
\bibitem [{\citenamefont {Itou}\ \emph {et~al.}(2009)\citenamefont {Itou},
  \citenamefont {Oyamada}, \citenamefont {Maegawa}, \citenamefont {Tamura},\
  and\ \citenamefont {Kato}}]{Itou2009}%
  \BibitemOpen
  \bibfield  {author} {\bibinfo {author} {\bibnamefont {Itou}, \bibfnamefont
  {T}}, \bibinfo {author} {\bibfnamefont {A}~\bibnamefont {Oyamada}}, \bibinfo
  {author} {\bibfnamefont {S}~\bibnamefont {Maegawa}}, \bibinfo {author}
  {\bibfnamefont {M}~\bibnamefont {Tamura}}, \ and\ \bibinfo {author}
  {\bibfnamefont {R}~\bibnamefont {Kato}}} (\bibinfo {year} {2009}),\ \bibfield
   {title} {\enquote {\bibinfo {title} {{$^{13}\mathrm{Cl}$ {NMR} study of the
  spin-liquid state in the triangular quantum antiferromagnet
  ${\mathrm{EtMe}}_{3}\mathrm{Sb}\mathbf{[}\mathrm{Pd}\mathbf{(}\mathrm{dmit}{\mathbf{)}}_{2}{\mathbf{]}}_{2}$}},}\
  }\href {\doibase 10.1088/1742-6596/145/1/012039} {\bibfield  {journal}
  {\bibinfo  {journal} {Journal of Physics: Conference Series}\ }\textbf
  {\bibinfo {volume} {145}},\ \bibinfo {pages} {012039}}\BibitemShut {NoStop}%
\bibitem [{\citenamefont {Jackeli}\ and\ \citenamefont
  {Khaliullin}(2009)}]{Jackeli2009}%
  \BibitemOpen
  \bibfield  {author} {\bibinfo {author} {\bibnamefont {Jackeli}, \bibfnamefont
  {G}}, \ and\ \bibinfo {author} {\bibfnamefont {G.}~\bibnamefont
  {Khaliullin}}} (\bibinfo {year} {2009}),\ \bibfield  {title} {\enquote
  {\bibinfo {title} {{Mott Insulators in the Strong Spin-Orbit Coupling Limit:
  From Heisenberg to a Quantum Compass and Kitaev Models}},}\ }\href {\doibase
  10.1103/PhysRevLett.102.017205} {\bibfield  {journal} {\bibinfo  {journal}
  {Phys. Rev. Lett.}\ }\textbf {\bibinfo {volume} {102}},\ \bibinfo {pages}
  {017205}}\BibitemShut {NoStop}%
\bibitem [{\citenamefont {Jang}\ \emph {et~al.}(2019)\citenamefont {Jang},
  \citenamefont {Sano}, \citenamefont {Kato},\ and\ \citenamefont
  {Motome}}]{Motome2019}%
  \BibitemOpen
  \bibfield  {author} {\bibinfo {author} {\bibnamefont {Jang}, \bibfnamefont
  {Seong-Hoon}}, \bibinfo {author} {\bibfnamefont {Ryoya}\ \bibnamefont
  {Sano}}, \bibinfo {author} {\bibfnamefont {Yasuyuki}\ \bibnamefont {Kato}}, \
  and\ \bibinfo {author} {\bibfnamefont {Yukitoshi}\ \bibnamefont {Motome}}}
  (\bibinfo {year} {2019}),\ \bibfield  {title} {\enquote {\bibinfo {title}
  {{Antiferromagnetic Kitaev interaction in $f$-electron based honeycomb
  magnets}},}\ }\href {\doibase 10.1103/PhysRevB.99.241106} {\bibfield
  {journal} {\bibinfo  {journal} {Phys. Rev. B}\ }\textbf {\bibinfo {volume}
  {99}},\ \bibinfo {pages} {241106}}\BibitemShut {NoStop}%
\bibitem [{\citenamefont {Janson}\ \emph {et~al.}(2010)\citenamefont {Janson},
  \citenamefont {Richter}, \citenamefont {Sindzingre},\ and\ \citenamefont
  {Rosner}}]{Janson2010}%
  \BibitemOpen
  \bibfield  {author} {\bibinfo {author} {\bibnamefont {Janson}, \bibfnamefont
  {O}}, \bibinfo {author} {\bibfnamefont {J.}~\bibnamefont {Richter}}, \bibinfo
  {author} {\bibfnamefont {P.}~\bibnamefont {Sindzingre}}, \ and\ \bibinfo
  {author} {\bibfnamefont {H.}~\bibnamefont {Rosner}}} (\bibinfo {year}
  {2010}),\ \bibfield  {title} {\enquote {\bibinfo {title} {{Coupled frustrated
  quantum spin-$\frac{1}{2}$ chains with orbital order in volborthite
  ${\text{Cu}}_{3}{\text{V}}_{2}{\text{O}}_{7}{(\text{OH})}_{2}\ensuremath{\cdot}2{\text{H}}_{2}\text{O}$}},}\
  }\href {\doibase 10.1103/PhysRevB.82.104434} {\bibfield  {journal} {\bibinfo
  {journal} {Phys. Rev. B}\ }\textbf {\bibinfo {volume} {82}},\ \bibinfo
  {pages} {104434}}\BibitemShut {NoStop}%
\bibitem [{\citenamefont {Jiang}\ \emph {et~al.}(2019)\citenamefont {Jiang},
  \citenamefont {Kim}, \citenamefont {Han},\ and\ \citenamefont
  {Ran}}]{Jiang_2019}%
  \BibitemOpen
  \bibfield  {author} {\bibinfo {author} {\bibnamefont {Jiang}, \bibfnamefont
  {Shenghan}}, \bibinfo {author} {\bibfnamefont {Panjin}\ \bibnamefont {Kim}},
  \bibinfo {author} {\bibfnamefont {Jung~Hoon}\ \bibnamefont {Han}}, \ and\
  \bibinfo {author} {\bibfnamefont {Ying}\ \bibnamefont {Ran}}} (\bibinfo
  {year} {2019}),\ \bibfield  {title} {\enquote {\bibinfo {title} {{Competing
  Spin Liquid Phases in the S=1/2 Heisenberg Model on the Kagome Lattice}},}\
  }\href {\doibase 10.21468/scipostphys.7.1.006} {\bibfield  {journal}
  {\bibinfo  {journal} {{SciPost} Physics}\ }\textbf {\bibinfo {volume}
  {7}}~(\bibinfo {number} {1}),\ 10.21468/scipostphys.7.1.006}\BibitemShut
  {NoStop}%
\bibitem [{\citenamefont {Johnson}\ \emph {et~al.}(2015)\citenamefont
  {Johnson}, \citenamefont {Williams}, \citenamefont {Haghighirad},
  \citenamefont {Singleton}, \citenamefont {Zapf}, \citenamefont {Manuel},
  \citenamefont {Mazin}, \citenamefont {Li}, \citenamefont {Jeschke},
  \citenamefont {Valent\'{\i}},\ and\ \citenamefont {Coldea}}]{Coldea2015}%
  \BibitemOpen
  \bibfield  {author} {\bibinfo {author} {\bibnamefont {Johnson}, \bibfnamefont
  {R~D}}, \bibinfo {author} {\bibfnamefont {S.~C.}\ \bibnamefont {Williams}},
  \bibinfo {author} {\bibfnamefont {A.~A.}\ \bibnamefont {Haghighirad}},
  \bibinfo {author} {\bibfnamefont {J.}~\bibnamefont {Singleton}}, \bibinfo
  {author} {\bibfnamefont {V.}~\bibnamefont {Zapf}}, \bibinfo {author}
  {\bibfnamefont {P.}~\bibnamefont {Manuel}}, \bibinfo {author} {\bibfnamefont
  {I.~I.}\ \bibnamefont {Mazin}}, \bibinfo {author} {\bibfnamefont
  {Y.}~\bibnamefont {Li}}, \bibinfo {author} {\bibfnamefont {H.~O.}\
  \bibnamefont {Jeschke}}, \bibinfo {author} {\bibfnamefont {R.}~\bibnamefont
  {Valent\'{\i}}}, \ and\ \bibinfo {author} {\bibfnamefont {R.}~\bibnamefont
  {Coldea}}} (\bibinfo {year} {2015}),\ \bibfield  {title} {\enquote {\bibinfo
  {title} {{Monoclinic crystal structure of
  $\ensuremath{\alpha}\ensuremath{-}{\mathrm{RuCl}}_{3}$ and the zigzag
  antiferromagnetic ground state}},}\ }\href {\doibase
  10.1103/PhysRevB.92.235119} {\bibfield  {journal} {\bibinfo  {journal} {Phys.
  Rev. B}\ }\textbf {\bibinfo {volume} {92}},\ \bibinfo {pages}
  {235119}}\BibitemShut {NoStop}%
\bibitem [{\citenamefont {Joshi}\ \emph {et~al.}(1999)\citenamefont {Joshi},
  \citenamefont {Ma}, \citenamefont {Mila}, \citenamefont {Shi},\ and\
  \citenamefont {Zhang}}]{PhysRevB.60.6584}%
  \BibitemOpen
  \bibfield  {author} {\bibinfo {author} {\bibnamefont {Joshi}, \bibfnamefont
  {A}}, \bibinfo {author} {\bibfnamefont {M.}~\bibnamefont {Ma}}, \bibinfo
  {author} {\bibfnamefont {F.}~\bibnamefont {Mila}}, \bibinfo {author}
  {\bibfnamefont {D.~N.}\ \bibnamefont {Shi}}, \ and\ \bibinfo {author}
  {\bibfnamefont {F.~C.}\ \bibnamefont {Zhang}}} (\bibinfo {year} {1999}),\
  \bibfield  {title} {\enquote {\bibinfo {title} {Elementary excitations in
  magnetically ordered systems with orbital degeneracy},}\ }\href {\doibase
  10.1103/PhysRevB.60.6584} {\bibfield  {journal} {\bibinfo  {journal} {Phys.
  Rev. B}\ }\textbf {\bibinfo {volume} {60}},\ \bibinfo {pages}
  {6584--6587}}\BibitemShut {NoStop}%
\bibitem [{\citenamefont {Joshi}(2018)}]{PhysRevB.98.060405}%
  \BibitemOpen
  \bibfield  {author} {\bibinfo {author} {\bibnamefont {Joshi}, \bibfnamefont
  {Darshan~G}}} (\bibinfo {year} {2018}),\ \bibfield  {title} {\enquote
  {\bibinfo {title} {{Topological excitations in the ferromagnetic
  Kitaev-Heisenberg model}},}\ }\href {\doibase 10.1103/PhysRevB.98.060405}
  {\bibfield  {journal} {\bibinfo  {journal} {Phys. Rev. B}\ }\textbf {\bibinfo
  {volume} {98}},\ \bibinfo {pages} {060405}}\BibitemShut {NoStop}%
\bibitem [{\citenamefont {Joy}\ and\ \citenamefont
  {Rosch}(2022)}]{PhysRevX.12.041004}%
  \BibitemOpen
  \bibfield  {author} {\bibinfo {author} {\bibnamefont {Joy}, \bibfnamefont
  {Aprem~P}}, \ and\ \bibinfo {author} {\bibfnamefont {Achim}\ \bibnamefont
  {Rosch}}} (\bibinfo {year} {2022}),\ \bibfield  {title} {\enquote {\bibinfo
  {title} {{Dynamics of Visons and Thermal Hall Effect in Perturbed Kitaev
  Models}},}\ }\href {\doibase 10.1103/PhysRevX.12.041004} {\bibfield
  {journal} {\bibinfo  {journal} {Phys. Rev. X}\ }\textbf {\bibinfo {volume}
  {12}},\ \bibinfo {pages} {041004}}\BibitemShut {NoStop}%
\bibitem [{\citenamefont {Kageyama}\ \emph {et~al.}(1999)\citenamefont
  {Kageyama}, \citenamefont {Yoshimura}, \citenamefont {Stern}, \citenamefont
  {Mushnikov}, \citenamefont {Onizuka}, \citenamefont {Kato}, \citenamefont
  {Kosuge}, \citenamefont {Slichter}, \citenamefont {Goto},\ and\ \citenamefont
  {Ueda}}]{Kageyama1999}%
  \BibitemOpen
  \bibfield  {author} {\bibinfo {author} {\bibnamefont {Kageyama},
  \bibfnamefont {H}}, \bibinfo {author} {\bibfnamefont {K.}~\bibnamefont
  {Yoshimura}}, \bibinfo {author} {\bibfnamefont {R.}~\bibnamefont {Stern}},
  \bibinfo {author} {\bibfnamefont {N.~V.}\ \bibnamefont {Mushnikov}}, \bibinfo
  {author} {\bibfnamefont {K.}~\bibnamefont {Onizuka}}, \bibinfo {author}
  {\bibfnamefont {M.}~\bibnamefont {Kato}}, \bibinfo {author} {\bibfnamefont
  {K.}~\bibnamefont {Kosuge}}, \bibinfo {author} {\bibfnamefont {C.~P.}\
  \bibnamefont {Slichter}}, \bibinfo {author} {\bibfnamefont {T.}~\bibnamefont
  {Goto}}, \ and\ \bibinfo {author} {\bibfnamefont {Y.}~\bibnamefont {Ueda}}}
  (\bibinfo {year} {1999}),\ \bibfield  {title} {\enquote {\bibinfo {title}
  {{Exact Dimer Ground State and Quantized Magnetization Plateaus in the
  Two-Dimensional Spin System
  ${\mathrm{SrCu}}_{2}({\mathrm{BO}}_{3}){}_{2}$}},}\ }\href {\doibase
  10.1103/PhysRevLett.82.3168} {\bibfield  {journal} {\bibinfo  {journal}
  {Phys. Rev. Lett.}\ }\textbf {\bibinfo {volume} {82}},\ \bibinfo {pages}
  {3168--3171}}\BibitemShut {NoStop}%
\bibitem [{\citenamefont {Kane}\ and\ \citenamefont
  {Fisher}(1997)}]{PhysRevB.55.15832}%
  \BibitemOpen
  \bibfield  {author} {\bibinfo {author} {\bibnamefont {Kane}, \bibfnamefont
  {C~L}}, \ and\ \bibinfo {author} {\bibfnamefont {Matthew P.~A.}\ \bibnamefont
  {Fisher}}} (\bibinfo {year} {1997}),\ \bibfield  {title} {\enquote {\bibinfo
  {title} {{Quantized thermal transport in the fractional quantum Hall
  effect}},}\ }\href {\doibase 10.1103/PhysRevB.55.15832} {\bibfield  {journal}
  {\bibinfo  {journal} {Phys. Rev. B}\ }\textbf {\bibinfo {volume} {55}},\
  \bibinfo {pages} {15832--15837}}\BibitemShut {NoStop}%
\bibitem [{\citenamefont {Kane}\ and\ \citenamefont {Mele}(2005)}]{Kane2005}%
  \BibitemOpen
  \bibfield  {author} {\bibinfo {author} {\bibnamefont {Kane}, \bibfnamefont
  {C~L}}, \ and\ \bibinfo {author} {\bibfnamefont {E.~J.}\ \bibnamefont
  {Mele}}} (\bibinfo {year} {2005}),\ \bibfield  {title} {\enquote {\bibinfo
  {title} {{Quantum spin Hall effect in graphene}},}\ }\href {\doibase
  10.1103/physrevlett.95.226801} {\bibfield  {journal} {\bibinfo  {journal}
  {Physical Review Letters}\ }\textbf {\bibinfo {volume} {95}}~(\bibinfo
  {number} {22}),\ \bibinfo {pages} {226801}}\BibitemShut {NoStop}%
\bibitem [{\citenamefont {Kao}\ \emph {et~al.}(2003)\citenamefont {Kao},
  \citenamefont {Enjalran}, \citenamefont {Del~Maestro}, \citenamefont
  {Molavian},\ and\ \citenamefont {Gingras}}]{PhysRevB.68.172407}%
  \BibitemOpen
  \bibfield  {author} {\bibinfo {author} {\bibnamefont {Kao}, \bibfnamefont
  {Ying-Jer}}, \bibinfo {author} {\bibfnamefont {Matthew}\ \bibnamefont
  {Enjalran}}, \bibinfo {author} {\bibfnamefont {Adrian}\ \bibnamefont
  {Del~Maestro}}, \bibinfo {author} {\bibfnamefont {Hamid~R.}\ \bibnamefont
  {Molavian}}, \ and\ \bibinfo {author} {\bibfnamefont {Michel J.~P.}\
  \bibnamefont {Gingras}}} (\bibinfo {year} {2003}),\ \bibfield  {title}
  {\enquote {\bibinfo {title} {{Understanding paramagnetic spin correlations in
  the spin-liquid pyrochlore
  ${\mathrm{Tb}}_{2}{\mathrm{Ti}}_{2}{\mathrm{O}}_{7}$}},}\ }\href {\doibase
  10.1103/PhysRevB.68.172407} {\bibfield  {journal} {\bibinfo  {journal} {Phys.
  Rev. B}\ }\textbf {\bibinfo {volume} {68}},\ \bibinfo {pages}
  {172407}}\BibitemShut {NoStop}%
\bibitem [{\citenamefont {Kasahara}\ \emph
  {et~al.}(2018{\natexlab{a}})\citenamefont {Kasahara}, \citenamefont
  {Ohnishi}, \citenamefont {Mizukami}, \citenamefont {Tanaka}, \citenamefont
  {Ma}, \citenamefont {Sugii}, \citenamefont {Kurita}, \citenamefont {Tanaka},
  \citenamefont {Nasu}, \citenamefont {Motome}, \citenamefont {Shibauchi},\
  and\ \citenamefont {Matsuda}}]{Kasahara2018}%
  \BibitemOpen
  \bibfield  {author} {\bibinfo {author} {\bibnamefont {Kasahara},
  \bibfnamefont {Y}}, \bibinfo {author} {\bibfnamefont {T.}~\bibnamefont
  {Ohnishi}}, \bibinfo {author} {\bibfnamefont {Y.}~\bibnamefont {Mizukami}},
  \bibinfo {author} {\bibfnamefont {O.}~\bibnamefont {Tanaka}}, \bibinfo
  {author} {\bibfnamefont {Sixiao}\ \bibnamefont {Ma}}, \bibinfo {author}
  {\bibfnamefont {K.}~\bibnamefont {Sugii}}, \bibinfo {author} {\bibfnamefont
  {N.}~\bibnamefont {Kurita}}, \bibinfo {author} {\bibfnamefont
  {H.}~\bibnamefont {Tanaka}}, \bibinfo {author} {\bibfnamefont
  {J.}~\bibnamefont {Nasu}}, \bibinfo {author} {\bibfnamefont {Y.}~\bibnamefont
  {Motome}}, \bibinfo {author} {\bibfnamefont {T.}~\bibnamefont {Shibauchi}}, \
  and\ \bibinfo {author} {\bibfnamefont {Y.}~\bibnamefont {Matsuda}}} (\bibinfo
  {year} {2018}{\natexlab{a}}),\ \bibfield  {title} {\enquote {\bibinfo {title}
  {{Majorana quantization and half-integer thermal quantum Hall effect in a
  Kitaev spin liquid}},}\ }\href {\doibase 10.1038/s41586-018-0274-0}
  {\bibfield  {journal} {\bibinfo  {journal} {Nature}\ }\textbf {\bibinfo
  {volume} {559}}~(\bibinfo {number} {7713}),\ \bibinfo {pages}
  {227--231}}\BibitemShut {NoStop}%
\bibitem [{\citenamefont {Kasahara}\ \emph
  {et~al.}(2018{\natexlab{b}})\citenamefont {Kasahara}, \citenamefont
  {Ohnishi}, \citenamefont {Mizukami}, \citenamefont {Tanaka}, \citenamefont
  {Ma}, \citenamefont {Sugii}, \citenamefont {Kurita}, \citenamefont {Tanaka},
  \citenamefont {Nasu}, \citenamefont {Motome}, \citenamefont {Shibauchi},\
  and\ \citenamefont {Matsuda}}]{Matsuda2018}%
  \BibitemOpen
  \bibfield  {author} {\bibinfo {author} {\bibnamefont {Kasahara},
  \bibfnamefont {Y}}, \bibinfo {author} {\bibfnamefont {T.}~\bibnamefont
  {Ohnishi}}, \bibinfo {author} {\bibfnamefont {Y.}~\bibnamefont {Mizukami}},
  \bibinfo {author} {\bibfnamefont {O.}~\bibnamefont {Tanaka}}, \bibinfo
  {author} {\bibfnamefont {Sixiao}\ \bibnamefont {Ma}}, \bibinfo {author}
  {\bibfnamefont {K.}~\bibnamefont {Sugii}}, \bibinfo {author} {\bibfnamefont
  {N.}~\bibnamefont {Kurita}}, \bibinfo {author} {\bibfnamefont
  {H.}~\bibnamefont {Tanaka}}, \bibinfo {author} {\bibfnamefont
  {J.}~\bibnamefont {Nasu}}, \bibinfo {author} {\bibfnamefont {Y.}~\bibnamefont
  {Motome}}, \bibinfo {author} {\bibfnamefont {T.}~\bibnamefont {Shibauchi}}, \
  and\ \bibinfo {author} {\bibfnamefont {Y.}~\bibnamefont {Matsuda}}} (\bibinfo
  {year} {2018}{\natexlab{b}}),\ \bibfield  {title} {\enquote {\bibinfo {title}
  {{Majorana quantization and half-integer thermal quantum Hall effect in a
  Kitaev spin liquid}},}\ }\href {\doibase 10.1038/s41586-018-0274-0}
  {\bibfield  {journal} {\bibinfo  {journal} {Nature}\ }\textbf {\bibinfo
  {volume} {559}}~(\bibinfo {number} {7713}),\ \bibinfo {pages}
  {227--231}}\BibitemShut {NoStop}%
\bibitem [{\citenamefont {Kasahara}\ \emph
  {et~al.}(2018{\natexlab{c}})\citenamefont {Kasahara}, \citenamefont {Sugii},
  \citenamefont {Ohnishi}, \citenamefont {Shimozawa}, \citenamefont
  {Yamashita}, \citenamefont {Kurita}, \citenamefont {Tanaka}, \citenamefont
  {Nasu}, \citenamefont {Motome}, \citenamefont {Shibauchi},\ and\
  \citenamefont {Matsuda}}]{Matsuda2018B}%
  \BibitemOpen
  \bibfield  {author} {\bibinfo {author} {\bibnamefont {Kasahara},
  \bibfnamefont {Y}}, \bibinfo {author} {\bibfnamefont {K.}~\bibnamefont
  {Sugii}}, \bibinfo {author} {\bibfnamefont {T.}~\bibnamefont {Ohnishi}},
  \bibinfo {author} {\bibfnamefont {M.}~\bibnamefont {Shimozawa}}, \bibinfo
  {author} {\bibfnamefont {M.}~\bibnamefont {Yamashita}}, \bibinfo {author}
  {\bibfnamefont {N.}~\bibnamefont {Kurita}}, \bibinfo {author} {\bibfnamefont
  {H.}~\bibnamefont {Tanaka}}, \bibinfo {author} {\bibfnamefont
  {J.}~\bibnamefont {Nasu}}, \bibinfo {author} {\bibfnamefont {Y.}~\bibnamefont
  {Motome}}, \bibinfo {author} {\bibfnamefont {T.}~\bibnamefont {Shibauchi}}, \
  and\ \bibinfo {author} {\bibfnamefont {Y.}~\bibnamefont {Matsuda}}} (\bibinfo
  {year} {2018}{\natexlab{c}}),\ \bibfield  {title} {\enquote {\bibinfo {title}
  {{Unusual Thermal Hall Effect in a Kitaev Spin Liquid Candidate
  {$\ensuremath{\alpha}\text{\ensuremath{-}}{\mathrm{RuCl}}_{3}$}}},}\ }\href
  {\doibase 10.1103/PhysRevLett.120.217205} {\bibfield  {journal} {\bibinfo
  {journal} {Phys. Rev. Lett.}\ }\textbf {\bibinfo {volume} {120}},\ \bibinfo
  {pages} {217205}}\BibitemShut {NoStop}%
\bibitem [{\citenamefont {Katsura}\ \emph {et~al.}(2010)\citenamefont
  {Katsura}, \citenamefont {Nagaosa},\ and\ \citenamefont {Lee}}]{Katsura2010}%
  \BibitemOpen
  \bibfield  {author} {\bibinfo {author} {\bibnamefont {Katsura}, \bibfnamefont
  {Hosho}}, \bibinfo {author} {\bibfnamefont {Naoto}\ \bibnamefont {Nagaosa}},
  \ and\ \bibinfo {author} {\bibfnamefont {Patrick~A.}\ \bibnamefont {Lee}}}
  (\bibinfo {year} {2010}),\ \bibfield  {title} {\enquote {\bibinfo {title}
  {{Theory of the thermal Hall effect in quantum magnets}},}\ }\href {\doibase
  10.1103/PhysRevLett.104.066403} {\bibfield  {journal} {\bibinfo  {journal}
  {Phys. Rev. Lett.}\ }\textbf {\bibinfo {volume} {104}},\ \bibinfo {pages}
  {066403}}\BibitemShut {NoStop}%
\bibitem [{\citenamefont {Khaliullin}(2013)}]{PhysRevLett.111.197201}%
  \BibitemOpen
  \bibfield  {author} {\bibinfo {author} {\bibnamefont {Khaliullin},
  \bibfnamefont {Giniyat}}} (\bibinfo {year} {2013}),\ \bibfield  {title}
  {\enquote {\bibinfo {title} {{Excitonic Magnetism in Van Vleck--type
  ${d}^{4}$ Mott Insulators}},}\ }\href {\doibase
  10.1103/PhysRevLett.111.197201} {\bibfield  {journal} {\bibinfo  {journal}
  {Phys. Rev. Lett.}\ }\textbf {\bibinfo {volume} {111}},\ \bibinfo {pages}
  {197201}}\BibitemShut {NoStop}%
\bibitem [{\citenamefont {Kim}\ \emph {et~al.}(2024)\citenamefont {Kim},
  \citenamefont {Saito}, \citenamefont {Yang}, \citenamefont {Ishizuka},
  \citenamefont {Coak}, \citenamefont {Lee}, \citenamefont {Sim}, \citenamefont
  {Oh}, \citenamefont {Nagaosa},\ and\ \citenamefont {Park}}]{s41467}%
  \BibitemOpen
  \bibfield  {author} {\bibinfo {author} {\bibnamefont {Kim}, \bibfnamefont
  {Ha-Leem}}, \bibinfo {author} {\bibfnamefont {Takuma}\ \bibnamefont {Saito}},
  \bibinfo {author} {\bibfnamefont {Heejun}\ \bibnamefont {Yang}}, \bibinfo
  {author} {\bibfnamefont {Hiroaki}\ \bibnamefont {Ishizuka}}, \bibinfo
  {author} {\bibfnamefont {Matthew}\ \bibnamefont {Coak}}, \bibinfo {author}
  {\bibfnamefont {Jun}\ \bibnamefont {Lee}}, \bibinfo {author} {\bibfnamefont
  {Hasung}\ \bibnamefont {Sim}}, \bibinfo {author} {\bibfnamefont {Yoon~Seok}\
  \bibnamefont {Oh}}, \bibinfo {author} {\bibfnamefont {Naoto}\ \bibnamefont
  {Nagaosa}}, \ and\ \bibinfo {author} {\bibfnamefont {Je-Geun}\ \bibnamefont
  {Park}}} (\bibinfo {year} {2024}),\ \bibfield  {title} {\enquote {\bibinfo
  {title} {{Thermal Hall effects due to topological spin fluctuations in
  YMnO$_3$}},}\ }\href {\doibase 10.1038/s41467-023-44448-9} {\bibfield
  {journal} {\bibinfo  {journal} {Nature Communications}\ }\textbf {\bibinfo
  {volume} {15}},\ \bibinfo {pages} {243}}\BibitemShut {NoStop}%
\bibitem [{\citenamefont {Kim}(2005)}]{Kim2005}%
  \BibitemOpen
  \bibfield  {author} {\bibinfo {author} {\bibnamefont {Kim}, \bibfnamefont
  {Ki-Seok}}} (\bibinfo {year} {2005}),\ \bibfield  {title} {\enquote {\bibinfo
  {title} {{Deconfinement in the presence of a Fermi surface}},}\ }\href
  {\doibase 10.1103/PhysRevB.72.245106} {\bibfield  {journal} {\bibinfo
  {journal} {Phys. Rev. B}\ }\textbf {\bibinfo {volume} {72}},\ \bibinfo
  {pages} {245106}}\BibitemShut {NoStop}%
\bibitem [{\citenamefont {Kim}\ \emph {et~al.}(2019)\citenamefont {Kim},
  \citenamefont {Nakata}, \citenamefont {Loss},\ and\ \citenamefont
  {Tserkovnyak}}]{PhysRevLett.122.057204}%
  \BibitemOpen
  \bibfield  {author} {\bibinfo {author} {\bibnamefont {Kim}, \bibfnamefont
  {Se~Kwon}}, \bibinfo {author} {\bibfnamefont {Kouki}\ \bibnamefont {Nakata}},
  \bibinfo {author} {\bibfnamefont {Daniel}\ \bibnamefont {Loss}}, \ and\
  \bibinfo {author} {\bibfnamefont {Yaroslav}\ \bibnamefont {Tserkovnyak}}}
  (\bibinfo {year} {2019}),\ \bibfield  {title} {\enquote {\bibinfo {title}
  {{Tunable Magnonic Thermal Hall Effect in Skyrmion Crystal Phases of
  Ferrimagnets}},}\ }\href {\doibase 10.1103/PhysRevLett.122.057204} {\bibfield
   {journal} {\bibinfo  {journal} {Phys. Rev. Lett.}\ }\textbf {\bibinfo
  {volume} {122}},\ \bibinfo {pages} {057204}}\BibitemShut {NoStop}%
\bibitem [{\citenamefont {Kim}\ \emph {et~al.}(2016)\citenamefont {Kim},
  \citenamefont {Ochoa}, \citenamefont {Zarzuela},\ and\ \citenamefont
  {Tserkovnyak}}]{PhysRevLett.117.227201}%
  \BibitemOpen
  \bibfield  {author} {\bibinfo {author} {\bibnamefont {Kim}, \bibfnamefont
  {Se~Kwon}}, \bibinfo {author} {\bibfnamefont {H\'ector}\ \bibnamefont
  {Ochoa}}, \bibinfo {author} {\bibfnamefont {Ricardo}\ \bibnamefont
  {Zarzuela}}, \ and\ \bibinfo {author} {\bibfnamefont {Yaroslav}\ \bibnamefont
  {Tserkovnyak}}} (\bibinfo {year} {2016}),\ \bibfield  {title} {\enquote
  {\bibinfo {title} {{Realization of the Haldane-Kane-Mele Model in a System of
  Localized Spins}},}\ }\href {\doibase 10.1103/PhysRevLett.117.227201}
  {\bibfield  {journal} {\bibinfo  {journal} {Phys. Rev. Lett.}\ }\textbf
  {\bibinfo {volume} {117}},\ \bibinfo {pages} {227201}}\BibitemShut {NoStop}%
\bibitem [{\citenamefont {Kitaev}(2003)}]{Kitaev2003}%
  \BibitemOpen
  \bibfield  {author} {\bibinfo {author} {\bibnamefont {Kitaev}, \bibfnamefont
  {A~Yu}}} (\bibinfo {year} {2003}),\ \bibfield  {title} {\enquote {\bibinfo
  {title} {Fault-tolerant quantum computation by anyons},}\ }\href {\doibase
  10.1016/s0003-4916(02)00018-0} {\bibfield  {journal} {\bibinfo  {journal}
  {Annals of Physics}\ }\textbf {\bibinfo {volume} {303}}~(\bibinfo {number}
  {1}),\ \bibinfo {pages} {2--30}}\BibitemShut {NoStop}%
\bibitem [{\citenamefont {Kitaev}(2006)}]{Kitaev2006}%
  \BibitemOpen
  \bibfield  {author} {\bibinfo {author} {\bibnamefont {Kitaev}, \bibfnamefont
  {Alexei}}} (\bibinfo {year} {2006}),\ \bibfield  {title} {\enquote {\bibinfo
  {title} {Anyons in an exactly solved model and beyond},}\ }\href {\doibase
  10.1016/j.aop.2005.10.005} {\bibfield  {journal} {\bibinfo  {journal} {Annals
  of Physics}\ }\textbf {\bibinfo {volume} {321}}~(\bibinfo {number} {1}),\
  \bibinfo {pages} {2--111}}\BibitemShut {NoStop}%
\bibitem [{\citenamefont {Knolle}\ and\ \citenamefont
  {Moessner}(2019)}]{Knolle2019}%
  \BibitemOpen
  \bibfield  {author} {\bibinfo {author} {\bibnamefont {Knolle}, \bibfnamefont
  {J}}, \ and\ \bibinfo {author} {\bibfnamefont {R.}~\bibnamefont {Moessner}}}
  (\bibinfo {year} {2019}),\ \bibfield  {title} {\enquote {\bibinfo {title} {A
  field guide to spin liquids},}\ }\href {\doibase
  10.1146/annurev-conmatphys-031218-013401} {\bibfield  {journal} {\bibinfo
  {journal} {Annual Review of Condensed Matter Physics}\ }\textbf {\bibinfo
  {volume} {10}}~(\bibinfo {number} {1}),\ \bibinfo {pages}
  {451--472}}\BibitemShut {NoStop}%
\bibitem [{\citenamefont {{Kosuke Fujiwara, Sota Kitamura, Takahiro
  Morimoto}}(2022)}]{Morimoto2022}%
  \BibitemOpen
  \bibfield  {author} {\bibinfo {author} {\bibnamefont {{Kosuke Fujiwara, Sota
  Kitamura, Takahiro Morimoto}},}} (\bibinfo {year} {2022}),\ \href@noop {}
  {\enquote {\bibinfo {title} {{{Thermal Hall responses in frustrated honeycomb
  spin systems}}},}\ }\Eprint {http://arxiv.org/abs/2203.16853}
  {arXiv:2203.16853} \BibitemShut {NoStop}%
\bibitem [{\citenamefont {Kurosaki}\ \emph {et~al.}(2005)\citenamefont
  {Kurosaki}, \citenamefont {Shimizu}, \citenamefont {Miyagawa}, \citenamefont
  {Kanoda},\ and\ \citenamefont {Saito}}]{Saito2005}%
  \BibitemOpen
  \bibfield  {author} {\bibinfo {author} {\bibnamefont {Kurosaki},
  \bibfnamefont {Y}}, \bibinfo {author} {\bibfnamefont {Y.}~\bibnamefont
  {Shimizu}}, \bibinfo {author} {\bibfnamefont {K.}~\bibnamefont {Miyagawa}},
  \bibinfo {author} {\bibfnamefont {K.}~\bibnamefont {Kanoda}}, \ and\ \bibinfo
  {author} {\bibfnamefont {G.}~\bibnamefont {Saito}}} (\bibinfo {year}
  {2005}),\ \bibfield  {title} {\enquote {\bibinfo {title} {{Mott transition
  from a spin liquid to a Fermi liquid in the spin-frustrated organic conductor
  $\ensuremath{\kappa}\mathrm{\text{\ensuremath{-}}}(\mathrm{ET}{)}_{2}{\mathrm{Cu}}_{2}(\mathrm{CN}{)}_{3}$}},}\
  }\href {\doibase 10.1103/PhysRevLett.95.177001} {\bibfield  {journal}
  {\bibinfo  {journal} {Phys. Rev. Lett.}\ }\textbf {\bibinfo {volume} {95}},\
  \bibinfo {pages} {177001}}\BibitemShut {NoStop}%
\bibitem [{\citenamefont {Lacroix}\ \emph {et~al.}(2011)\citenamefont
  {Lacroix}, \citenamefont {Mendels},\ and\ \citenamefont
  {Mila}}]{lacroix2011introduction}%
  \BibitemOpen
  \bibfield  {author} {\bibinfo {author} {\bibnamefont {Lacroix}, \bibfnamefont
  {C}}, \bibinfo {author} {\bibfnamefont {P.}~\bibnamefont {Mendels}}, \ and\
  \bibinfo {author} {\bibfnamefont {F.}~\bibnamefont {Mila}}} (\bibinfo {year}
  {2011}),\ \href {https://books.google.com.hk/books?id=utSV09ZuhOkC} {\emph
  {\bibinfo {title} {{Introduction to Frustrated Magnetism: Materials,
  Experiments, Theory}}}},\ Springer Series in Solid-State Sciences\ (\bibinfo
  {publisher} {Springer Berlin Heidelberg})\BibitemShut {NoStop}%
\bibitem [{\citenamefont {Larsen}\ \emph {et~al.}(2019)\citenamefont {Larsen},
  \citenamefont {R\o{}mer}, \citenamefont {Janas}, \citenamefont {Treue},
  \citenamefont {M\o{}nsted}, \citenamefont {Shaik}, \citenamefont
  {R\o{}nnow},\ and\ \citenamefont {Lefmann}}]{PhysRevB.99.054432}%
  \BibitemOpen
  \bibfield  {author} {\bibinfo {author} {\bibnamefont {Larsen}, \bibfnamefont
  {C~B}}, \bibinfo {author} {\bibfnamefont {A.~T.}\ \bibnamefont {R\o{}mer}},
  \bibinfo {author} {\bibfnamefont {S.}~\bibnamefont {Janas}}, \bibinfo
  {author} {\bibfnamefont {F.}~\bibnamefont {Treue}}, \bibinfo {author}
  {\bibfnamefont {B.}~\bibnamefont {M\o{}nsted}}, \bibinfo {author}
  {\bibfnamefont {N.~E.}\ \bibnamefont {Shaik}}, \bibinfo {author}
  {\bibfnamefont {H.~M.}\ \bibnamefont {R\o{}nnow}}, \ and\ \bibinfo {author}
  {\bibfnamefont {K.}~\bibnamefont {Lefmann}}} (\bibinfo {year} {2019}),\
  \bibfield  {title} {\enquote {\bibinfo {title} {{Exact diagonalization study
  of the Hubbard-parametrized four-spin ring exchange model on a square
  lattice}},}\ }\href {\doibase 10.1103/PhysRevB.99.054432} {\bibfield
  {journal} {\bibinfo  {journal} {Phys. Rev. B}\ }\textbf {\bibinfo {volume}
  {99}},\ \bibinfo {pages} {054432}}\BibitemShut {NoStop}%
\bibitem [{\citenamefont {Laurell}\ and\ \citenamefont
  {Fiete}(2017)}]{PhysRevLett.118.177201}%
  \BibitemOpen
  \bibfield  {author} {\bibinfo {author} {\bibnamefont {Laurell}, \bibfnamefont
  {Pontus}}, \ and\ \bibinfo {author} {\bibfnamefont {Gregory~A.}\ \bibnamefont
  {Fiete}}} (\bibinfo {year} {2017}),\ \bibfield  {title} {\enquote {\bibinfo
  {title} {Topological magnon bands and unconventional superconductivity in
  pyrochlore iridate thin films},}\ }\href {\doibase
  10.1103/PhysRevLett.118.177201} {\bibfield  {journal} {\bibinfo  {journal}
  {Phys. Rev. Lett.}\ }\textbf {\bibinfo {volume} {118}},\ \bibinfo {pages}
  {177201}}\BibitemShut {NoStop}%
\bibitem [{\citenamefont {Laurell}\ and\ \citenamefont
  {Fiete}(2018)}]{PhysRevB.98.094419}%
  \BibitemOpen
  \bibfield  {author} {\bibinfo {author} {\bibnamefont {Laurell}, \bibfnamefont
  {Pontus}}, \ and\ \bibinfo {author} {\bibfnamefont {Gregory~A.}\ \bibnamefont
  {Fiete}}} (\bibinfo {year} {2018}),\ \bibfield  {title} {\enquote {\bibinfo
  {title} {{Magnon thermal Hall effect in kagome antiferromagnets with
  Dzyaloshinskii-Moriya interactions}},}\ }\href {\doibase
  10.1103/PhysRevB.98.094419} {\bibfield  {journal} {\bibinfo  {journal} {Phys.
  Rev. B}\ }\textbf {\bibinfo {volume} {98}},\ \bibinfo {pages}
  {094419}}\BibitemShut {NoStop}%
\bibitem [{\citenamefont {Leahy}\ \emph {et~al.}(2017)\citenamefont {Leahy},
  \citenamefont {Pocs}, \citenamefont {Siegfried}, \citenamefont {Graf},
  \citenamefont {Do}, \citenamefont {Choi}, \citenamefont {Normand},\ and\
  \citenamefont {Lee}}]{Minhyea2017}%
  \BibitemOpen
  \bibfield  {author} {\bibinfo {author} {\bibnamefont {Leahy}, \bibfnamefont
  {Ian~A}}, \bibinfo {author} {\bibfnamefont {Christopher~A.}\ \bibnamefont
  {Pocs}}, \bibinfo {author} {\bibfnamefont {Peter~E.}\ \bibnamefont
  {Siegfried}}, \bibinfo {author} {\bibfnamefont {David}\ \bibnamefont {Graf}},
  \bibinfo {author} {\bibfnamefont {S.-H.}\ \bibnamefont {Do}}, \bibinfo
  {author} {\bibfnamefont {Kwang-Yong}\ \bibnamefont {Choi}}, \bibinfo {author}
  {\bibfnamefont {B.}~\bibnamefont {Normand}}, \ and\ \bibinfo {author}
  {\bibfnamefont {Minhyea}\ \bibnamefont {Lee}}} (\bibinfo {year} {2017}),\
  \bibfield  {title} {\enquote {\bibinfo {title} {{Anomalous Thermal
  Conductivity and Magnetic Torque Response in the Honeycomb Magnet
  {$\ensuremath{\alpha}\text{\ensuremath{-}}{\mathrm{RuCl}}_{3}$}}},}\ }\href
  {\doibase 10.1103/PhysRevLett.118.187203} {\bibfield  {journal} {\bibinfo
  {journal} {Phys. Rev. Lett.}\ }\textbf {\bibinfo {volume} {118}},\ \bibinfo
  {pages} {187203}}\BibitemShut {NoStop}%
\bibitem [{\citenamefont {Lee}\ \emph {et~al.}(2015)\citenamefont {Lee},
  \citenamefont {Han},\ and\ \citenamefont {Lee}}]{Lee2015}%
  \BibitemOpen
  \bibfield  {author} {\bibinfo {author} {\bibnamefont {Lee}, \bibfnamefont
  {Hyunyong}}, \bibinfo {author} {\bibfnamefont {Jung~Hoon}\ \bibnamefont
  {Han}}, \ and\ \bibinfo {author} {\bibfnamefont {Patrick~A.}\ \bibnamefont
  {Lee}}} (\bibinfo {year} {2015}),\ \bibfield  {title} {\enquote {\bibinfo
  {title} {{Thermal Hall effect of spins in a paramagnet}},}\ }\href {\doibase
  10.1103/PhysRevB.91.125413} {\bibfield  {journal} {\bibinfo  {journal} {Phys.
  Rev. B}\ }\textbf {\bibinfo {volume} {91}},\ \bibinfo {pages}
  {125413}}\BibitemShut {NoStop}%
\bibitem [{\citenamefont {Lee}(2008{\natexlab{a}})}]{Lee2008}%
  \BibitemOpen
  \bibfield  {author} {\bibinfo {author} {\bibnamefont {Lee}, \bibfnamefont
  {P~A}}} (\bibinfo {year} {2008}{\natexlab{a}}),\ \bibfield  {title} {\enquote
  {\bibinfo {title} {{An End to the Drought of Quantum Spin Liquids}},}\ }\href
  {\doibase 10.1126/science.1163196} {\bibfield  {journal} {\bibinfo  {journal}
  {Science}\ }\textbf {\bibinfo {volume} {321}}~(\bibinfo {number} {5894}),\
  \bibinfo {pages} {1306--1307}}\BibitemShut {NoStop}%
\bibitem [{\citenamefont {Lee}\ and\ \citenamefont
  {Nagaosa}(1992)}]{LeeNagaosa1992}%
  \BibitemOpen
  \bibfield  {author} {\bibinfo {author} {\bibnamefont {Lee}, \bibfnamefont
  {Patrick~A}}, \ and\ \bibinfo {author} {\bibfnamefont {Naoto}\ \bibnamefont
  {Nagaosa}}} (\bibinfo {year} {1992}),\ \bibfield  {title} {\enquote {\bibinfo
  {title} {{Gauge theory of the normal state of
  high-${\mathit{T}}_{\mathit{c}}$ superconductors}},}\ }\href {\doibase
  10.1103/PhysRevB.46.5621} {\bibfield  {journal} {\bibinfo  {journal} {Phys.
  Rev. B}\ }\textbf {\bibinfo {volume} {46}},\ \bibinfo {pages}
  {5621--5639}}\BibitemShut {NoStop}%
\bibitem [{\citenamefont {Lee}\ and\ \citenamefont
  {Nagaosa}(2013)}]{LeeNagaosa2013}%
  \BibitemOpen
  \bibfield  {author} {\bibinfo {author} {\bibnamefont {Lee}, \bibfnamefont
  {Patrick~A}}, \ and\ \bibinfo {author} {\bibfnamefont {Naoto}\ \bibnamefont
  {Nagaosa}}} (\bibinfo {year} {2013}),\ \bibfield  {title} {\enquote {\bibinfo
  {title} {{Proposal to use neutron scattering to access scalar spin chirality
  fluctuations in kagom{\'{e}} lattices}},}\ }\href {\doibase
  10.1103/PhysRevB.87.064423} {\bibfield  {journal} {\bibinfo  {journal} {Phys.
  Rev. B}\ }\textbf {\bibinfo {volume} {87}},\ \bibinfo {pages}
  {064423}}\BibitemShut {NoStop}%
\bibitem [{\citenamefont {Lee}(2008{\natexlab{b}})}]{SSLee2008}%
  \BibitemOpen
  \bibfield  {author} {\bibinfo {author} {\bibnamefont {Lee}, \bibfnamefont
  {Sung-Sik}}} (\bibinfo {year} {2008}{\natexlab{b}}),\ \bibfield  {title}
  {\enquote {\bibinfo {title} {{Stability of the {U(1)} spin liquid with a
  spinon Fermi surface in $2+1$ dimensions}},}\ }\href {\doibase
  10.1103/PhysRevB.78.085129} {\bibfield  {journal} {\bibinfo  {journal} {Phys.
  Rev. B}\ }\textbf {\bibinfo {volume} {78}},\ \bibinfo {pages}
  {085129}}\BibitemShut {NoStop}%
\bibitem [{\citenamefont {Lee}\ and\ \citenamefont {Lee}(2005)}]{Lee2005}%
  \BibitemOpen
  \bibfield  {author} {\bibinfo {author} {\bibnamefont {Lee}, \bibfnamefont
  {Sung-Sik}}, \ and\ \bibinfo {author} {\bibfnamefont {Patrick~A.}\
  \bibnamefont {Lee}}} (\bibinfo {year} {2005}),\ \bibfield  {title} {\enquote
  {\bibinfo {title} {{U(1) gauge theory of the Hubbard model: Spin liquid
  states and possible application to
  $\ensuremath{\kappa}\mathrm{\text{\ensuremath{-}}}(\mathrm{BEDT}\mathrm{\text{\ensuremath{-}}}\mathrm{TTF}{)}_{2}{\mathrm{Cu}}_{2}(\mathrm{CN}{)}_{3}$}},}\
  }\href {\doibase 10.1103/PhysRevLett.95.036403} {\bibfield  {journal}
  {\bibinfo  {journal} {Phys. Rev. Lett.}\ }\textbf {\bibinfo {volume} {95}},\
  \bibinfo {pages} {036403}}\BibitemShut {NoStop}%
\bibitem [{\citenamefont {Lee}\ \emph {et~al.}(2012)\citenamefont {Lee},
  \citenamefont {Onoda},\ and\ \citenamefont {Balents}}]{Balents2012}%
  \BibitemOpen
  \bibfield  {author} {\bibinfo {author} {\bibnamefont {Lee}, \bibfnamefont
  {SungBin}}, \bibinfo {author} {\bibfnamefont {Shigeki}\ \bibnamefont
  {Onoda}}, \ and\ \bibinfo {author} {\bibfnamefont {Leon}\ \bibnamefont
  {Balents}}} (\bibinfo {year} {2012}),\ \bibfield  {title} {\enquote {\bibinfo
  {title} {Generic quantum spin ice},}\ }\href {\doibase
  10.1103/PhysRevB.86.104412} {\bibfield  {journal} {\bibinfo  {journal} {Phys.
  Rev. B}\ }\textbf {\bibinfo {volume} {86}},\ \bibinfo {pages}
  {104412}}\BibitemShut {NoStop}%
\bibitem [{\citenamefont {Li}\ and\ \citenamefont {Chen}(2022)}]{Li_2022}%
  \BibitemOpen
  \bibfield  {author} {\bibinfo {author} {\bibnamefont {Li}, \bibfnamefont
  {Chao-Kai}}, \ and\ \bibinfo {author} {\bibfnamefont {Gang}\ \bibnamefont
  {Chen}}} (\bibinfo {year} {2022}),\ \bibfield  {title} {\enquote {\bibinfo
  {title} {Universal excitonic superexchange in spin-orbit-coupled mott
  insulators},}\ }\href {\doibase 10.1209/0295-5075/ac8612} {\bibfield
  {journal} {\bibinfo  {journal} {Europhysics Letters}\ }\textbf {\bibinfo
  {volume} {139}}~(\bibinfo {number} {5}),\ \bibinfo {pages}
  {56001}}\BibitemShut {NoStop}%
\bibitem [{\citenamefont {Li}\ \emph {et~al.}(2021)\citenamefont {Li},
  \citenamefont {Yao},\ and\ \citenamefont {Chen}}]{PhysRevResearch.3.033156}%
  \BibitemOpen
  \bibfield  {author} {\bibinfo {author} {\bibnamefont {Li}, \bibfnamefont
  {Chao-Kai}}, \bibinfo {author} {\bibfnamefont {Xu-Ping}\ \bibnamefont {Yao}},
  \ and\ \bibinfo {author} {\bibfnamefont {Gang}\ \bibnamefont {Chen}}}
  (\bibinfo {year} {2021}),\ \bibfield  {title} {\enquote {\bibinfo {title}
  {Twisted magnetic topological insulators},}\ }\href {\doibase
  10.1103/PhysRevResearch.3.033156} {\bibfield  {journal} {\bibinfo  {journal}
  {Phys. Rev. Res.}\ }\textbf {\bibinfo {volume} {3}},\ \bibinfo {pages}
  {033156}}\BibitemShut {NoStop}%
\bibitem [{\citenamefont {Li}\ \emph {et~al.}(2022)\citenamefont {Li},
  \citenamefont {Yao}, \citenamefont {Liu},\ and\ \citenamefont
  {Chen}}]{PhysRevLett.129.017202}%
  \BibitemOpen
  \bibfield  {author} {\bibinfo {author} {\bibnamefont {Li}, \bibfnamefont
  {Chao-Kai}}, \bibinfo {author} {\bibfnamefont {Xu-Ping}\ \bibnamefont {Yao}},
  \bibinfo {author} {\bibfnamefont {Jianpeng}\ \bibnamefont {Liu}}, \ and\
  \bibinfo {author} {\bibfnamefont {Gang}\ \bibnamefont {Chen}}} (\bibinfo
  {year} {2022}),\ \bibfield  {title} {\enquote {\bibinfo {title}
  {{Fractionalization on the Surface: Is Type-II Terminated
  $1T\text{\ensuremath{-}}{\mathrm{TaS}}_{2}$ Surface an Anomalously Realized
  Spin Liquid?}}}\ }\href {\doibase 10.1103/PhysRevLett.129.017202} {\bibfield
  {journal} {\bibinfo  {journal} {Phys. Rev. Lett.}\ }\textbf {\bibinfo
  {volume} {129}},\ \bibinfo {pages} {017202}}\BibitemShut {NoStop}%
\bibitem [{\citenamefont {Li}\ and\ \citenamefont
  {Chen}(2018)}]{PhysRevB.98.045109}%
  \BibitemOpen
  \bibfield  {author} {\bibinfo {author} {\bibnamefont {Li}, \bibfnamefont
  {Fei-Ye}}, \ and\ \bibinfo {author} {\bibfnamefont {Gang}\ \bibnamefont
  {Chen}}} (\bibinfo {year} {2018}),\ \bibfield  {title} {\enquote {\bibinfo
  {title} {Competing phases and topological excitations of spin-1 pyrochlore
  antiferromagnets},}\ }\href {\doibase 10.1103/PhysRevB.98.045109} {\bibfield
  {journal} {\bibinfo  {journal} {Phys. Rev. B}\ }\textbf {\bibinfo {volume}
  {98}},\ \bibinfo {pages} {045109}}\BibitemShut {NoStop}%
\bibitem [{\citenamefont {Li}\ and\ \citenamefont
  {Chen}(2019)}]{PhysRevB.100.045103}%
  \BibitemOpen
  \bibfield  {author} {\bibinfo {author} {\bibnamefont {Li}, \bibfnamefont
  {Fei-Ye}}, \ and\ \bibinfo {author} {\bibfnamefont {Gang}\ \bibnamefont
  {Chen}}} (\bibinfo {year} {2019}),\ \bibfield  {title} {\enquote {\bibinfo
  {title} {{Spin-orbital entanglement in ${d}^{8}$ Mott insulators: Possible
  excitonic magnetism in diamond lattice antiferromagnets}},}\ }\href {\doibase
  10.1103/PhysRevB.100.045103} {\bibfield  {journal} {\bibinfo  {journal}
  {Phys. Rev. B}\ }\textbf {\bibinfo {volume} {100}},\ \bibinfo {pages}
  {045103}}\BibitemShut {NoStop}%
\bibitem [{\citenamefont {Li}\ \emph {et~al.}(2017{\natexlab{a}})\citenamefont
  {Li}, \citenamefont {Li}, \citenamefont {Yu}, \citenamefont {Paramekanti},\
  and\ \citenamefont {Chen}}]{Feiye2017}%
  \BibitemOpen
  \bibfield  {author} {\bibinfo {author} {\bibnamefont {Li}, \bibfnamefont
  {Fei-Ye}}, \bibinfo {author} {\bibfnamefont {Yao-Dong}\ \bibnamefont {Li}},
  \bibinfo {author} {\bibfnamefont {Yue}\ \bibnamefont {Yu}}, \bibinfo {author}
  {\bibfnamefont {Arun}\ \bibnamefont {Paramekanti}}, \ and\ \bibinfo {author}
  {\bibfnamefont {Gang}\ \bibnamefont {Chen}}} (\bibinfo {year}
  {2017}{\natexlab{a}}),\ \bibfield  {title} {\enquote {\bibinfo {title}
  {{Kitaev materials beyond iridates: Order by quantum disorder and Weyl
  magnons in rare-earth double perovskites}},}\ }\href {\doibase
  10.1103/PhysRevB.95.085132} {\bibfield  {journal} {\bibinfo  {journal} {Phys.
  Rev. B}\ }\textbf {\bibinfo {volume} {95}},\ \bibinfo {pages}
  {085132}}\BibitemShut {NoStop}%
\bibitem [{\citenamefont {Li}\ \emph {et~al.}(2020)\citenamefont {Li},
  \citenamefont {Fauqu\'e}, \citenamefont {Zhu},\ and\ \citenamefont
  {Behnia}}]{PhysRevLett.124.105901}%
  \BibitemOpen
  \bibfield  {author} {\bibinfo {author} {\bibnamefont {Li}, \bibfnamefont
  {Xiaokang}}, \bibinfo {author} {\bibfnamefont {Beno\^{\i}t}\ \bibnamefont
  {Fauqu\'e}}, \bibinfo {author} {\bibfnamefont {Zengwei}\ \bibnamefont {Zhu}},
  \ and\ \bibinfo {author} {\bibfnamefont {Kamran}\ \bibnamefont {Behnia}}}
  (\bibinfo {year} {2020}),\ \bibfield  {title} {\enquote {\bibinfo {title}
  {{Phonon Thermal Hall Effect in Strontium Titanate}},}\ }\href {\doibase
  10.1103/PhysRevLett.124.105901} {\bibfield  {journal} {\bibinfo  {journal}
  {Phys. Rev. Lett.}\ }\textbf {\bibinfo {volume} {124}},\ \bibinfo {pages}
  {105901}}\BibitemShut {NoStop}%
\bibitem [{\citenamefont {Li}\ and\ \citenamefont
  {Chen}(2017{\natexlab{a}})}]{Gang2017B}%
  \BibitemOpen
  \bibfield  {author} {\bibinfo {author} {\bibnamefont {Li}, \bibfnamefont
  {Yao-Dong}}, \ and\ \bibinfo {author} {\bibfnamefont {Gang}\ \bibnamefont
  {Chen}}} (\bibinfo {year} {2017}{\natexlab{a}}),\ \bibfield  {title}
  {\enquote {\bibinfo {title} {{Detecting spin fractionalization in a spinon
  Fermi surface spin liquid}},}\ }\href {\doibase 10.1103/PhysRevB.96.075105}
  {\bibfield  {journal} {\bibinfo  {journal} {Phys. Rev. B}\ }\textbf {\bibinfo
  {volume} {96}},\ \bibinfo {pages} {075105}}\BibitemShut {NoStop}%
\bibitem [{\citenamefont {Li}\ and\ \citenamefont
  {Chen}(2017{\natexlab{b}})}]{Gang2017C}%
  \BibitemOpen
  \bibfield  {author} {\bibinfo {author} {\bibnamefont {Li}, \bibfnamefont
  {Yao-Dong}}, \ and\ \bibinfo {author} {\bibfnamefont {Gang}\ \bibnamefont
  {Chen}}} (\bibinfo {year} {2017}{\natexlab{b}}),\ \bibfield  {title}
  {\enquote {\bibinfo {title} {{Symmetry enriched {U(1)} topological orders for
  dipole-octupole doublets on a pyrochlore lattice}},}\ }\href {\doibase
  10.1103/PhysRevB.95.041106} {\bibfield  {journal} {\bibinfo  {journal} {Phys.
  Rev. B}\ }\textbf {\bibinfo {volume} {95}},\ \bibinfo {pages}
  {041106}}\BibitemShut {NoStop}%
\bibitem [{\citenamefont {Li}\ \emph {et~al.}(2017{\natexlab{b}})\citenamefont
  {Li}, \citenamefont {Lu},\ and\ \citenamefont {Chen}}]{Gang2017}%
  \BibitemOpen
  \bibfield  {author} {\bibinfo {author} {\bibnamefont {Li}, \bibfnamefont
  {Yao-Dong}}, \bibinfo {author} {\bibfnamefont {Yuan-Ming}\ \bibnamefont
  {Lu}}, \ and\ \bibinfo {author} {\bibfnamefont {Gang}\ \bibnamefont {Chen}}}
  (\bibinfo {year} {2017}{\natexlab{b}}),\ \bibfield  {title} {\enquote
  {\bibinfo {title} {{Spinon Fermi surface U(1) spin liquid in the
  spin-orbit-coupled triangular-lattice Mott insulator
  ${\mathrm{YbMgGaO}}_{4}$}},}\ }\href {\doibase 10.1103/PhysRevB.96.054445}
  {\bibfield  {journal} {\bibinfo  {journal} {Phys. Rev. B}\ }\textbf {\bibinfo
  {volume} {96}},\ \bibinfo {pages} {054445}}\BibitemShut {NoStop}%
\bibitem [{\citenamefont {Li}\ \emph {et~al.}(2016)\citenamefont {Li},
  \citenamefont {Wang},\ and\ \citenamefont {Chen}}]{Gang2016B}%
  \BibitemOpen
  \bibfield  {author} {\bibinfo {author} {\bibnamefont {Li}, \bibfnamefont
  {Yao-Dong}}, \bibinfo {author} {\bibfnamefont {Xiaoqun}\ \bibnamefont
  {Wang}}, \ and\ \bibinfo {author} {\bibfnamefont {Gang}\ \bibnamefont
  {Chen}}} (\bibinfo {year} {2016}),\ \bibfield  {title} {\enquote {\bibinfo
  {title} {Hidden multipolar orders of dipole-octupole doublets on a triangular
  lattice},}\ }\href {\doibase 10.1103/PhysRevB.94.201114} {\bibfield
  {journal} {\bibinfo  {journal} {Phys. Rev. B}\ }\textbf {\bibinfo {volume}
  {94}},\ \bibinfo {pages} {201114}}\BibitemShut {NoStop}%
\bibitem [{\citenamefont {Li}\ \emph {et~al.}(2019)\citenamefont {Li},
  \citenamefont {Yang}, \citenamefont {Zhou},\ and\ \citenamefont
  {Chen}}]{Gang2019}%
  \BibitemOpen
  \bibfield  {author} {\bibinfo {author} {\bibnamefont {Li}, \bibfnamefont
  {Yao-Dong}}, \bibinfo {author} {\bibfnamefont {Xu}~\bibnamefont {Yang}},
  \bibinfo {author} {\bibfnamefont {Yi}~\bibnamefont {Zhou}}, \ and\ \bibinfo
  {author} {\bibfnamefont {Gang}\ \bibnamefont {Chen}}} (\bibinfo {year}
  {2019}),\ \bibfield  {title} {\enquote {\bibinfo {title} {{Non-Kitaev spin
  liquids in Kitaev materials}},}\ }\href {\doibase 10.1103/PhysRevB.99.205119}
  {\bibfield  {journal} {\bibinfo  {journal} {Phys. Rev. B}\ }\textbf {\bibinfo
  {volume} {99}},\ \bibinfo {pages} {205119}}\BibitemShut {NoStop}%
\bibitem [{\citenamefont {Liu}\ \emph {et~al.}(2020{\natexlab{a}})\citenamefont
  {Liu}, \citenamefont {Huang},\ and\ \citenamefont
  {Chen}}]{PhysRevResearch.2.043013}%
  \BibitemOpen
  \bibfield  {author} {\bibinfo {author} {\bibnamefont {Liu}, \bibfnamefont
  {Changle}}, \bibinfo {author} {\bibfnamefont {Chun-Jiong}\ \bibnamefont
  {Huang}}, \ and\ \bibinfo {author} {\bibfnamefont {Gang}\ \bibnamefont
  {Chen}}} (\bibinfo {year} {2020}{\natexlab{a}}),\ \bibfield  {title}
  {\enquote {\bibinfo {title} {{Intrinsic quantum Ising model on a triangular
  lattice magnet $\mathrm{Tm}\mathrm{Mg}\mathrm{Ga}{\mathrm{O}}_{4}$}},}\
  }\href {\doibase 10.1103/PhysRevResearch.2.043013} {\bibfield  {journal}
  {\bibinfo  {journal} {Phys. Rev. Res.}\ }\textbf {\bibinfo {volume} {2}},\
  \bibinfo {pages} {043013}}\BibitemShut {NoStop}%
\bibitem [{\citenamefont {Liu}\ \emph {et~al.}(2019)\citenamefont {Liu},
  \citenamefont {Li},\ and\ \citenamefont {Chen}}]{PhysRevB.99.224407}%
  \BibitemOpen
  \bibfield  {author} {\bibinfo {author} {\bibnamefont {Liu}, \bibfnamefont
  {Changle}}, \bibinfo {author} {\bibfnamefont {Fei-Ye}\ \bibnamefont {Li}}, \
  and\ \bibinfo {author} {\bibfnamefont {Gang}\ \bibnamefont {Chen}}} (\bibinfo
  {year} {2019}),\ \bibfield  {title} {\enquote {\bibinfo {title} {Upper branch
  magnetism in quantum magnets: Collapses of excited levels and emergent
  selection rules},}\ }\href {\doibase 10.1103/PhysRevB.99.224407} {\bibfield
  {journal} {\bibinfo  {journal} {Phys. Rev. B}\ }\textbf {\bibinfo {volume}
  {99}},\ \bibinfo {pages} {224407}}\BibitemShut {NoStop}%
\bibitem [{\citenamefont {Liu}\ and\ \citenamefont
  {Khaliullin}(2018)}]{PhysRevB.97.014407}%
  \BibitemOpen
  \bibfield  {author} {\bibinfo {author} {\bibnamefont {Liu}, \bibfnamefont
  {Huimei}}, \ and\ \bibinfo {author} {\bibfnamefont {Giniyat}\ \bibnamefont
  {Khaliullin}}} (\bibinfo {year} {2018}),\ \bibfield  {title} {\enquote
  {\bibinfo {title} {{Pseudospin exchange interactions in ${d}^{7}$ cobalt
  compounds: Possible realization of the Kitaev model}},}\ }\href {\doibase
  10.1103/PhysRevB.97.014407} {\bibfield  {journal} {\bibinfo  {journal} {Phys.
  Rev. B}\ }\textbf {\bibinfo {volume} {97}},\ \bibinfo {pages}
  {014407}}\BibitemShut {NoStop}%
\bibitem [{\citenamefont {Liu}\ \emph {et~al.}(2020{\natexlab{b}})\citenamefont
  {Liu}, \citenamefont {Li}, \citenamefont {Chen},\ and\ \citenamefont
  {Wang}}]{PhysRevResearch.2.033260}%
  \BibitemOpen
  \bibfield  {author} {\bibinfo {author} {\bibnamefont {Liu}, \bibfnamefont
  {Jian~Qiao}}, \bibinfo {author} {\bibfnamefont {Fei-Ye}\ \bibnamefont {Li}},
  \bibinfo {author} {\bibfnamefont {Gang}\ \bibnamefont {Chen}}, \ and\
  \bibinfo {author} {\bibfnamefont {Ziqiang}\ \bibnamefont {Wang}}} (\bibinfo
  {year} {2020}{\natexlab{b}}),\ \bibfield  {title} {\enquote {\bibinfo {title}
  {Featureless quantum paramagnet with frustrated criticality and competing
  spiral magnetism on spin-1 honeycomb lattice magnet},}\ }\href {\doibase
  10.1103/PhysRevResearch.2.033260} {\bibfield  {journal} {\bibinfo  {journal}
  {Phys. Rev. Res.}\ }\textbf {\bibinfo {volume} {2}},\ \bibinfo {pages}
  {033260}}\BibitemShut {NoStop}%
\bibitem [{\citenamefont {Liu}\ \emph {et~al.}(2020{\natexlab{c}})\citenamefont
  {Liu}, \citenamefont {Quan}, \citenamefont {Lin},\ and\ \citenamefont
  {Zou}}]{Liu2020}%
  \BibitemOpen
  \bibfield  {author} {\bibinfo {author} {\bibnamefont {Liu}, \bibfnamefont
  {Jing}}, \bibinfo {author} {\bibfnamefont {Ya-Min}\ \bibnamefont {Quan}},
  \bibinfo {author} {\bibfnamefont {H.Q.}\ \bibnamefont {Lin}}, \ and\ \bibinfo
  {author} {\bibfnamefont {Liang-Jian}\ \bibnamefont {Zou}}} (\bibinfo {year}
  {2020}{\natexlab{c}}),\ \bibfield  {title} {\enquote {\bibinfo {title}
  {{Topologically different spin disorder phases of the $J_1$-$J_2$ Heisenberg
  model on the honeycomb lattice}},}\ }\href {\doibase
  10.1016/j.physe.2020.114037} {\bibfield  {journal} {\bibinfo  {journal}
  {Physica E: Low-dimensional Systems and Nanostructures}\ }\textbf {\bibinfo
  {volume} {120}},\ \bibinfo {pages} {114037}}\BibitemShut {NoStop}%
\bibitem [{\citenamefont {Liu}\ and\ \citenamefont {Normand}(2018)}]{Liu2018}%
  \BibitemOpen
  \bibfield  {author} {\bibinfo {author} {\bibnamefont {Liu}, \bibfnamefont
  {Zheng-Xin}}, \ and\ \bibinfo {author} {\bibfnamefont {B.}~\bibnamefont
  {Normand}}} (\bibinfo {year} {2018}),\ \bibfield  {title} {\enquote {\bibinfo
  {title} {Dirac and chiral quantum spin liquids on the honeycomb lattice in a
  magnetic field},}\ }\href {\doibase 10.1103/PhysRevLett.120.187201}
  {\bibfield  {journal} {\bibinfo  {journal} {Phys. Rev. Lett.}\ }\textbf
  {\bibinfo {volume} {120}},\ \bibinfo {pages} {187201}}\BibitemShut {NoStop}%
\bibitem [{\citenamefont {Lu}\ \emph {et~al.}(2019)\citenamefont {Lu},
  \citenamefont {Guo}, \citenamefont {Koval},\ and\ \citenamefont
  {Jia}}]{PhysRevB.99.054409}%
  \BibitemOpen
  \bibfield  {author} {\bibinfo {author} {\bibnamefont {Lu}, \bibfnamefont
  {Yalei}}, \bibinfo {author} {\bibfnamefont {Xing}\ \bibnamefont {Guo}},
  \bibinfo {author} {\bibfnamefont {Vladimir}\ \bibnamefont {Koval}}, \ and\
  \bibinfo {author} {\bibfnamefont {Chenglong}\ \bibnamefont {Jia}}} (\bibinfo
  {year} {2019}),\ \bibfield  {title} {\enquote {\bibinfo {title} {Topological
  thermal hall effect driven by spin-chirality fluctuations in frustrated
  antiferromagnets},}\ }\href {\doibase 10.1103/PhysRevB.99.054409} {\bibfield
  {journal} {\bibinfo  {journal} {Phys. Rev. B}\ }\textbf {\bibinfo {volume}
  {99}},\ \bibinfo {pages} {054409}}\BibitemShut {NoStop}%
\bibitem [{\citenamefont {Lu}\ \emph {et~al.}(2021)\citenamefont {Lu},
  \citenamefont {Li},\ and\ \citenamefont {Wu}}]{PhysRevLett.127.217202}%
  \BibitemOpen
  \bibfield  {author} {\bibinfo {author} {\bibnamefont {Lu}, \bibfnamefont
  {Yu-Shan}}, \bibinfo {author} {\bibfnamefont {Jian-Lin}\ \bibnamefont {Li}},
  \ and\ \bibinfo {author} {\bibfnamefont {Chien-Te}\ \bibnamefont {Wu}}}
  (\bibinfo {year} {2021}),\ \bibfield  {title} {\enquote {\bibinfo {title}
  {{Topological Phase Transitions of Dirac Magnons in Honeycomb
  Ferromagnets}},}\ }\href {\doibase 10.1103/PhysRevLett.127.217202} {\bibfield
   {journal} {\bibinfo  {journal} {Phys. Rev. Lett.}\ }\textbf {\bibinfo
  {volume} {127}},\ \bibinfo {pages} {217202}}\BibitemShut {NoStop}%
\bibitem [{\citenamefont {Luttinger}(1964)}]{Luttinger1964}%
  \BibitemOpen
  \bibfield  {author} {\bibinfo {author} {\bibnamefont {Luttinger},
  \bibfnamefont {J~M}}} (\bibinfo {year} {1964}),\ \bibfield  {title} {\enquote
  {\bibinfo {title} {Theory of thermal transport coefficients},}\ }\href
  {\doibase 10.1103/physrev.135.a1505} {\bibfield  {journal} {\bibinfo
  {journal} {Physical Review}\ }\textbf {\bibinfo {volume} {135}}~(\bibinfo
  {number} {6A}),\ \bibinfo {pages} {A1505--A1514}}\BibitemShut {NoStop}%
\bibitem [{\citenamefont {Lyu}\ and\ \citenamefont
  {Witczak-Krempa}(2023)}]{lyu2023phonons}%
  \BibitemOpen
  \bibfield  {author} {\bibinfo {author} {\bibnamefont {Lyu}, \bibfnamefont
  {Liuke}}, \ and\ \bibinfo {author} {\bibfnamefont {William}\ \bibnamefont
  {Witczak-Krempa}}} (\bibinfo {year} {2023}),\ \href@noop {} {\enquote
  {\bibinfo {title} {{Phonons behave like Electrons in the Thermal Hall Effect
  of the Cuprates}},}\ }\Eprint {http://arxiv.org/abs/2207.02240}
  {arXiv:2207.02240 [cond-mat.str-el]} \BibitemShut {NoStop}%
\bibitem [{\citenamefont {{M. Hirschberger, P. Czajka, S. Koohpayeh, W. Wang,
  and N. P. Ong}}(2019)}]{Hirschberger2019}%
  \BibitemOpen
  \bibfield  {author} {\bibinfo {author} {\bibnamefont {{M. Hirschberger, P.
  Czajka, S. Koohpayeh, W. Wang, and N. P. Ong}},}} (\bibinfo {year} {2019}),\
  \href@noop {} {\enquote {\bibinfo {title} {{{Enhanced thermal Hall
  conductivity below 1 Kelvin in the pyrochlore magnet Yb$_2$Ti$_2$O$_7$}}},}\
  }\Eprint {http://arxiv.org/abs/1903.00595} {arXiv:1903.00595} \BibitemShut
  {NoStop}%
\bibitem [{\citenamefont {Ma}\ and\ \citenamefont
  {Fiete}(2022)}]{PhysRevB.105.L100402}%
  \BibitemOpen
  \bibfield  {author} {\bibinfo {author} {\bibnamefont {Ma}, \bibfnamefont
  {Bowen}}, \ and\ \bibinfo {author} {\bibfnamefont {Gregory~A.}\ \bibnamefont
  {Fiete}}} (\bibinfo {year} {2022}),\ \bibfield  {title} {\enquote {\bibinfo
  {title} {{Antiferromagnetic insulators with tunable magnon-polaron Chern
  numbers induced by in-plane optical phonons}},}\ }\href {\doibase
  10.1103/PhysRevB.105.L100402} {\bibfield  {journal} {\bibinfo  {journal}
  {Phys. Rev. B}\ }\textbf {\bibinfo {volume} {105}},\ \bibinfo {pages}
  {L100402}}\BibitemShut {NoStop}%
\bibitem [{\citenamefont {Ma}\ \emph {et~al.}(2023)\citenamefont {Ma},
  \citenamefont {Wang},\ and\ \citenamefont {Chen}}]{ma2023chiral}%
  \BibitemOpen
  \bibfield  {author} {\bibinfo {author} {\bibnamefont {Ma}, \bibfnamefont
  {Bowen}}, \bibinfo {author} {\bibfnamefont {Z.~D.}\ \bibnamefont {Wang}}, \
  and\ \bibinfo {author} {\bibfnamefont {Gang}\ \bibnamefont {Chen}}} (\bibinfo
  {year} {2023}),\ \href@noop {} {\enquote {\bibinfo {title} {Chiral
  magneto-phonons with tunable topology in anisotropic quantum magnets},}\
  }\Eprint {http://arxiv.org/abs/2309.04064} {arXiv:2309.04064
  [cond-mat.mes-hall]} \BibitemShut {NoStop}%
\bibitem [{\citenamefont {Machida}\ \emph {et~al.}(2007)\citenamefont
  {Machida}, \citenamefont {Nakatsuji}, \citenamefont {Maeno}, \citenamefont
  {Tayama}, \citenamefont {Sakakibara},\ and\ \citenamefont
  {Onoda}}]{PhysRevLett.98.057203}%
  \BibitemOpen
  \bibfield  {author} {\bibinfo {author} {\bibnamefont {Machida}, \bibfnamefont
  {Y}}, \bibinfo {author} {\bibfnamefont {S.}~\bibnamefont {Nakatsuji}},
  \bibinfo {author} {\bibfnamefont {Y.}~\bibnamefont {Maeno}}, \bibinfo
  {author} {\bibfnamefont {T.}~\bibnamefont {Tayama}}, \bibinfo {author}
  {\bibfnamefont {T.}~\bibnamefont {Sakakibara}}, \ and\ \bibinfo {author}
  {\bibfnamefont {S.}~\bibnamefont {Onoda}}} (\bibinfo {year} {2007}),\
  \bibfield  {title} {\enquote {\bibinfo {title} {{Unconventional Anomalous
  Hall Effect Enhanced by a Noncoplanar Spin Texture in the Frustrated Kondo
  Lattice ${\mathrm{Pr}}_{2}{\mathrm{Ir}}_{2}{\mathrm{O}}_{7}$}},}\ }\href
  {\doibase 10.1103/PhysRevLett.98.057203} {\bibfield  {journal} {\bibinfo
  {journal} {Phys. Rev. Lett.}\ }\textbf {\bibinfo {volume} {98}},\ \bibinfo
  {pages} {057203}}\BibitemShut {NoStop}%
\bibitem [{\citenamefont {Machida}\ \emph {et~al.}(2009)\citenamefont
  {Machida}, \citenamefont {Nakatsuji}, \citenamefont {Onoda}, \citenamefont
  {Tayama},\ and\ \citenamefont {Sakakibara}}]{Machida2009}%
  \BibitemOpen
  \bibfield  {author} {\bibinfo {author} {\bibnamefont {Machida}, \bibfnamefont
  {Yo}}, \bibinfo {author} {\bibfnamefont {Satoru}\ \bibnamefont {Nakatsuji}},
  \bibinfo {author} {\bibfnamefont {Shigeki}\ \bibnamefont {Onoda}}, \bibinfo
  {author} {\bibfnamefont {Takashi}\ \bibnamefont {Tayama}}, \ and\ \bibinfo
  {author} {\bibfnamefont {Toshiro}\ \bibnamefont {Sakakibara}}} (\bibinfo
  {year} {2009}),\ \bibfield  {title} {\enquote {\bibinfo {title}
  {{Time-reversal symmetry breaking and spontaneous Hall effect without
  magnetic dipole order}},}\ }\href {\doibase 10.1038/nature08680} {\bibfield
  {journal} {\bibinfo  {journal} {Nature}\ }\textbf {\bibinfo {volume}
  {463}}~(\bibinfo {number} {7278}),\ \bibinfo {pages} {210--213}}\BibitemShut
  {NoStop}%
\bibitem [{\citenamefont {Malki}\ and\ \citenamefont
  {Schmidt}(2017)}]{PhysRevB.95.195137}%
  \BibitemOpen
  \bibfield  {author} {\bibinfo {author} {\bibnamefont {Malki}, \bibfnamefont
  {M}}, \ and\ \bibinfo {author} {\bibfnamefont {K.~P.}\ \bibnamefont
  {Schmidt}}} (\bibinfo {year} {2017}),\ \bibfield  {title} {\enquote {\bibinfo
  {title} {{Magnetic Chern bands and triplon Hall effect in an extended
  Shastry-Sutherland model}},}\ }\href {\doibase 10.1103/PhysRevB.95.195137}
  {\bibfield  {journal} {\bibinfo  {journal} {Phys. Rev. B}\ }\textbf {\bibinfo
  {volume} {95}},\ \bibinfo {pages} {195137}}\BibitemShut {NoStop}%
\bibitem [{\citenamefont {Mangeolle}\ \emph
  {et~al.}(2022{\natexlab{a}})\citenamefont {Mangeolle}, \citenamefont
  {Balents},\ and\ \citenamefont {Savary}}]{PhysRevX.12.041031}%
  \BibitemOpen
  \bibfield  {author} {\bibinfo {author} {\bibnamefont {Mangeolle},
  \bibfnamefont {L\'eo}}, \bibinfo {author} {\bibfnamefont {Leon}\ \bibnamefont
  {Balents}}, \ and\ \bibinfo {author} {\bibfnamefont {Lucile}\ \bibnamefont
  {Savary}}} (\bibinfo {year} {2022}{\natexlab{a}}),\ \bibfield  {title}
  {\enquote {\bibinfo {title} {{Phonon Thermal Hall Conductivity from
  Scattering with Collective Fluctuations}},}\ }\href {\doibase
  10.1103/PhysRevX.12.041031} {\bibfield  {journal} {\bibinfo  {journal} {Phys.
  Rev. X}\ }\textbf {\bibinfo {volume} {12}},\ \bibinfo {pages}
  {041031}}\BibitemShut {NoStop}%
\bibitem [{\citenamefont {Mangeolle}\ \emph
  {et~al.}(2022{\natexlab{b}})\citenamefont {Mangeolle}, \citenamefont
  {Balents},\ and\ \citenamefont {Savary}}]{PhysRevB.106.245139}%
  \BibitemOpen
  \bibfield  {author} {\bibinfo {author} {\bibnamefont {Mangeolle},
  \bibfnamefont {L\'eo}}, \bibinfo {author} {\bibfnamefont {Leon}\ \bibnamefont
  {Balents}}, \ and\ \bibinfo {author} {\bibfnamefont {Lucile}\ \bibnamefont
  {Savary}}} (\bibinfo {year} {2022}{\natexlab{b}}),\ \bibfield  {title}
  {\enquote {\bibinfo {title} {Thermal conductivity and theory of inelastic
  scattering of phonons by collective fluctuations},}\ }\href {\doibase
  10.1103/PhysRevB.106.245139} {\bibfield  {journal} {\bibinfo  {journal}
  {Phys. Rev. B}\ }\textbf {\bibinfo {volume} {106}},\ \bibinfo {pages}
  {245139}}\BibitemShut {NoStop}%
\bibitem [{\citenamefont {Marston}(1990)}]{Marston1990}%
  \BibitemOpen
  \bibfield  {author} {\bibinfo {author} {\bibnamefont {Marston}, \bibfnamefont
  {J~B}}} (\bibinfo {year} {1990}),\ \bibfield  {title} {\enquote {\bibinfo
  {title} {{Instantons and massless fermions in (2+1)-dimensional lattice QED
  and antiferromagnets}},}\ }\href {\doibase 10.1103/PhysRevLett.64.1166}
  {\bibfield  {journal} {\bibinfo  {journal} {Phys. Rev. Lett.}\ }\textbf
  {\bibinfo {volume} {64}},\ \bibinfo {pages} {1166--1169}}\BibitemShut
  {NoStop}%
\bibitem [{\citenamefont {Matan}\ \emph {et~al.}(2006)\citenamefont {Matan},
  \citenamefont {Grohol}, \citenamefont {Nocera}, \citenamefont {Yildirim},
  \citenamefont {Harris}, \citenamefont {Lee}, \citenamefont {Nagler},\ and\
  \citenamefont {Lee}}]{PhysRevLett.96.247201}%
  \BibitemOpen
  \bibfield  {author} {\bibinfo {author} {\bibnamefont {Matan}, \bibfnamefont
  {K}}, \bibinfo {author} {\bibfnamefont {D.}~\bibnamefont {Grohol}}, \bibinfo
  {author} {\bibfnamefont {D.~G.}\ \bibnamefont {Nocera}}, \bibinfo {author}
  {\bibfnamefont {T.}~\bibnamefont {Yildirim}}, \bibinfo {author}
  {\bibfnamefont {A.~B.}\ \bibnamefont {Harris}}, \bibinfo {author}
  {\bibfnamefont {S.~H.}\ \bibnamefont {Lee}}, \bibinfo {author} {\bibfnamefont
  {S.~E.}\ \bibnamefont {Nagler}}, \ and\ \bibinfo {author} {\bibfnamefont
  {Y.~S.}\ \bibnamefont {Lee}}} (\bibinfo {year} {2006}),\ \bibfield  {title}
  {\enquote {\bibinfo {title} {{Spin Waves in the Frustrated Kagom\'e Lattice
  Antiferromagnet
  ${\mathrm{KFe}}_{3}(\mathrm{OH}{)}_{6}({\mathrm{SO}}_{4}{)}_{2}$}},}\ }\href
  {\doibase 10.1103/PhysRevLett.96.247201} {\bibfield  {journal} {\bibinfo
  {journal} {Phys. Rev. Lett.}\ }\textbf {\bibinfo {volume} {96}},\ \bibinfo
  {pages} {247201}}\BibitemShut {NoStop}%
\bibitem [{\citenamefont {Matsumoto}\ and\ \citenamefont
  {Murakami}(2011{\natexlab{a}})}]{Matsumoto2011B}%
  \BibitemOpen
  \bibfield  {author} {\bibinfo {author} {\bibnamefont {Matsumoto},
  \bibfnamefont {Ryo}}, \ and\ \bibinfo {author} {\bibfnamefont {Shuichi}\
  \bibnamefont {Murakami}}} (\bibinfo {year} {2011}{\natexlab{a}}),\ \bibfield
  {title} {\enquote {\bibinfo {title} {{Rotational motion of magnons and the
  thermal Hall effect}},}\ }\href {\doibase 10.1103/PhysRevB.84.184406}
  {\bibfield  {journal} {\bibinfo  {journal} {Phys. Rev. B}\ }\textbf {\bibinfo
  {volume} {84}},\ \bibinfo {pages} {184406}}\BibitemShut {NoStop}%
\bibitem [{\citenamefont {Matsumoto}\ and\ \citenamefont
  {Murakami}(2011{\natexlab{b}})}]{Matsumoto2011}%
  \BibitemOpen
  \bibfield  {author} {\bibinfo {author} {\bibnamefont {Matsumoto},
  \bibfnamefont {Ryo}}, \ and\ \bibinfo {author} {\bibfnamefont {Shuichi}\
  \bibnamefont {Murakami}}} (\bibinfo {year} {2011}{\natexlab{b}}),\ \bibfield
  {title} {\enquote {\bibinfo {title} {Theoretical prediction of a rotating
  magnon wave packet in ferromagnets},}\ }\href {\doibase
  10.1103/PhysRevLett.106.197202} {\bibfield  {journal} {\bibinfo  {journal}
  {Phys. Rev. Lett.}\ }\textbf {\bibinfo {volume} {106}},\ \bibinfo {pages}
  {197202}}\BibitemShut {NoStop}%
\bibitem [{\citenamefont {Matsumoto}\ \emph {et~al.}(2014)\citenamefont
  {Matsumoto}, \citenamefont {Shindou},\ and\ \citenamefont
  {Murakami}}]{Shindou2014}%
  \BibitemOpen
  \bibfield  {author} {\bibinfo {author} {\bibnamefont {Matsumoto},
  \bibfnamefont {Ryo}}, \bibinfo {author} {\bibfnamefont {Ryuichi}\
  \bibnamefont {Shindou}}, \ and\ \bibinfo {author} {\bibfnamefont {Shuichi}\
  \bibnamefont {Murakami}}} (\bibinfo {year} {2014}),\ \bibfield  {title}
  {\enquote {\bibinfo {title} {{Thermal Hall effect of magnons in magnets with
  dipolar interaction}},}\ }\href {\doibase 10.1103/PhysRevB.89.054420}
  {\bibfield  {journal} {\bibinfo  {journal} {Phys. Rev. B}\ }\textbf {\bibinfo
  {volume} {89}},\ \bibinfo {pages} {054420}}\BibitemShut {NoStop}%
\bibitem [{\citenamefont {McClarty}\ \emph {et~al.}(2018)\citenamefont
  {McClarty}, \citenamefont {Dong}, \citenamefont {Gohlke}, \citenamefont
  {Rau}, \citenamefont {Pollmann}, \citenamefont {Moessner},\ and\
  \citenamefont {Penc}}]{PhysRevB.98.060404}%
  \BibitemOpen
  \bibfield  {author} {\bibinfo {author} {\bibnamefont {McClarty},
  \bibfnamefont {P~A}}, \bibinfo {author} {\bibfnamefont {X.-Y.}\ \bibnamefont
  {Dong}}, \bibinfo {author} {\bibfnamefont {M.}~\bibnamefont {Gohlke}},
  \bibinfo {author} {\bibfnamefont {J.~G.}\ \bibnamefont {Rau}}, \bibinfo
  {author} {\bibfnamefont {F.}~\bibnamefont {Pollmann}}, \bibinfo {author}
  {\bibfnamefont {R.}~\bibnamefont {Moessner}}, \ and\ \bibinfo {author}
  {\bibfnamefont {K.}~\bibnamefont {Penc}}} (\bibinfo {year} {2018}),\
  \bibfield  {title} {\enquote {\bibinfo {title} {{Topological magnons in
  Kitaev magnets at high fields}},}\ }\href {\doibase
  10.1103/PhysRevB.98.060404} {\bibfield  {journal} {\bibinfo  {journal} {Phys.
  Rev. B}\ }\textbf {\bibinfo {volume} {98}},\ \bibinfo {pages}
  {060404}}\BibitemShut {NoStop}%
\bibitem [{\citenamefont {McClarty}\ \emph {et~al.}(2017)\citenamefont
  {McClarty}, \citenamefont {Kr\"{u}ger}, \citenamefont {Guidi}, \citenamefont
  {Parker}, \citenamefont {Refson}, \citenamefont {Parker}, \citenamefont
  {Prabhakaran},\ and\ \citenamefont {Coldea}}]{McClarty2017}%
  \BibitemOpen
  \bibfield  {author} {\bibinfo {author} {\bibnamefont {McClarty},
  \bibfnamefont {P~A}}, \bibinfo {author} {\bibfnamefont {F.}~\bibnamefont
  {Kr\"{u}ger}}, \bibinfo {author} {\bibfnamefont {T.}~\bibnamefont {Guidi}},
  \bibinfo {author} {\bibfnamefont {S.~F.}\ \bibnamefont {Parker}}, \bibinfo
  {author} {\bibfnamefont {K.}~\bibnamefont {Refson}}, \bibinfo {author}
  {\bibfnamefont {A.~W.}\ \bibnamefont {Parker}}, \bibinfo {author}
  {\bibfnamefont {D.}~\bibnamefont {Prabhakaran}}, \ and\ \bibinfo {author}
  {\bibfnamefont {R.}~\bibnamefont {Coldea}}} (\bibinfo {year} {2017}),\
  \bibfield  {title} {\enquote {\bibinfo {title} {{Topological triplon modes
  and bound states in a Shastry{\textendash}Sutherland magnet}},}\ }\href
  {\doibase 10.1038/nphys4117} {\bibfield  {journal} {\bibinfo  {journal}
  {Nature Physics}\ }\textbf {\bibinfo {volume} {13}}~(\bibinfo {number} {8}),\
  \bibinfo {pages} {736--741}}\BibitemShut {NoStop}%
\bibitem [{\citenamefont {McClarty}(2022)}]{McClarty2022}%
  \BibitemOpen
  \bibfield  {author} {\bibinfo {author} {\bibnamefont {McClarty},
  \bibfnamefont {Paul~A}}} (\bibinfo {year} {2022}),\ \bibfield  {title}
  {\enquote {\bibinfo {title} {Topological magnons: A review},}\ }\href
  {\doibase 10.1146/annurev-conmatphys-031620-104715} {\bibfield  {journal}
  {\bibinfo  {journal} {Annual Review of Condensed Matter Physics}\ }\textbf
  {\bibinfo {volume} {13}}~(\bibinfo {number} {1}),\ \bibinfo {pages}
  {171--190}}\BibitemShut {NoStop}%
\bibitem [{\citenamefont {Messio}\ \emph {et~al.}(2010)\citenamefont {Messio},
  \citenamefont {C\'epas},\ and\ \citenamefont
  {Lhuillier}}]{PhysRevB.81.064428}%
  \BibitemOpen
  \bibfield  {author} {\bibinfo {author} {\bibnamefont {Messio}, \bibfnamefont
  {L}}, \bibinfo {author} {\bibfnamefont {O.}~\bibnamefont {C\'epas}}, \ and\
  \bibinfo {author} {\bibfnamefont {C.}~\bibnamefont {Lhuillier}}} (\bibinfo
  {year} {2010}),\ \bibfield  {title} {\enquote {\bibinfo {title}
  {{Schwinger-boson approach to the kagome antiferromagnet with
  Dzyaloshinskii-Moriya interactions: Phase diagram and dynamical structure
  factors}},}\ }\href {\doibase 10.1103/PhysRevB.81.064428} {\bibfield
  {journal} {\bibinfo  {journal} {Phys. Rev. B}\ }\textbf {\bibinfo {volume}
  {81}},\ \bibinfo {pages} {064428}}\BibitemShut {NoStop}%
\bibitem [{\citenamefont {Messio}\ \emph {et~al.}(2017)\citenamefont {Messio},
  \citenamefont {Bieri}, \citenamefont {Lhuillier},\ and\ \citenamefont
  {Bernu}}]{PhysRevLett.118.267201}%
  \BibitemOpen
  \bibfield  {author} {\bibinfo {author} {\bibnamefont {Messio}, \bibfnamefont
  {Laura}}, \bibinfo {author} {\bibfnamefont {Samuel}\ \bibnamefont {Bieri}},
  \bibinfo {author} {\bibfnamefont {Claire}\ \bibnamefont {Lhuillier}}, \ and\
  \bibinfo {author} {\bibfnamefont {Bernard}\ \bibnamefont {Bernu}}} (\bibinfo
  {year} {2017}),\ \bibfield  {title} {\enquote {\bibinfo {title} {{Chiral Spin
  Liquid on a Kagome Antiferromagnet Induced by the Dzyaloshinskii-Moriya
  Interaction}},}\ }\href {\doibase 10.1103/PhysRevLett.118.267201} {\bibfield
  {journal} {\bibinfo  {journal} {Phys. Rev. Lett.}\ }\textbf {\bibinfo
  {volume} {118}},\ \bibinfo {pages} {267201}}\BibitemShut {NoStop}%
\bibitem [{\citenamefont {Mezzacapo}\ and\ \citenamefont
  {Boninsegni}(2012)}]{Massimo2012}%
  \BibitemOpen
  \bibfield  {author} {\bibinfo {author} {\bibnamefont {Mezzacapo},
  \bibfnamefont {Fabio}}, \ and\ \bibinfo {author} {\bibfnamefont {Massimo}\
  \bibnamefont {Boninsegni}}} (\bibinfo {year} {2012}),\ \bibfield  {title}
  {\enquote {\bibinfo {title} {{Ground-state phase diagram of the quantum
  ${J}_{1}\ensuremath{-}{J}_{2}$ model on the honeycomb lattice}},}\ }\href
  {\doibase 10.1103/PhysRevB.85.060402} {\bibfield  {journal} {\bibinfo
  {journal} {Phys. Rev. B}\ }\textbf {\bibinfo {volume} {85}},\ \bibinfo
  {pages} {060402}}\BibitemShut {NoStop}%
\bibitem [{\citenamefont {Miyahara}\ and\ \citenamefont
  {Ueda}(1999)}]{Miyahara1999}%
  \BibitemOpen
  \bibfield  {author} {\bibinfo {author} {\bibnamefont {Miyahara},
  \bibfnamefont {Shin}}, \ and\ \bibinfo {author} {\bibfnamefont {Kazuo}\
  \bibnamefont {Ueda}}} (\bibinfo {year} {1999}),\ \bibfield  {title} {\enquote
  {\bibinfo {title} {{Exact Dimer Ground State of the Two Dimensional
  Heisenberg Spin System ${\mathrm{SrCu}}_{2}({\mathrm{BO}}_{3}){}_{2}$}},}\
  }\href {\doibase 10.1103/PhysRevLett.82.3701} {\bibfield  {journal} {\bibinfo
   {journal} {Phys. Rev. Lett.}\ }\textbf {\bibinfo {volume} {82}},\ \bibinfo
  {pages} {3701--3704}}\BibitemShut {NoStop}%
\bibitem [{\citenamefont {Molavian}\ \emph {et~al.}(2007)\citenamefont
  {Molavian}, \citenamefont {Gingras},\ and\ \citenamefont
  {Canals}}]{Canals2007}%
  \BibitemOpen
  \bibfield  {author} {\bibinfo {author} {\bibnamefont {Molavian},
  \bibfnamefont {Hamid~R}}, \bibinfo {author} {\bibfnamefont {Michel J.~P.}\
  \bibnamefont {Gingras}}, \ and\ \bibinfo {author} {\bibfnamefont {Benjamin}\
  \bibnamefont {Canals}}} (\bibinfo {year} {2007}),\ \bibfield  {title}
  {\enquote {\bibinfo {title} {{Dynamically induced frustration as a route to a
  quantum spin ice state in
  ${\mathrm{Tb}}_{2}{\mathrm{Ti}}_{2}{\mathrm{O}}_{7}$ via virtual crystal
  field excitations and quantum many-body effects}},}\ }\href {\doibase
  10.1103/PhysRevLett.98.157204} {\bibfield  {journal} {\bibinfo  {journal}
  {Phys. Rev. Lett.}\ }\textbf {\bibinfo {volume} {98}},\ \bibinfo {pages}
  {157204}}\BibitemShut {NoStop}%
\bibitem [{\citenamefont {Mook}\ \emph {et~al.}(2014)\citenamefont {Mook},
  \citenamefont {Henk},\ and\ \citenamefont {Mertig}}]{PhysRevB.89.134409}%
  \BibitemOpen
  \bibfield  {author} {\bibinfo {author} {\bibnamefont {Mook}, \bibfnamefont
  {Alexander}}, \bibinfo {author} {\bibfnamefont {J\"urgen}\ \bibnamefont
  {Henk}}, \ and\ \bibinfo {author} {\bibfnamefont {Ingrid}\ \bibnamefont
  {Mertig}}} (\bibinfo {year} {2014}),\ \bibfield  {title} {\enquote {\bibinfo
  {title} {{Magnon Hall effect and topology in kagome lattices: A theoretical
  investigation}},}\ }\href {\doibase 10.1103/PhysRevB.89.134409} {\bibfield
  {journal} {\bibinfo  {journal} {Phys. Rev. B}\ }\textbf {\bibinfo {volume}
  {89}},\ \bibinfo {pages} {134409}}\BibitemShut {NoStop}%
\bibitem [{\citenamefont {Mook}\ \emph {et~al.}(2016)\citenamefont {Mook},
  \citenamefont {Henk},\ and\ \citenamefont {Mertig}}]{PhysRevLett.117.157204}%
  \BibitemOpen
  \bibfield  {author} {\bibinfo {author} {\bibnamefont {Mook}, \bibfnamefont
  {Alexander}}, \bibinfo {author} {\bibfnamefont {J\"urgen}\ \bibnamefont
  {Henk}}, \ and\ \bibinfo {author} {\bibfnamefont {Ingrid}\ \bibnamefont
  {Mertig}}} (\bibinfo {year} {2016}),\ \bibfield  {title} {\enquote {\bibinfo
  {title} {Tunable magnon weyl points in ferromagnetic pyrochlores},}\ }\href
  {\doibase 10.1103/PhysRevLett.117.157204} {\bibfield  {journal} {\bibinfo
  {journal} {Phys. Rev. Lett.}\ }\textbf {\bibinfo {volume} {117}},\ \bibinfo
  {pages} {157204}}\BibitemShut {NoStop}%
\bibitem [{\citenamefont {Mook}\ \emph {et~al.}(2019)\citenamefont {Mook},
  \citenamefont {Henk},\ and\ \citenamefont {Mertig}}]{Mook2019}%
  \BibitemOpen
  \bibfield  {author} {\bibinfo {author} {\bibnamefont {Mook}, \bibfnamefont
  {Alexander}}, \bibinfo {author} {\bibfnamefont {J\"urgen}\ \bibnamefont
  {Henk}}, \ and\ \bibinfo {author} {\bibfnamefont {Ingrid}\ \bibnamefont
  {Mertig}}} (\bibinfo {year} {2019}),\ \bibfield  {title} {\enquote {\bibinfo
  {title} {{Thermal Hall effect in noncollinear coplanar insulating
  antiferromagnets}},}\ }\href {\doibase 10.1103/PhysRevB.99.014427} {\bibfield
   {journal} {\bibinfo  {journal} {Phys. Rev. B}\ }\textbf {\bibinfo {volume}
  {99}},\ \bibinfo {pages} {014427}}\BibitemShut {NoStop}%
\bibitem [{\citenamefont {Morey}\ \emph {et~al.}(2019)\citenamefont {Morey},
  \citenamefont {Scheie}, \citenamefont {Sheckelton}, \citenamefont {Brown},\
  and\ \citenamefont {McQueen}}]{PhysRevMaterials.3.014410}%
  \BibitemOpen
  \bibfield  {author} {\bibinfo {author} {\bibnamefont {Morey}, \bibfnamefont
  {Jennifer~R}}, \bibinfo {author} {\bibfnamefont {Allen}\ \bibnamefont
  {Scheie}}, \bibinfo {author} {\bibfnamefont {John~P.}\ \bibnamefont
  {Sheckelton}}, \bibinfo {author} {\bibfnamefont {Craig~M.}\ \bibnamefont
  {Brown}}, \ and\ \bibinfo {author} {\bibfnamefont {Tyrel~M.}\ \bibnamefont
  {McQueen}}} (\bibinfo {year} {2019}),\ \bibfield  {title} {\enquote {\bibinfo
  {title}
  {{$\mathrm{N}{\mathrm{i}}_{2}\mathrm{M}{\mathrm{o}}_{3}{\mathrm{O}}_{8}$:
  Complex antiferromagnetic order on a honeycomb lattice}},}\ }\href {\doibase
  10.1103/PhysRevMaterials.3.014410} {\bibfield  {journal} {\bibinfo  {journal}
  {Phys. Rev. Mater.}\ }\textbf {\bibinfo {volume} {3}},\ \bibinfo {pages}
  {014410}}\BibitemShut {NoStop}%
\bibitem [{\citenamefont {Moriya}(1960)}]{Moriya1960}%
  \BibitemOpen
  \bibfield  {author} {\bibinfo {author} {\bibnamefont {Moriya}, \bibfnamefont
  {T\^oru}}} (\bibinfo {year} {1960}),\ \bibfield  {title} {\enquote {\bibinfo
  {title} {Anisotropic superexchange interaction and weak ferromagnetism},}\
  }\href {\doibase 10.1103/PhysRev.120.91} {\bibfield  {journal} {\bibinfo
  {journal} {Phys. Rev.}\ }\textbf {\bibinfo {volume} {120}},\ \bibinfo {pages}
  {91--98}}\BibitemShut {NoStop}%
\bibitem [{\citenamefont {Motrunich}\ and\ \citenamefont
  {Senthil}(2005)}]{PhysRevB.71.125102}%
  \BibitemOpen
  \bibfield  {author} {\bibinfo {author} {\bibnamefont {Motrunich},
  \bibfnamefont {O~I}}, \ and\ \bibinfo {author} {\bibfnamefont
  {T.}~\bibnamefont {Senthil}}} (\bibinfo {year} {2005}),\ \bibfield  {title}
  {\enquote {\bibinfo {title} {Origin of artificial electrodynamics in
  three-dimensional bosonic models},}\ }\href {\doibase
  10.1103/PhysRevB.71.125102} {\bibfield  {journal} {\bibinfo  {journal} {Phys.
  Rev. B}\ }\textbf {\bibinfo {volume} {71}},\ \bibinfo {pages}
  {125102}}\BibitemShut {NoStop}%
\bibitem [{\citenamefont {Motrunich}(2005)}]{Motrunich2005}%
  \BibitemOpen
  \bibfield  {author} {\bibinfo {author} {\bibnamefont {Motrunich},
  \bibfnamefont {Olexei~I}}} (\bibinfo {year} {2005}),\ \bibfield  {title}
  {\enquote {\bibinfo {title} {{Variational study of triangular lattice
  spin-$1/2$ model with ring exchanges and spin liquid state in
  $\ensuremath{\kappa}\text{\ensuremath{-}}{(\mathrm{ET})}_{2}{\mathrm{Cu}}_{2}{(\mathrm{CN})}_{3}$}},}\
  }\href {\doibase 10.1103/PhysRevB.72.045105} {\bibfield  {journal} {\bibinfo
  {journal} {Phys. Rev. B}\ }\textbf {\bibinfo {volume} {72}},\ \bibinfo
  {pages} {045105}}\BibitemShut {NoStop}%
\bibitem [{\citenamefont {Motrunich}(2006)}]{Motrunich2006}%
  \BibitemOpen
  \bibfield  {author} {\bibinfo {author} {\bibnamefont {Motrunich},
  \bibfnamefont {Olexei~I}}} (\bibinfo {year} {2006}),\ \bibfield  {title}
  {\enquote {\bibinfo {title} {{Orbital magnetic field effects in spin liquid
  with spinon Fermi sea: Possible application to
  $\ensuremath{\kappa}\text{\ensuremath{-}}{(\mathrm{ET})}_{2}{\mathrm{Cu}}_{2}{(\mathrm{C}\mathrm{N})}_{3}$}},}\
  }\href {\doibase 10.1103/PhysRevB.73.155115} {\bibfield  {journal} {\bibinfo
  {journal} {Phys. Rev. B}\ }\textbf {\bibinfo {volume} {73}},\ \bibinfo
  {pages} {155115}}\BibitemShut {NoStop}%
\bibitem [{\citenamefont {Murthy}(1991)}]{Murthy1991}%
  \BibitemOpen
  \bibfield  {author} {\bibinfo {author} {\bibnamefont {Murthy}, \bibfnamefont
  {Ganpathy}}} (\bibinfo {year} {1991}),\ \bibfield  {title} {\enquote
  {\bibinfo {title} {{Disordered SU($N$) antiferromagnets and the
  renormalization of charged instanton gases in three dimensions}},}\ }\href
  {\doibase 10.1103/PhysRevLett.67.911} {\bibfield  {journal} {\bibinfo
  {journal} {Phys. Rev. Lett.}\ }\textbf {\bibinfo {volume} {67}},\ \bibinfo
  {pages} {911--914}}\BibitemShut {NoStop}%
\bibitem [{\citenamefont {Nasu}\ \emph {et~al.}(2016)\citenamefont {Nasu},
  \citenamefont {Knolle}, \citenamefont {Kovrizhin}, \citenamefont {Motome},\
  and\ \citenamefont {Moessner}}]{Nasu2016}%
  \BibitemOpen
  \bibfield  {author} {\bibinfo {author} {\bibnamefont {Nasu}, \bibfnamefont
  {J}}, \bibinfo {author} {\bibfnamefont {J.}~\bibnamefont {Knolle}}, \bibinfo
  {author} {\bibfnamefont {D.~L.}\ \bibnamefont {Kovrizhin}}, \bibinfo {author}
  {\bibfnamefont {Y.}~\bibnamefont {Motome}}, \ and\ \bibinfo {author}
  {\bibfnamefont {R.}~\bibnamefont {Moessner}}} (\bibinfo {year} {2016}),\
  \bibfield  {title} {\enquote {\bibinfo {title} {{Fermionic response from
  fractionalization in an insulating two-dimensional~magnet}},}\ }\href
  {\doibase 10.1038/nphys3809} {\bibfield  {journal} {\bibinfo  {journal}
  {Nature Physics}\ }\textbf {\bibinfo {volume} {12}}~(\bibinfo {number}
  {10}),\ \bibinfo {pages} {912--915}}\BibitemShut {NoStop}%
\bibitem [{\citenamefont {Nasu}\ \emph {et~al.}(2017)\citenamefont {Nasu},
  \citenamefont {Yoshitake},\ and\ \citenamefont {Motome}}]{Motome2017}%
  \BibitemOpen
  \bibfield  {author} {\bibinfo {author} {\bibnamefont {Nasu}, \bibfnamefont
  {Joji}}, \bibinfo {author} {\bibfnamefont {Junki}\ \bibnamefont {Yoshitake}},
  \ and\ \bibinfo {author} {\bibfnamefont {Yukitoshi}\ \bibnamefont {Motome}}}
  (\bibinfo {year} {2017}),\ \bibfield  {title} {\enquote {\bibinfo {title}
  {{Thermal transport in the Kitaev model}},}\ }\href {\doibase
  10.1103/PhysRevLett.119.127204} {\bibfield  {journal} {\bibinfo  {journal}
  {Phys. Rev. Lett.}\ }\textbf {\bibinfo {volume} {119}},\ \bibinfo {pages}
  {127204}}\BibitemShut {NoStop}%
\bibitem [{\citenamefont {Neumann}\ \emph {et~al.}(2022)\citenamefont
  {Neumann}, \citenamefont {Mook}, \citenamefont {Henk},\ and\ \citenamefont
  {Mertig}}]{PhysRevLett.128.117201}%
  \BibitemOpen
  \bibfield  {author} {\bibinfo {author} {\bibnamefont {Neumann}, \bibfnamefont
  {Robin~R}}, \bibinfo {author} {\bibfnamefont {Alexander}\ \bibnamefont
  {Mook}}, \bibinfo {author} {\bibfnamefont {J\"urgen}\ \bibnamefont {Henk}}, \
  and\ \bibinfo {author} {\bibfnamefont {Ingrid}\ \bibnamefont {Mertig}}}
  (\bibinfo {year} {2022}),\ \bibfield  {title} {\enquote {\bibinfo {title}
  {{Thermal Hall Effect of Magnons in Collinear Antiferromagnetic Insulators:
  Signatures of Magnetic and Topological Phase Transitions}},}\ }\href
  {\doibase 10.1103/PhysRevLett.128.117201} {\bibfield  {journal} {\bibinfo
  {journal} {Phys. Rev. Lett.}\ }\textbf {\bibinfo {volume} {128}},\ \bibinfo
  {pages} {117201}}\BibitemShut {NoStop}%
\bibitem [{\citenamefont {Ni}\ \emph {et~al.}(2019)\citenamefont {Ni},
  \citenamefont {Pan}, \citenamefont {Song}, \citenamefont {Huang},
  \citenamefont {Zeng}, \citenamefont {Yu}, \citenamefont {Cheng},
  \citenamefont {Wang}, \citenamefont {Dai}, \citenamefont {Kato},\ and\
  \citenamefont {Li}}]{Shiyan2019}%
  \BibitemOpen
  \bibfield  {author} {\bibinfo {author} {\bibnamefont {Ni}, \bibfnamefont
  {J~M}}, \bibinfo {author} {\bibfnamefont {B.~L.}\ \bibnamefont {Pan}},
  \bibinfo {author} {\bibfnamefont {B.~Q.}\ \bibnamefont {Song}}, \bibinfo
  {author} {\bibfnamefont {Y.~Y.}\ \bibnamefont {Huang}}, \bibinfo {author}
  {\bibfnamefont {J.~Y.}\ \bibnamefont {Zeng}}, \bibinfo {author}
  {\bibfnamefont {Y.~J.}\ \bibnamefont {Yu}}, \bibinfo {author} {\bibfnamefont
  {E.~J.}\ \bibnamefont {Cheng}}, \bibinfo {author} {\bibfnamefont {L.~S.}\
  \bibnamefont {Wang}}, \bibinfo {author} {\bibfnamefont {D.~Z.}\ \bibnamefont
  {Dai}}, \bibinfo {author} {\bibfnamefont {R.}~\bibnamefont {Kato}}, \ and\
  \bibinfo {author} {\bibfnamefont {S.~Y.}\ \bibnamefont {Li}}} (\bibinfo
  {year} {2019}),\ \bibfield  {title} {\enquote {\bibinfo {title} {{Absence of
  magnetic thermal conductivity in the quantum spin liquid candidate
  ${\mathrm{EtMe}}_{3}\mathrm{Sb}[\mathrm{Pd}(\text{dmit}{)}_{2}{]}_{2}$}},}\
  }\href {\doibase 10.1103/PhysRevLett.123.247204} {\bibfield  {journal}
  {\bibinfo  {journal} {Phys. Rev. Lett.}\ }\textbf {\bibinfo {volume} {123}},\
  \bibinfo {pages} {247204}}\BibitemShut {NoStop}%
\bibitem [{\citenamefont {O'Brien}\ \emph {et~al.}(2016)\citenamefont
  {O'Brien}, \citenamefont {Hermanns},\ and\ \citenamefont
  {Trebst}}]{PhysRevB.93.085101}%
  \BibitemOpen
  \bibfield  {author} {\bibinfo {author} {\bibnamefont {O'Brien}, \bibfnamefont
  {Kevin}}, \bibinfo {author} {\bibfnamefont {Maria}\ \bibnamefont {Hermanns}},
  \ and\ \bibinfo {author} {\bibfnamefont {Simon}\ \bibnamefont {Trebst}}}
  (\bibinfo {year} {2016}),\ \bibfield  {title} {\enquote {\bibinfo {title}
  {{Classification of gapless ${\mathbb{Z}}_{2}$ spin liquids in
  three-dimensional Kitaev models}},}\ }\href {\doibase
  10.1103/PhysRevB.93.085101} {\bibfield  {journal} {\bibinfo  {journal} {Phys.
  Rev. B}\ }\textbf {\bibinfo {volume} {93}},\ \bibinfo {pages}
  {085101}}\BibitemShut {NoStop}%
\bibitem [{\citenamefont {Okamoto}(2013)}]{Okamoto2013}%
  \BibitemOpen
  \bibfield  {author} {\bibinfo {author} {\bibnamefont {Okamoto}, \bibfnamefont
  {Satoshi}}} (\bibinfo {year} {2013}),\ \bibfield  {title} {\enquote {\bibinfo
  {title} {{Global phase diagram of a doped Kitaev-Heisenberg model}},}\ }\href
  {\doibase 10.1103/PhysRevB.87.064508} {\bibfield  {journal} {\bibinfo
  {journal} {Phys. Rev. B}\ }\textbf {\bibinfo {volume} {87}},\ \bibinfo
  {pages} {064508}}\BibitemShut {NoStop}%
\bibitem [{\citenamefont {Okuma}\ \emph {et~al.}(2019)\citenamefont {Okuma},
  \citenamefont {Nakamura}, \citenamefont {Okubo}, \citenamefont {Miyake},
  \citenamefont {Matsuo}, \citenamefont {Kindo}, \citenamefont {Tokunaga},
  \citenamefont {Kawashima}, \citenamefont {Takeyama},\ and\ \citenamefont
  {Hiroi}}]{Okuma2019}%
  \BibitemOpen
  \bibfield  {author} {\bibinfo {author} {\bibnamefont {Okuma}, \bibfnamefont
  {R}}, \bibinfo {author} {\bibfnamefont {D.}~\bibnamefont {Nakamura}},
  \bibinfo {author} {\bibfnamefont {T.}~\bibnamefont {Okubo}}, \bibinfo
  {author} {\bibfnamefont {A.}~\bibnamefont {Miyake}}, \bibinfo {author}
  {\bibfnamefont {A.}~\bibnamefont {Matsuo}}, \bibinfo {author} {\bibfnamefont
  {K.}~\bibnamefont {Kindo}}, \bibinfo {author} {\bibfnamefont
  {M.}~\bibnamefont {Tokunaga}}, \bibinfo {author} {\bibfnamefont
  {N.}~\bibnamefont {Kawashima}}, \bibinfo {author} {\bibfnamefont
  {S.}~\bibnamefont {Takeyama}}, \ and\ \bibinfo {author} {\bibfnamefont
  {Z.}~\bibnamefont {Hiroi}}} (\bibinfo {year} {2019}),\ \bibfield  {title}
  {\enquote {\bibinfo {title} {A series of magnon crystals appearing under
  ultrahigh magnetic fields in a kagom{\'{e}} antiferromagnet},}\ }\href
  {\doibase 10.1038/s41467-019-09063-7} {\bibfield  {journal} {\bibinfo
  {journal} {Nature Communications}\ }\textbf {\bibinfo {volume}
  {10}}~(\bibinfo {number} {1}),\ 10.1038/s41467-019-09063-7}\BibitemShut
  {NoStop}%
\bibitem [{\citenamefont {Okuma}\ \emph {et~al.}(2017)\citenamefont {Okuma},
  \citenamefont {Yajima}, \citenamefont {Nishio-Hamane}, \citenamefont
  {Okubo},\ and\ \citenamefont {Hiroi}}]{Okuma2017}%
  \BibitemOpen
  \bibfield  {author} {\bibinfo {author} {\bibnamefont {Okuma}, \bibfnamefont
  {Ryutaro}}, \bibinfo {author} {\bibfnamefont {Takeshi}\ \bibnamefont
  {Yajima}}, \bibinfo {author} {\bibfnamefont {Daisuke}\ \bibnamefont
  {Nishio-Hamane}}, \bibinfo {author} {\bibfnamefont {Tsuyoshi}\ \bibnamefont
  {Okubo}}, \ and\ \bibinfo {author} {\bibfnamefont {Zenji}\ \bibnamefont
  {Hiroi}}} (\bibinfo {year} {2017}),\ \bibfield  {title} {\enquote {\bibinfo
  {title} {{Weak ferromagnetic order breaking the threefold rotational symmetry
  of the underlying kagome lattice in
  $\mathrm{CdC}{\mathrm{u}}_{3}{(\mathrm{OH})}_{6}{({\mathrm{NO}}_{3})}_{2}\ifmmode\cdot\else\textperiodcentered\fi{}{\mathrm{H}}_{2}\mathrm{O}$}},}\
  }\href {\doibase 10.1103/PhysRevB.95.094427} {\bibfield  {journal} {\bibinfo
  {journal} {Phys. Rev. B}\ }\textbf {\bibinfo {volume} {95}},\ \bibinfo
  {pages} {094427}}\BibitemShut {NoStop}%
\bibitem [{\citenamefont {Onoda}\ and\ \citenamefont
  {Tanaka}(2010)}]{Tanaka2010}%
  \BibitemOpen
  \bibfield  {author} {\bibinfo {author} {\bibnamefont {Onoda}, \bibfnamefont
  {Shigeki}}, \ and\ \bibinfo {author} {\bibfnamefont {Yoichi}\ \bibnamefont
  {Tanaka}}} (\bibinfo {year} {2010}),\ \bibfield  {title} {\enquote {\bibinfo
  {title} {{Quantum Melting of Spin Ice: Emergent Cooperative Quadrupole and
  Chirality}},}\ }\href {\doibase 10.1103/PhysRevLett.105.047201} {\bibfield
  {journal} {\bibinfo  {journal} {Phys. Rev. Lett.}\ }\textbf {\bibinfo
  {volume} {105}},\ \bibinfo {pages} {047201}}\BibitemShut {NoStop}%
\bibitem [{\citenamefont {Onoda}\ and\ \citenamefont
  {Tanaka}(2011)}]{PhysRevB.83.094411}%
  \BibitemOpen
  \bibfield  {author} {\bibinfo {author} {\bibnamefont {Onoda}, \bibfnamefont
  {Shigeki}}, \ and\ \bibinfo {author} {\bibfnamefont {Yoichi}\ \bibnamefont
  {Tanaka}}} (\bibinfo {year} {2011}),\ \bibfield  {title} {\enquote {\bibinfo
  {title} {{Quantum fluctuations in the effective pseudospin-$\frac{1}{2}$
  model for magnetic pyrochlore oxides}},}\ }\href {\doibase
  10.1103/PhysRevB.83.094411} {\bibfield  {journal} {\bibinfo  {journal} {Phys.
  Rev. B}\ }\textbf {\bibinfo {volume} {83}},\ \bibinfo {pages}
  {094411}}\BibitemShut {NoStop}%
\bibitem [{\citenamefont {Onose}\ \emph {et~al.}(2010)\citenamefont {Onose},
  \citenamefont {Ideue}, \citenamefont {Katsura}, \citenamefont {Shiomi},
  \citenamefont {Nagaosa},\ and\ \citenamefont {Tokura}}]{Onose2010}%
  \BibitemOpen
  \bibfield  {author} {\bibinfo {author} {\bibnamefont {Onose}, \bibfnamefont
  {Y}}, \bibinfo {author} {\bibfnamefont {T.}~\bibnamefont {Ideue}}, \bibinfo
  {author} {\bibfnamefont {H.}~\bibnamefont {Katsura}}, \bibinfo {author}
  {\bibfnamefont {Y.}~\bibnamefont {Shiomi}}, \bibinfo {author} {\bibfnamefont
  {N.}~\bibnamefont {Nagaosa}}, \ and\ \bibinfo {author} {\bibfnamefont
  {Y.}~\bibnamefont {Tokura}}} (\bibinfo {year} {2010}),\ \bibfield  {title}
  {\enquote {\bibinfo {title} {{Observation of the magnon Hall effect}},}\
  }\href {\doibase 10.1126/science.1188260} {\bibfield  {journal} {\bibinfo
  {journal} {Science}\ }\textbf {\bibinfo {volume} {329}}~(\bibinfo {number}
  {5989}),\ \bibinfo {pages} {297--299}}\BibitemShut {NoStop}%
\bibitem [{\citenamefont {Owerre}(2016{\natexlab{a}})}]{Owerre2016B}%
  \BibitemOpen
  \bibfield  {author} {\bibinfo {author} {\bibnamefont {Owerre}, \bibfnamefont
  {S~A}}} (\bibinfo {year} {2016}{\natexlab{a}}),\ \bibfield  {title} {\enquote
  {\bibinfo {title} {A first theoretical realization of honeycomb topological
  magnon insulator},}\ }\href {\doibase 10.1088/0953-8984/28/38/386001}
  {\bibfield  {journal} {\bibinfo  {journal} {Journal of Physics: Condensed
  Matter}\ }\textbf {\bibinfo {volume} {28}}~(\bibinfo {number} {38}),\
  \bibinfo {pages} {386001}}\BibitemShut {NoStop}%
\bibitem [{\citenamefont {Owerre}(2016{\natexlab{b}})}]{Owerre2016C}%
  \BibitemOpen
  \bibfield  {author} {\bibinfo {author} {\bibnamefont {Owerre}, \bibfnamefont
  {S~A}}} (\bibinfo {year} {2016}{\natexlab{b}}),\ \bibfield  {title} {\enquote
  {\bibinfo {title} {{Magnon Hall effect without
  Dzyaloshinskii{\textendash}Moriya interaction}},}\ }\href {\doibase
  10.1088/0953-8984/29/3/03lt01} {\bibfield  {journal} {\bibinfo  {journal}
  {Journal of Physics: Condensed Matter}\ }\textbf {\bibinfo {volume}
  {29}}~(\bibinfo {number} {3}),\ \bibinfo {pages} {03LT01}}\BibitemShut
  {NoStop}%
\bibitem [{\citenamefont {Owerre}(2017{\natexlab{a}})}]{Owerre2017B}%
  \BibitemOpen
  \bibfield  {author} {\bibinfo {author} {\bibnamefont {Owerre}, \bibfnamefont
  {S~A}}} (\bibinfo {year} {2017}{\natexlab{a}}),\ \bibfield  {title} {\enquote
  {\bibinfo {title} {Topological magnetic excitations on the distorted
  kagom{\'{e}} antiferromagnets: Applications to volborthite, vesignieite, and
  edwardsite},}\ }\href {\doibase 10.1209/0295-5075/117/37006} {\bibfield
  {journal} {\bibinfo  {journal} {{EPL} (Europhysics Letters)}\ }\textbf
  {\bibinfo {volume} {117}}~(\bibinfo {number} {3}),\ \bibinfo {pages}
  {37006}}\BibitemShut {NoStop}%
\bibitem [{\citenamefont {Owerre}(2017{\natexlab{b}})}]{Owerre2017}%
  \BibitemOpen
  \bibfield  {author} {\bibinfo {author} {\bibnamefont {Owerre}, \bibfnamefont
  {S~A}}} (\bibinfo {year} {2017}{\natexlab{b}}),\ \bibfield  {title} {\enquote
  {\bibinfo {title} {{Topological thermal Hall effect in frustrated
  kagom{\'{e}} antiferromagnets}},}\ }\href {\doibase
  10.1103/PhysRevB.95.014422} {\bibfield  {journal} {\bibinfo  {journal} {Phys.
  Rev. B}\ }\textbf {\bibinfo {volume} {95}},\ \bibinfo {pages}
  {014422}}\BibitemShut {NoStop}%
\bibitem [{\citenamefont {Park}\ and\ \citenamefont
  {Yang}(2020)}]{PhysRevB.102.214421}%
  \BibitemOpen
  \bibfield  {author} {\bibinfo {author} {\bibnamefont {Park}, \bibfnamefont
  {Sungjoon}}, \ and\ \bibinfo {author} {\bibfnamefont {Bohm-Jung}\
  \bibnamefont {Yang}}} (\bibinfo {year} {2020}),\ \bibfield  {title} {\enquote
  {\bibinfo {title} {{Thermal Hall effect from a two-dimensional Schwinger
  boson gas with Rashba spin-orbit interaction: Application to ferromagnets
  with in-plane Dzyaloshinskii-Moriya interaction}},}\ }\href {\doibase
  10.1103/PhysRevB.102.214421} {\bibfield  {journal} {\bibinfo  {journal}
  {Phys. Rev. B}\ }\textbf {\bibinfo {volume} {102}},\ \bibinfo {pages}
  {214421}}\BibitemShut {NoStop}%
\bibitem [{\citenamefont {Patel}\ and\ \citenamefont
  {Trivedi}(2019)}]{Patel2019}%
  \BibitemOpen
  \bibfield  {author} {\bibinfo {author} {\bibnamefont {Patel}, \bibfnamefont
  {Niravkumar~D}}, \ and\ \bibinfo {author} {\bibfnamefont {Nandini}\
  \bibnamefont {Trivedi}}} (\bibinfo {year} {2019}),\ \bibfield  {title}
  {\enquote {\bibinfo {title} {Magnetic field-induced intermediate quantum spin
  liquid with a spinon fermi surface},}\ }\href {\doibase
  10.1073/pnas.1821406116} {\bibfield  {journal} {\bibinfo  {journal}
  {Proceedings of the National Academy of Sciences}\ }\textbf {\bibinfo
  {volume} {116}}~(\bibinfo {number} {25}),\ \bibinfo {pages}
  {12199--12203}}\BibitemShut {NoStop}%
\bibitem [{\citenamefont {Pauling}(1935)}]{Pauling1935}%
  \BibitemOpen
  \bibfield  {author} {\bibinfo {author} {\bibnamefont {Pauling}, \bibfnamefont
  {Linus}}} (\bibinfo {year} {1935}),\ \bibfield  {title} {\enquote {\bibinfo
  {title} {The structure and entropy of ice and of other crystals with some
  randomness of atomic arrangement},}\ }\href {\doibase 10.1021/ja01315a102}
  {\bibfield  {journal} {\bibinfo  {journal} {Journal of the American Chemical
  Society}\ }\textbf {\bibinfo {volume} {57}}~(\bibinfo {number} {12}),\
  \bibinfo {pages} {2680--2684}}\BibitemShut {NoStop}%
\bibitem [{\citenamefont {Pawlak}\ \emph {et~al.}(1980)\citenamefont {Pawlak},
  \citenamefont {Duczmal}, \citenamefont {Pokrzywnicki},\ and\ \citenamefont
  {Czopnik}}]{PAWLAK1980195}%
  \BibitemOpen
  \bibfield  {author} {\bibinfo {author} {\bibnamefont {Pawlak}, \bibfnamefont
  {L}}, \bibinfo {author} {\bibfnamefont {M.}~\bibnamefont {Duczmal}}, \bibinfo
  {author} {\bibfnamefont {S.}~\bibnamefont {Pokrzywnicki}}, \ and\ \bibinfo
  {author} {\bibfnamefont {A.}~\bibnamefont {Czopnik}}} (\bibinfo {year}
  {1980}),\ \bibfield  {title} {\enquote {\bibinfo {title} {Magnetic
  susceptibility and crystal field parameters of ytterbium sesquiselenide},}\
  }\href {\doibase https://doi.org/10.1016/0038-1098(80)91145-X} {\bibfield
  {journal} {\bibinfo  {journal} {Solid State Communications}\ }\textbf
  {\bibinfo {volume} {34}}~(\bibinfo {number} {3}),\ \bibinfo {pages}
  {195--197}}\BibitemShut {NoStop}%
\bibitem [{\citenamefont {Pesin}\ and\ \citenamefont
  {Balents}(2010)}]{Pesin_2010}%
  \BibitemOpen
  \bibfield  {author} {\bibinfo {author} {\bibnamefont {Pesin}, \bibfnamefont
  {Dmytro}}, \ and\ \bibinfo {author} {\bibfnamefont {Leon}\ \bibnamefont
  {Balents}}} (\bibinfo {year} {2010}),\ \bibfield  {title} {\enquote {\bibinfo
  {title} {Mott physics and band topology in materials with strong
  spin{\textendash}orbit interaction},}\ }\href {\doibase 10.1038/nphys1606}
  {\bibfield  {journal} {\bibinfo  {journal} {Nature Physics}\ }\textbf
  {\bibinfo {volume} {6}}~(\bibinfo {number} {5}),\ \bibinfo {pages}
  {376--381}}\BibitemShut {NoStop}%
\bibitem [{\citenamefont {Petit}\ \emph {et~al.}(2016)\citenamefont {Petit},
  \citenamefont {Lhotel}, \citenamefont {Canals}, \citenamefont {Hatnean},
  \citenamefont {Ollivier}, \citenamefont {Mutka}, \citenamefont {Ressouche},
  \citenamefont {Wildes}, \citenamefont {Lees},\ and\ \citenamefont
  {Balakrishnan}}]{Petit2016}%
  \BibitemOpen
  \bibfield  {author} {\bibinfo {author} {\bibnamefont {Petit}, \bibfnamefont
  {S}}, \bibinfo {author} {\bibfnamefont {E.}~\bibnamefont {Lhotel}}, \bibinfo
  {author} {\bibfnamefont {B.}~\bibnamefont {Canals}}, \bibinfo {author}
  {\bibfnamefont {M.~Ciomaga}\ \bibnamefont {Hatnean}}, \bibinfo {author}
  {\bibfnamefont {J.}~\bibnamefont {Ollivier}}, \bibinfo {author}
  {\bibfnamefont {H.}~\bibnamefont {Mutka}}, \bibinfo {author} {\bibfnamefont
  {E.}~\bibnamefont {Ressouche}}, \bibinfo {author} {\bibfnamefont {A.~R.}\
  \bibnamefont {Wildes}}, \bibinfo {author} {\bibfnamefont {M.~R.}\
  \bibnamefont {Lees}}, \ and\ \bibinfo {author} {\bibfnamefont
  {G.}~\bibnamefont {Balakrishnan}}} (\bibinfo {year} {2016}),\ \bibfield
  {title} {\enquote {\bibinfo {title} {Observation of magnetic fragmentation in
  spin ice},}\ }\href {\doibase 10.1038/nphys3710} {\bibfield  {journal}
  {\bibinfo  {journal} {Nature Physics}\ }\textbf {\bibinfo {volume}
  {12}}~(\bibinfo {number} {8}),\ \bibinfo {pages} {746--750}}\BibitemShut
  {NoStop}%
\bibitem [{\citenamefont {Reig-i Plessis}\ \emph {et~al.}(2019)\citenamefont
  {Reig-i Plessis}, \citenamefont {Geldern}, \citenamefont {Aczel},
  \citenamefont {Kochkov}, \citenamefont {Clark},\ and\ \citenamefont
  {MacDougall}}]{PhysRevB.99.134438}%
  \BibitemOpen
  \bibfield  {author} {\bibinfo {author} {\bibnamefont {Reig-i Plessis},
  \bibfnamefont {D}}, \bibinfo {author} {\bibfnamefont {S.~V.}\ \bibnamefont
  {Geldern}}, \bibinfo {author} {\bibfnamefont {A.~A.}\ \bibnamefont {Aczel}},
  \bibinfo {author} {\bibfnamefont {D.}~\bibnamefont {Kochkov}}, \bibinfo
  {author} {\bibfnamefont {B.~K.}\ \bibnamefont {Clark}}, \ and\ \bibinfo
  {author} {\bibfnamefont {G.~J.}\ \bibnamefont {MacDougall}}} (\bibinfo {year}
  {2019}),\ \bibfield  {title} {\enquote {\bibinfo {title} {{Deviation from the
  dipole-ice model in the spinel spin-ice candidate
  ${\mathrm{MgEr}}_{2}{\mathrm{Se}}_{4}$}},}\ }\href {\doibase
  10.1103/PhysRevB.99.134438} {\bibfield  {journal} {\bibinfo  {journal} {Phys.
  Rev. B}\ }\textbf {\bibinfo {volume} {99}},\ \bibinfo {pages}
  {134438}}\BibitemShut {NoStop}%
\bibitem [{\citenamefont {Plumb}\ \emph {et~al.}(2014)\citenamefont {Plumb},
  \citenamefont {Clancy}, \citenamefont {Sandilands}, \citenamefont {Shankar},
  \citenamefont {Hu}, \citenamefont {Burch}, \citenamefont {Kee},\ and\
  \citenamefont {Kim}}]{Plumb2014}%
  \BibitemOpen
  \bibfield  {author} {\bibinfo {author} {\bibnamefont {Plumb}, \bibfnamefont
  {K~W}}, \bibinfo {author} {\bibfnamefont {J.~P.}\ \bibnamefont {Clancy}},
  \bibinfo {author} {\bibfnamefont {L.~J.}\ \bibnamefont {Sandilands}},
  \bibinfo {author} {\bibfnamefont {V.~Vijay}\ \bibnamefont {Shankar}},
  \bibinfo {author} {\bibfnamefont {Y.~F.}\ \bibnamefont {Hu}}, \bibinfo
  {author} {\bibfnamefont {K.~S.}\ \bibnamefont {Burch}}, \bibinfo {author}
  {\bibfnamefont {Hae-Young}\ \bibnamefont {Kee}}, \ and\ \bibinfo {author}
  {\bibfnamefont {Young-June}\ \bibnamefont {Kim}}} (\bibinfo {year} {2014}),\
  \bibfield  {title} {\enquote {\bibinfo {title}
  {{$\ensuremath{\alpha}\ensuremath{-}{\mathrm{RuCl}}_{3}$: A spin-orbit
  assisted Mott insulator on a honeycomb lattice}},}\ }\href {\doibase
  10.1103/PhysRevB.90.041112} {\bibfield  {journal} {\bibinfo  {journal} {Phys.
  Rev. B}\ }\textbf {\bibinfo {volume} {90}},\ \bibinfo {pages}
  {041112}}\BibitemShut {NoStop}%
\bibitem [{\citenamefont {Pollini}(1996)}]{PhysRevB.53.12769}%
  \BibitemOpen
  \bibfield  {author} {\bibinfo {author} {\bibnamefont {Pollini}, \bibfnamefont
  {I}}} (\bibinfo {year} {1996}),\ \bibfield  {title} {\enquote {\bibinfo
  {title} {{Electronic properties of the narrow-band material
  $\ensuremath{\alpha}\ensuremath{-}{\mathrm{RuCl}}_{3}$}},}\ }\href {\doibase
  10.1103/PhysRevB.53.12769} {\bibfield  {journal} {\bibinfo  {journal} {Phys.
  Rev. B}\ }\textbf {\bibinfo {volume} {53}},\ \bibinfo {pages}
  {12769--12776}}\BibitemShut {NoStop}%
\bibitem [{\citenamefont {Polyakov}(1977)}]{Polyakov1977}%
  \BibitemOpen
  \bibfield  {author} {\bibinfo {author} {\bibnamefont {Polyakov},
  \bibfnamefont {AM}}} (\bibinfo {year} {1977}),\ \bibfield  {title} {\enquote
  {\bibinfo {title} {Quark confinement and topology of gauge theories},}\
  }\href {\doibase 10.1016/0550-3213(77)90086-4} {\bibfield  {journal}
  {\bibinfo  {journal} {Nuclear Physics B}\ }\textbf {\bibinfo {volume}
  {120}}~(\bibinfo {number} {3}),\ \bibinfo {pages} {429--458}}\BibitemShut
  {NoStop}%
\bibitem [{\citenamefont {Porée}\ \emph {et~al.}(2023)\citenamefont {Porée},
  \citenamefont {Yan}, \citenamefont {Desrochers}, \citenamefont {Petit},
  \citenamefont {Lhotel}, \citenamefont {Appel}, \citenamefont {Ollivier},
  \citenamefont {Kim}, \citenamefont {Nevidomskyy},\ and\ \citenamefont
  {Sibille}}]{poree2023fractional}%
  \BibitemOpen
  \bibfield  {author} {\bibinfo {author} {\bibnamefont {Porée}, \bibfnamefont
  {Victor}}, \bibinfo {author} {\bibfnamefont {Han}\ \bibnamefont {Yan}},
  \bibinfo {author} {\bibfnamefont {Félix}\ \bibnamefont {Desrochers}},
  \bibinfo {author} {\bibfnamefont {Sylvain}\ \bibnamefont {Petit}}, \bibinfo
  {author} {\bibfnamefont {Elsa}\ \bibnamefont {Lhotel}}, \bibinfo {author}
  {\bibfnamefont {Markus}\ \bibnamefont {Appel}}, \bibinfo {author}
  {\bibfnamefont {Jacques}\ \bibnamefont {Ollivier}}, \bibinfo {author}
  {\bibfnamefont {Yong~Baek}\ \bibnamefont {Kim}}, \bibinfo {author}
  {\bibfnamefont {Andriy~H.}\ \bibnamefont {Nevidomskyy}}, \ and\ \bibinfo
  {author} {\bibfnamefont {Romain}\ \bibnamefont {Sibille}}} (\bibinfo {year}
  {2023}),\ \href@noop {} {\enquote {\bibinfo {title} {Fractional matter
  coupled to the emergent gauge field in a quantum spin ice},}\ }\Eprint
  {http://arxiv.org/abs/2304.05452} {arXiv:2304.05452 [cond-mat.str-el]}
  \BibitemShut {NoStop}%
\bibitem [{\citenamefont {Qin}\ \emph {et~al.}(2011)\citenamefont {Qin},
  \citenamefont {Niu},\ and\ \citenamefont {Shi}}]{Qin2011}%
  \BibitemOpen
  \bibfield  {author} {\bibinfo {author} {\bibnamefont {Qin}, \bibfnamefont
  {Tao}}, \bibinfo {author} {\bibfnamefont {Qian}\ \bibnamefont {Niu}}, \ and\
  \bibinfo {author} {\bibfnamefont {Junren}\ \bibnamefont {Shi}}} (\bibinfo
  {year} {2011}),\ \bibfield  {title} {\enquote {\bibinfo {title} {{Energy
  magnetization and the thermal Hall effect}},}\ }\href {\doibase
  10.1103/PhysRevLett.107.236601} {\bibfield  {journal} {\bibinfo  {journal}
  {Phys. Rev. Lett.}\ }\textbf {\bibinfo {volume} {107}},\ \bibinfo {pages}
  {236601}}\BibitemShut {NoStop}%
\bibitem [{\citenamefont {Ramirez}\ \emph {et~al.}(1999)\citenamefont
  {Ramirez}, \citenamefont {Hayashi}, \citenamefont {Cava}, \citenamefont
  {Siddharthan},\ and\ \citenamefont {Shastry}}]{Ramirez1999}%
  \BibitemOpen
  \bibfield  {author} {\bibinfo {author} {\bibnamefont {Ramirez}, \bibfnamefont
  {A~P}}, \bibinfo {author} {\bibfnamefont {A.}~\bibnamefont {Hayashi}},
  \bibinfo {author} {\bibfnamefont {R.~J.}\ \bibnamefont {Cava}}, \bibinfo
  {author} {\bibfnamefont {R.}~\bibnamefont {Siddharthan}}, \ and\ \bibinfo
  {author} {\bibfnamefont {B.~S.}\ \bibnamefont {Shastry}}} (\bibinfo {year}
  {1999}),\ \bibfield  {title} {\enquote {\bibinfo {title} {Zero-point entropy
  in `spin ice'},}\ }\href {\doibase 10.1038/20619} {\bibfield  {journal}
  {\bibinfo  {journal} {Nature}\ }\textbf {\bibinfo {volume} {399}}~(\bibinfo
  {number} {6734}),\ \bibinfo {pages} {333--335}}\BibitemShut {NoStop}%
\bibitem [{\citenamefont {Ran}\ \emph {et~al.}(2007)\citenamefont {Ran},
  \citenamefont {Hermele}, \citenamefont {Lee},\ and\ \citenamefont
  {Wen}}]{Ran2007}%
  \BibitemOpen
  \bibfield  {author} {\bibinfo {author} {\bibnamefont {Ran}, \bibfnamefont
  {Ying}}, \bibinfo {author} {\bibfnamefont {Michael}\ \bibnamefont {Hermele}},
  \bibinfo {author} {\bibfnamefont {Patrick~A.}\ \bibnamefont {Lee}}, \ and\
  \bibinfo {author} {\bibfnamefont {Xiao-Gang}\ \bibnamefont {Wen}}} (\bibinfo
  {year} {2007}),\ \bibfield  {title} {\enquote {\bibinfo {title}
  {{Projected-wave-function study of the spin-$1/2$ Heisenberg model on the
  kagom\'e lattice}},}\ }\href {\doibase 10.1103/PhysRevLett.98.117205}
  {\bibfield  {journal} {\bibinfo  {journal} {Phys. Rev. Lett.}\ }\textbf
  {\bibinfo {volume} {98}},\ \bibinfo {pages} {117205}}\BibitemShut {NoStop}%
\bibitem [{\citenamefont {Rantner}\ and\ \citenamefont {Wen}(2002)}]{Wen2002B}%
  \BibitemOpen
  \bibfield  {author} {\bibinfo {author} {\bibnamefont {Rantner}, \bibfnamefont
  {Walter}}, \ and\ \bibinfo {author} {\bibfnamefont {Xiao-Gang}\ \bibnamefont
  {Wen}}} (\bibinfo {year} {2002}),\ \bibfield  {title} {\enquote {\bibinfo
  {title} {{Spin correlations in the algebraic spin liquid: Implications for
  high-${T}_{c}$ superconductors}},}\ }\href {\doibase
  10.1103/PhysRevB.66.144501} {\bibfield  {journal} {\bibinfo  {journal} {Phys.
  Rev. B}\ }\textbf {\bibinfo {volume} {66}},\ \bibinfo {pages}
  {144501}}\BibitemShut {NoStop}%
\bibitem [{\citenamefont {Rau}\ and\ \citenamefont {Gingras}(2018)}]{Rau2018}%
  \BibitemOpen
  \bibfield  {author} {\bibinfo {author} {\bibnamefont {Rau}, \bibfnamefont
  {Jeffrey~G}}, \ and\ \bibinfo {author} {\bibfnamefont {Michel J.~P.}\
  \bibnamefont {Gingras}}} (\bibinfo {year} {2018}),\ \bibfield  {title}
  {\enquote {\bibinfo {title} {Frustration and anisotropic exchange in
  ytterbium magnets with edge-shared octahedra},}\ }\href {\doibase
  10.1103/PhysRevB.98.054408} {\bibfield  {journal} {\bibinfo  {journal} {Phys.
  Rev. B}\ }\textbf {\bibinfo {volume} {98}},\ \bibinfo {pages}
  {054408}}\BibitemShut {NoStop}%
\bibitem [{\citenamefont {Rau}\ \emph {et~al.}(2014)\citenamefont {Rau},
  \citenamefont {Lee},\ and\ \citenamefont {Kee}}]{Rau2014}%
  \BibitemOpen
  \bibfield  {author} {\bibinfo {author} {\bibnamefont {Rau}, \bibfnamefont
  {Jeffrey~G}}, \bibinfo {author} {\bibfnamefont {Eric Kin-Ho}\ \bibnamefont
  {Lee}}, \ and\ \bibinfo {author} {\bibfnamefont {Hae-Young}\ \bibnamefont
  {Kee}}} (\bibinfo {year} {2014}),\ \bibfield  {title} {\enquote {\bibinfo
  {title} {{Generic spin model for the honeycomb Iridates beyond the Kitaev
  limit}},}\ }\href {\doibase 10.1103/PhysRevLett.112.077204} {\bibfield
  {journal} {\bibinfo  {journal} {Phys. Rev. Lett.}\ }\textbf {\bibinfo
  {volume} {112}},\ \bibinfo {pages} {077204}}\BibitemShut {NoStop}%
\bibitem [{\citenamefont {Rau}\ \emph {et~al.}(2016)\citenamefont {Rau},
  \citenamefont {Lee},\ and\ \citenamefont {Kee}}]{Rau2016}%
  \BibitemOpen
  \bibfield  {author} {\bibinfo {author} {\bibnamefont {Rau}, \bibfnamefont
  {Jeffrey~G}}, \bibinfo {author} {\bibfnamefont {Eric Kin-Ho}\ \bibnamefont
  {Lee}}, \ and\ \bibinfo {author} {\bibfnamefont {Hae-Young}\ \bibnamefont
  {Kee}}} (\bibinfo {year} {2016}),\ \bibfield  {title} {\enquote {\bibinfo
  {title} {{Spin-orbit physics giving rise to novel phases in correlated
  systems: Iridates and related materials}},}\ }\href {\doibase
  10.1146/annurev-conmatphys-031115-011319} {\bibfield  {journal} {\bibinfo
  {journal} {Annual Review of Condensed Matter Physics}\ }\textbf {\bibinfo
  {volume} {7}}~(\bibinfo {number} {1}),\ \bibinfo {pages}
  {195--221}}\BibitemShut {NoStop}%
\bibitem [{\citenamefont {Romh{\'{a}}nyi}\ \emph {et~al.}(2015)\citenamefont
  {Romh{\'{a}}nyi}, \citenamefont {Penc},\ and\ \citenamefont
  {Ganesh}}]{Romhnyi2015}%
  \BibitemOpen
  \bibfield  {author} {\bibinfo {author} {\bibnamefont {Romh{\'{a}}nyi},
  \bibfnamefont {Judit}}, \bibinfo {author} {\bibfnamefont {Karlo}\
  \bibnamefont {Penc}}, \ and\ \bibinfo {author} {\bibfnamefont
  {R.}~\bibnamefont {Ganesh}}} (\bibinfo {year} {2015}),\ \bibfield  {title}
  {\enquote {\bibinfo {title} {{Hall effect of triplons in a dimerized quantum
  magnet}},}\ }\href {\doibase 10.1038/ncomms7805} {\bibfield  {journal}
  {\bibinfo  {journal} {Nature Communications}\ }\textbf {\bibinfo {volume}
  {6}}~(\bibinfo {number} {1}),\ \bibinfo {pages} {6805}}\BibitemShut {NoStop}%
\bibitem [{\citenamefont {Romh\'anyi}\ \emph {et~al.}(2011)\citenamefont
  {Romh\'anyi}, \citenamefont {Totsuka},\ and\ \citenamefont
  {Penc}}]{Romhanyi2011}%
  \BibitemOpen
  \bibfield  {author} {\bibinfo {author} {\bibnamefont {Romh\'anyi},
  \bibfnamefont {Judit}}, \bibinfo {author} {\bibfnamefont {Keisuke}\
  \bibnamefont {Totsuka}}, \ and\ \bibinfo {author} {\bibfnamefont {Karlo}\
  \bibnamefont {Penc}}} (\bibinfo {year} {2011}),\ \bibfield  {title} {\enquote
  {\bibinfo {title} {{Effect of Dzyaloshinskii-Moriya interactions on the phase
  diagram and magnetic excitations of SrCu${}_{2}$(BO${}_{3}$)${}_{2}$}},}\
  }\href {\doibase 10.1103/PhysRevB.83.024413} {\bibfield  {journal} {\bibinfo
  {journal} {Phys. Rev. B}\ }\textbf {\bibinfo {volume} {83}},\ \bibinfo
  {pages} {024413}}\BibitemShut {NoStop}%
\bibitem [{\citenamefont {Ross}\ \emph {et~al.}(2009)\citenamefont {Ross},
  \citenamefont {Ruff}, \citenamefont {Adams}, \citenamefont {Gardner},
  \citenamefont {Dabkowska}, \citenamefont {Qiu}, \citenamefont {Copley},\ and\
  \citenamefont {Gaulin}}]{Ross2009}%
  \BibitemOpen
  \bibfield  {author} {\bibinfo {author} {\bibnamefont {Ross}, \bibfnamefont
  {K~A}}, \bibinfo {author} {\bibfnamefont {J.~P.~C.}\ \bibnamefont {Ruff}},
  \bibinfo {author} {\bibfnamefont {C.~P.}\ \bibnamefont {Adams}}, \bibinfo
  {author} {\bibfnamefont {J.~S.}\ \bibnamefont {Gardner}}, \bibinfo {author}
  {\bibfnamefont {H.~A.}\ \bibnamefont {Dabkowska}}, \bibinfo {author}
  {\bibfnamefont {Y.}~\bibnamefont {Qiu}}, \bibinfo {author} {\bibfnamefont
  {J.~R.~D.}\ \bibnamefont {Copley}}, \ and\ \bibinfo {author} {\bibfnamefont
  {B.~D.}\ \bibnamefont {Gaulin}}} (\bibinfo {year} {2009}),\ \bibfield
  {title} {\enquote {\bibinfo {title} {{Two-dimensional kagom{\'{e}}
  correlations and field induced order in the ferromagnetic $XY$ pyrochlore
  ${\mathrm{Yb}}_{2}{\mathrm{Ti}}_{2}{\mathbf{O}}_{7}$}},}\ }\href {\doibase
  10.1103/PhysRevLett.103.227202} {\bibfield  {journal} {\bibinfo  {journal}
  {Phys. Rev. Lett.}\ }\textbf {\bibinfo {volume} {103}},\ \bibinfo {pages}
  {227202}}\BibitemShut {NoStop}%
\bibitem [{\citenamefont {Ross}\ \emph {et~al.}(2011)\citenamefont {Ross},
  \citenamefont {Savary}, \citenamefont {Gaulin},\ and\ \citenamefont
  {Balents}}]{Ross2011}%
  \BibitemOpen
  \bibfield  {author} {\bibinfo {author} {\bibnamefont {Ross}, \bibfnamefont
  {Kate~A}}, \bibinfo {author} {\bibfnamefont {Lucile}\ \bibnamefont {Savary}},
  \bibinfo {author} {\bibfnamefont {Bruce~D.}\ \bibnamefont {Gaulin}}, \ and\
  \bibinfo {author} {\bibfnamefont {Leon}\ \bibnamefont {Balents}}} (\bibinfo
  {year} {2011}),\ \bibfield  {title} {\enquote {\bibinfo {title} {Quantum
  excitations in quantum spin ice},}\ }\href {\doibase
  10.1103/PhysRevX.1.021002} {\bibfield  {journal} {\bibinfo  {journal} {Phys.
  Rev. X}\ }\textbf {\bibinfo {volume} {1}},\ \bibinfo {pages}
  {021002}}\BibitemShut {NoStop}%
\bibitem [{\citenamefont {R\"uckriegel}\ \emph {et~al.}(2018)\citenamefont
  {R\"uckriegel}, \citenamefont {Brataas},\ and\ \citenamefont
  {Duine}}]{PhysRevB.97.081106}%
  \BibitemOpen
  \bibfield  {author} {\bibinfo {author} {\bibnamefont {R\"uckriegel},
  \bibfnamefont {Andreas}}, \bibinfo {author} {\bibfnamefont {Arne}\
  \bibnamefont {Brataas}}, \ and\ \bibinfo {author} {\bibfnamefont
  {Rembert~A.}\ \bibnamefont {Duine}}} (\bibinfo {year} {2018}),\ \bibfield
  {title} {\enquote {\bibinfo {title} {Bulk and edge spin transport in
  topological magnon insulators},}\ }\href {\doibase
  10.1103/PhysRevB.97.081106} {\bibfield  {journal} {\bibinfo  {journal} {Phys.
  Rev. B}\ }\textbf {\bibinfo {volume} {97}},\ \bibinfo {pages}
  {081106}}\BibitemShut {NoStop}%
\bibitem [{\citenamefont {Ryu}\ \emph {et~al.}(2012)\citenamefont {Ryu},
  \citenamefont {Moore},\ and\ \citenamefont {Ludwig}}]{PhysRevB.85.045104}%
  \BibitemOpen
  \bibfield  {author} {\bibinfo {author} {\bibnamefont {Ryu}, \bibfnamefont
  {Shinsei}}, \bibinfo {author} {\bibfnamefont {Joel~E.}\ \bibnamefont
  {Moore}}, \ and\ \bibinfo {author} {\bibfnamefont {Andreas W.~W.}\
  \bibnamefont {Ludwig}}} (\bibinfo {year} {2012}),\ \bibfield  {title}
  {\enquote {\bibinfo {title} {Electromagnetic and gravitational responses and
  anomalies in topological insulators and superconductors},}\ }\href {\doibase
  10.1103/PhysRevB.85.045104} {\bibfield  {journal} {\bibinfo  {journal} {Phys.
  Rev. B}\ }\textbf {\bibinfo {volume} {85}},\ \bibinfo {pages}
  {045104}}\BibitemShut {NoStop}%
\bibitem [{\citenamefont {Sachdev}(1992)}]{PhysRevB.45.12377}%
  \BibitemOpen
  \bibfield  {author} {\bibinfo {author} {\bibnamefont {Sachdev}, \bibfnamefont
  {Subir}}} (\bibinfo {year} {1992}),\ \bibfield  {title} {\enquote {\bibinfo
  {title} {Kagome- and triangular-lattice heisenberg antiferromagnets: Ordering
  from quantum fluctuations and quantum-disordered ground states with
  unconfined bosonic spinons},}\ }\href {\doibase 10.1103/PhysRevB.45.12377}
  {\bibfield  {journal} {\bibinfo  {journal} {Phys. Rev. B}\ }\textbf {\bibinfo
  {volume} {45}},\ \bibinfo {pages} {12377--12396}}\BibitemShut {NoStop}%
\bibitem [{\citenamefont {Sachdev}\ and\ \citenamefont
  {Bhatt}(1990)}]{Sachdev1990}%
  \BibitemOpen
  \bibfield  {author} {\bibinfo {author} {\bibnamefont {Sachdev}, \bibfnamefont
  {Subir}}, \ and\ \bibinfo {author} {\bibfnamefont {R.~N.}\ \bibnamefont
  {Bhatt}}} (\bibinfo {year} {1990}),\ \bibfield  {title} {\enquote {\bibinfo
  {title} {Bond-operator representation of quantum spins: Mean-field theory of
  frustrated quantum heisenberg antiferromagnets},}\ }\href {\doibase
  10.1103/PhysRevB.41.9323} {\bibfield  {journal} {\bibinfo  {journal} {Phys.
  Rev. B}\ }\textbf {\bibinfo {volume} {41}},\ \bibinfo {pages}
  {9323--9329}}\BibitemShut {NoStop}%
\bibitem [{\citenamefont {Samajdar}\ \emph
  {et~al.}(2019{\natexlab{a}})\citenamefont {Samajdar}, \citenamefont
  {Chatterjee}, \citenamefont {Sachdev},\ and\ \citenamefont
  {Scheurer}}]{Sachdev2019}%
  \BibitemOpen
  \bibfield  {author} {\bibinfo {author} {\bibnamefont {Samajdar},
  \bibfnamefont {Rhine}}, \bibinfo {author} {\bibfnamefont {Shubhayu}\
  \bibnamefont {Chatterjee}}, \bibinfo {author} {\bibfnamefont {Subir}\
  \bibnamefont {Sachdev}}, \ and\ \bibinfo {author} {\bibfnamefont
  {Mathias~S.}\ \bibnamefont {Scheurer}}} (\bibinfo {year}
  {2019}{\natexlab{a}}),\ \bibfield  {title} {\enquote {\bibinfo {title}
  {{Thermal Hall effect in square-lattice spin liquids: A Schwinger boson
  mean-field study}},}\ }\href {\doibase 10.1103/PhysRevB.99.165126} {\bibfield
   {journal} {\bibinfo  {journal} {Phys. Rev. B}\ }\textbf {\bibinfo {volume}
  {99}},\ \bibinfo {pages} {165126}}\BibitemShut {NoStop}%
\bibitem [{\citenamefont {Samajdar}\ \emph
  {et~al.}(2019{\natexlab{b}})\citenamefont {Samajdar}, \citenamefont
  {Scheurer}, \citenamefont {Chatterjee}, \citenamefont {Guo}, \citenamefont
  {Xu},\ and\ \citenamefont {Sachdev}}]{Samajdar2019}%
  \BibitemOpen
  \bibfield  {author} {\bibinfo {author} {\bibnamefont {Samajdar},
  \bibfnamefont {Rhine}}, \bibinfo {author} {\bibfnamefont {Mathias~S.}\
  \bibnamefont {Scheurer}}, \bibinfo {author} {\bibfnamefont {Shubhayu}\
  \bibnamefont {Chatterjee}}, \bibinfo {author} {\bibfnamefont {Haoyu}\
  \bibnamefont {Guo}}, \bibinfo {author} {\bibfnamefont {Cenke}\ \bibnamefont
  {Xu}}, \ and\ \bibinfo {author} {\bibfnamefont {Subir}\ \bibnamefont
  {Sachdev}}} (\bibinfo {year} {2019}{\natexlab{b}}),\ \bibfield  {title}
  {\enquote {\bibinfo {title} {{Enhanced thermal Hall effect in the
  square-lattice N{\'{e}}el state}},}\ }\href {\doibase
  10.1038/s41567-019-0669-3} {\bibfield  {journal} {\bibinfo  {journal} {Nature
  Physics}\ }\textbf {\bibinfo {volume} {15}}~(\bibinfo {number} {12}),\
  \bibinfo {pages} {1290--1294}}\BibitemShut {NoStop}%
\bibitem [{\citenamefont {Sandilands}\ \emph {et~al.}(2015)\citenamefont
  {Sandilands}, \citenamefont {Tian}, \citenamefont {Plumb}, \citenamefont
  {Kim},\ and\ \citenamefont {Burch}}]{Sandilands2015}%
  \BibitemOpen
  \bibfield  {author} {\bibinfo {author} {\bibnamefont {Sandilands},
  \bibfnamefont {Luke~J}}, \bibinfo {author} {\bibfnamefont {Yao}\ \bibnamefont
  {Tian}}, \bibinfo {author} {\bibfnamefont {Kemp~W.}\ \bibnamefont {Plumb}},
  \bibinfo {author} {\bibfnamefont {Young-June}\ \bibnamefont {Kim}}, \ and\
  \bibinfo {author} {\bibfnamefont {Kenneth~S.}\ \bibnamefont {Burch}}}
  (\bibinfo {year} {2015}),\ \bibfield  {title} {\enquote {\bibinfo {title}
  {{Scattering continuum and possible fractionalized excitations in
  $\ensuremath{\alpha}\ensuremath{-}{\mathrm{RuCl}}_{3}$}},}\ }\href {\doibase
  10.1103/PhysRevLett.114.147201} {\bibfield  {journal} {\bibinfo  {journal}
  {Phys. Rev. Lett.}\ }\textbf {\bibinfo {volume} {114}},\ \bibinfo {pages}
  {147201}}\BibitemShut {NoStop}%
\bibitem [{\citenamefont {Sano}\ \emph {et~al.}(2018)\citenamefont {Sano},
  \citenamefont {Kato},\ and\ \citenamefont {Motome}}]{PhysRevB.97.014408}%
  \BibitemOpen
  \bibfield  {author} {\bibinfo {author} {\bibnamefont {Sano}, \bibfnamefont
  {Ryoya}}, \bibinfo {author} {\bibfnamefont {Yasuyuki}\ \bibnamefont {Kato}},
  \ and\ \bibinfo {author} {\bibfnamefont {Yukitoshi}\ \bibnamefont {Motome}}}
  (\bibinfo {year} {2018}),\ \bibfield  {title} {\enquote {\bibinfo {title}
  {{Kitaev-Heisenberg Hamiltonian for high-spin ${d}^{7}$ Mott insulators}},}\
  }\href {\doibase 10.1103/PhysRevB.97.014408} {\bibfield  {journal} {\bibinfo
  {journal} {Phys. Rev. B}\ }\textbf {\bibinfo {volume} {97}},\ \bibinfo
  {pages} {014408}}\BibitemShut {NoStop}%
\bibitem [{\citenamefont {Savary}\ and\ \citenamefont
  {Balents}(2012)}]{Savary2012}%
  \BibitemOpen
  \bibfield  {author} {\bibinfo {author} {\bibnamefont {Savary}, \bibfnamefont
  {Lucile}}, \ and\ \bibinfo {author} {\bibfnamefont {Leon}\ \bibnamefont
  {Balents}}} (\bibinfo {year} {2012}),\ \bibfield  {title} {\enquote {\bibinfo
  {title} {{Coulombic quantum liquids in spin-$1/2$ pyrochlores}},}\ }\href
  {\doibase 10.1103/PhysRevLett.108.037202} {\bibfield  {journal} {\bibinfo
  {journal} {Phys. Rev. Lett.}\ }\textbf {\bibinfo {volume} {108}},\ \bibinfo
  {pages} {037202}}\BibitemShut {NoStop}%
\bibitem [{\citenamefont {Savary}\ and\ \citenamefont
  {Balents}(2013)}]{Savary2013}%
  \BibitemOpen
  \bibfield  {author} {\bibinfo {author} {\bibnamefont {Savary}, \bibfnamefont
  {Lucile}}, \ and\ \bibinfo {author} {\bibfnamefont {Leon}\ \bibnamefont
  {Balents}}} (\bibinfo {year} {2013}),\ \bibfield  {title} {\enquote {\bibinfo
  {title} {Spin liquid regimes at nonzero temperature in quantum spin ice},}\
  }\href {\doibase 10.1103/PhysRevB.87.205130} {\bibfield  {journal} {\bibinfo
  {journal} {Phys. Rev. B}\ }\textbf {\bibinfo {volume} {87}},\ \bibinfo
  {pages} {205130}}\BibitemShut {NoStop}%
\bibitem [{\citenamefont {Savary}\ and\ \citenamefont
  {Balents}(2016)}]{Savary2016}%
  \BibitemOpen
  \bibfield  {author} {\bibinfo {author} {\bibnamefont {Savary}, \bibfnamefont
  {Lucile}}, \ and\ \bibinfo {author} {\bibfnamefont {Leon}\ \bibnamefont
  {Balents}}} (\bibinfo {year} {2016}),\ \bibfield  {title} {\enquote {\bibinfo
  {title} {Quantum spin liquids: a review},}\ }\href {\doibase
  10.1088/0034-4885/80/1/016502} {\bibfield  {journal} {\bibinfo  {journal}
  {Reports on Progress in Physics}\ }\textbf {\bibinfo {volume} {80}}~(\bibinfo
  {number} {1}),\ \bibinfo {pages} {016502}}\BibitemShut {NoStop}%
\bibitem [{\citenamefont {Schaffer}\ \emph {et~al.}(2012)\citenamefont
  {Schaffer}, \citenamefont {Bhattacharjee},\ and\ \citenamefont
  {Kim}}]{YongBaek2012}%
  \BibitemOpen
  \bibfield  {author} {\bibinfo {author} {\bibnamefont {Schaffer},
  \bibfnamefont {Robert}}, \bibinfo {author} {\bibfnamefont {Subhro}\
  \bibnamefont {Bhattacharjee}}, \ and\ \bibinfo {author} {\bibfnamefont
  {Yong~Baek}\ \bibnamefont {Kim}}} (\bibinfo {year} {2012}),\ \bibfield
  {title} {\enquote {\bibinfo {title} {{Quantum phase transition in
  Heisenberg-Kitaev model}},}\ }\href {\doibase 10.1103/PhysRevB.86.224417}
  {\bibfield  {journal} {\bibinfo  {journal} {Phys. Rev. B}\ }\textbf {\bibinfo
  {volume} {86}},\ \bibinfo {pages} {224417}}\BibitemShut {NoStop}%
\bibitem [{\citenamefont {Seemann}\ \emph {et~al.}(2015)\citenamefont
  {Seemann}, \citenamefont {K\"odderitzsch}, \citenamefont {Wimmer},\ and\
  \citenamefont {Ebert}}]{PhysRevB.92.155138}%
  \BibitemOpen
  \bibfield  {author} {\bibinfo {author} {\bibnamefont {Seemann}, \bibfnamefont
  {M}}, \bibinfo {author} {\bibfnamefont {D.}~\bibnamefont {K\"odderitzsch}},
  \bibinfo {author} {\bibfnamefont {S.}~\bibnamefont {Wimmer}}, \ and\ \bibinfo
  {author} {\bibfnamefont {H.}~\bibnamefont {Ebert}}} (\bibinfo {year}
  {2015}),\ \bibfield  {title} {\enquote {\bibinfo {title} {Symmetry-imposed
  shape of linear response tensors},}\ }\href {\doibase
  10.1103/PhysRevB.92.155138} {\bibfield  {journal} {\bibinfo  {journal} {Phys.
  Rev. B}\ }\textbf {\bibinfo {volume} {92}},\ \bibinfo {pages}
  {155138}}\BibitemShut {NoStop}%
\bibitem [{\citenamefont {Senthil}\ and\ \citenamefont
  {Fisher}(2000)}]{Fisher2000}%
  \BibitemOpen
  \bibfield  {author} {\bibinfo {author} {\bibnamefont {Senthil}, \bibfnamefont
  {T}}, \ and\ \bibinfo {author} {\bibfnamefont {Matthew P.~A.}\ \bibnamefont
  {Fisher}}} (\bibinfo {year} {2000}),\ \bibfield  {title} {\enquote {\bibinfo
  {title} {{${Z}_{2}$ gauge theory of electron fractionalization in strongly
  correlated systems}},}\ }\href {\doibase 10.1103/PhysRevB.62.7850} {\bibfield
   {journal} {\bibinfo  {journal} {Phys. Rev. B}\ }\textbf {\bibinfo {volume}
  {62}},\ \bibinfo {pages} {7850--7881}}\BibitemShut {NoStop}%
\bibitem [{\citenamefont {Senthil}\ and\ \citenamefont
  {Fisher}(2001{\natexlab{a}})}]{Senthil2001}%
  \BibitemOpen
  \bibfield  {author} {\bibinfo {author} {\bibnamefont {Senthil}, \bibfnamefont
  {T}}, \ and\ \bibinfo {author} {\bibfnamefont {Matthew P.~A.}\ \bibnamefont
  {Fisher}}} (\bibinfo {year} {2001}{\natexlab{a}}),\ \bibfield  {title}
  {\enquote {\bibinfo {title} {{Detecting fractions of electrons in the
  high-${T}_{c}$ cuprates}},}\ }\href {\doibase 10.1103/PhysRevB.64.214511}
  {\bibfield  {journal} {\bibinfo  {journal} {Phys. Rev. B}\ }\textbf {\bibinfo
  {volume} {64}},\ \bibinfo {pages} {214511}}\BibitemShut {NoStop}%
\bibitem [{\citenamefont {Senthil}\ and\ \citenamefont
  {Fisher}(2001{\natexlab{b}})}]{Fisher2001}%
  \BibitemOpen
  \bibfield  {author} {\bibinfo {author} {\bibnamefont {Senthil}, \bibfnamefont
  {T}}, \ and\ \bibinfo {author} {\bibfnamefont {Matthew P.~A.}\ \bibnamefont
  {Fisher}}} (\bibinfo {year} {2001}{\natexlab{b}}),\ \bibfield  {title}
  {\enquote {\bibinfo {title} {{Fractionalization in the cuprates: Detecting
  the topological order}},}\ }\href {\doibase 10.1103/PhysRevLett.86.292}
  {\bibfield  {journal} {\bibinfo  {journal} {Phys. Rev. Lett.}\ }\textbf
  {\bibinfo {volume} {86}},\ \bibinfo {pages} {292--295}}\BibitemShut {NoStop}%
\bibitem [{\citenamefont {Senthil}\ and\ \citenamefont
  {Fisher}(2001{\natexlab{c}})}]{Senthil2001B}%
  \BibitemOpen
  \bibfield  {author} {\bibinfo {author} {\bibnamefont {Senthil}, \bibfnamefont
  {T}}, \ and\ \bibinfo {author} {\bibfnamefont {Matthew P.~A.}\ \bibnamefont
  {Fisher}}} (\bibinfo {year} {2001}{\natexlab{c}}),\ \bibfield  {title}
  {\enquote {\bibinfo {title} {{Fractionalization, topological order, and
  cuprate superconductivity}},}\ }\href {\doibase 10.1103/PhysRevB.63.134521}
  {\bibfield  {journal} {\bibinfo  {journal} {Phys. Rev. B}\ }\textbf {\bibinfo
  {volume} {63}},\ \bibinfo {pages} {134521}}\BibitemShut {NoStop}%
\bibitem [{\citenamefont {Senthil}\ and\ \citenamefont
  {Motrunich}(2002)}]{Senthil2002}%
  \BibitemOpen
  \bibfield  {author} {\bibinfo {author} {\bibnamefont {Senthil}, \bibfnamefont
  {T}}, \ and\ \bibinfo {author} {\bibfnamefont {O.}~\bibnamefont {Motrunich}}}
  (\bibinfo {year} {2002}),\ \bibfield  {title} {\enquote {\bibinfo {title}
  {{Microscopic models for fractionalized phases in strongly correlated
  systems}},}\ }\href {\doibase 10.1103/PhysRevB.66.205104} {\bibfield
  {journal} {\bibinfo  {journal} {Phys. Rev. B}\ }\textbf {\bibinfo {volume}
  {66}},\ \bibinfo {pages} {205104}}\BibitemShut {NoStop}%
\bibitem [{\citenamefont {Shastry}\ and\ \citenamefont
  {Sutherland}(1981)}]{SS1981}%
  \BibitemOpen
  \bibfield  {author} {\bibinfo {author} {\bibnamefont {Shastry}, \bibfnamefont
  {B~Sriram}}, \ and\ \bibinfo {author} {\bibfnamefont {Bill}\ \bibnamefont
  {Sutherland}}} (\bibinfo {year} {1981}),\ \bibfield  {title} {\enquote
  {\bibinfo {title} {Exact ground state of a quantum mechanical
  antiferromagnet},}\ }\href {\doibase 10.1016/0378-4363(81)90838-x} {\bibfield
   {journal} {\bibinfo  {journal} {{Physica B+C}}\ }\textbf {\bibinfo {volume}
  {108}}~(\bibinfo {number} {1-3}),\ \bibinfo {pages} {1069--1070}}\BibitemShut
  {NoStop}%
\bibitem [{\citenamefont {Shimizu}\ \emph {et~al.}(2003)\citenamefont
  {Shimizu}, \citenamefont {Miyagawa}, \citenamefont {Kanoda}, \citenamefont
  {Maesato},\ and\ \citenamefont {Saito}}]{Saito2003}%
  \BibitemOpen
  \bibfield  {author} {\bibinfo {author} {\bibnamefont {Shimizu}, \bibfnamefont
  {Y}}, \bibinfo {author} {\bibfnamefont {K.}~\bibnamefont {Miyagawa}},
  \bibinfo {author} {\bibfnamefont {K.}~\bibnamefont {Kanoda}}, \bibinfo
  {author} {\bibfnamefont {M.}~\bibnamefont {Maesato}}, \ and\ \bibinfo
  {author} {\bibfnamefont {G.}~\bibnamefont {Saito}}} (\bibinfo {year}
  {2003}),\ \bibfield  {title} {\enquote {\bibinfo {title} {{Spin liquid state
  in an organic Mott insulator with a triangular lattice}},}\ }\href {\doibase
  10.1103/PhysRevLett.91.107001} {\bibfield  {journal} {\bibinfo  {journal}
  {Phys. Rev. Lett.}\ }\textbf {\bibinfo {volume} {91}},\ \bibinfo {pages}
  {107001}}\BibitemShut {NoStop}%
\bibitem [{\citenamefont {Shindou}\ \emph
  {et~al.}(2013{\natexlab{a}})\citenamefont {Shindou}, \citenamefont
  {Matsumoto}, \citenamefont {Murakami},\ and\ \citenamefont
  {Ohe}}]{Shindou2013B}%
  \BibitemOpen
  \bibfield  {author} {\bibinfo {author} {\bibnamefont {Shindou}, \bibfnamefont
  {Ryuichi}}, \bibinfo {author} {\bibfnamefont {Ryo}\ \bibnamefont
  {Matsumoto}}, \bibinfo {author} {\bibfnamefont {Shuichi}\ \bibnamefont
  {Murakami}}, \ and\ \bibinfo {author} {\bibfnamefont {Jun-ichiro}\
  \bibnamefont {Ohe}}} (\bibinfo {year} {2013}{\natexlab{a}}),\ \bibfield
  {title} {\enquote {\bibinfo {title} {Topological chiral magnonic edge mode in
  a magnonic crystal},}\ }\href {\doibase 10.1103/PhysRevB.87.174427}
  {\bibfield  {journal} {\bibinfo  {journal} {Phys. Rev. B}\ }\textbf {\bibinfo
  {volume} {87}},\ \bibinfo {pages} {174427}}\BibitemShut {NoStop}%
\bibitem [{\citenamefont {Shindou}\ \emph
  {et~al.}(2013{\natexlab{b}})\citenamefont {Shindou}, \citenamefont {Ohe},
  \citenamefont {Matsumoto}, \citenamefont {Murakami},\ and\ \citenamefont
  {Saitoh}}]{Shindou2013A}%
  \BibitemOpen
  \bibfield  {author} {\bibinfo {author} {\bibnamefont {Shindou}, \bibfnamefont
  {Ryuichi}}, \bibinfo {author} {\bibfnamefont {Jun-ichiro}\ \bibnamefont
  {Ohe}}, \bibinfo {author} {\bibfnamefont {Ryo}\ \bibnamefont {Matsumoto}},
  \bibinfo {author} {\bibfnamefont {Shuichi}\ \bibnamefont {Murakami}}, \ and\
  \bibinfo {author} {\bibfnamefont {Eiji}\ \bibnamefont {Saitoh}}} (\bibinfo
  {year} {2013}{\natexlab{b}}),\ \bibfield  {title} {\enquote {\bibinfo {title}
  {{Chiral spin-wave edge modes in dipolar magnetic thin films}},}\ }\href
  {\doibase 10.1103/PhysRevB.87.174402} {\bibfield  {journal} {\bibinfo
  {journal} {Phys. Rev. B}\ }\textbf {\bibinfo {volume} {87}},\ \bibinfo
  {pages} {174402}}\BibitemShut {NoStop}%
\bibitem [{\citenamefont {Sibille}\ \emph {et~al.}(2020)\citenamefont
  {Sibille}, \citenamefont {Gauthier}, \citenamefont {Lhotel}, \citenamefont
  {Por{\'{e}}e}, \citenamefont {Pomjakushin}, \citenamefont {Ewings},
  \citenamefont {Perring}, \citenamefont {Ollivier}, \citenamefont {Wildes},
  \citenamefont {Ritter}, \citenamefont {Hansen}, \citenamefont {Keen},
  \citenamefont {Nilsen}, \citenamefont {Keller}, \citenamefont {Petit},\ and\
  \citenamefont {Fennell}}]{Sibille_2020}%
  \BibitemOpen
  \bibfield  {author} {\bibinfo {author} {\bibnamefont {Sibille}, \bibfnamefont
  {Romain}}, \bibinfo {author} {\bibfnamefont {Nicolas}\ \bibnamefont
  {Gauthier}}, \bibinfo {author} {\bibfnamefont {Elsa}\ \bibnamefont {Lhotel}},
  \bibinfo {author} {\bibfnamefont {Victor}\ \bibnamefont {Por{\'{e}}e}},
  \bibinfo {author} {\bibfnamefont {Vladimir}\ \bibnamefont {Pomjakushin}},
  \bibinfo {author} {\bibfnamefont {Russell~A.}\ \bibnamefont {Ewings}},
  \bibinfo {author} {\bibfnamefont {Toby~G.}\ \bibnamefont {Perring}}, \bibinfo
  {author} {\bibfnamefont {Jacques}\ \bibnamefont {Ollivier}}, \bibinfo
  {author} {\bibfnamefont {Andrew}\ \bibnamefont {Wildes}}, \bibinfo {author}
  {\bibfnamefont {Clemens}\ \bibnamefont {Ritter}}, \bibinfo {author}
  {\bibfnamefont {Thomas~C.}\ \bibnamefont {Hansen}}, \bibinfo {author}
  {\bibfnamefont {David~A.}\ \bibnamefont {Keen}}, \bibinfo {author}
  {\bibfnamefont {G{\o}ran~J.}\ \bibnamefont {Nilsen}}, \bibinfo {author}
  {\bibfnamefont {Lukas}\ \bibnamefont {Keller}}, \bibinfo {author}
  {\bibfnamefont {Sylvain}\ \bibnamefont {Petit}}, \ and\ \bibinfo {author}
  {\bibfnamefont {Tom}\ \bibnamefont {Fennell}}} (\bibinfo {year} {2020}),\
  \bibfield  {title} {\enquote {\bibinfo {title} {A quantum liquid of magnetic
  octupoles on the pyrochlore lattice},}\ }\href {\doibase
  10.1038/s41567-020-0827-7} {\bibfield  {journal} {\bibinfo  {journal} {Nature
  Physics}\ }\textbf {\bibinfo {volume} {16}}~(\bibinfo {number} {5}),\
  \bibinfo {pages} {546--552}}\BibitemShut {NoStop}%
\bibitem [{\citenamefont {Sibille}\ \emph {et~al.}(2015)\citenamefont
  {Sibille}, \citenamefont {Lhotel}, \citenamefont {Pomjakushin}, \citenamefont
  {Baines}, \citenamefont {Fennell},\ and\ \citenamefont
  {Kenzelmann}}]{Sibille2015}%
  \BibitemOpen
  \bibfield  {author} {\bibinfo {author} {\bibnamefont {Sibille}, \bibfnamefont
  {Romain}}, \bibinfo {author} {\bibfnamefont {Elsa}\ \bibnamefont {Lhotel}},
  \bibinfo {author} {\bibfnamefont {Vladimir}\ \bibnamefont {Pomjakushin}},
  \bibinfo {author} {\bibfnamefont {Chris}\ \bibnamefont {Baines}}, \bibinfo
  {author} {\bibfnamefont {Tom}\ \bibnamefont {Fennell}}, \ and\ \bibinfo
  {author} {\bibfnamefont {Michel}\ \bibnamefont {Kenzelmann}}} (\bibinfo
  {year} {2015}),\ \bibfield  {title} {\enquote {\bibinfo {title} {{Candidate
  quantum spin liquid in the ${\mathrm{Ce}}^{3+}$ pyrochlore stannate
  ${\mathrm{Ce}}_{2}{\mathrm{Sn}}_{2}{\mathrm{O}}_{7}$}},}\ }\href {\doibase
  10.1103/PhysRevLett.115.097202} {\bibfield  {journal} {\bibinfo  {journal}
  {Phys. Rev. Lett.}\ }\textbf {\bibinfo {volume} {115}},\ \bibinfo {pages}
  {097202}}\BibitemShut {NoStop}%
\bibitem [{\citenamefont {Singh}(2011)}]{Singh2011}%
  \BibitemOpen
  \bibfield  {author} {\bibinfo {author} {\bibnamefont {Singh}, \bibfnamefont
  {Rajiv}}} (\bibinfo {year} {2011}),\ \bibfield  {title} {\enquote {\bibinfo
  {title} {Spinning on ice},}\ }\href {\doibase 10.1103/physics.4.77}
  {\bibfield  {journal} {\bibinfo  {journal} {Physics}\ }\textbf {\bibinfo
  {volume} {4}},\ \bibinfo {pages} {77}}\BibitemShut {NoStop}%
\bibitem [{\citenamefont {Singh}\ and\ \citenamefont
  {Gegenwart}(2010)}]{Singh2010}%
  \BibitemOpen
  \bibfield  {author} {\bibinfo {author} {\bibnamefont {Singh}, \bibfnamefont
  {Yogesh}}, \ and\ \bibinfo {author} {\bibfnamefont {P.}~\bibnamefont
  {Gegenwart}}} (\bibinfo {year} {2010}),\ \bibfield  {title} {\enquote
  {\bibinfo {title} {{Antiferromagnetic Mott insulating state in single
  crystals of the honeycomb lattice material
  ${\text{Na}}_{2}{\text{IrO}}_{3}$}},}\ }\href {\doibase
  10.1103/PhysRevB.82.064412} {\bibfield  {journal} {\bibinfo  {journal} {Phys.
  Rev. B}\ }\textbf {\bibinfo {volume} {82}},\ \bibinfo {pages}
  {064412}}\BibitemShut {NoStop}%
\bibitem [{\citenamefont {Singh}\ \emph {et~al.}(2012)\citenamefont {Singh},
  \citenamefont {Manni}, \citenamefont {Reuther}, \citenamefont {Berlijn},
  \citenamefont {Thomale}, \citenamefont {Ku}, \citenamefont {Trebst},\ and\
  \citenamefont {Gegenwart}}]{Singh2012}%
  \BibitemOpen
  \bibfield  {author} {\bibinfo {author} {\bibnamefont {Singh}, \bibfnamefont
  {Yogesh}}, \bibinfo {author} {\bibfnamefont {S.}~\bibnamefont {Manni}},
  \bibinfo {author} {\bibfnamefont {J.}~\bibnamefont {Reuther}}, \bibinfo
  {author} {\bibfnamefont {T.}~\bibnamefont {Berlijn}}, \bibinfo {author}
  {\bibfnamefont {R.}~\bibnamefont {Thomale}}, \bibinfo {author} {\bibfnamefont
  {W.}~\bibnamefont {Ku}}, \bibinfo {author} {\bibfnamefont {S.}~\bibnamefont
  {Trebst}}, \ and\ \bibinfo {author} {\bibfnamefont {P.}~\bibnamefont
  {Gegenwart}}} (\bibinfo {year} {2012}),\ \bibfield  {title} {\enquote
  {\bibinfo {title} {{Relevance of the Heisenberg-Kitaev Model for the
  Honeycomb Lattice Iridates ${A}_{2}{\mathrm{IrO}}_{3}$}},}\ }\href {\doibase
  10.1103/PhysRevLett.108.127203} {\bibfield  {journal} {\bibinfo  {journal}
  {Phys. Rev. Lett.}\ }\textbf {\bibinfo {volume} {108}},\ \bibinfo {pages}
  {127203}}\BibitemShut {NoStop}%
\bibitem [{\citenamefont {Smith}\ \emph {et~al.}(2022)\citenamefont {Smith},
  \citenamefont {Benton}, \citenamefont {Yahne}, \citenamefont {Placke},
  \citenamefont {Sch\"afer}, \citenamefont {Gaudet}, \citenamefont {Dudemaine},
  \citenamefont {Fitterman}, \citenamefont {Beare}, \citenamefont {Wildes},
  \citenamefont {Bhattacharya}, \citenamefont {DeLazzer}, \citenamefont
  {Buhariwalla}, \citenamefont {Butch}, \citenamefont {Movshovich},
  \citenamefont {Garrett}, \citenamefont {Marjerrison}, \citenamefont {Clancy},
  \citenamefont {Kermarrec}, \citenamefont {Luke}, \citenamefont {Bianchi},
  \citenamefont {Ross},\ and\ \citenamefont {Gaulin}}]{PhysRevX.12.021015}%
  \BibitemOpen
  \bibfield  {author} {\bibinfo {author} {\bibnamefont {Smith}, \bibfnamefont
  {E~M}}, \bibinfo {author} {\bibfnamefont {O.}~\bibnamefont {Benton}},
  \bibinfo {author} {\bibfnamefont {D.~R.}\ \bibnamefont {Yahne}}, \bibinfo
  {author} {\bibfnamefont {B.}~\bibnamefont {Placke}}, \bibinfo {author}
  {\bibfnamefont {R.}~\bibnamefont {Sch\"afer}}, \bibinfo {author}
  {\bibfnamefont {J.}~\bibnamefont {Gaudet}}, \bibinfo {author} {\bibfnamefont
  {J.}~\bibnamefont {Dudemaine}}, \bibinfo {author} {\bibfnamefont
  {A.}~\bibnamefont {Fitterman}}, \bibinfo {author} {\bibfnamefont
  {J.}~\bibnamefont {Beare}}, \bibinfo {author} {\bibfnamefont {A.~R.}\
  \bibnamefont {Wildes}}, \bibinfo {author} {\bibfnamefont {S.}~\bibnamefont
  {Bhattacharya}}, \bibinfo {author} {\bibfnamefont {T.}~\bibnamefont
  {DeLazzer}}, \bibinfo {author} {\bibfnamefont {C.~R.~C.}\ \bibnamefont
  {Buhariwalla}}, \bibinfo {author} {\bibfnamefont {N.~P.}\ \bibnamefont
  {Butch}}, \bibinfo {author} {\bibfnamefont {R.}~\bibnamefont {Movshovich}},
  \bibinfo {author} {\bibfnamefont {J.~D.}\ \bibnamefont {Garrett}}, \bibinfo
  {author} {\bibfnamefont {C.~A.}\ \bibnamefont {Marjerrison}}, \bibinfo
  {author} {\bibfnamefont {J.~P.}\ \bibnamefont {Clancy}}, \bibinfo {author}
  {\bibfnamefont {E.}~\bibnamefont {Kermarrec}}, \bibinfo {author}
  {\bibfnamefont {G.~M.}\ \bibnamefont {Luke}}, \bibinfo {author}
  {\bibfnamefont {A.~D.}\ \bibnamefont {Bianchi}}, \bibinfo {author}
  {\bibfnamefont {K.~A.}\ \bibnamefont {Ross}}, \ and\ \bibinfo {author}
  {\bibfnamefont {B.~D.}\ \bibnamefont {Gaulin}}} (\bibinfo {year} {2022}),\
  \bibfield  {title} {\enquote {\bibinfo {title} {{Case for a
  ${\mathrm{U}(1)}_{\ensuremath{\pi}}$ Quantum Spin Liquid Ground State in the
  Dipole-Octupole Pyrochlore
  ${\mathrm{Ce}}_{2}{\mathrm{Zr}}_{2}{\mathrm{O}}_{7}$}},}\ }\href {\doibase
  10.1103/PhysRevX.12.021015} {\bibfield  {journal} {\bibinfo  {journal} {Phys.
  Rev. X}\ }\textbf {\bibinfo {volume} {12}},\ \bibinfo {pages}
  {021015}}\BibitemShut {NoStop}%
\bibitem [{\citenamefont {Sodemann}\ \emph {et~al.}(2018)\citenamefont
  {Sodemann}, \citenamefont {Chowdhury},\ and\ \citenamefont
  {Senthil}}]{PhysRevB.97.045152}%
  \BibitemOpen
  \bibfield  {author} {\bibinfo {author} {\bibnamefont {Sodemann},
  \bibfnamefont {Inti}}, \bibinfo {author} {\bibfnamefont {Debanjan}\
  \bibnamefont {Chowdhury}}, \ and\ \bibinfo {author} {\bibfnamefont
  {T.}~\bibnamefont {Senthil}}} (\bibinfo {year} {2018}),\ \bibfield  {title}
  {\enquote {\bibinfo {title} {{Quantum oscillations in insulators with neutral
  Fermi surfaces}},}\ }\href {\doibase 10.1103/PhysRevB.97.045152} {\bibfield
  {journal} {\bibinfo  {journal} {Phys. Rev. B}\ }\textbf {\bibinfo {volume}
  {97}},\ \bibinfo {pages} {045152}}\BibitemShut {NoStop}%
\bibitem [{\citenamefont {Song}\ and\ \citenamefont
  {Senthil}(2022)}]{song2022translationenriched}%
  \BibitemOpen
  \bibfield  {author} {\bibinfo {author} {\bibnamefont {Song}, \bibfnamefont
  {Xue-Yang}}, \ and\ \bibinfo {author} {\bibfnamefont {T.}~\bibnamefont
  {Senthil}}} (\bibinfo {year} {2022}),\ \href@noop {} {\enquote {\bibinfo
  {title} {{Translation-enriched $Z_2$ spin liquids and topological vison
  bands: Possible application to $\alpha$-RuCl$_3$}},}\ }\Eprint
  {http://arxiv.org/abs/2206.14197} {arXiv:2206.14197 [cond-mat.str-el]}
  \BibitemShut {NoStop}%
\bibitem [{\citenamefont {Song}\ \emph {et~al.}(2019)\citenamefont {Song},
  \citenamefont {Wang}, \citenamefont {Vishwanath},\ and\ \citenamefont
  {He}}]{DSLMotherState}%
  \BibitemOpen
  \bibfield  {author} {\bibinfo {author} {\bibnamefont {Song}, \bibfnamefont
  {Xue-Yang}}, \bibinfo {author} {\bibfnamefont {Chong}\ \bibnamefont {Wang}},
  \bibinfo {author} {\bibfnamefont {Ashvin}\ \bibnamefont {Vishwanath}}, \ and\
  \bibinfo {author} {\bibfnamefont {Yin-Chen}\ \bibnamefont {He}}} (\bibinfo
  {year} {2019}),\ \bibfield  {title} {\enquote {\bibinfo {title} {Unifying
  description of competing orders in two-dimensional quantum magnets},}\ }\href
  {https://doi.org/10.1038%2Fs41467-019-11727-3} {\bibfield  {journal}
  {\bibinfo  {journal} {Nature Communications}\ }\textbf {\bibinfo {volume}
  {10}}~(\bibinfo {number} {1}),\ \bibinfo {pages} {4254}}\BibitemShut
  {NoStop}%
\bibitem [{\citenamefont {Spivak}\ \emph {et~al.}(2010)\citenamefont {Spivak},
  \citenamefont {Kravchenko}, \citenamefont {Kivelson},\ and\ \citenamefont
  {Gao}}]{RevModPhys.82.1743}%
  \BibitemOpen
  \bibfield  {author} {\bibinfo {author} {\bibnamefont {Spivak}, \bibfnamefont
  {B}}, \bibinfo {author} {\bibfnamefont {S.~V.}\ \bibnamefont {Kravchenko}},
  \bibinfo {author} {\bibfnamefont {S.~A.}\ \bibnamefont {Kivelson}}, \ and\
  \bibinfo {author} {\bibfnamefont {X.~P.~A.}\ \bibnamefont {Gao}}} (\bibinfo
  {year} {2010}),\ \bibfield  {title} {\enquote {\bibinfo {title} {Colloquium:
  Transport in strongly correlated two dimensional electron fluids},}\ }\href
  {\doibase 10.1103/RevModPhys.82.1743} {\bibfield  {journal} {\bibinfo
  {journal} {Rev. Mod. Phys.}\ }\textbf {\bibinfo {volume} {82}},\ \bibinfo
  {pages} {1743--1766}}\BibitemShut {NoStop}%
\bibitem [{\citenamefont {Starykh}(2015)}]{Starykh_2015}%
  \BibitemOpen
  \bibfield  {author} {\bibinfo {author} {\bibnamefont {Starykh}, \bibfnamefont
  {Oleg~A}}} (\bibinfo {year} {2015}),\ \bibfield  {title} {\enquote {\bibinfo
  {title} {Unusual ordered phases of highly frustrated magnets: a review},}\
  }\href {\doibase 10.1088/0034-4885/78/5/052502} {\bibfield  {journal}
  {\bibinfo  {journal} {Reports on Progress in Physics}\ }\textbf {\bibinfo
  {volume} {78}}~(\bibinfo {number} {5}),\ \bibinfo {pages}
  {052502}}\BibitemShut {NoStop}%
\bibitem [{\citenamefont {Stone}(2012)}]{PhysRevB.85.184503}%
  \BibitemOpen
  \bibfield  {author} {\bibinfo {author} {\bibnamefont {Stone}, \bibfnamefont
  {Michael}}} (\bibinfo {year} {2012}),\ \bibfield  {title} {\enquote {\bibinfo
  {title} {{Gravitational anomalies and thermal Hall effect in topological
  insulators}},}\ }\href {\doibase 10.1103/PhysRevB.85.184503} {\bibfield
  {journal} {\bibinfo  {journal} {Phys. Rev. B}\ }\textbf {\bibinfo {volume}
  {85}},\ \bibinfo {pages} {184503}}\BibitemShut {NoStop}%
\bibitem [{\citenamefont {Strohm}\ \emph {et~al.}(2005)\citenamefont {Strohm},
  \citenamefont {Rikken},\ and\ \citenamefont {Wyder}}]{PhysRevLett.95.155901}%
  \BibitemOpen
  \bibfield  {author} {\bibinfo {author} {\bibnamefont {Strohm}, \bibfnamefont
  {C}}, \bibinfo {author} {\bibfnamefont {G.~L. J.~A.}\ \bibnamefont {Rikken}},
  \ and\ \bibinfo {author} {\bibfnamefont {P.}~\bibnamefont {Wyder}}} (\bibinfo
  {year} {2005}),\ \bibfield  {title} {\enquote {\bibinfo {title}
  {{Phenomenological Evidence for the Phonon Hall Effect}},}\ }\href {\doibase
  10.1103/PhysRevLett.95.155901} {\bibfield  {journal} {\bibinfo  {journal}
  {Phys. Rev. Lett.}\ }\textbf {\bibinfo {volume} {95}},\ \bibinfo {pages}
  {155901}}\BibitemShut {NoStop}%
\bibitem [{\citenamefont {Su}\ \emph {et~al.}(2019)\citenamefont {Su},
  \citenamefont {Li}, \citenamefont {Jiao}, \citenamefont {Liu}, \citenamefont
  {Sun}, \citenamefont {Wang}, \citenamefont {Sui}, \citenamefont {Zhou},
  \citenamefont {Chen},\ and\ \citenamefont {Cheng}}]{Su_2019}%
  \BibitemOpen
  \bibfield  {author} {\bibinfo {author} {\bibnamefont {Su}, \bibfnamefont
  {Na}}, \bibinfo {author} {\bibfnamefont {Feiye}\ \bibnamefont {Li}}, \bibinfo
  {author} {\bibfnamefont {Yuanyuan}\ \bibnamefont {Jiao}}, \bibinfo {author}
  {\bibfnamefont {Ziyi}\ \bibnamefont {Liu}}, \bibinfo {author} {\bibfnamefont
  {Jianping}\ \bibnamefont {Sun}}, \bibinfo {author} {\bibfnamefont {Bosen}\
  \bibnamefont {Wang}}, \bibinfo {author} {\bibfnamefont {Yu}~\bibnamefont
  {Sui}}, \bibinfo {author} {\bibfnamefont {Haidong}\ \bibnamefont {Zhou}},
  \bibinfo {author} {\bibfnamefont {Gang}\ \bibnamefont {Chen}}, \ and\
  \bibinfo {author} {\bibfnamefont {Jinguang}\ \bibnamefont {Cheng}}} (\bibinfo
  {year} {2019}),\ \bibfield  {title} {\enquote {\bibinfo {title} {{Asymmetric
  ferromagnetic criticality in pyrochlore ferromagnet Lu$_2$V$_2$O$_7$}},}\
  }\href {\doibase 10.1016/j.scib.2019.06.025} {\bibfield  {journal} {\bibinfo
  {journal} {Science Bulletin}\ }\textbf {\bibinfo {volume} {64}}~(\bibinfo
  {number} {17}),\ \bibinfo {pages} {1222--1227}}\BibitemShut {NoStop}%
\bibitem [{\citenamefont {Sugii}\ \emph
  {et~al.}(2017{\natexlab{a}})\citenamefont {Sugii}, \citenamefont {Shimozawa},
  \citenamefont {Watanabe}, \citenamefont {Suzuki}, \citenamefont {Halim},
  \citenamefont {Kimata}, \citenamefont {Matsumoto}, \citenamefont
  {Nakatsuji},\ and\ \citenamefont {Yamashita}}]{Yamashita2017}%
  \BibitemOpen
  \bibfield  {author} {\bibinfo {author} {\bibnamefont {Sugii}, \bibfnamefont
  {K}}, \bibinfo {author} {\bibfnamefont {M.}~\bibnamefont {Shimozawa}},
  \bibinfo {author} {\bibfnamefont {D.}~\bibnamefont {Watanabe}}, \bibinfo
  {author} {\bibfnamefont {Y.}~\bibnamefont {Suzuki}}, \bibinfo {author}
  {\bibfnamefont {M.}~\bibnamefont {Halim}}, \bibinfo {author} {\bibfnamefont
  {M.}~\bibnamefont {Kimata}}, \bibinfo {author} {\bibfnamefont
  {Y.}~\bibnamefont {Matsumoto}}, \bibinfo {author} {\bibfnamefont
  {S.}~\bibnamefont {Nakatsuji}}, \ and\ \bibinfo {author} {\bibfnamefont
  {M.}~\bibnamefont {Yamashita}}} (\bibinfo {year} {2017}{\natexlab{a}}),\
  \bibfield  {title} {\enquote {\bibinfo {title} {{Thermal Hall effect in a
  phonon-glass ${\mathrm{Ba}}_{3}{\mathrm{CuSb}}_{2}{\mathrm{O}}_{9}$}},}\
  }\href {\doibase 10.1103/PhysRevLett.118.145902} {\bibfield  {journal}
  {\bibinfo  {journal} {Phys. Rev. Lett.}\ }\textbf {\bibinfo {volume} {118}},\
  \bibinfo {pages} {145902}}\BibitemShut {NoStop}%
\bibitem [{\citenamefont {Sugii}\ \emph
  {et~al.}(2017{\natexlab{b}})\citenamefont {Sugii}, \citenamefont {Shimozawa},
  \citenamefont {Watanabe}, \citenamefont {Suzuki}, \citenamefont {Halim},
  \citenamefont {Kimata}, \citenamefont {Matsumoto}, \citenamefont
  {Nakatsuji},\ and\ \citenamefont {Yamashita}}]{PhysRevLett.118.145902}%
  \BibitemOpen
  \bibfield  {author} {\bibinfo {author} {\bibnamefont {Sugii}, \bibfnamefont
  {K}}, \bibinfo {author} {\bibfnamefont {M.}~\bibnamefont {Shimozawa}},
  \bibinfo {author} {\bibfnamefont {D.}~\bibnamefont {Watanabe}}, \bibinfo
  {author} {\bibfnamefont {Y.}~\bibnamefont {Suzuki}}, \bibinfo {author}
  {\bibfnamefont {M.}~\bibnamefont {Halim}}, \bibinfo {author} {\bibfnamefont
  {M.}~\bibnamefont {Kimata}}, \bibinfo {author} {\bibfnamefont
  {Y.}~\bibnamefont {Matsumoto}}, \bibinfo {author} {\bibfnamefont
  {S.}~\bibnamefont {Nakatsuji}}, \ and\ \bibinfo {author} {\bibfnamefont
  {M.}~\bibnamefont {Yamashita}}} (\bibinfo {year} {2017}{\natexlab{b}}),\
  \bibfield  {title} {\enquote {\bibinfo {title} {{Thermal Hall Effect in a
  Phonon-Glass ${\mathrm{Ba}}_{3}{\mathrm{CuSb}}_{2}{\mathrm{O}}_{9}$}},}\
  }\href {\doibase 10.1103/PhysRevLett.118.145902} {\bibfield  {journal}
  {\bibinfo  {journal} {Phys. Rev. Lett.}\ }\textbf {\bibinfo {volume} {118}},\
  \bibinfo {pages} {145902}}\BibitemShut {NoStop}%
\bibitem [{\citenamefont {Sun}\ \emph {et~al.}(2021)\citenamefont {Sun},
  \citenamefont {Sengupta}, \citenamefont {Nam},\ and\ \citenamefont
  {Yang}}]{Sun2021}%
  \BibitemOpen
  \bibfield  {author} {\bibinfo {author} {\bibnamefont {Sun}, \bibfnamefont
  {Hao}}, \bibinfo {author} {\bibfnamefont {Pinaki}\ \bibnamefont {Sengupta}},
  \bibinfo {author} {\bibfnamefont {Donguk}\ \bibnamefont {Nam}}, \ and\
  \bibinfo {author} {\bibfnamefont {Bo}~\bibnamefont {Yang}}} (\bibinfo {year}
  {2021}),\ \bibfield  {title} {\enquote {\bibinfo {title} {{Negative thermal
  Hall conductance in a two-dimer Shastry-Sutherland model with a
  $\ensuremath{\pi}$-flux Dirac triplon}},}\ }\href {\doibase
  10.1103/PhysRevB.103.L140404} {\bibfield  {journal} {\bibinfo  {journal}
  {Phys. Rev. B}\ }\textbf {\bibinfo {volume} {103}},\ \bibinfo {pages}
  {L140404}}\BibitemShut {NoStop}%
\bibitem [{\citenamefont {Suzuki}\ \emph {et~al.}(2017)\citenamefont {Suzuki},
  \citenamefont {Koretsune}, \citenamefont {Ochi},\ and\ \citenamefont
  {Arita}}]{PhysRevB.95.094406}%
  \BibitemOpen
  \bibfield  {author} {\bibinfo {author} {\bibnamefont {Suzuki}, \bibfnamefont
  {M-T}}, \bibinfo {author} {\bibfnamefont {T.}~\bibnamefont {Koretsune}},
  \bibinfo {author} {\bibfnamefont {M.}~\bibnamefont {Ochi}}, \ and\ \bibinfo
  {author} {\bibfnamefont {R.}~\bibnamefont {Arita}}} (\bibinfo {year}
  {2017}),\ \bibfield  {title} {\enquote {\bibinfo {title} {{Cluster multipole
  theory for anomalous Hall effect in antiferromagnets}},}\ }\href {\doibase
  10.1103/PhysRevB.95.094406} {\bibfield  {journal} {\bibinfo  {journal} {Phys.
  Rev. B}\ }\textbf {\bibinfo {volume} {95}},\ \bibinfo {pages}
  {094406}}\BibitemShut {NoStop}%
\bibitem [{\citenamefont {Szasz}\ \emph {et~al.}(2020)\citenamefont {Szasz},
  \citenamefont {Motruk}, \citenamefont {Zaletel},\ and\ \citenamefont
  {Moore}}]{PhysRevX.10.021042}%
  \BibitemOpen
  \bibfield  {author} {\bibinfo {author} {\bibnamefont {Szasz}, \bibfnamefont
  {Aaron}}, \bibinfo {author} {\bibfnamefont {Johannes}\ \bibnamefont
  {Motruk}}, \bibinfo {author} {\bibfnamefont {Michael~P.}\ \bibnamefont
  {Zaletel}}, \ and\ \bibinfo {author} {\bibfnamefont {Joel~E.}\ \bibnamefont
  {Moore}}} (\bibinfo {year} {2020}),\ \bibfield  {title} {\enquote {\bibinfo
  {title} {{Chiral Spin Liquid Phase of the Triangular Lattice Hubbard Model: A
  Density Matrix Renormalization Group Study}},}\ }\href {\doibase
  10.1103/PhysRevX.10.021042} {\bibfield  {journal} {\bibinfo  {journal} {Phys.
  Rev. X}\ }\textbf {\bibinfo {volume} {10}},\ \bibinfo {pages}
  {021042}}\BibitemShut {NoStop}%
\bibitem [{\citenamefont {Taguchi}\ \emph {et~al.}(2001)\citenamefont
  {Taguchi}, \citenamefont {Oohara}, \citenamefont {Yoshizawa}, \citenamefont
  {Nagaosa},\ and\ \citenamefont {Tokura}}]{Taguchi2001}%
  \BibitemOpen
  \bibfield  {author} {\bibinfo {author} {\bibnamefont {Taguchi}, \bibfnamefont
  {Y}}, \bibinfo {author} {\bibfnamefont {Y.}~\bibnamefont {Oohara}}, \bibinfo
  {author} {\bibfnamefont {H.}~\bibnamefont {Yoshizawa}}, \bibinfo {author}
  {\bibfnamefont {N.}~\bibnamefont {Nagaosa}}, \ and\ \bibinfo {author}
  {\bibfnamefont {Y.}~\bibnamefont {Tokura}}} (\bibinfo {year} {2001}),\
  \bibfield  {title} {\enquote {\bibinfo {title} {{Spin Chirality, Berry Phase,
  and Anomalous Hall Effect in a Frustrated Ferromagnet}},}\ }\href {\doibase
  10.1126/science.1058161} {\bibfield  {journal} {\bibinfo  {journal}
  {Science}\ }\textbf {\bibinfo {volume} {291}}~(\bibinfo {number} {5513}),\
  \bibinfo {pages} {2573--2576}}\BibitemShut {NoStop}%
\bibitem [{\citenamefont {Takagi}\ \emph {et~al.}(2019)\citenamefont {Takagi},
  \citenamefont {Takayama}, \citenamefont {Jackeli}, \citenamefont
  {Khaliullin},\ and\ \citenamefont {Nagler}}]{Takagi2019}%
  \BibitemOpen
  \bibfield  {author} {\bibinfo {author} {\bibnamefont {Takagi}, \bibfnamefont
  {Hidenori}}, \bibinfo {author} {\bibfnamefont {Tomohiro}\ \bibnamefont
  {Takayama}}, \bibinfo {author} {\bibfnamefont {George}\ \bibnamefont
  {Jackeli}}, \bibinfo {author} {\bibfnamefont {Giniyat}\ \bibnamefont
  {Khaliullin}}, \ and\ \bibinfo {author} {\bibfnamefont {Stephen~E.}\
  \bibnamefont {Nagler}}} (\bibinfo {year} {2019}),\ \bibfield  {title}
  {\enquote {\bibinfo {title} {{Concept and realization of Kitaev quantum spin
  liquids}},}\ }\href {\doibase 10.1038/s42254-019-0038-2} {\bibfield
  {journal} {\bibinfo  {journal} {Nature Reviews Physics}\ }\textbf {\bibinfo
  {volume} {1}}~(\bibinfo {number} {4}),\ \bibinfo {pages}
  {264--280}}\BibitemShut {NoStop}%
\bibitem [{\citenamefont {Takahashi}\ and\ \citenamefont
  {Nagaosa}(2016)}]{PhysRevLett.117.217205}%
  \BibitemOpen
  \bibfield  {author} {\bibinfo {author} {\bibnamefont {Takahashi},
  \bibfnamefont {Ryuji}}, \ and\ \bibinfo {author} {\bibfnamefont {Naoto}\
  \bibnamefont {Nagaosa}}} (\bibinfo {year} {2016}),\ \bibfield  {title}
  {\enquote {\bibinfo {title} {{Berry Curvature in Magnon-Phonon Hybrid
  Systems}},}\ }\href {\doibase 10.1103/PhysRevLett.117.217205} {\bibfield
  {journal} {\bibinfo  {journal} {Phys. Rev. Lett.}\ }\textbf {\bibinfo
  {volume} {117}},\ \bibinfo {pages} {217205}}\BibitemShut {NoStop}%
\bibitem [{\citenamefont {Takeda}\ \emph {et~al.}(2023)\citenamefont {Takeda},
  \citenamefont {Kawano}, \citenamefont {Tamura}, \citenamefont {Akazawa},
  \citenamefont {Yan}, \citenamefont {Waki}, \citenamefont {Nakamura},
  \citenamefont {Sato}, \citenamefont {Narumi}, \citenamefont {Hagiwara},
  \citenamefont {Yamashita},\ and\ \citenamefont {Hotta}}]{takeda2023emergent}%
  \BibitemOpen
  \bibfield  {author} {\bibinfo {author} {\bibnamefont {Takeda}, \bibfnamefont
  {Hikaru}}, \bibinfo {author} {\bibfnamefont {Masataka}\ \bibnamefont
  {Kawano}}, \bibinfo {author} {\bibfnamefont {Kyo}\ \bibnamefont {Tamura}},
  \bibinfo {author} {\bibfnamefont {Masatoshi}\ \bibnamefont {Akazawa}},
  \bibinfo {author} {\bibfnamefont {Jian}\ \bibnamefont {Yan}}, \bibinfo
  {author} {\bibfnamefont {Takeshi}\ \bibnamefont {Waki}}, \bibinfo {author}
  {\bibfnamefont {Hiroyuki}\ \bibnamefont {Nakamura}}, \bibinfo {author}
  {\bibfnamefont {Kazuki}\ \bibnamefont {Sato}}, \bibinfo {author}
  {\bibfnamefont {Yasuo}\ \bibnamefont {Narumi}}, \bibinfo {author}
  {\bibfnamefont {Masayuki}\ \bibnamefont {Hagiwara}}, \bibinfo {author}
  {\bibfnamefont {Minoru}\ \bibnamefont {Yamashita}}, \ and\ \bibinfo {author}
  {\bibfnamefont {Chisa}\ \bibnamefont {Hotta}}} (\bibinfo {year} {2023}),\
  \href@noop {} {\enquote {\bibinfo {title} {{Emergent SU(3) magnons and
  thermal Hall effect in the antiferromagnetic skyrmion lattice}},}\ }\Eprint
  {http://arxiv.org/abs/2304.08029} {arXiv:2304.08029 [cond-mat.str-el]}
  \BibitemShut {NoStop}%
\bibitem [{\citenamefont {Takikawa}\ and\ \citenamefont
  {Fujimoto}(2019)}]{Fujimoto2019}%
  \BibitemOpen
  \bibfield  {author} {\bibinfo {author} {\bibnamefont {Takikawa},
  \bibfnamefont {Daichi}}, \ and\ \bibinfo {author} {\bibfnamefont {Satoshi}\
  \bibnamefont {Fujimoto}}} (\bibinfo {year} {2019}),\ \bibfield  {title}
  {\enquote {\bibinfo {title} {{Impact of off-diagonal exchange interactions on
  the Kitaev spin-liquid state of
  $\ensuremath{\alpha}\text{\ensuremath{-}}{\mathrm{RuCl}}_{3}$}},}\ }\href
  {\doibase 10.1103/PhysRevB.99.224409} {\bibfield  {journal} {\bibinfo
  {journal} {Phys. Rev. B}\ }\textbf {\bibinfo {volume} {99}},\ \bibinfo
  {pages} {224409}}\BibitemShut {NoStop}%
\bibitem [{\citenamefont {Teng}\ \emph {et~al.}(2020)\citenamefont {Teng},
  \citenamefont {Zhang}, \citenamefont {Samajdar}, \citenamefont {Scheurer},\
  and\ \citenamefont {Sachdev}}]{Sachdev2020}%
  \BibitemOpen
  \bibfield  {author} {\bibinfo {author} {\bibnamefont {Teng}, \bibfnamefont
  {Yanting}}, \bibinfo {author} {\bibfnamefont {Yunchao}\ \bibnamefont
  {Zhang}}, \bibinfo {author} {\bibfnamefont {Rhine}\ \bibnamefont {Samajdar}},
  \bibinfo {author} {\bibfnamefont {Mathias~S.}\ \bibnamefont {Scheurer}}, \
  and\ \bibinfo {author} {\bibfnamefont {Subir}\ \bibnamefont {Sachdev}}}
  (\bibinfo {year} {2020}),\ \bibfield  {title} {\enquote {\bibinfo {title}
  {{Unquantized thermal Hall effect in quantum spin liquids with spinon Fermi
  surfaces}},}\ }\href {\doibase 10.1103/PhysRevResearch.2.033283} {\bibfield
  {journal} {\bibinfo  {journal} {Phys. Rev. Research}\ }\textbf {\bibinfo
  {volume} {2}},\ \bibinfo {pages} {033283}}\BibitemShut {NoStop}%
\bibitem [{\citenamefont {Tokiwa}\ \emph {et~al.}(2018)\citenamefont {Tokiwa},
  \citenamefont {Yamashita}, \citenamefont {Terazawa}, \citenamefont {Kimura},
  \citenamefont {Kasahara}, \citenamefont {Onishi}, \citenamefont {Kato},
  \citenamefont {Halim}, \citenamefont {Gegenwart}, \citenamefont {Shibauchi},
  \citenamefont {Nakatsuji}, \citenamefont {Moon},\ and\ \citenamefont
  {Matsuda}}]{doi:10.7566/JPSJ.87.064702}%
  \BibitemOpen
  \bibfield  {author} {\bibinfo {author} {\bibnamefont {Tokiwa}, \bibfnamefont
  {Yoshifumi}}, \bibinfo {author} {\bibfnamefont {Takuya}\ \bibnamefont
  {Yamashita}}, \bibinfo {author} {\bibfnamefont {Daiki}\ \bibnamefont
  {Terazawa}}, \bibinfo {author} {\bibfnamefont {Kenta}\ \bibnamefont
  {Kimura}}, \bibinfo {author} {\bibfnamefont {Yuichi}\ \bibnamefont
  {Kasahara}}, \bibinfo {author} {\bibfnamefont {Takafumi}\ \bibnamefont
  {Onishi}}, \bibinfo {author} {\bibfnamefont {Yasuyuki}\ \bibnamefont {Kato}},
  \bibinfo {author} {\bibfnamefont {Mario}\ \bibnamefont {Halim}}, \bibinfo
  {author} {\bibfnamefont {Philipp}\ \bibnamefont {Gegenwart}}, \bibinfo
  {author} {\bibfnamefont {Takasada}\ \bibnamefont {Shibauchi}}, \bibinfo
  {author} {\bibfnamefont {Satoru}\ \bibnamefont {Nakatsuji}}, \bibinfo
  {author} {\bibfnamefont {Eun-Gook}\ \bibnamefont {Moon}}, \ and\ \bibinfo
  {author} {\bibfnamefont {Yuji}\ \bibnamefont {Matsuda}}} (\bibinfo {year}
  {2018}),\ \bibfield  {title} {\enquote {\bibinfo {title} {{Discovery of
  Emergent Photon and Monopoles in a Quantum Spin Liquid}},}\ }\href {\doibase
  10.7566/JPSJ.87.064702} {\bibfield  {journal} {\bibinfo  {journal} {Journal
  of the Physical Society of Japan}\ }\textbf {\bibinfo {volume}
  {87}}~(\bibinfo {number} {6}),\ \bibinfo {pages} {064702}},\ \Eprint
  {http://arxiv.org/abs/https://doi.org/10.7566/JPSJ.87.064702}
  {https://doi.org/10.7566/JPSJ.87.064702} \BibitemShut {NoStop}%
\bibitem [{\citenamefont {Trebst}\ and\ \citenamefont
  {Hickey}(2022)}]{Trebst2022}%
  \BibitemOpen
  \bibfield  {author} {\bibinfo {author} {\bibnamefont {Trebst}, \bibfnamefont
  {Simon}}, \ and\ \bibinfo {author} {\bibfnamefont {Ciar{\'{a}}n}\
  \bibnamefont {Hickey}}} (\bibinfo {year} {2022}),\ \bibfield  {title}
  {\enquote {\bibinfo {title} {Kitaev materials},}\ }\href {\doibase
  10.1016/j.physrep.2021.11.003} {\bibfield  {journal} {\bibinfo  {journal}
  {Physics Reports}\ }\textbf {\bibinfo {volume} {950}},\ \bibinfo {pages}
  {1--37}}\BibitemShut {NoStop}%
\bibitem [{\citenamefont {Trumper}\ \emph {et~al.}(1997)\citenamefont
  {Trumper}, \citenamefont {Manuel}, \citenamefont {Gazza},\ and\ \citenamefont
  {Ceccatto}}]{PhysRevLett.78.2216}%
  \BibitemOpen
  \bibfield  {author} {\bibinfo {author} {\bibnamefont {Trumper}, \bibfnamefont
  {A~E}}, \bibinfo {author} {\bibfnamefont {L.~O.}\ \bibnamefont {Manuel}},
  \bibinfo {author} {\bibfnamefont {C.~J.}\ \bibnamefont {Gazza}}, \ and\
  \bibinfo {author} {\bibfnamefont {H.~A.}\ \bibnamefont {Ceccatto}}} (\bibinfo
  {year} {1997}),\ \bibfield  {title} {\enquote {\bibinfo {title}
  {{Schwinger-Boson Approach to Quantum Spin Systems: Gaussian Fluctuations in
  the ``Natural'' Gauge}},}\ }\href {\doibase 10.1103/PhysRevLett.78.2216}
  {\bibfield  {journal} {\bibinfo  {journal} {Phys. Rev. Lett.}\ }\textbf
  {\bibinfo {volume} {78}},\ \bibinfo {pages} {2216--2219}}\BibitemShut
  {NoStop}%
\bibitem [{\citenamefont {Vinkler-Aviv}\ and\ \citenamefont
  {Rosch}(2018)}]{Rosch2018}%
  \BibitemOpen
  \bibfield  {author} {\bibinfo {author} {\bibnamefont {Vinkler-Aviv},
  \bibfnamefont {Yuval}}, \ and\ \bibinfo {author} {\bibfnamefont {Achim}\
  \bibnamefont {Rosch}}} (\bibinfo {year} {2018}),\ \bibfield  {title}
  {\enquote {\bibinfo {title} {{Approximately quantized thermal Hall effect of
  chiral liquids coupled to phonons}},}\ }\href {\doibase
  10.1103/PhysRevX.8.031032} {\bibfield  {journal} {\bibinfo  {journal} {Phys.
  Rev. X}\ }\textbf {\bibinfo {volume} {8}},\ \bibinfo {pages}
  {031032}}\BibitemShut {NoStop}%
\bibitem [{\citenamefont {Wang}\ and\ \citenamefont
  {Senthil}(2016)}]{PhysRevX.6.011034}%
  \BibitemOpen
  \bibfield  {author} {\bibinfo {author} {\bibnamefont {Wang}, \bibfnamefont
  {Chong}}, \ and\ \bibinfo {author} {\bibfnamefont {T.}~\bibnamefont
  {Senthil}}} (\bibinfo {year} {2016}),\ \bibfield  {title} {\enquote {\bibinfo
  {title} {{Time-Reversal Symmetric $U(1)$ Quantum Spin Liquids}},}\ }\href
  {\doibase 10.1103/PhysRevX.6.011034} {\bibfield  {journal} {\bibinfo
  {journal} {Phys. Rev. X}\ }\textbf {\bibinfo {volume} {6}},\ \bibinfo {pages}
  {011034}}\BibitemShut {NoStop}%
\bibitem [{\citenamefont {Wang}\ and\ \citenamefont
  {Vishwanath}(2006)}]{PhysRevB.74.174423}%
  \BibitemOpen
  \bibfield  {author} {\bibinfo {author} {\bibnamefont {Wang}, \bibfnamefont
  {Fa}}, \ and\ \bibinfo {author} {\bibfnamefont {Ashvin}\ \bibnamefont
  {Vishwanath}}} (\bibinfo {year} {2006}),\ \bibfield  {title} {\enquote
  {\bibinfo {title} {{Spin-liquid states on the triangular and Kagom\'e
  lattices: A projective-symmetry-group analysis of Schwinger boson states}},}\
  }\href {\doibase 10.1103/PhysRevB.74.174423} {\bibfield  {journal} {\bibinfo
  {journal} {Phys. Rev. B}\ }\textbf {\bibinfo {volume} {74}},\ \bibinfo
  {pages} {174423}}\BibitemShut {NoStop}%
\bibitem [{\citenamefont {Wang}\ \emph {et~al.}(2007)\citenamefont {Wang},
  \citenamefont {Vishwanath},\ and\ \citenamefont {Kim}}]{PhysRevB.76.094421}%
  \BibitemOpen
  \bibfield  {author} {\bibinfo {author} {\bibnamefont {Wang}, \bibfnamefont
  {Fa}}, \bibinfo {author} {\bibfnamefont {Ashvin}\ \bibnamefont {Vishwanath}},
  \ and\ \bibinfo {author} {\bibfnamefont {Yong~Baek}\ \bibnamefont {Kim}}}
  (\bibinfo {year} {2007}),\ \bibfield  {title} {\enquote {\bibinfo {title}
  {{Quantum and classical spins on the spatially distorted kagom\'e lattice:
  Applications to volborthite
  {Cu}$_{3}$V$_{2}$O$_{7}$OH$_{2}$$\cdot$2H$_{2}$O}},}\ }\href {\doibase
  10.1103/PhysRevB.76.094421} {\bibfield  {journal} {\bibinfo  {journal} {Phys.
  Rev. B}\ }\textbf {\bibinfo {volume} {76}},\ \bibinfo {pages}
  {094421}}\BibitemShut {NoStop}%
\bibitem [{\citenamefont {Watanabe}\ \emph {et~al.}(2016)\citenamefont
  {Watanabe}, \citenamefont {Sugii}, \citenamefont {Shimozawa}, \citenamefont
  {Suzuki}, \citenamefont {Yajima}, \citenamefont {Ishikawa}, \citenamefont
  {Hiroi}, \citenamefont {Shibauchi}, \citenamefont {Matsuda},\ and\
  \citenamefont {Yamashita}}]{Watanabe2016}%
  \BibitemOpen
  \bibfield  {author} {\bibinfo {author} {\bibnamefont {Watanabe},
  \bibfnamefont {Daiki}}, \bibinfo {author} {\bibfnamefont {Kaori}\
  \bibnamefont {Sugii}}, \bibinfo {author} {\bibfnamefont {Masaaki}\
  \bibnamefont {Shimozawa}}, \bibinfo {author} {\bibfnamefont {Yoshitaka}\
  \bibnamefont {Suzuki}}, \bibinfo {author} {\bibfnamefont {Takeshi}\
  \bibnamefont {Yajima}}, \bibinfo {author} {\bibfnamefont {Hajime}\
  \bibnamefont {Ishikawa}}, \bibinfo {author} {\bibfnamefont {Zenji}\
  \bibnamefont {Hiroi}}, \bibinfo {author} {\bibfnamefont {Takasada}\
  \bibnamefont {Shibauchi}}, \bibinfo {author} {\bibfnamefont {Yuji}\
  \bibnamefont {Matsuda}}, \ and\ \bibinfo {author} {\bibfnamefont {Minoru}\
  \bibnamefont {Yamashita}}} (\bibinfo {year} {2016}),\ \bibfield  {title}
  {\enquote {\bibinfo {title} {Emergence of nontrivial magnetic excitations in
  a spin-liquid state of kagom{\'{e}} volborthite},}\ }\href {\doibase
  10.1073/pnas.1524076113} {\bibfield  {journal} {\bibinfo  {journal}
  {Proceedings of the National Academy of Sciences}\ }\textbf {\bibinfo
  {volume} {113}}~(\bibinfo {number} {31}),\ \bibinfo {pages}
  {8653--8657}}\BibitemShut {NoStop}%
\bibitem [{\citenamefont {Wen}\ \emph {et~al.}(2017)\citenamefont {Wen},
  \citenamefont {Koohpayeh}, \citenamefont {Ross}, \citenamefont {Trump},
  \citenamefont {McQueen}, \citenamefont {Kimura}, \citenamefont {Nakatsuji},
  \citenamefont {Qiu}, \citenamefont {Pajerowski}, \citenamefont {Copley},\
  and\ \citenamefont {Broholm}}]{Broholm2017}%
  \BibitemOpen
  \bibfield  {author} {\bibinfo {author} {\bibnamefont {Wen}, \bibfnamefont
  {J-J}}, \bibinfo {author} {\bibfnamefont {S.~M.}\ \bibnamefont {Koohpayeh}},
  \bibinfo {author} {\bibfnamefont {K.~A.}\ \bibnamefont {Ross}}, \bibinfo
  {author} {\bibfnamefont {B.~A.}\ \bibnamefont {Trump}}, \bibinfo {author}
  {\bibfnamefont {T.~M.}\ \bibnamefont {McQueen}}, \bibinfo {author}
  {\bibfnamefont {K.}~\bibnamefont {Kimura}}, \bibinfo {author} {\bibfnamefont
  {S.}~\bibnamefont {Nakatsuji}}, \bibinfo {author} {\bibfnamefont
  {Y.}~\bibnamefont {Qiu}}, \bibinfo {author} {\bibfnamefont {D.~M.}\
  \bibnamefont {Pajerowski}}, \bibinfo {author} {\bibfnamefont {J.~R.~D.}\
  \bibnamefont {Copley}}, \ and\ \bibinfo {author} {\bibfnamefont {C.~L.}\
  \bibnamefont {Broholm}}} (\bibinfo {year} {2017}),\ \bibfield  {title}
  {\enquote {\bibinfo {title} {{Disordered route to the coulomb quantum spin
  liquid: Random transverse fields on spin ice in
  ${\mathrm{Pr}}_{2}{\mathrm{Zr}}_{2}{\mathrm{O}}_{7}$}},}\ }\href {\doibase
  10.1103/PhysRevLett.118.107206} {\bibfield  {journal} {\bibinfo  {journal}
  {Phys. Rev. Lett.}\ }\textbf {\bibinfo {volume} {118}},\ \bibinfo {pages}
  {107206}}\BibitemShut {NoStop}%
\bibitem [{\citenamefont {Wen}\ \emph {et~al.}(1989{\natexlab{a}})\citenamefont
  {Wen}, \citenamefont {Wilczek},\ and\ \citenamefont {Zee}}]{Zee1989}%
  \BibitemOpen
  \bibfield  {author} {\bibinfo {author} {\bibnamefont {Wen}, \bibfnamefont
  {X~G}}, \bibinfo {author} {\bibfnamefont {Frank}\ \bibnamefont {Wilczek}}, \
  and\ \bibinfo {author} {\bibfnamefont {A.}~\bibnamefont {Zee}}} (\bibinfo
  {year} {1989}{\natexlab{a}}),\ \bibfield  {title} {\enquote {\bibinfo {title}
  {Chiral spin states and superconductivity},}\ }\href {\doibase
  10.1103/PhysRevB.39.11413} {\bibfield  {journal} {\bibinfo  {journal} {Phys.
  Rev. B}\ }\textbf {\bibinfo {volume} {39}},\ \bibinfo {pages}
  {11413--11423}}\BibitemShut {NoStop}%
\bibitem [{\citenamefont {Wen}\ \emph {et~al.}(1989{\natexlab{b}})\citenamefont
  {Wen}, \citenamefont {Wilczek},\ and\ \citenamefont {Zee}}]{Wen1989B}%
  \BibitemOpen
  \bibfield  {author} {\bibinfo {author} {\bibnamefont {Wen}, \bibfnamefont
  {X~G}}, \bibinfo {author} {\bibfnamefont {Frank}\ \bibnamefont {Wilczek}}, \
  and\ \bibinfo {author} {\bibfnamefont {A.}~\bibnamefont {Zee}}} (\bibinfo
  {year} {1989}{\natexlab{b}}),\ \bibfield  {title} {\enquote {\bibinfo {title}
  {Chiral spin states and superconductivity},}\ }\href {\doibase
  10.1103/PhysRevB.39.11413} {\bibfield  {journal} {\bibinfo  {journal} {Phys.
  Rev. B}\ }\textbf {\bibinfo {volume} {39}},\ \bibinfo {pages}
  {11413--11423}}\BibitemShut {NoStop}%
\bibitem [{\citenamefont {Wen}(2002)}]{Wen2002}%
  \BibitemOpen
  \bibfield  {author} {\bibinfo {author} {\bibnamefont {Wen}, \bibfnamefont
  {Xiao-Gang}}} (\bibinfo {year} {2002}),\ \bibfield  {title} {\enquote
  {\bibinfo {title} {Quantum orders and symmetric spin liquids},}\ }\href
  {\doibase 10.1103/PhysRevB.65.165113} {\bibfield  {journal} {\bibinfo
  {journal} {Phys. Rev. B}\ }\textbf {\bibinfo {volume} {65}},\ \bibinfo
  {pages} {165113}}\BibitemShut {NoStop}%
\bibitem [{\citenamefont {Widmann}\ \emph {et~al.}(2019)\citenamefont
  {Widmann}, \citenamefont {Tsurkan}, \citenamefont {Prishchenko},
  \citenamefont {Mazurenko}, \citenamefont {Tsirlin},\ and\ \citenamefont
  {Loidl}}]{Loidl2019}%
  \BibitemOpen
  \bibfield  {author} {\bibinfo {author} {\bibnamefont {Widmann}, \bibfnamefont
  {S}}, \bibinfo {author} {\bibfnamefont {V.}~\bibnamefont {Tsurkan}}, \bibinfo
  {author} {\bibfnamefont {D.~A.}\ \bibnamefont {Prishchenko}}, \bibinfo
  {author} {\bibfnamefont {V.~G.}\ \bibnamefont {Mazurenko}}, \bibinfo {author}
  {\bibfnamefont {A.~A.}\ \bibnamefont {Tsirlin}}, \ and\ \bibinfo {author}
  {\bibfnamefont {A.}~\bibnamefont {Loidl}}} (\bibinfo {year} {2019}),\
  \bibfield  {title} {\enquote {\bibinfo {title} {{Thermodynamic evidence of
  fractionalized excitations in
  $\ensuremath{\alpha}\ensuremath{-}\mathrm{RuC}{\mathrm{l}}_{3}$}},}\ }\href
  {\doibase 10.1103/PhysRevB.99.094415} {\bibfield  {journal} {\bibinfo
  {journal} {Phys. Rev. B}\ }\textbf {\bibinfo {volume} {99}},\ \bibinfo
  {pages} {094415}}\BibitemShut {NoStop}%
\bibitem [{\citenamefont {Winter}\ \emph {et~al.}(2016)\citenamefont {Winter},
  \citenamefont {Li}, \citenamefont {Jeschke},\ and\ \citenamefont
  {Valent\'{\i}}}]{Valenti2016}%
  \BibitemOpen
  \bibfield  {author} {\bibinfo {author} {\bibnamefont {Winter}, \bibfnamefont
  {Stephen~M}}, \bibinfo {author} {\bibfnamefont {Ying}\ \bibnamefont {Li}},
  \bibinfo {author} {\bibfnamefont {Harald~O.}\ \bibnamefont {Jeschke}}, \ and\
  \bibinfo {author} {\bibfnamefont {Roser}\ \bibnamefont {Valent\'{\i}}}}
  (\bibinfo {year} {2016}),\ \bibfield  {title} {\enquote {\bibinfo {title}
  {{Challenges in design of Kitaev materials: Magnetic interactions from
  competing energy scales}},}\ }\href {\doibase 10.1103/PhysRevB.93.214431}
  {\bibfield  {journal} {\bibinfo  {journal} {Phys. Rev. B}\ }\textbf {\bibinfo
  {volume} {93}},\ \bibinfo {pages} {214431}}\BibitemShut {NoStop}%
\bibitem [{\citenamefont {Winter}\ \emph
  {et~al.}(2017{\natexlab{a}})\citenamefont {Winter}, \citenamefont {Riedl},
  \citenamefont {Maksimov}, \citenamefont {Chernyshev}, \citenamefont
  {Honecker},\ and\ \citenamefont {Valent{\'{\i}}}}]{Winter2017}%
  \BibitemOpen
  \bibfield  {author} {\bibinfo {author} {\bibnamefont {Winter}, \bibfnamefont
  {Stephen~M}}, \bibinfo {author} {\bibfnamefont {Kira}\ \bibnamefont {Riedl}},
  \bibinfo {author} {\bibfnamefont {Pavel~A.}\ \bibnamefont {Maksimov}},
  \bibinfo {author} {\bibfnamefont {Alexander~L.}\ \bibnamefont {Chernyshev}},
  \bibinfo {author} {\bibfnamefont {Andreas}\ \bibnamefont {Honecker}}, \ and\
  \bibinfo {author} {\bibfnamefont {Roser}\ \bibnamefont {Valent{\'{\i}}}}}
  (\bibinfo {year} {2017}{\natexlab{a}}),\ \bibfield  {title} {\enquote
  {\bibinfo {title} {Breakdown of magnons in a strongly spin-orbital coupled
  magnet},}\ }\href {\doibase 10.1038/s41467-017-01177-0} {\bibfield  {journal}
  {\bibinfo  {journal} {Nature Communications}\ }\textbf {\bibinfo {volume}
  {8}}~(\bibinfo {number} {1}),\ \bibinfo {pages} {1152}}\BibitemShut {NoStop}%
\bibitem [{\citenamefont {Winter}\ \emph
  {et~al.}(2017{\natexlab{b}})\citenamefont {Winter}, \citenamefont {Tsirlin},
  \citenamefont {Daghofer}, \citenamefont {van~den Brink}, \citenamefont
  {Singh}, \citenamefont {Gegenwart},\ and\ \citenamefont
  {Valent{\'{\i}}}}]{Valenti2017}%
  \BibitemOpen
  \bibfield  {author} {\bibinfo {author} {\bibnamefont {Winter}, \bibfnamefont
  {Stephen~M}}, \bibinfo {author} {\bibfnamefont {Alexander~A}\ \bibnamefont
  {Tsirlin}}, \bibinfo {author} {\bibfnamefont {Maria}\ \bibnamefont
  {Daghofer}}, \bibinfo {author} {\bibfnamefont {Jeroen}\ \bibnamefont {van~den
  Brink}}, \bibinfo {author} {\bibfnamefont {Yogesh}\ \bibnamefont {Singh}},
  \bibinfo {author} {\bibfnamefont {Philipp}\ \bibnamefont {Gegenwart}}, \ and\
  \bibinfo {author} {\bibfnamefont {Roser}\ \bibnamefont {Valent{\'{\i}}}}}
  (\bibinfo {year} {2017}{\natexlab{b}}),\ \bibfield  {title} {\enquote
  {\bibinfo {title} {{Models and materials for generalized Kitaev
  magnetism}},}\ }\href {\doibase 10.1088/1361-648x/aa8cf5} {\bibfield
  {journal} {\bibinfo  {journal} {Journal of Physics: Condensed Matter}\
  }\textbf {\bibinfo {volume} {29}}~(\bibinfo {number} {49}),\ \bibinfo {pages}
  {493002}}\BibitemShut {NoStop}%
\bibitem [{\citenamefont {Witczak-Krempa}\ \emph {et~al.}(2014)\citenamefont
  {Witczak-Krempa}, \citenamefont {Chen}, \citenamefont {Kim},\ and\
  \citenamefont {Balents}}]{Balents2014}%
  \BibitemOpen
  \bibfield  {author} {\bibinfo {author} {\bibnamefont {Witczak-Krempa},
  \bibfnamefont {William}}, \bibinfo {author} {\bibfnamefont {Gang}\
  \bibnamefont {Chen}}, \bibinfo {author} {\bibfnamefont {Yong~Baek}\
  \bibnamefont {Kim}}, \ and\ \bibinfo {author} {\bibfnamefont {Leon}\
  \bibnamefont {Balents}}} (\bibinfo {year} {2014}),\ \bibfield  {title}
  {\enquote {\bibinfo {title} {{Correlated Quantum Phenomena in the Strong
  Spin-Orbit Regime}},}\ }\href {\doibase
  10.1146/annurev-conmatphys-020911-125138} {\bibfield  {journal} {\bibinfo
  {journal} {Annual Review of Condensed Matter Physics}\ }\textbf {\bibinfo
  {volume} {5}}~(\bibinfo {number} {1}),\ \bibinfo {pages}
  {57--82}}\BibitemShut {NoStop}%
\bibitem [{\citenamefont {Xu}\ \emph {et~al.}(2023)\citenamefont {Xu},
  \citenamefont {Carnahan}, \citenamefont {Zhang}, \citenamefont {Sretenovic},
  \citenamefont {Zhang}, \citenamefont {Xiao},\ and\ \citenamefont
  {Ke}}]{PhysRevB.107.L060404}%
  \BibitemOpen
  \bibfield  {author} {\bibinfo {author} {\bibnamefont {Xu}, \bibfnamefont
  {Chunqiang}}, \bibinfo {author} {\bibfnamefont {Caitlin}\ \bibnamefont
  {Carnahan}}, \bibinfo {author} {\bibfnamefont {Heda}\ \bibnamefont {Zhang}},
  \bibinfo {author} {\bibfnamefont {Milos}\ \bibnamefont {Sretenovic}},
  \bibinfo {author} {\bibfnamefont {Pengpeng}\ \bibnamefont {Zhang}}, \bibinfo
  {author} {\bibfnamefont {Di}~\bibnamefont {Xiao}}, \ and\ \bibinfo {author}
  {\bibfnamefont {Xianglin}\ \bibnamefont {Ke}}} (\bibinfo {year} {2023}),\
  \bibfield  {title} {\enquote {\bibinfo {title} {Thermal hall effect in a van
  der waals triangular magnet ${\mathrm{fecl}}_{2}$},}\ }\href {\doibase
  10.1103/PhysRevB.107.L060404} {\bibfield  {journal} {\bibinfo  {journal}
  {Phys. Rev. B}\ }\textbf {\bibinfo {volume} {107}},\ \bibinfo {pages}
  {L060404}}\BibitemShut {NoStop}%
\bibitem [{\citenamefont {Xu}\ \emph {et~al.}(2015)\citenamefont {Xu},
  \citenamefont {Anand}, \citenamefont {Bera}, \citenamefont {Frontzek},
  \citenamefont {Abernathy}, \citenamefont {Casati}, \citenamefont
  {Siemensmeyer},\ and\ \citenamefont {Lake}}]{PhysRevB.92.224430}%
  \BibitemOpen
  \bibfield  {author} {\bibinfo {author} {\bibnamefont {Xu}, \bibfnamefont
  {J}}, \bibinfo {author} {\bibfnamefont {V.~K.}\ \bibnamefont {Anand}},
  \bibinfo {author} {\bibfnamefont {A.~K.}\ \bibnamefont {Bera}}, \bibinfo
  {author} {\bibfnamefont {M.}~\bibnamefont {Frontzek}}, \bibinfo {author}
  {\bibfnamefont {D.~L.}\ \bibnamefont {Abernathy}}, \bibinfo {author}
  {\bibfnamefont {N.}~\bibnamefont {Casati}}, \bibinfo {author} {\bibfnamefont
  {K.}~\bibnamefont {Siemensmeyer}}, \ and\ \bibinfo {author} {\bibfnamefont
  {B.}~\bibnamefont {Lake}}} (\bibinfo {year} {2015}),\ \bibfield  {title}
  {\enquote {\bibinfo {title} {{Magnetic structure and crystal-field states of
  the pyrochlore antiferromagnet
  ${\mathrm{Nd}}_{2}{\mathrm{Zr}}_{2}{\mathrm{O}}_{7}$}},}\ }\href {\doibase
  10.1103/PhysRevB.92.224430} {\bibfield  {journal} {\bibinfo  {journal} {Phys.
  Rev. B}\ }\textbf {\bibinfo {volume} {92}},\ \bibinfo {pages}
  {224430}}\BibitemShut {NoStop}%
\bibitem [{\citenamefont {Xu}\ \emph {et~al.}(2020)\citenamefont {Xu},
  \citenamefont {Benton}, \citenamefont {Islam}, \citenamefont {Guidi},
  \citenamefont {Ehlers},\ and\ \citenamefont {Lake}}]{PhysRevLett.124.097203}%
  \BibitemOpen
  \bibfield  {author} {\bibinfo {author} {\bibnamefont {Xu}, \bibfnamefont
  {J}}, \bibinfo {author} {\bibfnamefont {Owen}\ \bibnamefont {Benton}},
  \bibinfo {author} {\bibfnamefont {A.~T. M.~N.}\ \bibnamefont {Islam}},
  \bibinfo {author} {\bibfnamefont {T.}~\bibnamefont {Guidi}}, \bibinfo
  {author} {\bibfnamefont {G.}~\bibnamefont {Ehlers}}, \ and\ \bibinfo {author}
  {\bibfnamefont {B.}~\bibnamefont {Lake}}} (\bibinfo {year} {2020}),\
  \bibfield  {title} {\enquote {\bibinfo {title} {{Order out of a Coulomb Phase
  and Higgs Transition: Frustrated Transverse Interactions of
  ${\mathrm{Nd}}_{2}{\mathrm{Zr}}_{2}{\mathrm{O}}_{7}$}},}\ }\href {\doibase
  10.1103/PhysRevLett.124.097203} {\bibfield  {journal} {\bibinfo  {journal}
  {Phys. Rev. Lett.}\ }\textbf {\bibinfo {volume} {124}},\ \bibinfo {pages}
  {097203}}\BibitemShut {NoStop}%
\bibitem [{\citenamefont {Yamashita}\ \emph {et~al.}(2020)\citenamefont
  {Yamashita}, \citenamefont {Gouchi}, \citenamefont {Uwatoko}, \citenamefont
  {Kurita},\ and\ \citenamefont {Tanaka}}]{Yamashita2020B}%
  \BibitemOpen
  \bibfield  {author} {\bibinfo {author} {\bibnamefont {Yamashita},
  \bibfnamefont {M}}, \bibinfo {author} {\bibfnamefont {J.}~\bibnamefont
  {Gouchi}}, \bibinfo {author} {\bibfnamefont {Y.}~\bibnamefont {Uwatoko}},
  \bibinfo {author} {\bibfnamefont {N.}~\bibnamefont {Kurita}}, \ and\ \bibinfo
  {author} {\bibfnamefont {H.}~\bibnamefont {Tanaka}}} (\bibinfo {year}
  {2020}),\ \bibfield  {title} {\enquote {\bibinfo {title} {{Sample dependence
  of half-integer quantized thermal Hall effect in the Kitaev spin-liquid
  candidate $\ensuremath{\alpha}\ensuremath{-}{\mathrm{RuCl}}_{3}$}},}\ }\href
  {\doibase 10.1103/PhysRevB.102.220404} {\bibfield  {journal} {\bibinfo
  {journal} {Phys. Rev. B}\ }\textbf {\bibinfo {volume} {102}},\ \bibinfo
  {pages} {220404}}\BibitemShut {NoStop}%
\bibitem [{\citenamefont {Yamashita}\ \emph {et~al.}(2010)\citenamefont
  {Yamashita}, \citenamefont {Nakata}, \citenamefont {Senshu}, \citenamefont
  {Nagata}, \citenamefont {Yamamoto}, \citenamefont {Kato}, \citenamefont
  {Shibauchi},\ and\ \citenamefont {Matsuda}}]{Yamashita2010}%
  \BibitemOpen
  \bibfield  {author} {\bibinfo {author} {\bibnamefont {Yamashita},
  \bibfnamefont {M}}, \bibinfo {author} {\bibfnamefont {N.}~\bibnamefont
  {Nakata}}, \bibinfo {author} {\bibfnamefont {Y.}~\bibnamefont {Senshu}},
  \bibinfo {author} {\bibfnamefont {M.}~\bibnamefont {Nagata}}, \bibinfo
  {author} {\bibfnamefont {H.~M.}\ \bibnamefont {Yamamoto}}, \bibinfo {author}
  {\bibfnamefont {R.}~\bibnamefont {Kato}}, \bibinfo {author} {\bibfnamefont
  {T.}~\bibnamefont {Shibauchi}}, \ and\ \bibinfo {author} {\bibfnamefont
  {Y.}~\bibnamefont {Matsuda}}} (\bibinfo {year} {2010}),\ \bibfield  {title}
  {\enquote {\bibinfo {title} {{Highly Mobile Gapless Excitations in a
  Two-Dimensional Candidate Quantum Spin Liquid}},}\ }\href {\doibase
  10.1126/science.1188200} {\bibfield  {journal} {\bibinfo  {journal}
  {Science}\ }\textbf {\bibinfo {volume} {328}}~(\bibinfo {number} {5983}),\
  \bibinfo {pages} {1246--1248}}\BibitemShut {NoStop}%
\bibitem [{\citenamefont {Yamashita}\ \emph
  {et~al.}(2008{\natexlab{a}})\citenamefont {Yamashita}, \citenamefont
  {Nakata}, \citenamefont {Kasahara}, \citenamefont {Sasaki}, \citenamefont
  {Yoneyama}, \citenamefont {Kobayashi}, \citenamefont {Fujimoto},
  \citenamefont {Shibauchi},\ and\ \citenamefont {Matsuda}}]{Yamashita2008}%
  \BibitemOpen
  \bibfield  {author} {\bibinfo {author} {\bibnamefont {Yamashita},
  \bibfnamefont {Minoru}}, \bibinfo {author} {\bibfnamefont {Norihito}\
  \bibnamefont {Nakata}}, \bibinfo {author} {\bibfnamefont {Yuichi}\
  \bibnamefont {Kasahara}}, \bibinfo {author} {\bibfnamefont {Takahiko}\
  \bibnamefont {Sasaki}}, \bibinfo {author} {\bibfnamefont {Naoki}\
  \bibnamefont {Yoneyama}}, \bibinfo {author} {\bibfnamefont {Norio}\
  \bibnamefont {Kobayashi}}, \bibinfo {author} {\bibfnamefont {Satoshi}\
  \bibnamefont {Fujimoto}}, \bibinfo {author} {\bibfnamefont {Takasada}\
  \bibnamefont {Shibauchi}}, \ and\ \bibinfo {author} {\bibfnamefont {Yuji}\
  \bibnamefont {Matsuda}}} (\bibinfo {year} {2008}{\natexlab{a}}),\ \bibfield
  {title} {\enquote {\bibinfo {title} {{Thermal-transport measurements in a
  quantum spin-liquid state of the frustrated triangular magnet
  $\alpha$-({BEDT}-{TTF})$_2$Cu$_2$({CN})$_3$}},}\ }\href {\doibase
  10.1038/nphys1134} {\bibfield  {journal} {\bibinfo  {journal} {Nature
  Physics}\ }\textbf {\bibinfo {volume} {5}}~(\bibinfo {number} {1}),\ \bibinfo
  {pages} {44--47}}\BibitemShut {NoStop}%
\bibitem [{\citenamefont {Yamashita}\ \emph
  {et~al.}(2008{\natexlab{b}})\citenamefont {Yamashita}, \citenamefont
  {Nakazawa}, \citenamefont {Oguni}, \citenamefont {Oshima}, \citenamefont
  {Nojiri}, \citenamefont {Shimizu}, \citenamefont {Miyagawa},\ and\
  \citenamefont {Kanoda}}]{Yamashita2008B}%
  \BibitemOpen
  \bibfield  {author} {\bibinfo {author} {\bibnamefont {Yamashita},
  \bibfnamefont {Satoshi}}, \bibinfo {author} {\bibfnamefont {Yasuhiro}\
  \bibnamefont {Nakazawa}}, \bibinfo {author} {\bibfnamefont {Masaharu}\
  \bibnamefont {Oguni}}, \bibinfo {author} {\bibfnamefont {Yugo}\ \bibnamefont
  {Oshima}}, \bibinfo {author} {\bibfnamefont {Hiroyuki}\ \bibnamefont
  {Nojiri}}, \bibinfo {author} {\bibfnamefont {Yasuhiro}\ \bibnamefont
  {Shimizu}}, \bibinfo {author} {\bibfnamefont {Kazuya}\ \bibnamefont
  {Miyagawa}}, \ and\ \bibinfo {author} {\bibfnamefont {Kazushi}\ \bibnamefont
  {Kanoda}}} (\bibinfo {year} {2008}{\natexlab{b}}),\ \bibfield  {title}
  {\enquote {\bibinfo {title} {{Thermodynamic properties of a spin-1/2
  spin-liquid state in a $\kappa$-type organic salt}},}\ }\href {\doibase
  10.1038/nphys942} {\bibfield  {journal} {\bibinfo  {journal} {Nature
  Physics}\ }\textbf {\bibinfo {volume} {4}}~(\bibinfo {number} {6}),\ \bibinfo
  {pages} {459--462}}\BibitemShut {NoStop}%
\bibitem [{\citenamefont {Yamaura}\ \emph {et~al.}(2012)\citenamefont
  {Yamaura}, \citenamefont {Ohgushi}, \citenamefont {Ohsumi}, \citenamefont
  {Hasegawa}, \citenamefont {Yamauchi}, \citenamefont {Sugimoto}, \citenamefont
  {Takeshita}, \citenamefont {Tokuda}, \citenamefont {Takata}, \citenamefont
  {Udagawa}, \citenamefont {Takigawa}, \citenamefont {Harima}, \citenamefont
  {Arima},\ and\ \citenamefont {Hiroi}}]{PhysRevLett.108.247205}%
  \BibitemOpen
  \bibfield  {author} {\bibinfo {author} {\bibnamefont {Yamaura}, \bibfnamefont
  {J}}, \bibinfo {author} {\bibfnamefont {K.}~\bibnamefont {Ohgushi}}, \bibinfo
  {author} {\bibfnamefont {H.}~\bibnamefont {Ohsumi}}, \bibinfo {author}
  {\bibfnamefont {T.}~\bibnamefont {Hasegawa}}, \bibinfo {author}
  {\bibfnamefont {I.}~\bibnamefont {Yamauchi}}, \bibinfo {author}
  {\bibfnamefont {K.}~\bibnamefont {Sugimoto}}, \bibinfo {author}
  {\bibfnamefont {S.}~\bibnamefont {Takeshita}}, \bibinfo {author}
  {\bibfnamefont {A.}~\bibnamefont {Tokuda}}, \bibinfo {author} {\bibfnamefont
  {M.}~\bibnamefont {Takata}}, \bibinfo {author} {\bibfnamefont
  {M.}~\bibnamefont {Udagawa}}, \bibinfo {author} {\bibfnamefont
  {M.}~\bibnamefont {Takigawa}}, \bibinfo {author} {\bibfnamefont
  {H.}~\bibnamefont {Harima}}, \bibinfo {author} {\bibfnamefont
  {T.}~\bibnamefont {Arima}}, \ and\ \bibinfo {author} {\bibfnamefont
  {Z.}~\bibnamefont {Hiroi}}} (\bibinfo {year} {2012}),\ \bibfield  {title}
  {\enquote {\bibinfo {title} {{Tetrahedral Magnetic Order and the
  Metal-Insulator Transition in the Pyrochlore Lattice of
  ${\mathrm{Cd}}_{2}{\mathrm{Os}}_{2}{\mathbf{O}}_{7}$}},}\ }\href {\doibase
  10.1103/PhysRevLett.108.247205} {\bibfield  {journal} {\bibinfo  {journal}
  {Phys. Rev. Lett.}\ }\textbf {\bibinfo {volume} {108}},\ \bibinfo {pages}
  {247205}}\BibitemShut {NoStop}%
\bibitem [{\citenamefont {Yan}\ and\ \citenamefont
  {Felser}(2017)}]{doi:10.1146/annurev-conmatphys-031016-025458}%
  \BibitemOpen
  \bibfield  {author} {\bibinfo {author} {\bibnamefont {Yan}, \bibfnamefont
  {Binghai}}, \ and\ \bibinfo {author} {\bibfnamefont {Claudia}\ \bibnamefont
  {Felser}}} (\bibinfo {year} {2017}),\ \bibfield  {title} {\enquote {\bibinfo
  {title} {Topological materials: Weyl semimetals},}\ }\href {\doibase
  10.1146/annurev-conmatphys-031016-025458} {\bibfield  {journal} {\bibinfo
  {journal} {Annual Review of Condensed Matter Physics}\ }\textbf {\bibinfo
  {volume} {8}}~(\bibinfo {number} {1}),\ \bibinfo {pages}
  {337--354}}\BibitemShut {NoStop}%
\bibitem [{\citenamefont {Yan}\ \emph {et~al.}(2011)\citenamefont {Yan},
  \citenamefont {Huse},\ and\ \citenamefont {White}}]{Yan_2011}%
  \BibitemOpen
  \bibfield  {author} {\bibinfo {author} {\bibnamefont {Yan}, \bibfnamefont
  {Simeng}}, \bibinfo {author} {\bibfnamefont {David~A.}\ \bibnamefont {Huse}},
  \ and\ \bibinfo {author} {\bibfnamefont {Steven~R.}\ \bibnamefont {White}}}
  (\bibinfo {year} {2011}),\ \bibfield  {title} {\enquote {\bibinfo {title}
  {{Spin-Liquid Ground State of the S = 1/2 Kagome Heisenberg
  Antiferromagnet}},}\ }\href {\doibase 10.1126/science.1201080} {\bibfield
  {journal} {\bibinfo  {journal} {Science}\ }\textbf {\bibinfo {volume}
  {332}}~(\bibinfo {number} {6034}),\ \bibinfo {pages}
  {1173--1176}}\BibitemShut {NoStop}%
\bibitem [{\citenamefont {Yang}\ \emph
  {et~al.}(2020{\natexlab{a}})\citenamefont {Yang}, \citenamefont {Kim},\ and\
  \citenamefont {Lee}}]{Sungbin2020}%
  \BibitemOpen
  \bibfield  {author} {\bibinfo {author} {\bibnamefont {Yang}, \bibfnamefont
  {Hyeok-Jun}}, \bibinfo {author} {\bibfnamefont {Hee~Seung}\ \bibnamefont
  {Kim}}, \ and\ \bibinfo {author} {\bibfnamefont {SungBin}\ \bibnamefont
  {Lee}}} (\bibinfo {year} {2020}{\natexlab{a}}),\ \bibfield  {title} {\enquote
  {\bibinfo {title} {{Magnetic field and thermal Hall effect in a pyrochlore
  U(1) quantum spin liquid}},}\ }\href {\doibase 10.1103/PhysRevB.102.060405}
  {\bibfield  {journal} {\bibinfo  {journal} {Phys. Rev. B}\ }\textbf {\bibinfo
  {volume} {102}},\ \bibinfo {pages} {060405}}\BibitemShut {NoStop}%
\bibitem [{\citenamefont {Yang}\ \emph
  {et~al.}(2020{\natexlab{b}})\citenamefont {Yang}, \citenamefont {Zhang},\
  and\ \citenamefont {Zhang}}]{Fuchun2020}%
  \BibitemOpen
  \bibfield  {author} {\bibinfo {author} {\bibnamefont {Yang}, \bibfnamefont
  {Yi-Feng}}, \bibinfo {author} {\bibfnamefont {Guang-Ming}\ \bibnamefont
  {Zhang}}, \ and\ \bibinfo {author} {\bibfnamefont {Fu-Chun}\ \bibnamefont
  {Zhang}}} (\bibinfo {year} {2020}{\natexlab{b}}),\ \bibfield  {title}
  {\enquote {\bibinfo {title} {{Universal behavior of the thermal Hall
  conductivity}},}\ }\href {\doibase 10.1103/PhysRevLett.124.186602} {\bibfield
   {journal} {\bibinfo  {journal} {Phys. Rev. Lett.}\ }\textbf {\bibinfo
  {volume} {124}},\ \bibinfo {pages} {186602}}\BibitemShut {NoStop}%
\bibitem [{\citenamefont {Yao}\ \emph {et~al.}(2020)\citenamefont {Yao},
  \citenamefont {Li},\ and\ \citenamefont {Chen}}]{Yao2020}%
  \BibitemOpen
  \bibfield  {author} {\bibinfo {author} {\bibnamefont {Yao}, \bibfnamefont
  {Xu-Ping}}, \bibinfo {author} {\bibfnamefont {Yao-Dong}\ \bibnamefont {Li}},
  \ and\ \bibinfo {author} {\bibfnamefont {Gang}\ \bibnamefont {Chen}}}
  (\bibinfo {year} {2020}),\ \bibfield  {title} {\enquote {\bibinfo {title}
  {Pyrochlore {U(1)} spin liquid of mixed-symmetry enrichments in magnetic
  fields},}\ }\href {\doibase 10.1103/PhysRevResearch.2.013334} {\bibfield
  {journal} {\bibinfo  {journal} {Phys. Rev. Research}\ }\textbf {\bibinfo
  {volume} {2}},\ \bibinfo {pages} {013334}}\BibitemShut {NoStop}%
\bibitem [{\citenamefont {Yao}\ \emph {et~al.}(2022)\citenamefont {Yao},
  \citenamefont {Luo},\ and\ \citenamefont {Chen}}]{PhysRevB.105.024401}%
  \BibitemOpen
  \bibfield  {author} {\bibinfo {author} {\bibnamefont {Yao}, \bibfnamefont
  {Xu-Ping}}, \bibinfo {author} {\bibfnamefont {Rui~Leonard}\ \bibnamefont
  {Luo}}, \ and\ \bibinfo {author} {\bibfnamefont {Gang}\ \bibnamefont {Chen}}}
  (\bibinfo {year} {2022}),\ \bibfield  {title} {\enquote {\bibinfo {title}
  {{Intertwining SU($N$) symmetry and frustration on a honeycomb lattice}},}\
  }\href {\doibase 10.1103/PhysRevB.105.024401} {\bibfield  {journal} {\bibinfo
   {journal} {Phys. Rev. B}\ }\textbf {\bibinfo {volume} {105}},\ \bibinfo
  {pages} {024401}}\BibitemShut {NoStop}%
\bibitem [{\citenamefont {Ye}\ \emph {et~al.}(1999)\citenamefont {Ye},
  \citenamefont {Kim}, \citenamefont {Millis}, \citenamefont {Shraiman},
  \citenamefont {Majumdar},\ and\ \citenamefont {Te\ifmmode \check{s}\else
  \v{s}\fi{}anovi\ifmmode~\acute{c}\else \'{c}\fi{}}}]{PhysRevLett.83.3737}%
  \BibitemOpen
  \bibfield  {author} {\bibinfo {author} {\bibnamefont {Ye}, \bibfnamefont
  {Jinwu}}, \bibinfo {author} {\bibfnamefont {Yong~Baek}\ \bibnamefont {Kim}},
  \bibinfo {author} {\bibfnamefont {A.~J.}\ \bibnamefont {Millis}}, \bibinfo
  {author} {\bibfnamefont {B.~I.}\ \bibnamefont {Shraiman}}, \bibinfo {author}
  {\bibfnamefont {P.}~\bibnamefont {Majumdar}}, \ and\ \bibinfo {author}
  {\bibfnamefont {Z.}~\bibnamefont {Te\ifmmode \check{s}\else
  \v{s}\fi{}anovi\ifmmode~\acute{c}\else \'{c}\fi{}}}} (\bibinfo {year}
  {1999}),\ \bibfield  {title} {\enquote {\bibinfo {title} {{Berry Phase Theory
  of the Anomalous Hall Effect: Application to Colossal Magnetoresistance
  Manganites}},}\ }\href {\doibase 10.1103/PhysRevLett.83.3737} {\bibfield
  {journal} {\bibinfo  {journal} {Phys. Rev. Lett.}\ }\textbf {\bibinfo
  {volume} {83}},\ \bibinfo {pages} {3737--3740}}\BibitemShut {NoStop}%
\bibitem [{\citenamefont {Ye}\ \emph {et~al.}(2018{\natexlab{a}})\citenamefont
  {Ye}, \citenamefont {Kang}, \citenamefont {Liu}, \citenamefont {von Cube},
  \citenamefont {Wicker}, \citenamefont {Suzuki}, \citenamefont {Jozwiak},
  \citenamefont {Bostwick}, \citenamefont {Rotenberg}, \citenamefont {Bell},
  \citenamefont {Fu}, \citenamefont {Comin},\ and\ \citenamefont
  {Checkelsky}}]{Ye_2018}%
  \BibitemOpen
  \bibfield  {author} {\bibinfo {author} {\bibnamefont {Ye}, \bibfnamefont
  {Linda}}, \bibinfo {author} {\bibfnamefont {Mingu}\ \bibnamefont {Kang}},
  \bibinfo {author} {\bibfnamefont {Junwei}\ \bibnamefont {Liu}}, \bibinfo
  {author} {\bibfnamefont {Felix}\ \bibnamefont {von Cube}}, \bibinfo {author}
  {\bibfnamefont {Christina~R.}\ \bibnamefont {Wicker}}, \bibinfo {author}
  {\bibfnamefont {Takehito}\ \bibnamefont {Suzuki}}, \bibinfo {author}
  {\bibfnamefont {Chris}\ \bibnamefont {Jozwiak}}, \bibinfo {author}
  {\bibfnamefont {Aaron}\ \bibnamefont {Bostwick}}, \bibinfo {author}
  {\bibfnamefont {Eli}\ \bibnamefont {Rotenberg}}, \bibinfo {author}
  {\bibfnamefont {David~C.}\ \bibnamefont {Bell}}, \bibinfo {author}
  {\bibfnamefont {Liang}\ \bibnamefont {Fu}}, \bibinfo {author} {\bibfnamefont
  {Riccardo}\ \bibnamefont {Comin}}, \ and\ \bibinfo {author} {\bibfnamefont
  {Joseph~G.}\ \bibnamefont {Checkelsky}}} (\bibinfo {year}
  {2018}{\natexlab{a}}),\ \bibfield  {title} {\enquote {\bibinfo {title}
  {Massive dirac fermions in a ferromagnetic kagome metal},}\ }\href {\doibase
  10.1038/nature25987} {\bibfield  {journal} {\bibinfo  {journal} {Nature}\
  }\textbf {\bibinfo {volume} {555}}~(\bibinfo {number} {7698}),\ \bibinfo
  {pages} {638--642}}\BibitemShut {NoStop}%
\bibitem [{\citenamefont {Ye}\ \emph {et~al.}(2018{\natexlab{b}})\citenamefont
  {Ye}, \citenamefont {Hal\'asz}, \citenamefont {Savary},\ and\ \citenamefont
  {Balents}}]{Balents2018}%
  \BibitemOpen
  \bibfield  {author} {\bibinfo {author} {\bibnamefont {Ye}, \bibfnamefont
  {Mengxing}}, \bibinfo {author} {\bibfnamefont {G\'abor~B.}\ \bibnamefont
  {Hal\'asz}}, \bibinfo {author} {\bibfnamefont {Lucile}\ \bibnamefont
  {Savary}}, \ and\ \bibinfo {author} {\bibfnamefont {Leon}\ \bibnamefont
  {Balents}}} (\bibinfo {year} {2018}{\natexlab{b}}),\ \bibfield  {title}
  {\enquote {\bibinfo {title} {{Quantization of the thermal Hall conductivity
  at small Hall angles}},}\ }\href {\doibase 10.1103/PhysRevLett.121.147201}
  {\bibfield  {journal} {\bibinfo  {journal} {Phys. Rev. Lett.}\ }\textbf
  {\bibinfo {volume} {121}},\ \bibinfo {pages} {147201}}\BibitemShut {NoStop}%
\bibitem [{\citenamefont {Yin}\ \emph {et~al.}(2020)\citenamefont {Yin},
  \citenamefont {Ma}, \citenamefont {Cochran}, \citenamefont {Xu},
  \citenamefont {Zhang}, \citenamefont {Tien}, \citenamefont {Shumiya},
  \citenamefont {Cheng}, \citenamefont {Jiang}, \citenamefont {Lian},
  \citenamefont {Song}, \citenamefont {Chang}, \citenamefont {Belopolski},
  \citenamefont {Multer}, \citenamefont {Litskevich}, \citenamefont {Cheng},
  \citenamefont {Yang}, \citenamefont {Swidler}, \citenamefont {Zhou},
  \citenamefont {Lin}, \citenamefont {Neupert}, \citenamefont {Wang},
  \citenamefont {Yao}, \citenamefont {Chang}, \citenamefont {Jia},\ and\
  \citenamefont {Hasan}}]{Yin_2020}%
  \BibitemOpen
  \bibfield  {author} {\bibinfo {author} {\bibnamefont {Yin}, \bibfnamefont
  {Jia-Xin}}, \bibinfo {author} {\bibfnamefont {Wenlong}\ \bibnamefont {Ma}},
  \bibinfo {author} {\bibfnamefont {Tyler~A.}\ \bibnamefont {Cochran}},
  \bibinfo {author} {\bibfnamefont {Xitong}\ \bibnamefont {Xu}}, \bibinfo
  {author} {\bibfnamefont {Songtian~S.}\ \bibnamefont {Zhang}}, \bibinfo
  {author} {\bibfnamefont {Hung-Ju}\ \bibnamefont {Tien}}, \bibinfo {author}
  {\bibfnamefont {Nana}\ \bibnamefont {Shumiya}}, \bibinfo {author}
  {\bibfnamefont {Guangming}\ \bibnamefont {Cheng}}, \bibinfo {author}
  {\bibfnamefont {Kun}\ \bibnamefont {Jiang}}, \bibinfo {author} {\bibfnamefont
  {Biao}\ \bibnamefont {Lian}}, \bibinfo {author} {\bibfnamefont {Zhida}\
  \bibnamefont {Song}}, \bibinfo {author} {\bibfnamefont {Guoqing}\
  \bibnamefont {Chang}}, \bibinfo {author} {\bibfnamefont {Ilya}\ \bibnamefont
  {Belopolski}}, \bibinfo {author} {\bibfnamefont {Daniel}\ \bibnamefont
  {Multer}}, \bibinfo {author} {\bibfnamefont {Maksim}\ \bibnamefont
  {Litskevich}}, \bibinfo {author} {\bibfnamefont {Zi-Jia}\ \bibnamefont
  {Cheng}}, \bibinfo {author} {\bibfnamefont {Xian~P.}\ \bibnamefont {Yang}},
  \bibinfo {author} {\bibfnamefont {Bianca}\ \bibnamefont {Swidler}}, \bibinfo
  {author} {\bibfnamefont {Huibin}\ \bibnamefont {Zhou}}, \bibinfo {author}
  {\bibfnamefont {Hsin}\ \bibnamefont {Lin}}, \bibinfo {author} {\bibfnamefont
  {Titus}\ \bibnamefont {Neupert}}, \bibinfo {author} {\bibfnamefont {Ziqiang}\
  \bibnamefont {Wang}}, \bibinfo {author} {\bibfnamefont {Nan}\ \bibnamefont
  {Yao}}, \bibinfo {author} {\bibfnamefont {Tay-Rong}\ \bibnamefont {Chang}},
  \bibinfo {author} {\bibfnamefont {Shuang}\ \bibnamefont {Jia}}, \ and\
  \bibinfo {author} {\bibfnamefont {M.~Zahid}\ \bibnamefont {Hasan}}} (\bibinfo
  {year} {2020}),\ \bibfield  {title} {\enquote {\bibinfo {title}
  {{Quantum-limit Chern topological magnetism in {TbMn}$_6$Sn$_6$}},}\ }\href
  {\doibase 10.1038/s41586-020-2482-7} {\bibfield  {journal} {\bibinfo
  {journal} {Nature}\ }\textbf {\bibinfo {volume} {583}}~(\bibinfo {number}
  {7817}),\ \bibinfo {pages} {533--536}}\BibitemShut {NoStop}%
\bibitem [{\citenamefont {Yokoi}\ \emph {et~al.}(2021)\citenamefont {Yokoi},
  \citenamefont {Ma}, \citenamefont {Kasahara}, \citenamefont {Kasahara},
  \citenamefont {Shibauchi}, \citenamefont {Kurita}, \citenamefont {Tanaka},
  \citenamefont {Nasu}, \citenamefont {Motome}, \citenamefont {Hickey},
  \citenamefont {Trebst},\ and\ \citenamefont {Matsuda}}]{Matsuda2021}%
  \BibitemOpen
  \bibfield  {author} {\bibinfo {author} {\bibnamefont {Yokoi}, \bibfnamefont
  {T}}, \bibinfo {author} {\bibfnamefont {S.}~\bibnamefont {Ma}}, \bibinfo
  {author} {\bibfnamefont {Y.}~\bibnamefont {Kasahara}}, \bibinfo {author}
  {\bibfnamefont {S.}~\bibnamefont {Kasahara}}, \bibinfo {author}
  {\bibfnamefont {T.}~\bibnamefont {Shibauchi}}, \bibinfo {author}
  {\bibfnamefont {N.}~\bibnamefont {Kurita}}, \bibinfo {author} {\bibfnamefont
  {H.}~\bibnamefont {Tanaka}}, \bibinfo {author} {\bibfnamefont
  {J.}~\bibnamefont {Nasu}}, \bibinfo {author} {\bibfnamefont {Y.}~\bibnamefont
  {Motome}}, \bibinfo {author} {\bibfnamefont {C.}~\bibnamefont {Hickey}},
  \bibinfo {author} {\bibfnamefont {S.}~\bibnamefont {Trebst}}, \ and\ \bibinfo
  {author} {\bibfnamefont {Y.}~\bibnamefont {Matsuda}}} (\bibinfo {year}
  {2021}),\ \bibfield  {title} {\enquote {\bibinfo {title} {{Half-integer
  quantized anomalous thermal Hall effect in the Kitaev material candidate
  $\ensuremath{\alpha}\ensuremath{-}{\mathrm{RuCl}}_{3}$}},}\ }\href {\doibase
  10.1126/science.aay5551} {\bibfield  {journal} {\bibinfo  {journal}
  {Science}\ }\textbf {\bibinfo {volume} {373}}~(\bibinfo {number} {6554}),\
  \bibinfo {pages} {568--572}}\BibitemShut {NoStop}%
\bibitem [{\citenamefont {Yoshida}\ \emph {et~al.}(2017)\citenamefont
  {Yoshida}, \citenamefont {Noguchi}, \citenamefont {Matsushita}, \citenamefont
  {Ishii}, \citenamefont {Ihara}, \citenamefont {Oda}, \citenamefont {Okabe},
  \citenamefont {Yamashita}, \citenamefont {Nakazawa}, \citenamefont {Takata},
  \citenamefont {Kida}, \citenamefont {Narumi},\ and\ \citenamefont
  {Hagiwara}}]{Yoshida2017}%
  \BibitemOpen
  \bibfield  {author} {\bibinfo {author} {\bibnamefont {Yoshida}, \bibfnamefont
  {Hiroyuki}}, \bibinfo {author} {\bibfnamefont {Naoya}\ \bibnamefont
  {Noguchi}}, \bibinfo {author} {\bibfnamefont {Yoshitaka}\ \bibnamefont
  {Matsushita}}, \bibinfo {author} {\bibfnamefont {Yuto}\ \bibnamefont
  {Ishii}}, \bibinfo {author} {\bibfnamefont {Yoshihiko}\ \bibnamefont
  {Ihara}}, \bibinfo {author} {\bibfnamefont {Migaku}\ \bibnamefont {Oda}},
  \bibinfo {author} {\bibfnamefont {Hirotaka}\ \bibnamefont {Okabe}}, \bibinfo
  {author} {\bibfnamefont {Satoshi}\ \bibnamefont {Yamashita}}, \bibinfo
  {author} {\bibfnamefont {Yasuhiro}\ \bibnamefont {Nakazawa}}, \bibinfo
  {author} {\bibfnamefont {Atsushi}\ \bibnamefont {Takata}}, \bibinfo {author}
  {\bibfnamefont {Takanori}\ \bibnamefont {Kida}}, \bibinfo {author}
  {\bibfnamefont {Yasuo}\ \bibnamefont {Narumi}}, \ and\ \bibinfo {author}
  {\bibfnamefont {Masayuki}\ \bibnamefont {Hagiwara}}} (\bibinfo {year}
  {2017}),\ \bibfield  {title} {\enquote {\bibinfo {title} {{Unusual Magnetic
  State with Dual Magnetic Excitations in the Single Crystal of $S$ = 1/2
  Kagome Lattice Antiferromagnet
  {CaCu}$_3${(OH)}$_6${Cl}$_2\cdot$0.6{H}$_2${O}}},}\ }\href {\doibase
  10.7566/jpsj.86.033704} {\bibfield  {journal} {\bibinfo  {journal} {Journal
  of the Physical Society of Japan}\ }\textbf {\bibinfo {volume}
  {86}}~(\bibinfo {number} {3}),\ \bibinfo {pages} {033704}}\BibitemShut
  {NoStop}%
\bibitem [{\citenamefont {Yoshida}\ \emph {et~al.}(2012)\citenamefont
  {Yoshida}, \citenamefont {ichi Yamaura}, \citenamefont {Isobe}, \citenamefont
  {Okamoto}, \citenamefont {Nilsen},\ and\ \citenamefont
  {Hiroi}}]{Yoshida2012}%
  \BibitemOpen
  \bibfield  {author} {\bibinfo {author} {\bibnamefont {Yoshida}, \bibfnamefont
  {Hiroyuki}}, \bibinfo {author} {\bibfnamefont {Jun}\ \bibnamefont {ichi
  Yamaura}}, \bibinfo {author} {\bibfnamefont {Masaaki}\ \bibnamefont {Isobe}},
  \bibinfo {author} {\bibfnamefont {Yoshihiko}\ \bibnamefont {Okamoto}},
  \bibinfo {author} {\bibfnamefont {G{\o}ran~J.}\ \bibnamefont {Nilsen}}, \
  and\ \bibinfo {author} {\bibfnamefont {Zenji}\ \bibnamefont {Hiroi}}}
  (\bibinfo {year} {2012}),\ \bibfield  {title} {\enquote {\bibinfo {title}
  {Orbital switching in a frustrated magnet},}\ }\href {\doibase
  10.1038/ncomms1875} {\bibfield  {journal} {\bibinfo  {journal} {Nature
  Communications}\ }\textbf {\bibinfo {volume} {3}}~(\bibinfo {number} {1}),\
  \bibinfo {pages} {860}}\BibitemShut {NoStop}%
\bibitem [{\citenamefont {Yoshitake}\ \emph {et~al.}(2016)\citenamefont
  {Yoshitake}, \citenamefont {Nasu},\ and\ \citenamefont
  {Motome}}]{Yoshitake_2016}%
  \BibitemOpen
  \bibfield  {author} {\bibinfo {author} {\bibnamefont {Yoshitake},
  \bibfnamefont {Junki}}, \bibinfo {author} {\bibfnamefont {Joji}\ \bibnamefont
  {Nasu}}, \ and\ \bibinfo {author} {\bibfnamefont {Yukitoshi}\ \bibnamefont
  {Motome}}} (\bibinfo {year} {2016}),\ \bibfield  {title} {\enquote {\bibinfo
  {title} {{Fractional Spin Fluctuations as a Precursor of Quantum Spin
  Liquids: Majorana Dynamical Mean-Field Study for the Kitaev Model}},}\ }\href
  {\doibase 10.1103/physrevlett.117.157203} {\bibfield  {journal} {\bibinfo
  {journal} {Physical Review Letters}\ }\textbf {\bibinfo {volume}
  {117}}~(\bibinfo {number} {15}),\ 10.1103/physrevlett.117.157203}\BibitemShut
  {NoStop}%
\bibitem [{\citenamefont {Zayed}\ \emph {et~al.}(2017)\citenamefont {Zayed},
  \citenamefont {Rüegg}, \citenamefont {J.}, \citenamefont {Läuchli},
  \citenamefont {Panagopoulos}, \citenamefont {Saxena}, \citenamefont
  {Ellerby}, \citenamefont {McMorrow}, \citenamefont {Strässle}, \citenamefont
  {Klotz}, \citenamefont {Hamel}, \citenamefont {Sadykov}, \citenamefont
  {Pomjakushin}, \citenamefont {Boehm}, \citenamefont
  {Jim{\'{e}}nez{\textendash}Ruiz}, \citenamefont {Schneidewind}, \citenamefont
  {Pomjakushina}, \citenamefont {Stingaciu}, \citenamefont {Conder},\ and\
  \citenamefont {R{\o}nnow}}]{Zayed_2017}%
  \BibitemOpen
  \bibfield  {author} {\bibinfo {author} {\bibnamefont {Zayed}, \bibfnamefont
  {M~E}}, \bibinfo {author} {\bibfnamefont {Ch.}\ \bibnamefont {Rüegg}},
  \bibinfo {author} {\bibfnamefont {J.~Larrea}\ \bibnamefont {J.}}, \bibinfo
  {author} {\bibfnamefont {A.~M.}\ \bibnamefont {Läuchli}}, \bibinfo {author}
  {\bibfnamefont {C.}~\bibnamefont {Panagopoulos}}, \bibinfo {author}
  {\bibfnamefont {S.~S.}\ \bibnamefont {Saxena}}, \bibinfo {author}
  {\bibfnamefont {M.}~\bibnamefont {Ellerby}}, \bibinfo {author} {\bibfnamefont
  {D.~F.}\ \bibnamefont {McMorrow}}, \bibinfo {author} {\bibfnamefont {Th.}\
  \bibnamefont {Strässle}}, \bibinfo {author} {\bibfnamefont {S.}~\bibnamefont
  {Klotz}}, \bibinfo {author} {\bibfnamefont {G.}~\bibnamefont {Hamel}},
  \bibinfo {author} {\bibfnamefont {R.~A.}\ \bibnamefont {Sadykov}}, \bibinfo
  {author} {\bibfnamefont {V.}~\bibnamefont {Pomjakushin}}, \bibinfo {author}
  {\bibfnamefont {M.}~\bibnamefont {Boehm}}, \bibinfo {author} {\bibfnamefont
  {M.}~\bibnamefont {Jim{\'{e}}nez{\textendash}Ruiz}}, \bibinfo {author}
  {\bibfnamefont {A.}~\bibnamefont {Schneidewind}}, \bibinfo {author}
  {\bibfnamefont {E.}~\bibnamefont {Pomjakushina}}, \bibinfo {author}
  {\bibfnamefont {M.}~\bibnamefont {Stingaciu}}, \bibinfo {author}
  {\bibfnamefont {K.}~\bibnamefont {Conder}}, \ and\ \bibinfo {author}
  {\bibfnamefont {H.~M.}\ \bibnamefont {R{\o}nnow}}} (\bibinfo {year} {2017}),\
  \bibfield  {title} {\enquote {\bibinfo {title} {{4-spin plaquette singlet
  state in the Shastry{\textendash}Sutherland compound
  {SrCu}$_2$({BO}$_3$)$_2$}},}\ }\href {\doibase 10.1038/nphys4190} {\bibfield
  {journal} {\bibinfo  {journal} {Nature Physics}\ }\textbf {\bibinfo {volume}
  {13}}~(\bibinfo {number} {10}),\ \bibinfo {pages} {962--966}}\BibitemShut
  {NoStop}%
\bibitem [{\citenamefont {Zhang}\ \emph
  {et~al.}(2021{\natexlab{a}})\citenamefont {Zhang}, \citenamefont {Chern},\
  and\ \citenamefont {Kim}}]{PhysRevB.103.174402}%
  \BibitemOpen
  \bibfield  {author} {\bibinfo {author} {\bibnamefont {Zhang}, \bibfnamefont
  {Emily~Z}}, \bibinfo {author} {\bibfnamefont {Li~Ern}\ \bibnamefont {Chern}},
  \ and\ \bibinfo {author} {\bibfnamefont {Yong~Baek}\ \bibnamefont {Kim}}}
  (\bibinfo {year} {2021}{\natexlab{a}}),\ \bibfield  {title} {\enquote
  {\bibinfo {title} {{Topological magnons for thermal Hall transport in
  frustrated magnets with bond-dependent interactions}},}\ }\href {\doibase
  10.1103/PhysRevB.103.174402} {\bibfield  {journal} {\bibinfo  {journal}
  {Phys. Rev. B}\ }\textbf {\bibinfo {volume} {103}},\ \bibinfo {pages}
  {174402}}\BibitemShut {NoStop}%
\bibitem [{\citenamefont {Zhang}\ \emph
  {et~al.}(2021{\natexlab{b}})\citenamefont {Zhang}, \citenamefont {Xu},
  \citenamefont {Carnahan}, \citenamefont {Sretenovic}, \citenamefont {Suri},
  \citenamefont {Xiao},\ and\ \citenamefont {Ke}}]{PhysRevLett.127.247202}%
  \BibitemOpen
  \bibfield  {author} {\bibinfo {author} {\bibnamefont {Zhang}, \bibfnamefont
  {Heda}}, \bibinfo {author} {\bibfnamefont {Chunqiang}\ \bibnamefont {Xu}},
  \bibinfo {author} {\bibfnamefont {Caitlin}\ \bibnamefont {Carnahan}},
  \bibinfo {author} {\bibfnamefont {Milos}\ \bibnamefont {Sretenovic}},
  \bibinfo {author} {\bibfnamefont {Nishchay}\ \bibnamefont {Suri}}, \bibinfo
  {author} {\bibfnamefont {Di}~\bibnamefont {Xiao}}, \ and\ \bibinfo {author}
  {\bibfnamefont {X.}~\bibnamefont {Ke}}} (\bibinfo {year}
  {2021}{\natexlab{b}}),\ \bibfield  {title} {\enquote {\bibinfo {title}
  {{Anomalous Thermal Hall Effect in an Insulating van der Waals Magnet}},}\
  }\href {\doibase 10.1103/PhysRevLett.127.247202} {\bibfield  {journal}
  {\bibinfo  {journal} {Physical Review Letters}\ }\textbf {\bibinfo {volume}
  {127}},\ \bibinfo {pages} {247202}}\BibitemShut {NoStop}%
\bibitem [{\citenamefont {Zhang}\ \emph {et~al.}(2013)\citenamefont {Zhang},
  \citenamefont {Ren}, \citenamefont {Wang},\ and\ \citenamefont
  {Li}}]{Lifa2013}%
  \BibitemOpen
  \bibfield  {author} {\bibinfo {author} {\bibnamefont {Zhang}, \bibfnamefont
  {Lifa}}, \bibinfo {author} {\bibfnamefont {Jie}\ \bibnamefont {Ren}},
  \bibinfo {author} {\bibfnamefont {Jian-Sheng}\ \bibnamefont {Wang}}, \ and\
  \bibinfo {author} {\bibfnamefont {Baowen}\ \bibnamefont {Li}}} (\bibinfo
  {year} {2013}),\ \bibfield  {title} {\enquote {\bibinfo {title} {Topological
  magnon insulator in insulating ferromagnet},}\ }\href {\doibase
  10.1103/PhysRevB.87.144101} {\bibfield  {journal} {\bibinfo  {journal} {Phys.
  Rev. B}\ }\textbf {\bibinfo {volume} {87}},\ \bibinfo {pages}
  {144101}}\BibitemShut {NoStop}%
\bibitem [{\citenamefont {Zhang}\ \emph {et~al.}(2020)\citenamefont {Zhang},
  \citenamefont {Gao}, \citenamefont {Liu},\ and\ \citenamefont
  {Chen}}]{Zhang2020}%
  \BibitemOpen
  \bibfield  {author} {\bibinfo {author} {\bibnamefont {Zhang}, \bibfnamefont
  {Xiao-Tian}}, \bibinfo {author} {\bibfnamefont {Yong~Hao}\ \bibnamefont
  {Gao}}, \bibinfo {author} {\bibfnamefont {Chunxiao}\ \bibnamefont {Liu}}, \
  and\ \bibinfo {author} {\bibfnamefont {Gang}\ \bibnamefont {Chen}}} (\bibinfo
  {year} {2020}),\ \bibfield  {title} {\enquote {\bibinfo {title} {{Topological
  thermal Hall effect of magnetic monopoles in the pyrochlore {U(1)} spin
  liquid}},}\ }\href {\doibase 10.1103/PhysRevResearch.2.013066} {\bibfield
  {journal} {\bibinfo  {journal} {Phys. Rev. Research}\ }\textbf {\bibinfo
  {volume} {2}},\ \bibinfo {pages} {013066}}\BibitemShut {NoStop}%
\bibitem [{\citenamefont {Zhang}\ \emph {et~al.}(2019)\citenamefont {Zhang},
  \citenamefont {Zhang}, \citenamefont {Okamoto},\ and\ \citenamefont
  {Xiao}}]{PhysRevLett.123.167202}%
  \BibitemOpen
  \bibfield  {author} {\bibinfo {author} {\bibnamefont {Zhang}, \bibfnamefont
  {Xiaoou}}, \bibinfo {author} {\bibfnamefont {Yinhan}\ \bibnamefont {Zhang}},
  \bibinfo {author} {\bibfnamefont {Satoshi}\ \bibnamefont {Okamoto}}, \ and\
  \bibinfo {author} {\bibfnamefont {Di}~\bibnamefont {Xiao}}} (\bibinfo {year}
  {2019}),\ \bibfield  {title} {\enquote {\bibinfo {title} {{Thermal Hall
  Effect Induced by Magnon-Phonon Interactions}},}\ }\href {\doibase
  10.1103/PhysRevLett.123.167202} {\bibfield  {journal} {\bibinfo  {journal}
  {Phys. Rev. Lett.}\ }\textbf {\bibinfo {volume} {123}},\ \bibinfo {pages}
  {167202}}\BibitemShut {NoStop}%
\bibitem [{\citenamefont {Zhao}\ \emph {et~al.}(2016)\citenamefont {Zhao},
  \citenamefont {Calder}, \citenamefont {Aczel}, \citenamefont {McGuire},
  \citenamefont {Sales}, \citenamefont {Mandrus}, \citenamefont {Chen},
  \citenamefont {Trivedi}, \citenamefont {Zhou},\ and\ \citenamefont
  {Yan}}]{PhysRevB.93.134426}%
  \BibitemOpen
  \bibfield  {author} {\bibinfo {author} {\bibnamefont {Zhao}, \bibfnamefont
  {Z~Y}}, \bibinfo {author} {\bibfnamefont {S.}~\bibnamefont {Calder}},
  \bibinfo {author} {\bibfnamefont {A.~A.}\ \bibnamefont {Aczel}}, \bibinfo
  {author} {\bibfnamefont {M.~A.}\ \bibnamefont {McGuire}}, \bibinfo {author}
  {\bibfnamefont {B.~C.}\ \bibnamefont {Sales}}, \bibinfo {author}
  {\bibfnamefont {D.~G.}\ \bibnamefont {Mandrus}}, \bibinfo {author}
  {\bibfnamefont {G.}~\bibnamefont {Chen}}, \bibinfo {author} {\bibfnamefont
  {N.}~\bibnamefont {Trivedi}}, \bibinfo {author} {\bibfnamefont {H.~D.}\
  \bibnamefont {Zhou}}, \ and\ \bibinfo {author} {\bibfnamefont {J.-Q.}\
  \bibnamefont {Yan}}} (\bibinfo {year} {2016}),\ \bibfield  {title} {\enquote
  {\bibinfo {title} {{Fragile singlet ground-state magnetism in the pyrochlore
  osmates ${R}_{2}{\mathrm{Os}}_{2}{\mathrm{O}}_{7}$ ($R=\mathrm{Y}$ and
  Ho)}},}\ }\href {\doibase 10.1103/PhysRevB.93.134426} {\bibfield  {journal}
  {\bibinfo  {journal} {Phys. Rev. B}\ }\textbf {\bibinfo {volume} {93}},\
  \bibinfo {pages} {134426}}\BibitemShut {NoStop}%
\bibitem [{\citenamefont {Zhitomirsky}\ and\ \citenamefont
  {Ueda}(1996)}]{PhysRevB.54.9007}%
  \BibitemOpen
  \bibfield  {author} {\bibinfo {author} {\bibnamefont {Zhitomirsky},
  \bibfnamefont {M~E}}, \ and\ \bibinfo {author} {\bibfnamefont {Kazuo}\
  \bibnamefont {Ueda}}} (\bibinfo {year} {1996}),\ \bibfield  {title} {\enquote
  {\bibinfo {title} {Valence-bond crystal phase of a frustrated spin-1/2
  square-lattice antiferromagnet},}\ }\href {\doibase 10.1103/PhysRevB.54.9007}
  {\bibfield  {journal} {\bibinfo  {journal} {Phys. Rev. B}\ }\textbf {\bibinfo
  {volume} {54}},\ \bibinfo {pages} {9007--9010}}\BibitemShut {NoStop}%
\bibitem [{\citenamefont {Zhu}\ \emph {et~al.}(2019)\citenamefont {Zhu},
  \citenamefont {shu Gong},\ and\ \citenamefont {Sheng}}]{Zhu_2019}%
  \BibitemOpen
  \bibfield  {author} {\bibinfo {author} {\bibnamefont {Zhu}, \bibfnamefont
  {W}}, \bibinfo {author} {\bibfnamefont {Shou}\ \bibnamefont {shu Gong}}, \
  and\ \bibinfo {author} {\bibfnamefont {D.~N.}\ \bibnamefont {Sheng}}}
  (\bibinfo {year} {2019}),\ \bibfield  {title} {\enquote {\bibinfo {title}
  {Identifying spinon excitations from dynamic structure factor of spin-1/2
  heisenberg antiferromagnet on the kagome lattice},}\ }\href {\doibase
  10.1073/pnas.1807840116} {\bibfield  {journal} {\bibinfo  {journal}
  {Proceedings of the National Academy of Sciences}\ }\textbf {\bibinfo
  {volume} {116}}~(\bibinfo {number} {12}),\ \bibinfo {pages}
  {5437--5441}}\BibitemShut {NoStop}%
\bibitem [{\citenamefont {Zhu}\ \emph {et~al.}(2018)\citenamefont {Zhu},
  \citenamefont {Kimchi}, \citenamefont {Sheng},\ and\ \citenamefont
  {Fu}}]{Fu2018}%
  \BibitemOpen
  \bibfield  {author} {\bibinfo {author} {\bibnamefont {Zhu}, \bibfnamefont
  {Zheng}}, \bibinfo {author} {\bibfnamefont {Itamar}\ \bibnamefont {Kimchi}},
  \bibinfo {author} {\bibfnamefont {D.~N.}\ \bibnamefont {Sheng}}, \ and\
  \bibinfo {author} {\bibfnamefont {Liang}\ \bibnamefont {Fu}}} (\bibinfo
  {year} {2018}),\ \bibfield  {title} {\enquote {\bibinfo {title} {{Robust
  non-Abelian spin liquid and a possible intermediate phase in the
  antiferromagnetic Kitaev model with magnetic field}},}\ }\href {\doibase
  10.1103/PhysRevB.97.241110} {\bibfield  {journal} {\bibinfo  {journal} {Phys.
  Rev. B}\ }\textbf {\bibinfo {volume} {97}},\ \bibinfo {pages}
  {241110}}\BibitemShut {NoStop}%
\bibitem [{\citenamefont {Zhu}\ \emph {et~al.}(2013)\citenamefont {Zhu},
  \citenamefont {Huse},\ and\ \citenamefont {White}}]{Zhu2013}%
  \BibitemOpen
  \bibfield  {author} {\bibinfo {author} {\bibnamefont {Zhu}, \bibfnamefont
  {Zhenyue}}, \bibinfo {author} {\bibfnamefont {David~A.}\ \bibnamefont
  {Huse}}, \ and\ \bibinfo {author} {\bibfnamefont {Steven~R.}\ \bibnamefont
  {White}}} (\bibinfo {year} {2013}),\ \bibfield  {title} {\enquote {\bibinfo
  {title} {{Weak plaquette valence bond order in the $S\mathbf{=}1/2$ honeycomb
  ${J}_{1}\mathbf{\ensuremath{-}}{J}_{2}$ Heisenberg model}},}\ }\href
  {\doibase 10.1103/PhysRevLett.110.127205} {\bibfield  {journal} {\bibinfo
  {journal} {Phys. Rev. Lett.}\ }\textbf {\bibinfo {volume} {110}},\ \bibinfo
  {pages} {127205}}\BibitemShut {NoStop}%
\bibitem [{\citenamefont {Zhuo}\ \emph {et~al.}(2021)\citenamefont {Zhuo},
  \citenamefont {Li},\ and\ \citenamefont {Manchon}}]{PhysRevB.104.144422}%
  \BibitemOpen
  \bibfield  {author} {\bibinfo {author} {\bibnamefont {Zhuo}, \bibfnamefont
  {Fengjun}}, \bibinfo {author} {\bibfnamefont {Hang}\ \bibnamefont {Li}}, \
  and\ \bibinfo {author} {\bibfnamefont {Aur\'elien}\ \bibnamefont {Manchon}}}
  (\bibinfo {year} {2021}),\ \bibfield  {title} {\enquote {\bibinfo {title}
  {{Topological phase transition and thermal Hall effect in kagome
  ferromagnets}},}\ }\href {\doibase 10.1103/PhysRevB.104.144422} {\bibfield
  {journal} {\bibinfo  {journal} {Phys. Rev. B}\ }\textbf {\bibinfo {volume}
  {104}},\ \bibinfo {pages} {144422}}\BibitemShut {NoStop}%
\bibitem [{\citenamefont {Zhuo}\ \emph {et~al.}(2022)\citenamefont {Zhuo},
  \citenamefont {Li},\ and\ \citenamefont {Manchon}}]{Zhuo2022}%
  \BibitemOpen
  \bibfield  {author} {\bibinfo {author} {\bibnamefont {Zhuo}, \bibfnamefont
  {Fengjun}}, \bibinfo {author} {\bibfnamefont {Hang}\ \bibnamefont {Li}}, \
  and\ \bibinfo {author} {\bibfnamefont {Aur{\'{e}}lien}\ \bibnamefont
  {Manchon}}} (\bibinfo {year} {2022}),\ \bibfield  {title} {\enquote {\bibinfo
  {title} {{Topological thermal Hall effect and magnonic edge states in kagome
  ferromagnets with bond anisotropy}},}\ }\href {\doibase
  10.1088/1367-2630/ac51a8} {\bibfield  {journal} {\bibinfo  {journal} {New
  Journal of Physics}\ }\textbf {\bibinfo {volume} {24}}~(\bibinfo {number}
  {2}),\ \bibinfo {pages} {023033}}\BibitemShut {NoStop}%
\bibitem [{\citenamefont {Zou}\ and\ \citenamefont {He}(2020)}]{Zou2020}%
  \BibitemOpen
  \bibfield  {author} {\bibinfo {author} {\bibnamefont {Zou}, \bibfnamefont
  {Liujun}}, \ and\ \bibinfo {author} {\bibfnamefont {Yin-Chen}\ \bibnamefont
  {He}}} (\bibinfo {year} {2020}),\ \bibfield  {title} {\enquote {\bibinfo
  {title} {{Field-induced ${\mathrm{QCD}}_{3}$-Chern-Simons quantum
  criticalities in Kitaev materials}},}\ }\href {\doibase
  10.1103/PhysRevResearch.2.013072} {\bibfield  {journal} {\bibinfo  {journal}
  {Phys. Rev. Research}\ }\textbf {\bibinfo {volume} {2}},\ \bibinfo {pages}
  {013072}}\BibitemShut {NoStop}%
\end{thebibliography}%






\end{document}